\documentclass[aps,prd,superscriptaddress,showpacs,preprint]{revtex4}
\usepackage{graphicx, bm}
\usepackage[usenames]{color}
\usepackage{float}
\usepackage{subcaption}

\begin{document}

\title{Searching the anomalous electromagnetic and weak dipole moments of the top-quark  at the Bestest Little Higgs Model}

\author{E. Cruz-Albaro\footnote{elicruzalbaro88@gmail.com}}
\affiliation{\small Facultad de F\'{\i}sica, Universidad Aut\'onoma de Zacatecas\\
            Apartado Postal C-580, 98060 Zacatecas, M\'exico.\\}

\author{ A. Guti\'errez-Rodr\'{\i}guez\footnote{alexgu@fisica.uaz.edu.mx}}
\affiliation{\small Facultad de F\'{\i}sica, Universidad Aut\'onoma de Zacatecas\\
         Apartado Postal C-580, 98060 Zacatecas, M\'exico.\\}

\author{ M. A. Hern\'andez-Ru\'{\i}z \footnote{mahernan@uaz.edu.mx}
}
\affiliation{\small Unidad Acad\'emica de Ciencias Qu\'{\i}micas, Universidad Aut\'onoma de Zacatecas\\
         Apartado Postal C-585, 98060 Zacatecas, M\'exico.\\}

%
%

\author{T. Cisneros-P\'erez \footnote{tzihue@gmail.com}
}
\affiliation{\small Unidad Acad\'emica de Ciencias Qu\'{\i}micas, Universidad Aut\'onoma de Zacatecas\\
         Apartado Postal C-585, 98060 Zacatecas, M\'exico.\\}


\date{\today}

\begin{abstract}

In this paper, using a Bestest Little Higgs Model (BLHM) approach, we obtain bounds on the top-quark electromagnetic and weak dipole moments. These physical observables are sensitive to quantum corrections induced by virtual particles at the one-loop level. The contributions of the virtual particles are obtained from the vertices of scalar bosons, vector bosons, and heavy quarks:
 $tq_iS_i$, $S_i=h_0, H_0, A_0, \phi^{0}, \eta^{0}, \sigma, H^{\pm}, \phi^{\pm},
\eta^{\pm}$; $tq_iV_i$, $V_i= \gamma, Z, W, Z', W'$; $V q_i\bar q_i$, $V=\gamma,Z$, $q_i=b, t, B, T, T_5, T_6, T^{2/3}, T^{5/3}$; $  V W^{-}_{i}\Phi^{+}_i $, $W_{i} \equiv W, W'$, $\Phi^{\pm}_{i}= \phi^{\pm}, \eta^{\pm}$; $ZZH_i$, $H_i=h_0, H_0$; $ZZ' H_i$;  and $VW'W'$.
With these new contributions, we evaluated the electroweak dipole moments $  a_t$, $ a^{W}_t $, $ d_t $ and $ d^{W}_{t} $ of the top-quark in the diagonalization schemes $y_2 > y_3$ and $y_2 < y_3$ to the Yukawa couplings.
For the BLHM parameters, we choose the following values of $m_{A_0}= 1000$ GeV, $m_{\eta^0}= 100$ GeV, $f=[1000, 3000]$ GeV, $F=[3000, 6000]$ GeV and $\tan\beta=3$, which give the best sensitivity on the electroweak dipole moments of the top-quark in this context: $a_t=1.39 \times 10^{-4} + i\, 6.55 \times 10^{-5}$; $a^{W}_t=6.31 \times 10^{-5} - i\, 1.39 \times 10^{-5}$ for the $y_2 > y_3 $ scenario, while $a_t= 2.12 \times 10^{-4} + i\, 4.49 \times 10^{-5}$; $a^{W}_t= 2.13 \times 10^{-4} - i\, 1.34 \times 10^{-5}$ for the $y_2 < y_3 $ scenario.
\end{abstract}

\pacs{14.65.Ha, 12.15.Mm, 12.60.-i \\
Keywords: Top quarks, Neutral currents, Models beyond the Standard Model.}

\vspace{5mm}

\maketitle

\section{Introduction}

Given the unique features of the top-quark, the precision measurements of their properties are essential
and exciting because it is so far the heaviest fundamental particle discovered, and as such, it is expected to be more sensitive to new physics at higher energy scales.
Given its large mass, the top-quark may also play a unique role within the Standard Model (SM), and its precise characterization may shed light on the electroweak
symmetry breaking mechanism~\cite{PRL81-1998,PRD59-1999} or on yet unanswered questions of the SM. From a theoretical point of view, the top offers a unique probe for understanding the electroweak sector and physics beyond the SM (BSM).

The top-quark is a key particle in various extensions BSM and is considered a laboratory for many experimental or simulation aspects
in searches for new physics. In particular, the top-quark anomalous couplings to gauge bosons in the $t\bar t \gamma$ and $t\bar t Z$ vertices
have made the top-quark one of the most attractive particles for new physics searches. In this regard, the study of the physics of the
$t$-quark by the Tevatron collider at the Fermilab~\cite{PLB713-2012, PLB693-2010,PRL102-2009} and the ATLAS and CMS Collaborations
\cite{ATLAS-CONF-2012-126,PRL110-2013,EPJC79-2019,JHEP03-2020} at the Large Hadron Collider (LHC) has been developed
significantly in recent years and now represents a very active physics program.
There are also desirable programs to study top physics at future $e^{-} p$ and $e^{+}e^{-}$ colliders such as the Future Circular Collider-hadron electron (FCC-he)~\cite{Barletta:2014vea,Koratzinos:2015fya}, the International Linear Collider (ILC)~\cite{Behnke:2013xla,Baer:2013cma,Adolphsen:2013kya} and the Compact Linear Collider (CLIC)~\cite{deBlas:2018mhx,Robson:2018enq,Roloff:2018dqu,CLIC:2016zwp,Abramowicz:2016zbo}, among others. Future colliders will offer the opportunity to explore top-quark electroweak couplings with very competitive precision.

A far less explored aspect of top-quark physics is the interactions with neutral electroweak gauge bosons and the Higgs boson. In particular, the study of anomalous couplings in electroweak corrections is relatively unexplored. Measurements of these couplings provide an excellent opportunity to probe the interactions of new physics since they have small magnitudes in the SM (see Table~\ref{limites-SM}).
These interactions are described by so-called form factors, which include Electric Dipole Moment (EDM), Weak Electric Dipole Moment (WEDM), Anomalous Magnetic Dipole Moment (AMDM), and Anomalous Weak Magnetic Dipole Moment (AWMDM).
Experimental detection of these magnitudes at the current sensitivity of the LHC or future colliders  would be clear evidence of new BSM physics.

\begin{table}[H]
\caption{The sensitivity achievable on the electromagnetic and weak dipole moments of the top-quark in the SM.
\label{limites-SM}}
\center
\begin{tabular}{c c c c }
\hline
$ \text{Model} $  &  \hspace{0.8cm}   $ \text{Sensitivity of the model} $   & \hspace{0.8cm} $\text{C.L.}$  & \hspace{0.8cm}  $\text{Reference}$  \\
\hline
\hline
SM   & \hspace{0.8cm} $ a_t = 0.02,\ \ d_t <10^{-30} \ \text{e\ cm} $ &  \hspace{0.8cm} $ 68\, \% $  &  \hspace{0.8cm}  \cite{Bernreuther:2005gq,Soni:1992tn,Pospelov:1991zt,Hoogeveen:1990cb}  \\
 SM   & \hspace{0.8cm} $ a^{W}_t \sim 10^{-3} $ &  \hspace{0.8cm} $ 95\, \%   $  &  \hspace{0.8cm} \cite{Bernabeu:1995gs,Hollik:1998vz}\\
\hline
\end{tabular}
\end{table}

With these motivations, we carried out a study on the electromagnetic and weak dipole moments of the quark-top in the context of the
BLHM~\cite{JHEP09-2010,Cruz-Albaro:2022kty,Cruz-Albaro:2022lks}.
This extension of the SM predicts the existence of new physical scalar bosons neutral and charged $h_0, H_0, A_0, \phi^{0},\eta^{0},
\sigma, H^{\pm}, \phi^{\pm}, \eta^{\pm}$, new heavy gauge bosons $Z', W'$ and new heavy quarks $B, T,T_5, T_6, T^{2/3},T^{5/3}$.
The effects of the new physics are searched for indirectly through virtual corrections of the new particles predicted by the BLHM.
The  EDM $d_t$, AMDM $a_t$,  WEDM $d^W_t$ and AWMDM $a^W_t$ of the top-quark are induced via the Feynman diagrams depicted in Figs.~\ref{dipolo} and~\ref{dipoloweak},
where $S_i$, $ \Phi^{\pm}_i $ and $ H_i $ represent scalar bosons, $V_i$ and $ W^{\pm}_i $ neutral and charged gauge bosons, and  $b_i$ and $q_i$  the heavy quarks. Therefore, among the new
contributions of the model, there are those arising from the vertices of scalars bosons, vector bosons, and heavy quarks
contribution, that is to say, vertices of the form:  $tq_iS_i$, $S_i=h_0, H_0, A_0, \phi^{0}, \eta^{0}, \sigma, H^{\pm}, \phi^{\pm},
\eta^{\pm}$; $tq_iV_i$, $V_i= \gamma, Z, W, Z', W'$; $V q_i\bar q_i$, $V=\gamma,Z$, and $q_i=b, t, B, T, T_5, T_6, T^{2/3}, T^{5/3}$; $  V W^{-}_{i}\Phi^{+}_i $, $\Phi^{\pm}_{i}= \phi^{\pm}, \eta^{\pm}$, and $W_{i} \equiv W, W'$; $ZZH_i$, $H_i=h_0, H_0$; $ZZ' H_i$;  and $VW'W'$. With these vertices, we calculate the one-loop contributions to the electroweak dipole moments  of the top-quark for the diagonalization schemes  $y_2>y_3$ and $y_2<y_3$, with $m_{A_0}= 1000$ GeV, $m_{\eta^0}= 100$ GeV, $f=[1000, 3000]$ GeV, $F=[3000,6000]$ GeV and $\tan\beta=3$.

Regarding the two possible scenarios of analysis of the electroweak dipole moments of the top-quark, that is to say, the diagonalization schemes $y_2>y_3$ and $y_2<y_3$, which arise because in the scenario $y_2=y_3$, the masses of the heavy quarks $T$ and $T_5$ are degenerate at lowest order. Thus, the analytical expressions for the heavy quark masses represented by Eqs.~(\ref{mT})-(\ref{mB})  were obtained under the assumption that $y_2 \neq y_3$.  Consequently, different diagonalization schemes are required for the region where $y_2=y_3$ versus the region where $|y_2-y_3|>0$~\cite{Godfrey:2012tf,PhenomenologyBLH}.
In this study, we carry out our analysis considering the scenarios $y_2>y_3$ and $y_2<y_3$ (see Subsection~\ref{subsecfermion}).

The paper is structured as follows. In Section II, we give a brief review of the BLHM.
In Section III, we present the predictions of the BLHM on the electromagnetic and weak  dipole moments of the top-quark.
In Section~\ref{resultadosNum}, we present our numerical results.
Finally, we give our conclusions in Section~\ref{conclusions}.
In Appendix~\ref{numericasparciales},  we provide all the partial numerical contributions of the particles that induce the AMDM and AWMDM of the top-quark on the $y_2 > y_3$  scenario.
To avoid being redundant in the presentation of our results, in Appendix~\ref{y2menor}, we investigate the AMDM and AWMDM of the top-quark on the $y_2 < y_3$ scenario.
This analysis complements our previous work~\cite{Cruz-Albaro:2022kty} that discusses the AWMDM of the top in the mentioned scenario. In Appendix~\ref{reglasFeynman}, we present the complete set of Feynman rules for studying the electromagnetic and weak dipole moments of the top-quark in the
context of the BLHM.

\section{The Bestest Little Higgs Model}

Various extensions of the SM have been proposed to solve the problem of the mass hierarchy. One proposed
extension is the Little Higgs Model (LHM)~\cite{Arkani1,Arkani2}, which employs collective symmetry breaking.
Its main idea is to represent the SM Higgs boson as a pseudo-Nambu-Goldstone boson of an approximate global
symmetry spontaneously broken at a scale in the TeV range.
In these models, the collective symmetry-breaking mechanisms are implemented in the
gauge sector, fermion sector, and the Higgs sector, which predict new particles within the mass range of a few TeV. These new particles play the role of partners of the
top-quark, of the gauge bosons, and the Higgs boson, the effect of which is to generate radiative corrections for the mass
of the Higgs boson and thus cancel the divergent corrections induced by SM particles.
LHM~\cite{Arkani1,Arkani2,Arkani3}, on the other hand, already has strong constraints from electroweak
precision data. These constraints typically require the new gauge bosons of LHM to be quite heavy \cite{PRD67-2003,
PRD68-2003}. In most LHM, the top partners are heavier than the new gauge bosons, which can lead to significant fine-tuning
in the Higgs potential \cite{JHEP03-2005}.

An exciting and relatively recent model is the BLHM~\cite{JHEP09-2010,Cruz-Albaro:2022kty,Cruz-Albaro:2022lks} overcomes these difficulties by including separate symmetry breaking scales at which the heavy gauge boson and top partners obtain their masses. This model was formulated to resolve the theoretical inconsistencies present in most other LH models, generates heavy gauge boson partner masses above the excluded mass range, and has top light partners below the upper bound from fine-tuning~\cite{Godfrey:2012tf}.
The BLHM is based on two independent non-linear sigma models. With the first field $\Sigma$, the global symmetry
$SO(6)_A\times SO(6)_B$ is broken to the diagonal group $SO(6)_V$ at the energy scale $f$, while with the second field
$\Delta$, the global symmetry $SU(2)_C \times SU(2)_D$ to the diagonal subgroup $SU(2)$ to the scale $F> f$. In the first
stage are generated 15 pseudo-Nambu-Goldstone bosons that are parameterized as

\begin{equation}\label{Sigma}
\Sigma=e^{i\Pi/f}  e^{2i\Pi_{h}/f}e^{i\Pi/f},
\end{equation}

\noindent
where $\Pi$ and $\Pi_h$ are complex and antisymmetric matrices given in Ref.~\cite{JHEP09-2010}. Regarding the second stage of spontaneous symmetry-breaking, the pseudo-Nambu-Goldstone bosons of the field $\Delta$ are parameterized as follows

\begin{equation}\label{Delta}
\Delta=F e^{2i \Pi_d/F},\, \,\, \, \, \Pi_d=\chi_a \frac{\tau^{a}}{2} \ \ (a=1,2,3),
\end{equation}

\noindent
$\chi_a$ represents the Nambu-Goldstone fields and the $\tau_a$ correspond to the Pauli matrices~\cite{JHEP09-2010}, which are the
generators of the SU(2) group.

\subsection{The scalar sector}

The BLHM Higgs fields, $h_1$ and $h_2$, form the Higgs potential that undergoes spontaneous symmetry breaking~\cite{JHEP09-2010,Kalyniak,Erikson}:

\begin{equation}\label{Vhiggs}
V_{Higgs}=\frac{1}{2}m_{1}^{2}h^{T}_{1}h_1 + \frac{1}{2}m_{2}^{2}h^{T}_{2}h_2 -B_\mu h^{T}_{1} h_2 + \frac{\lambda_{0}}{2} (h^{T}_{1}h_2)^{2}.
\end{equation}

\noindent The potential reaches a minimum when $m_1, m_2 >0$, while to break the electroweak symmetry requires $B_\mu > m_1 m_2$. The symmetry-breaking mechanism is implemented in the BLHM when the Higgs doublets acquire
their vacuum expectation values (VEVs), $\langle h_1\rangle ^{T}=(v_1,0,0,0)$ and $ \langle h_2 \rangle ^{T}=(v_2,0,0,0)$. By demanding that
these VEVs minimize the Higgs potential of Eq.~(\ref{Vhiggs}), the following relations are obtained

\begin{eqnarray}\label{v12}
&&v^{2}_1=\frac{1}{\lambda_0}\frac{m_2}{m_1}(B_\mu-m_1 m_2),\\
&&v^{2}_2=\frac{1}{\lambda_0}\frac{m_1}{m_2}(B_\mu-m_1 m_2).
\end{eqnarray}

\noindent These parameters can be expressed as follows

\begin{equation}\label{vvacio}
v^{2}\equiv v^{2}_1 +v^{2}_2= \frac{1}{\lambda_0}\left( \frac{m^{2}_1 + m^{2}_2}{m_1 m_2} \right) \left(B_\mu - m_1 m_2\right)\simeq \left(246\ \ \text{GeV}\right)^{2},
\end{equation}

\begin{equation}\label{beta}
\text{tan}\, \beta=\frac{v_1}{v_2}=\frac{m_2}{m_1}.
\end{equation}

\noindent From the diagonalization of the mass matrix for the scalar sector,
three non-physical fields $G_0$ and $G^{\pm}$, two physical scalar fields $H^{\pm}$
and three neutral physical scalar fields $h_0$, $H_0$ and $A_0$ are generated~\cite{Kalyniak,PhenomenologyBLH}. The lightest state, $h_0$, is identified as the
scalar boson of the SM. The masses of these fields are given as

\begin{eqnarray}\label{masaAGH}
m_{G_0}&=&m_{G^{\pm}}=0,\\
m^{2}_{A_{0}}&=&m^{2}_{H^{\pm}} =m^{2}_1+m^{2}_2,\label{mHmas} \\
m^{2}_{H_{0}} &=& \frac{B_\mu}{\text{sin}\, 2\beta} + \sqrt{\frac{B^{2}_{\mu}}{\text{sin}^{2}\, 2\beta} -2\lambda_0 B_\mu v^{2} \text{sin}\, 2\beta +\lambda^{2}_{0} v^{4} \text{sin}^{2}\, 2\beta  } \label{mh0H0}.
\end{eqnarray}

\noindent The four parameters present in the Higgs potential $ m_1,  m_2, B_\mu$ and $\lambda_0$ can be replaced by another more phenomenologically accessible set. That is, the masses of the states $h_0$ and $A_0$, the angle $\beta$ and the VEV $v$~\cite{Kalyniak}:

\begin{eqnarray}\label{parametros}
B_\mu &=&\frac{1}{2}(\lambda_0  v^{2} + m^{2}_{A_{0}}  )\, \text{sin}\, 2\beta,\\
\lambda_0 &=& \frac{m^{2}_{h_{0}}}{v^{2}}\Big(\frac{  m^{2}_{h_{0}}- m^{2}_{A_{0}} }{m^{2}_{h_{0}}-m^{2}_{A_{0}} \text{sin}^{2}\, 2\beta }\Big),\\
\text{tan}\, \alpha &=& \frac{ B_\mu \text{cot}\, 2\beta+ \sqrt{(B^{2}_\mu/\text{sin}^{2}\, 2\beta)-2\lambda_0 B_\mu v^{2} \text{sin}\, 2\beta+ \lambda^{2}_{0} v^{4}\text{sin}^{2}\, 2\beta  }  }{B_\mu -\lambda_0 v^{2} \text{sin}\, 2\beta},\label{alpha}   \\
m^{2}_{H_{0}} &=& \frac{B_\mu}{\text{sin}\, 2\beta}+ \sqrt{\frac{B^{2}_{\mu}}{\text{sin}^{2}\, 2\beta} -2\lambda_0 B_\mu v^{2} \text{sin}\, 2\beta +\lambda^{2}_{0} v^{4} \text{sin}^{2}\, 2\beta  }, \label{mH0}\\
m^{2}_{\sigma}&=&(\lambda_{56} + \lambda_{65})f^{2}=2\lambda_0 f^{2} \text{K}_\sigma. \label{masaescalar}
\end{eqnarray}

\noindent The variables $\lambda_{56}$ and $\lambda_{65}$ in Eq.~(\ref{masaescalar}) represent the coefficients of the quartic potential defined
in~\cite{JHEP09-2010}, both variables take values different from zero to achieve the collective breaking of the symmetry
and generate a quartic coupling of the Higgs boson~\cite{JHEP09-2010,Kalyniak}.
The BLHM also contains scalar triplet fields that get a contribution to their mass from the explicit symmetry-breaking terms in the model, as defined in Ref.~\cite{JHEP09-2010}, that depends on the parameter $m_4$,

\begin{eqnarray}
m^{2}_{\phi^{0}}&=& \frac{16}{3}F^{2} \frac{3 g^{2}_{A} g^{2}_{B}}{32 \pi^{2}} \log \left( \frac{\Lambda^{2}}{m^{2}_{W'^{\pm}}}\right) + m^{2}_{4} \frac{f^{4}+ F^{4}}{F^{2}(f^{2}+F^{2})},\\
m^{2}_{\phi^{\pm}}&=& \frac{16}{3}F^{2} \frac{3 g^{2}_{A} g^{2}_{B}}{32 \pi^{2}} \log \left( \frac{\Lambda^{2}}{m^{2}_{W'^{\pm}}}\right) + m^{2}_{4} \frac{f^{4}+f^{2}F^{2}+F^{4}}{F^{2}(f^{2}+F^{2})},\\
m^{2}_{\eta^{\pm}}&=&  m^{2}_{4}+ \frac{3 f^{2} g^{2}_{Y}}{64 \pi^{2}}\frac{\Lambda^{2}}{F^{2}},\\
m^{2}_{\eta^{0}}&=&m^{2}_{4}.
\end{eqnarray}

\subsection{The gauge sector}

In the BLHM, the new gauge bosons develop masses proportional to $\sqrt{f^2+F^2}\sim F$. This makes the masses of the gauge bosons large relative to other particles that have masses proportional to $f$. The kinetic terms of the gauge fields in the BLHM
are given as follows:

\begin{equation}\label{Lcinetico}
\mathcal{L}=\frac{f^{2}}{8} \text{Tr}(D_{\mu} \Sigma^{\dagger} D^{\mu} \Sigma) + \frac{F^{2}}{4} \text{Tr}(D_\mu \Delta^{\dagger} D^{\mu} \Delta),
\end{equation}

\noindent where

\begin{eqnarray}\label{derivadasC}
D_{\mu}\Sigma&=&\partial_{\mu} \Sigma +i g_A A^{a}_{1\mu} T^{a}_L \Sigma- i g_B \Sigma A^{a}_{2\mu} T^{a}_L+ i g_{Y} B^{3}_{\mu}(T^{3}_{R}\Sigma-\Sigma T^{3}_{R}),\\
D_{\mu}\Delta&=&\partial_{\mu} \Delta +i g_A A^{a}_{1\mu} \frac{\tau^{a}}{2}  \Delta- i g_B \Delta A^{a}_{2\mu} \frac{\tau^{a}}{2}.
\end{eqnarray}

\noindent $T^{a}_{L}$ are the generators of the group $SO(6)_A$ corresponding to the subgroup $SU(2)_{LA}$, while $T^3_R$ represents
the third component of the $SO(6)_B$ generators corresponding to the $SU(2)_{LB} $ subgroup, these matrices are provided in~\cite{JHEP09-2010}.
$g_A$ and $A^{a}_{1\mu}$ denote the gauge coupling and field associated with the gauge bosons of $SU(2)_{LA}$. $g_B$ and $A^{a}_{2\mu}$
represent the gauge coupling and the field associated with $SU(2)_{LB}$, while $g_Y$ and $B^{3}_{\mu}$ denote the hypercharge and the field.
When $\Sigma$ and $\Delta$ get their VEVs, the gauge fields $A^{a}_{1\mu}$ and $A^{a}_{2\mu}$ are mixed to form a massless triplet
$A^{a}_{0\mu}$ and a massive triplet $A^{a}_{H\mu}$,

\begin{equation}\label{AA}
A^{a}_{0\mu}=\text{cos}\, \theta_g A^{a}_{1\mu} + \text{sin}\, \theta_g A^{a}_{2\mu}, \hspace{5mm} A^{a}_{H\mu}= \text{sin}\, \theta_g A^{a}_{1\mu}- \text{cos}\, \theta_g A^{a}_{2\mu},
\end{equation}

\noindent with the mixing angles

\begin{equation}\label{gagb}
s_g\equiv \sin \theta_g=\frac{g_A}{\sqrt{g_{A}^{2}+g_{B}^{2}} },\ \ c_g \equiv \cos \theta_g=\frac{g_B}{\sqrt{g_{A}^{2}+g_{B}^{2}} },
\end{equation}

\noindent
which are related to the electroweak gauge coupling $g$ through

\begin{equation}\label{g}
\frac{1}{g^{2}}=\frac{1}{g^{2}_A}+\frac{1}{g^{2}_B}.
\end{equation}

After breaking the electroweak symmetry, when the Higgs doublets, $h_1$, and $h_2$ acquire their VEVs, the masses
of the gauge bosons of the BLHM are generated. In terms of the model parameters, the masses are given by

\begin{eqnarray}\label{masaBoson}
m^{2}_{\gamma} &=0&, \\
m^{2}_{Z}&=&\frac{1}{4}\left(g^{2}+g^{2}_Y \right)v^{2} \left(1-\frac{v^{2}}{12 f^2} \left(2+\frac{3f^2}{f^2+F^2} \left( s^{2}_g -c^{2}_g \right)^{2} \right)  \right), \\
m^{2}_{W^{\pm}}&=& \frac{1}{4} g^{2} v^{2} \left(1-  \frac{v^{2}}{12 f^2} \left(2+  \frac{3f^2}{f^2+F^2} \left(s^{2}_g -c^{2}_g  \right)^{2}\right)  \right),\\
m^{2}_{Z'}&=&m^{2}_{W'^{\pm}} +  \frac{g^2 s^{2}_W v^4}{16 c^{2}_W (f^2+F^2)} \left(s^{2}_g -c^{2}_g \right)^{2}, \label{mzprima} \\
m^{2}_{W'^{\pm}}&=& \frac{g^2}{4 c^{2}_{g} s^{2}_{g}} \left(f^2+F^2 \right)  - m^{2}_{W^{\pm}}. \label{mwprima}
\end{eqnarray}

The weak mixing angle is defined as

\begin{eqnarray}\label{angulodebil}
s_W&&\equiv\sin \theta_W = \frac{g_Y}{\sqrt{g^2+ g^{2}_Y }}, \\
c_W&&\equiv\cos \theta_W= \frac{g}{\sqrt{g^2+ g^{2}_Y }}.
\end{eqnarray}

\subsection{The Yang-Mills sector}

The gauge boson self-interactions arise from the following Lagrangian terms:

\begin{eqnarray}\label{YM}
\mathcal{L}=F_{1\mu \nu} F^{\mu \nu}_{1}  + F_{2\mu \nu} F^{\mu \nu}_{2},
\end{eqnarray}

\noindent
where $ F^{\mu \nu}_{1,2}$ are given by:
\begin{eqnarray}
F^{\mu \nu}_{1} &=& \partial^{\mu} A^{\alpha \nu}_{1} - \partial^{\nu} A^{\alpha \mu}_{1} + g_{A} \sum_{b} \sum_{c} \epsilon^{a b c} A^{b \mu}_{1} A^{c \nu}_{1}, \\
F^{\mu \nu}_{2} &=& \partial^{\mu} A^{\alpha \nu}_{2} - \partial^{\nu} A^{\alpha \mu}_{2} + g_{B} \sum_{b} \sum_{c} \epsilon^{a b c} A^{b \mu}_{2} A^{c \nu}_{2}.
\end{eqnarray}

\noindent
In these equations, the indices $a$, $b$ and $c$ run over the three gauge fields~\cite{Martin:2012kqb}; $\epsilon^{a b c}$ is the anti-symmetric tensor.

\subsection{The fermion sector} \label{subsecfermion}

For the construction of the Yukawa interactions in the BLHM, the fermions must be transformed under the group $SO(6)_A$ or $SO(6)_B$.
In this model, the fermion sector is divided into two parts. First, the sector of massive fermions is represented by Eq.~(\ref{Ltop}).
This sector includes the top and bottom quarks of the SM and a series of new heavy quarks arranged in four multiplets, $Q$ and $Q'$,
which transform under $SO(6)_A$, while $U^c$ and $U^{'c}_5$  are transformed under the group $SO(6)_B$. Second, the sector of
light fermions contained in  Eq.~(\ref{Lligeros}), in this expression, all the interactions of the remaining fermions of the SM with the
exotic particles of the BLHM are generated.

For massive fermions, the Lagrangian that describes them is given by \cite{JHEP09-2010}

\begin{equation}\label{Ltop}
\mathcal{L}_t=y_1 f Q^{T} S \Sigma S U^{c} + y_2 f Q'^{T} \Sigma U^{c} +y_3 f Q^{T} \Sigma U'^{c}_{5} +y_b f q_{3}^{T}(-2 i T^{2}_{R} \Sigma) U^{c}_{b}+ h.c.,
\end{equation}

\noindent where $ S = \text{diag} (1,1,1,1, -1, -1) $. The multiplets are arranged as follows

\begin{eqnarray}\label{camposf}
Q^{T}&=&\frac{1}{\sqrt{2}}\left( \left(-Q_{a_1} -Q_{b_2}\right), i\left(Q_{a_1} -Q_{b_2} \right),  \left(Q_{a_2} -Q_{b_1}\right), i\left(Q_{a_2} -Q_{b_1}\right), Q_{5},Q_{6} \right),\\
Q'^{T}&=&\frac{1}{\sqrt{2}} (-Q'_{a_1}, iQ'_{a_1},Q'_{a_2},iQ'_{a_2},0,0 ),\\
q_{3}^{T}&=& \frac{1}{\sqrt{2}} (-\bar{t}_L, i\bar{t}_L,\bar{b}_L,i\bar{b}_L,0,0 ),\\
U^{cT}&=& \frac{1}{\sqrt{2}} \left( (-U^{c}_{b_1} -U^{c}_{a_2}), i (U^{c}_{b_1} -U^{c}_{a_2}),  (U^{c}_{b_2} -U^{c}_{a_1}), i (U^{c}_{b_2} -U^{c}_{a_1}), U^{c}_{5},U^{c}_{6} \right),\\
U'^{cT}&=&(0,0,0,0,U'^{c}_5,0),\\
U_{b}^{cT}&=&(0,0,0,0,b^{c},0).
\end{eqnarray}

\noindent The explicit forms of the components of the multiplets are defined in Ref.~\cite{JHEP09-2010}. For simplicity, the Yukawa couplings are assumed to be real $y_1, y_2, y_3$ $\in \mathcal{R}$. The Yukawa coupling of the top-quark is defined as

\begin{equation}\label{yt}
 y_t= \frac{3 y_1 y_2 y_3}{ \sqrt{ (y^{2}_1 +y^{2}_2)(y^{2}_1 +y^{2}_3)}}=\frac{m_{t}}{v \sin \beta}.
 \end{equation}

\noindent For light fermions, the corresponding Lagrangian is

\begin{equation}\label{Lligeros}
\mathcal{L}_{light}= \sum_{i=1,2} y_u f q^{T}_i \Sigma u^{c}_{i} + \sum_{i=1,2} y_{d} f q^{T}_{i}(-2i T^{2}_{R} \Sigma) d^{c}_i
+\sum_{i=1,2,3} y_e f l^{T}_i (-2i T^{2}_{R} \Sigma) e^{c}_i + h.c.,
\end{equation}

\noindent with

\begin{eqnarray}\label{qligeros}
 q^{T}_i &=&\frac{1}{\sqrt{2}} (-\bar{u}_{iL}, i \bar{u}_{iL}, \bar{d}_{iL}, i \bar{d}_{iL},0,0),\\
 l^{T}_i &=&\frac{1}{\sqrt{2}} (-\bar{\nu}_{iL}, i \bar{\nu}_{iL}, \bar{e}_{iL}, i \bar{e}_{iL},0,0),\\
 u^{cT}_i &=&(0,0,0,0,u^{c}_i,0),\\
 d^{cT}_i &=&(0,0,0,0,d^{c}_i,0),\\
 e^{cT}_i &=&(0,0,0,0,e^{c}_i,0).
\end{eqnarray}

After breaking the electroweak symmetry, the resulting mass terms are expanded by a power series up to $\frac{v^2}{f^2}$,
and the mass matrices are diagonalized using perturbation theory.
The fermion mass eigenstates are calculated under the assumption that $y_{2} \neq y_{3}$,
otherwise the masses of the heavy quarks $T$ and $T_{5}$ are degenerate at the lowest order. Consequently, different diagonalization schemes are required for the region where $y_2=y_3$ versus the region where $|y_2-y_3|>0$~\cite{Godfrey:2012tf,PhenomenologyBLH}

\begin{eqnarray}
  m^{2}_t &=&y^{2}_t v^{2}_1,\label{mt} \\
  m^{2}_T &=& (y^{2}_1 + y^{2}_2)f^2 + \frac{9 v^{2}_1 y^{2}_1  y^{2}_2  y^{2}_3 }{(y^{2}_1 + y^{2}_2) (y^{2}_2 - y^{2}_3)}, \label{mT} \\
  m^{2}_{T_5} &=& (y^{2}_1 + y^{2}_3)f^2 - \frac{9 v^{2}_1 y^{2}_1  y^{2}_2  y^{2}_3 }{(y^{2}_1 + y^{2}_3) (y^{2}_2 - y^{2}_3)},\label{MT5} \\
  m^{2}_{T_6} &=&m^{2}_{T^{2/3}_b}=m^{2}_{T^{5/3}_b} =y^{2}_1 f^2,\label{mT6} \\
  m^{2}_B & =&(y^{2}_1 + y^{2}_2)f^2,\label{mB}
\end{eqnarray}

\noindent with $v_1=v\sin \beta$ and $v_2=v\cos \beta$.

\subsection{The currents sector}

The Lagrangian that describes the interactions of fermions with the gauge bosons is~\cite{JHEP09-2010,PhenomenologyBLH}
\begin{eqnarray}\label{LbaseW}
 \mathcal{L} &=& \bar{Q} \bar{\tau}^{\mu} D_{\mu}Q + \bar{Q}' \bar{\tau}^{\mu} D_{\mu}Q'- U^{c\dagger} \tau^{\mu} D_{\mu}U^{c}-  U'^{c\dagger} \tau^{\mu} D_{\mu}U'^{c} -  U_{b}^{c\dagger} \tau^{\mu} D_{\mu}U_{b}^{c} +\sum_{i=1,2}  q^{\dagger}_i \tau^{\mu} D_{\mu} q_i   \nonumber \\
 &+& \sum_{i=1,2,3}  l^{\dagger}_i \tau^{\mu} D_{\mu} l_i
 - \sum_{i=1,2,3}  e_i^{c\dagger} \tau^{\mu} D_{\mu} e^{c}_i - \sum_{i=1,2}  u_{i}^{c\dagger} \tau^{\mu} D_{\mu} u^{c}_{i} - \sum_{i=1,2}  d_{i}^{c\dagger} \tau^{\mu} D_{\mu} d^{c}_i,
\end{eqnarray}

\noindent  where $\tau^{\mu}$ and $\bar{\tau}^{\mu}$ are defined according to~\cite{Spremier}. On the other hand, the respective covariant derivatives are~\cite{PhenomenologyBLH,Martin:2012kqb}

\begin{eqnarray}\label{dcovariantes}
  D_{\mu}Q & =&\partial_{\mu}Q+ \sum_{a} (i g_A A^{a}_{1\mu} T^{a}_{L} Q )+ ig_Y B_{3\mu} (T^{3}_{R} +T^{+}_{X} )Q, \\
  D_{\mu}Q'  & =&\partial_{\mu}Q'+\sum_{a} (i g_A A^{a}_{1\mu} T^{a}_{L} Q' )+ ig_Y B_{3\mu} \left(\frac{1}{6} \right)Q',\\
   D_{\mu}U^{c} & =&\partial_{\mu}U^{c}+ \sum_{a} (i g_B A^{a}_{2\mu} T^{a}_{L} U^{c} )+ ig_Y B_{3\mu} (T^{3}_{R} +T^{-}_{X} )U^{c},\\
 D_{\mu}U'^{c} & =&\partial_{\mu}U'^{c}+  ig_Y B_{3\mu} T^{-}_{X} U'^{c},  \\
 D_{\mu}U_{b}^{c} & =&\partial_{\mu}U_{b}^{c}+  ig_Y B_{3\mu} \left(\frac{1}{3} \right) U_{b}^{c},  \\
     D_{\mu}q_i & =&\partial_{\mu}q_i+ \sum_{a} (i g_A A^{a}_{1\mu} T^{a}_{L} q_i )+ ig_Y B_{3\mu} (T^{3}_{R} +T^{+}_{X} )q_i,\\
  D_{\mu}l_i & =&\partial_{\mu}l_i +\sum_{a} (i g_B A^{a}_{2\mu} T^{a}_{L} l_i )+ ig_Y B_{3\mu} T^{3}_{R} l_i,\\
   D_{\mu}e^{c}_i & =&\partial_{\mu}e_{i}^{c} + ig_Y B_{3\mu} T^{e}_{X} e^{c}_i,\\
 D_{\mu}u^{c}_i & =&\partial_{\mu}u_{i}^{c}+  ig_Y B_{3\mu} T^{-}_{X} u^{c}_i,\\
    D_{\mu}d^{c}_i & =& \partial_{\mu}d_{i}^{c}+ ig_Y B_{3\mu} T^{d}_{X} d^{c}_i.\label{dcovariantes70}
\end{eqnarray}

\section{AMDM and AWMDM of the top-quark in the BLHM}


The electroweak properties of fermions are characterized by physical magnitudes called form factors. These measure properties such as the electric charge, the AMDM, the EDM, the AWMDM, the WEDM, and others. Some of these quantities are already present in classical theory, while others arise for the first time as a quantum fluctuation of one-loop or higher orders.
In quantum field theory, the electromagnetic and weak properties of fermions arise through their interaction with the gauge bosons $V$, $V= \gamma, Z$.
The most general Lorentz-invariant vertex function describing the interaction of a gauge boson with two fermions can be written in terms of ten form factors \cite{NPB551-1999,NPB812-2009}, which are functions of the kinematic invariants.
In the low energy limit, these correspond to couplings that multiply dimension-four or-five operators in an effective
Lagrangian and may be complex. If the $V$ boson couples to effectively massless fermions, the number of independent form factors is reduced to eight. In addition, if the fermions are on-shell, the number is further reduced to four. In this way, the $V \bar{f} f$ vertex function can be written in the form

{\small
\begin{eqnarray}\label{verticeZtt}
ie\bar{u}(p') \Gamma^{\mu}_{ V\bar{f}f} u(p)=ie \bar{u}(p')\big\{ \gamma^{\mu}\left[ F^{V}_{V}(q^{2})-F^{V}_{A}(q^{2})\gamma^{5}\right]
+ i \sigma^{\mu \nu} q_{\nu} \left[ F^{V}_{M}(q^{2})- iF^{V}_{E}(q^{2})\gamma^{5}\right] \big\}u(p),
\end{eqnarray}
}

\noindent
where $e$ is the proton charge and $q=p'-p$ is the $ V $ gauge boson transferred four-momentum. The terms $F^{V}_{V}(0)$ and $F^{V}_{A}(0)$ in the low energy
limit are the $V\bar{f}f$ vector and axial-vector form factors in the SM, while $F^{V}_M(q^2)$ and $F^{V}_E(q^2)$ are associated with the form factors of the  electromagnetic or weak dipole moments. The latter appear due to quantum corrections and are a valuable tool to study the effects
of new physics indirectly through virtual corrections of new particles predicted by extensions of the SM.

In this work, to produce a pair of heavy fermions, particularly a pair of top-quarks, the gauge boson $V$ must be off mass shell.
For this case, the well-known gauge dependence
problem arises and occurs when one studies the radiative corrections to fermion-pair production at colliders with center-of-mass energy beyond the mass of the $V$ boson, that is,
$\sqrt{q^{2}}>2m_t$~\cite{Bernabeu:1995gs}. In this case, the pinch technique (PT)~\cite{Papavassiliou:1993qe,Cornwall:1989gv,Cornwall:1981zr,Alkofer:2000wg,Papavassiliou:1996zn}
can be used to remove the gauge dependence. Another alternative is the background-field method~\cite{Denner:1994xt}. The PT consists in reorganizing the usual Feynman-diagram expressions into manifestly gauge-independent portions. One extracts from the box diagrams those gauge-dependent pieces which are kinematically equivalent to the  $t\bar tV$ vertex corrections. Those pieces offset the gauge non-invariance of the vertex corrections.
One may calculate these quantities in any gauge as long as the pinch contributions are consistently identified in the box and
vertex graphs. However, in this particular case, the computation in the Feynman-´t Hooft  gauge is most convenient since, in that
gauge, the MDM’s (WMDM's) do not receive any contributions from the box diagrams.
Indeed, such contributions from the box diagram arise from the longitudinal terms in the gauge boson propagators, which are not present in this gauge~\cite{Bernabeu:1995gs,Papavassiliou:1993qe}.
Therefore, calculating the AMDM (AWMDM) form factors of the top-quark in the Feynman-'t Hooft gauge is more convenient.
In this study, the computations are performed in the Feynman-'t Hooft gauge.
Our results determine the effects of the new physics that could be potentially very important.

Concerning form factors, $F^{V}_M (q^2)$ and $F^{V}_E(q^2)$, these are related to the AMDM $a_t$, AWMDM $a^{W}_t$, EDM $d_t $ and WEDM $d^{W}_t $ as follows

\begin{eqnarray}\label{MDD}
  F^{\gamma}_{M}(q^{2})= -\frac{a_t}{2 m_t}, &  \qquad F^{Z}_{M}(q^{2})= -\frac{a^{W}_t}{2 m_t}, \\
  F^{\gamma}_{E}(q^{2})= -\frac{d_t}{ e}, &  \qquad  F^{Z}_{E}(q^{2})= -\frac{d^{W}_t}{ e}.
\end{eqnarray}

\subsection{Contribution of new scalar bosons, gauge bosons, and heavy quarks to the AMDM and AWMDM of the top-quark}

Investigation of the electromagnetic and weak dipole moments of the top-quark provide deep insight into the particle of interest.  A key motivation for the detailed study of the top is its large mass which implies that it is the SM particle  most strongly coupled to the electroweak symmetry-breaking sector.
Additionally, the study of top-quark couplings is one of the cornerstones of the analysis program at the LHC and future high-energy colliders.
Therefore, the top-quark is a suitable particle to look for signatures of new physics at the TeV scale.

\begin{figure}[H]
\center
\subfloat[]{\includegraphics[width=4.5cm]{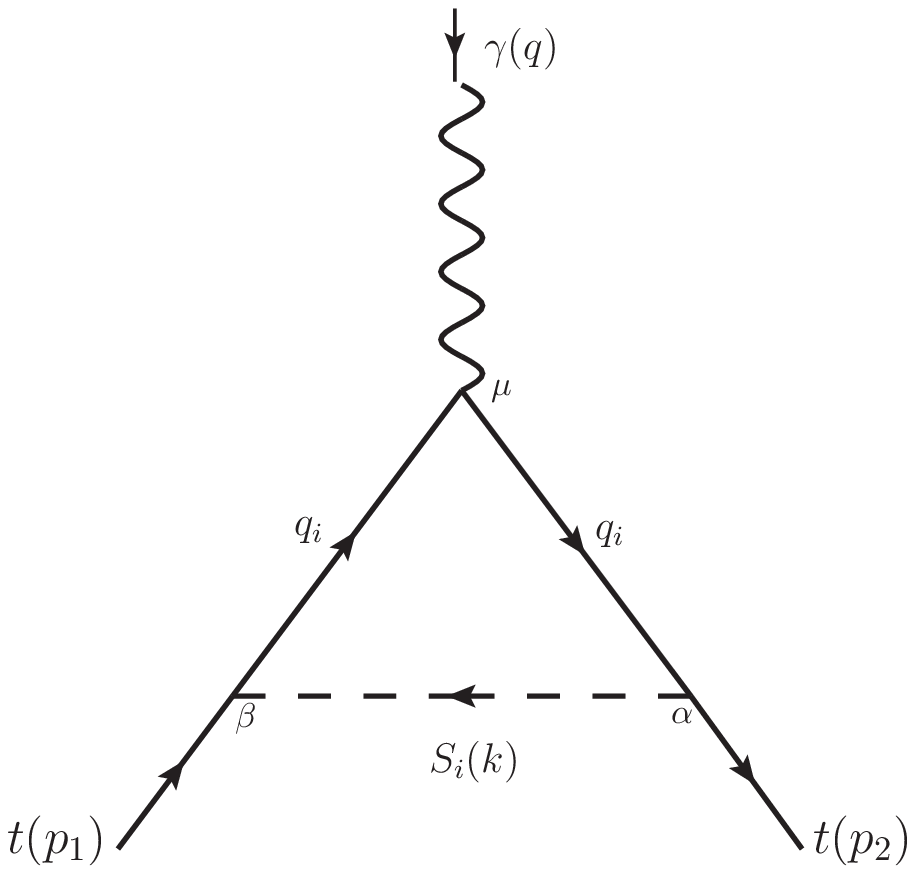}}
\subfloat[]{\includegraphics[width=4.5cm]{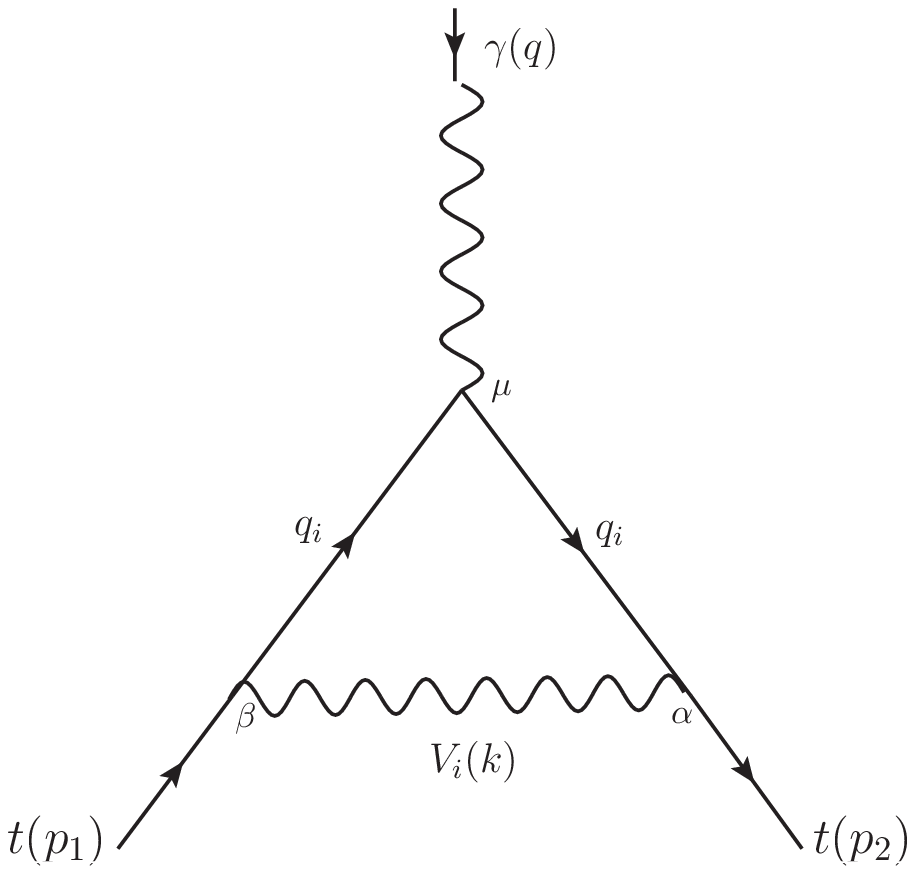}}
\subfloat[]{\includegraphics[width=4.5cm]{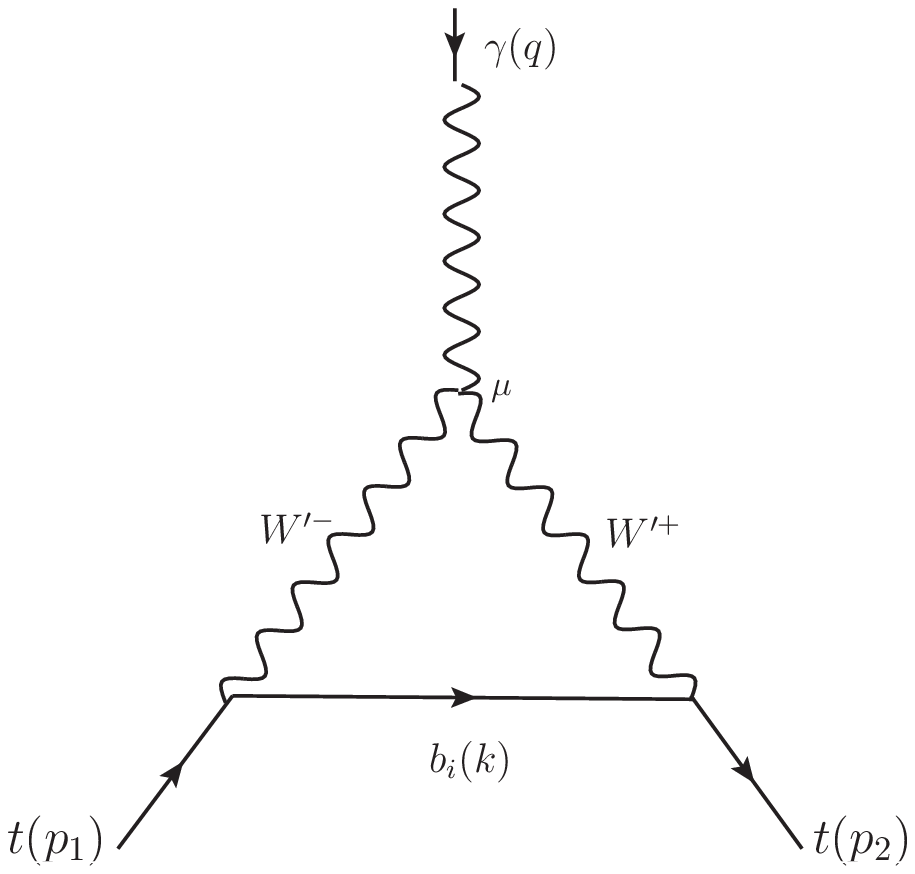}}\\
\subfloat[]{\includegraphics[width=4.5cm]{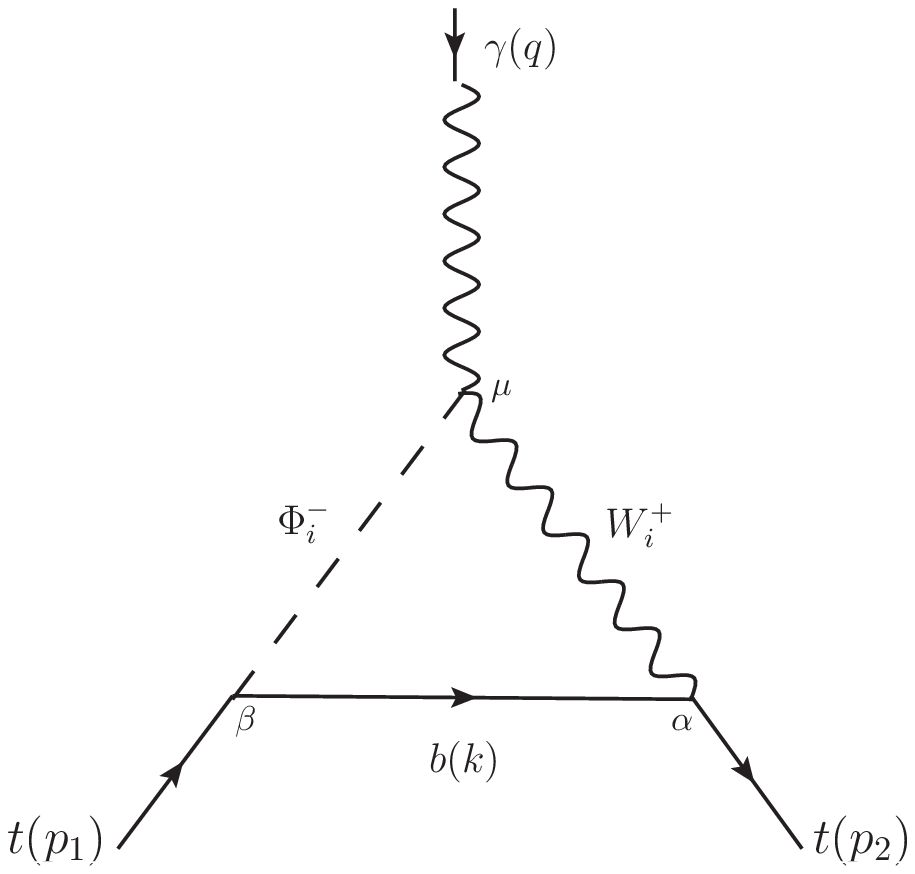}}
\subfloat[]{\includegraphics[width=4.5cm]{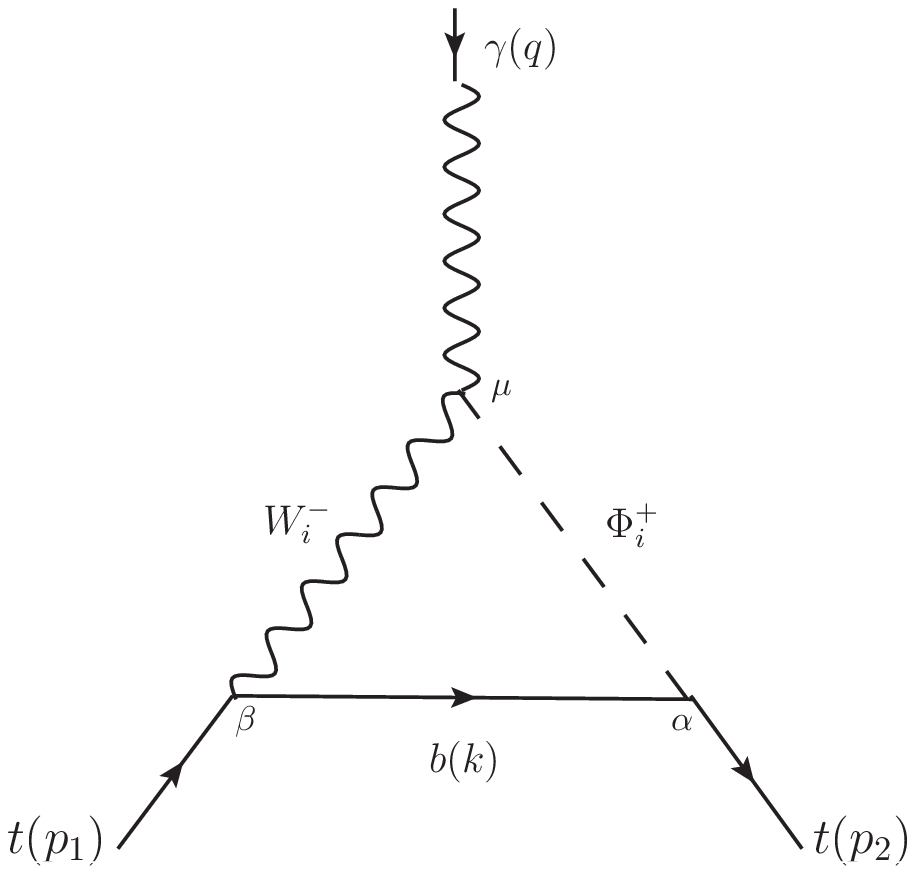}}
\caption{ \label{dipolo} Generic Feynman diagrams that contribute to the AMDM of the top-quark at one-loop, $q_{i}\equiv b, t, B, T, T_{5}, T_{6}, T^{2/3}, T^{5/3}$, and $b_{i}\equiv b, B$.
 a) Scalar contributions, $S_{i}\equiv h_{0}, H_{0}, A_{0}, \eta^{0},\phi^{0}, \sigma, H^{\pm}, \eta^{\pm},\phi^{\pm}$. b) and c) Vector contributions, $V_i \equiv \gamma, Z, W, Z', W'$. d) and e) Scalar-vector contributions, $\Phi^{\pm}_{i}= \phi^{\pm}, \eta^{\pm}$, and $W_{i} \equiv W, W'$.}
\end{figure}

\begin{figure}[H]
\center
\subfloat[]{\includegraphics[width=4.5cm]{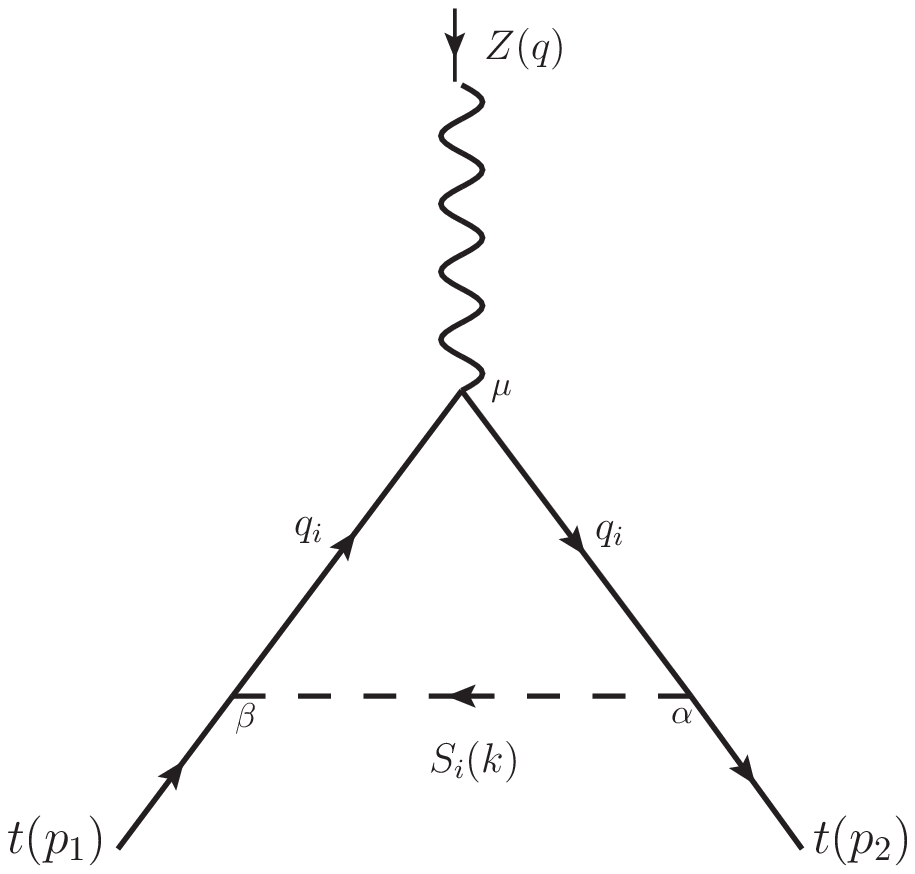}}
\subfloat[]{\includegraphics[width=4.5cm]{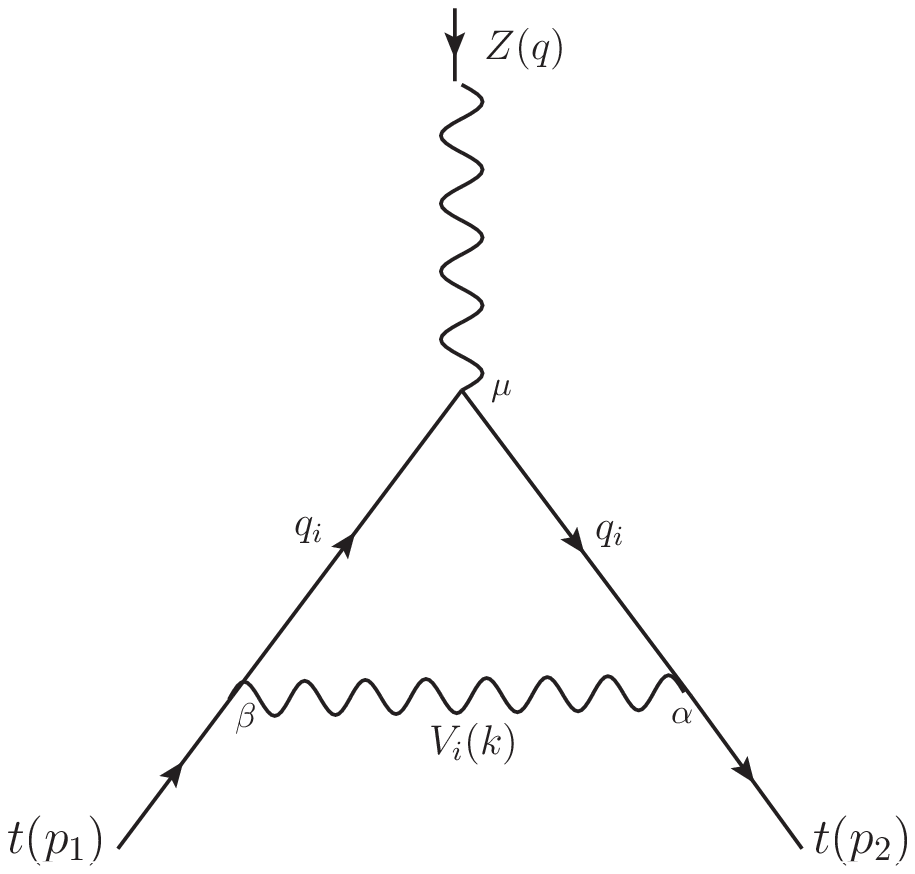}}
\subfloat[]{\includegraphics[width=4.5cm]{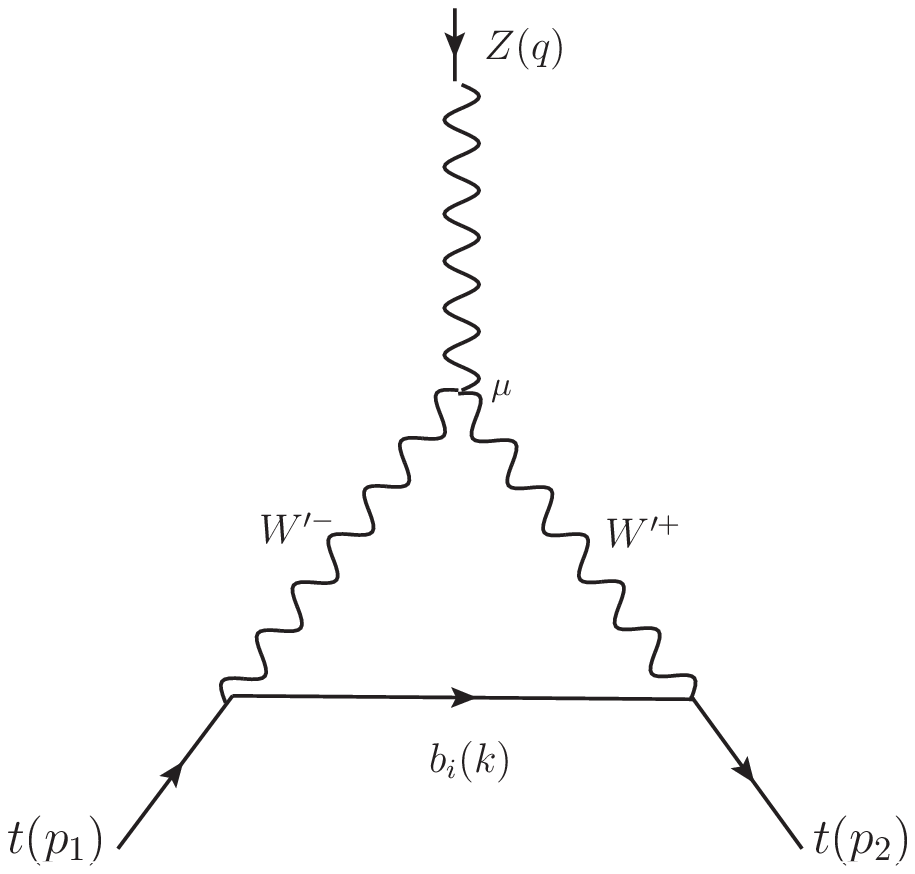}}\\
\subfloat[]{\includegraphics[width=4.5cm]{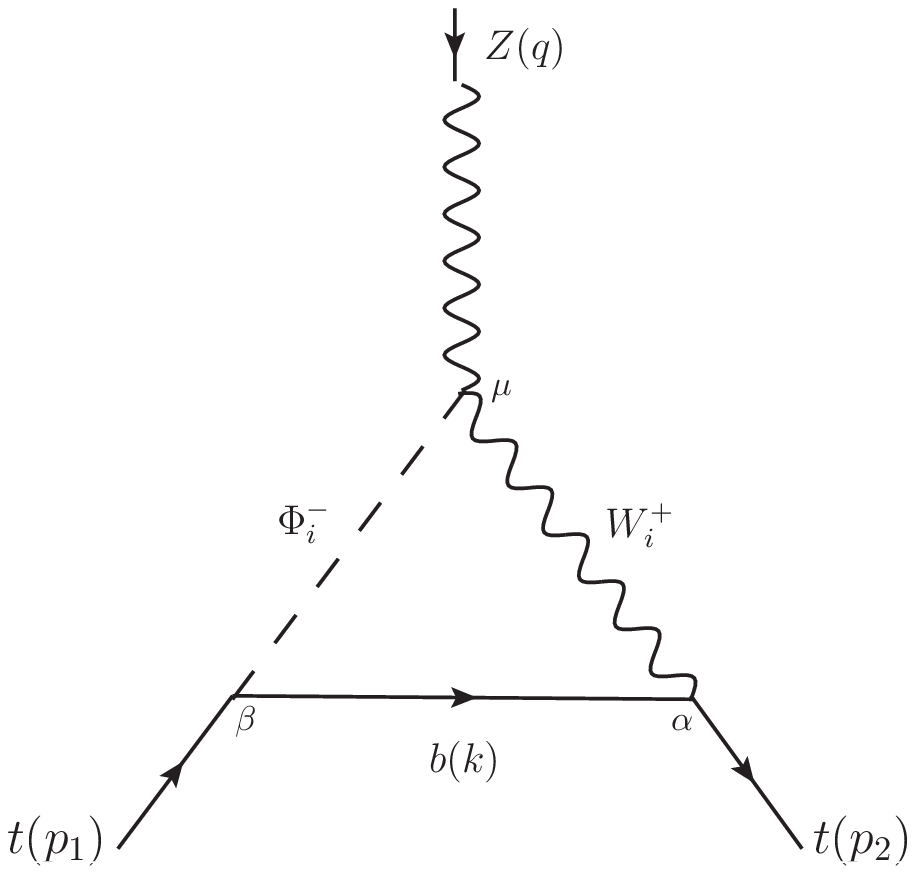}}
\subfloat[]{\includegraphics[width=4.5cm]{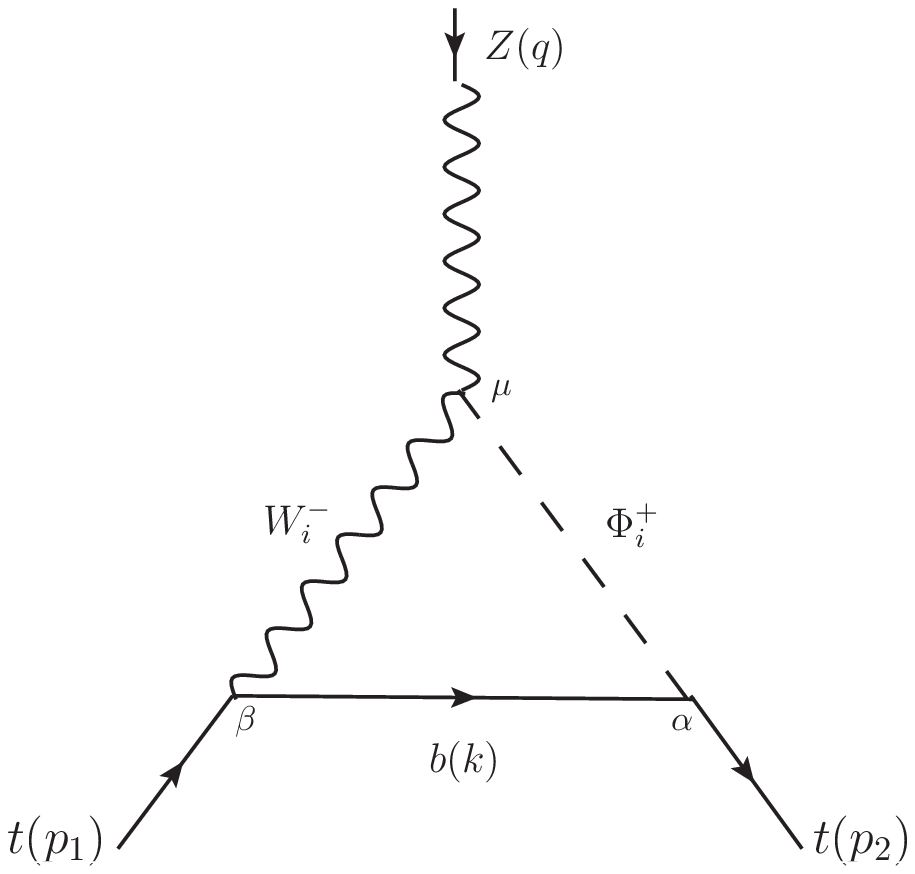}}\\
\subfloat[]{\includegraphics[width=4.5cm]{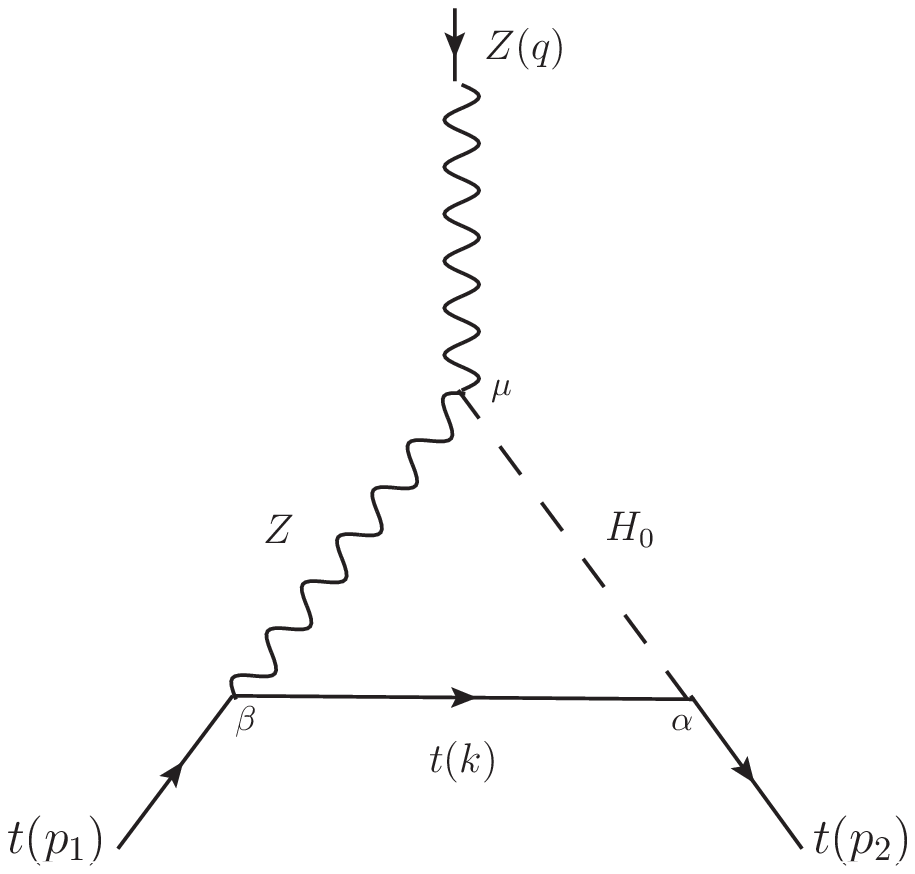}}
\subfloat[]{\includegraphics[width=4.5cm]{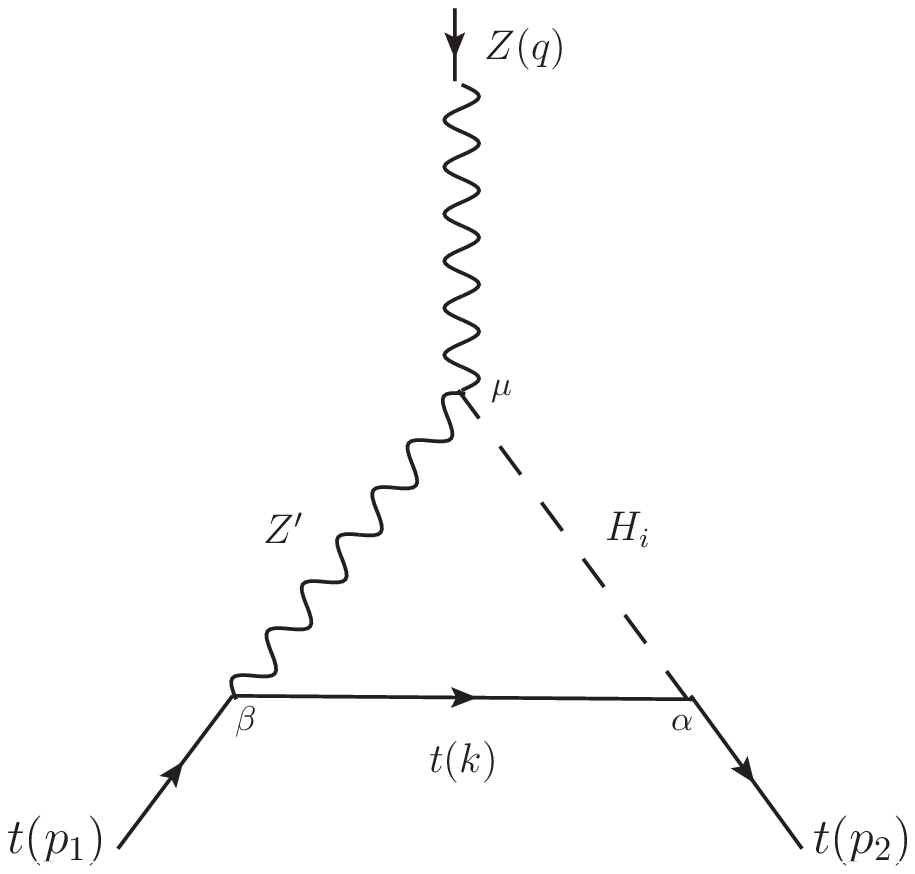}}
\caption{ \label{dipoloweak} Generic Feynman diagrams that contribute to the AWMDM of the top-quark at one-loop, $q_{i}\equiv b, t, B, T, T_{5}, T_{6}, T^{2/3}, T^{5/3}$, an $b_{i}\equiv b, B$.
 a) Scalar contributions, $S_{i}\equiv h_{0}, H_{0}, A_{0}, \eta^{0},\phi^{0}, \sigma, H^{\pm}, \eta^{\pm},\phi^{\pm}$. b) and c) Vector contributions, $V_i \equiv \gamma, Z, W, Z', W'$.
d)-g) Scalar-vector contributions, $W^{\pm}_{i}=W^{\pm}, W'^{\pm}$,  $\Phi^{\pm}_{i}= \phi^{\pm}, \eta^{\pm}$, and $ H_i=h_0, H_0 $.}
\end{figure}

In this subsection, we derive the AMDM and AWMDM of the top-quark at the one-loop level. These magnitudes carry essential information about their interactions with the new particles predicted by the BLHM. As for the EDM and WEDM, these do not receive one-loop-level contributions in the context of the BLHM.
In this way, we begin by finding all possible contributions to one-loop to the  $F^{V}_M (q^2)$ form factors that can be classified for study into three classes of triangular diagrams, i.e., Feynman diagrams are classified according to the contributions generated by the particles circulating in the loop of the $V\bar{t}t$ vertex: scalar contributions, vector contributions, and scalar-vector (s-v) contributions.
From  Figs.~\ref{dipolo} and~\ref{dipoloweak}, we can see
that $S_{i}$, $V_{i}$, $ \Phi^{\pm}_i $, $ W^{\pm}_i $, $ H_i $, $ q_{i}$ and $ b_i $ are  the particles circulating in the loop: $S_{i} \equiv h_0$
(SM Higgs boson), $H_0, A_0, \phi^{0}, \eta^{0}, \sigma, H^{\pm}, \phi^{\pm}, \eta^{\pm}$; $V_{i} \equiv  \gamma, Z, W, Z', W'$;  $\Phi^{\pm}_{i} \equiv \phi^{\pm}, \eta^{\pm}$;  $W^{\pm}_{i} \equiv W^{\pm}, W'^{\pm}$; $ H_i \equiv h_0, H_0 $; $q_i \equiv  b, t, B, T, T_5, T_6, T^{2/3}, T^{5/3}$; and $b_{i}\equiv b, B$.
To obtain the amplitude of each contribution, we need to know the Feynman rules involved in the diagrams shown in
Figs.~\ref{dipolo} and ~\ref{dipoloweak}, these vertices are provided in Appendix~\ref{reglasFeynman}~\cite{Cruz-Albaro:2022kty,Cruz-Albaro:2022lks}.

In the Feynman-´t Hooft gauge, about 60 Feynman diagrams that contribute to the one-loop corrections to the vertex  $V\bar{t}t$. It is relevant to mention that in the context of this gauge, diagrams involving Goldstone bosons ($G^{0}, G^{\pm}$) must also be considered. However, their contributions are much smaller than those where purely physical particles ($S_i, \ V_i$, $ \Phi^{\pm}_i $, $ W^{\pm}_i $, $ H_i $) are involved. This is due to the coupling constants that describe their corresponding vertices, which are suppressed.

Shown in Figs.~\ref{dipolo} and~\ref{dipoloweak} are all the generic Feynman diagrams that contribute to the AMDM and AWMDM of the top-quark, respectively. The amplitudes of these contributions can be written in the following compact form:

\begin{eqnarray}\label{amplitudesSV1}
\mathcal{M}^{\mu}_{t}(S_{i}q_i q_i)&=& \int \frac{d^{4}k}{(2\pi)^{4}} \bar{u}(p_{2}) \left(f^{*}_{S_i} + f^{*}_{P_i}\gamma^{5} \right) \left[i \frac{\not\! k + \not\!p_{2}+m_{q_i}  }{(k+p_{2})^{2}-m^{2}_{q_{i}}} \right] \left( \gamma^{\mu} (F_{V_i}+F_{A_{i}}\gamma^{5}) \right)  \nonumber \\
&\times & \left[i \frac{ \not\! k + \not\!p_{1}+m_{q_i}  }{(k+p_{1})^{2}-m^{2}_{q_{i}}} \right] \left(f_{S_i} + f_{P_i}\gamma^{5} \right) u(p_{1}) \left(\frac{i}{k^{2}- m^{2}_{S_{i}}} \right), \\
\mathcal{M}^{\mu}_{t}(V_{i} q_i q_i) &=&  \int \frac{d^{4}k}{(2\pi)^{4}} \bar{u}(p_{2})  \left( \gamma^{\alpha}\left(f^{*}_{V_i} + f^{*}_{A_i}\gamma^{5} \right) \right) \left[i \frac{\not\! k + \not\! p_{2}+m_{q_i}  }{(k+p_{2})^{2}-m^{2}_{q_{i}}} \right]  \left( \gamma^{\mu} (F_{V_{i}}+F_{A_{i}}\gamma^{5}) \right) \nonumber \\
&\times & \left[i \frac{ \not\! k + \not\! p_{1}+m_{q_i}  }{(k+p_{1})^{2}-m^{2}_{q_{i}}} \right] \left( \gamma^{\beta}\left(f_{V_i} + f_{A_i}\gamma^{5} \right)\right) u(p_{1})
\left[\frac{i}{k^{2}- m^{2}_{V_{i}}} \left(-g_{\alpha \beta} \right) \right], \label{amplitudesSV2} \\
\mathcal{M}^{\mu}_{t}(W'^{-} W'^{+} b_i) &=&  \int \frac{d^{4}k}{(2\pi)^{4}} \bar{u}(p_{2}) \left( \gamma^{\nu}\left(f^{*}_{V_i} + f^{*}_{A_i}\gamma^{5} \right) \right) \left[i \frac{\not\! k +m_{b_i} }{k^{2}-m^{2}_{b_i}} \right] \left( \gamma^{\eta}\left(f_{V_i} + f_{A_i}\gamma^{5} \right) \right)  u(p_1)\nonumber \\
&\times & \left[\frac{i}{(k-p_2)^{2}- m^{2}_{W'}} \left(-g_{\alpha \nu} \right) \right] A^{\alpha \mu \beta}_{  W' V W' }   \left[\frac{i}{(k-p_1)^{2}- m^{2}_{W'}} \left(-g_{ \beta \eta} \right) \right],\label{amplitudesSV3} \\
\mathcal{M}^{\mu}_{t}(  \Phi^{-}_{i} W^{+}_{i} b ) &=&  \int \frac{d^{4}k}{(2\pi)^{4}} \bar{u}(p_{2})  \left( \gamma^{\alpha}\left(f_{V_i} + f_{A_i}\gamma^{5} \right) \right)  \left[i \frac{\not\! k + m_{b} }{k^{2}-m^{2}_{b} } \right] \left(f_{S_i} + f_{P_i}\gamma^{5} \right) u(p_1)\nonumber \\
&\times & \left[\frac{i}{(k-p_2)^{2}- m^{2}_{W^{+}_{i}}} \left(-g_{\alpha \nu} \right) \right] \left( C_{ W^{+}_{i} V \Phi^{-}_{i} }\, g^{\mu \nu} \right)  \left[\frac{i}{(k-p_1)^{2}- m^{2}_{ \Phi^{-}_{i} }}  \right], \label{amplitudesSV4}\\
\mathcal{M}^{\mu}_{t}(  W^{-}_{i}  \Phi^{+}_{i} b ) &=&  \int \frac{d^{4}k}{(2\pi)^{4}} \bar{u}(p_{2}) \left(f^{*}_{S_i} + f^{*}_{P_i}\gamma^{5} \right)  \left[i \frac{\not\! k +m_{b}}{k^{2}-m^{2}_{b} } \right] \left( \gamma^{\beta}\left(f^{*}_{V_i} + f^{*}_{A_i}\gamma^{5} \right) \right) u(p_1)
\nonumber \\
&\times &   \left[\frac{i}{(k-p_2)^{2}- m^{2}_{\Phi^{+}_{i}}} \left(-g_{\beta \nu} \right) \right] \left( C_{ W^{-}_{i} V \Phi^{+}_{i} }\, g^{\mu \nu} \right) \left[\frac{i}{(k-p_1)^{2}- m^{2}_{ W^{-}_{i} }}  \right],
\label{amplitudesSV5}\\
 \mathcal{M}^{\mu}_{t}({\tt Z} H_i t) &=&  \int \frac{d^{4}k}{(2\pi)^{4}} \bar{u}(p_{2})  \left(f_{S_i} + f_{P_i}\gamma^{5} \right)  \left[i \frac{\not\! k + m_{t} }{k^{2}-m^{2}_{t}} \right]  \left( \gamma^{\beta} (F_{V_i}+F_{A_{i}}\gamma^{5}) \right) u(p_1)\nonumber \\
&\times & \left[\frac{i}{(k-p_2)^{2}- m^{2}_{H_i }} \right] \left( C_{ {\tt Z} Z H_i  }\, g^{\mu \nu} \right)   \left[\frac{i}{(k-p_1)^{2}- m^{2}_{{\tt Z} }} \left(-g_{\beta \nu} \right)  \right],\label{amplitudesSV6}
\end{eqnarray}

\noindent where $f_{S_{i}}, f_{P_{i}}, f_{V_{i}}$, and $ f_{A_{i}}$ denote the form factors of the scalars,
pseudoscalar, vector, and axial-vector.
On the other hand, $C$ represents the coupling constants of the corresponding vertices, while ${\tt Z}=Z,Z'$. The tensor $ A^{\alpha \mu \beta}_{    W' V W' }  $ is provided in Table~\ref{3boson} of Appendix~\ref{reglasFeynman}.
Once the amplitudes have been solved, we take on the task of picking only the coefficients of the  $\sigma^{\mu \nu} q_{\nu}$ and $\sigma^{\mu \nu} q_{\nu} \gamma^{5}$ tensors. These represent the form factors of our interest, $F^{V}_M$ and $F^{V}_E$ (see Eq.~(\ref{verticeZtt})).
As mentioned above, the $F^{V}_E$ form factor does not arise at the one-loop level in the BLHM scenario. Therefore, the form factors $F^{V}_M$ will be the magnitudes contained in coded form by the AMDM $a_t$ and AWMDM $a^{W}_t$ of the top-quark, which in turn represent the contributions of the new physics. $a_t$ and  $a^{W}_t$ are obtained as follows

\begin{eqnarray}\label{awt}
a_{t}\equiv a_{t}^{BLHM} &=& [a_{t}]^{S_i} + [a_{t}]^{V_i} +  [a_{t}]^{S_i - V_i},\\
a^{W}_{t }\equiv a^{W-BLHM}_{t} &=& [a^{W}_{t}]^{S_i} + [a^{W}_{t}]^{V_i} + [a^{W}_{t}]^{S_i - V_i}.
\end{eqnarray}

\subsection{ The top-quark at the ILC}

Future lepton colliders promise to precisely probe top-quark interactions because they offer a very clean and well-defined environment. In particular, the physics case for one or more $e^{+} e^{-}$ colliders is by now well established, and there are several such collider projects with sufficient energy to produce top-quark pairs, i.e., with center-of-mass energy $\sqrt{s}> 2 m_t$. However, a mature design exists for a linear $e^{+} e^{-}$ collider that can reach center-of-mass energies of up to 1000 GeV, namely the ILC.
At the ILC~\cite{Behnke:2013xla,Baer:2013cma,Adolphsen:2013kya}, the production of top-quarks in the $e^{+} e^{-} \rightarrow Z^{*}/\gamma \rightarrow t\bar{t}$  process will be a powerful tool to determine the scale of the new physics indirectly.  The process $e^{+} e^{-} \rightarrow Z^{*}/\gamma \rightarrow t\bar{t}$  is also extremely sensitive to top electroweak couplings.
\\
 The ILC is designed to operate in phase II at a center-of-mass energy of
$\sqrt{s}=500$ GeV. At this energy, top-quark pairs are produced numerously well above the threshold~\cite{Cao:2015qta}.
Thus, to give numerical results to $a_{t}$ and $a^{W}_{t}$, we adopt the same collider parameters
of the $e^{+} e^{-}$ linear collider, that is,  $\sqrt{s}=\sqrt{q^{2}} =500$ GeV.
We have calculated the contributions to the AMDM and AWMDM of the on-shell top-quark with the $V$ gauge boson at the expected center-of-mass energy for the ILC.

\section{Numerical results}
\label{resultadosNum}

For our numerical analysis of the electromagnetic and weak properties of the top-quark in the context of the BLHM, we briefly review the free parameters of the BLHM.
Subsequently, we discuss each study scenario's numerical contributions generated for the AMDM and AWMDM of the top-quark.

\subsection{Parameters space of the BLHM}

We consider the following BLHM input parameters: $ y_i $,  $m_{A_{0}}$, $m_{\eta_{0}}$, $\tan \beta$, $\tan \theta_{g}$, $f$ and $F$.

\noindent \textbf{Yukawa couplings, $y_{i}$, $i=1,2,3$}:
In the fermionic sector of the BLHM,  the most stringent theoretical constraint on the masses of the exotic quarks comes from
fine-tuning the Higgs potential due to fermion loops. It is, therefore, essential to determine realistic values of
the three Yukawa couplings, $y_{1,2,3}$, and the top-quark Yukawa coupling, $y_{t}$, that evade the fine-tuning
constraints. In this sense, a fit on the Yukawa coupling parameters is required. In the BLHM, the size of the
fine-tuning can be computed in the following way~\cite{JHEP09-2010,PhenomenologyBLH}

\begin{eqnarray} \label{fine-tuning}
\Psi=\frac{| \delta m^{2}_{1} |}{\lambda_0 v^{2} \cos^{2} \beta}, \  \delta m^{2}_{1} = -\frac{27 f^2}{8 \pi^{2} } \frac{ y^{2}_{1} y^{2}_{2} y^{2}_{3}  }{y^{2}_{2}-y^{2}_{3} }\, \text{log} \left( \frac{y^{2}_{1} + y^{2}_{2}}{y^{2}_{1} + y^{2}_{3}} \right).
\end{eqnarray}

\noindent  If $\Psi\sim 1$, this indicates no fine-tuning in the model. On the other hand, the top-quark Yukawa
coupling is determined by

\begin{eqnarray}
y_{t}=\frac{m_{t}}{v \sin \beta}=\frac{3 y_1 y_2 y_3}{\sqrt{\left(y^{2}_{1}+ y^{2}_{2} \right) \left(y^{2}_{1} + y^{2}_{3}  \right)   }},
\end{eqnarray}

\noindent where $m_{t}=172.76$ GeV is the top-quark mass, thus finding $y_{t}=0.74$. With this fixed value of $y_t$,
we can randomize perturbative values for the $y_{1,2,3}$ parameters. To obtain a numerical estimate of the fine-tuning
and an upper limit for the  $f$ scale where the new physics does not significantly require fine-tuning, we choose the following
values
\begin{itemize}
\item Scenario $y_2 > y_3$, $y_1=0.61$, $y_2=0.84$ and $y_3=0.35$~\cite{Cruz-Albaro:2022kty,Cruz-Albaro:2022lks}.
\item Scenario $y_2 < y_3$, $y_1=0.61$, $y_2=0.35$ and $y_3=0.84$~\cite{Cruz-Albaro:2022kty,Cruz-Albaro:2022lks}.
\end{itemize}

\noindent In these scenarios, we will analyze the AMDM and AWMDM of the top-quark in the context of the BLHM.
\vspace{0.1cm}

\noindent \textbf{ The pseudoscalar mass $A_{0}$}: This parameter is fixed around 1000 GeV.
Our choice is consistent with the current search results for new scalar bosons~\cite{ATLAS:2020gxx}. Data recorded by the ATLAS experiment at the LHC, corresponding to an integrated luminosity of 139 $\text{fb}^{-1}$ from proton-proton collisions at center-of-mass energy 13 TeV, were used to search for a heavy Higgs boson $A_{0}$, decaying into $ZH$, where $H$ denotes another heavy Higgs boson with mass $m_{H}>125$ GeV.
\vspace{0.1cm}

\noindent   \textbf{ The scalar mass $\eta_{0}$}: In the BLHM scenario, the free parameters $m_{4, 5, 6}$~\cite{JHEP09-2010} are
introduced to break all the axial symmetries in the Higgs potential, giving positive masses to all scalars. Specifically,
the $\eta_{0}$ scalar receives a mass equal to $m_{4}=m_{\eta_{0}}=100$ GeV, according to the BLHM, and the restriction
$m_{4}\gtrsim 10$ GeV must be considered~\cite{JHEP09-2010}.
\vspace{0.1cm}

\noindent  \textbf{ The ratio of the VEVs of the two Higgs doublets, $\tan \beta$}: Several theoretical constraints can be applied to this parameter, primarily due to perturbativity requirements. Two constraints limit the value of $\tan \beta$, the first of which is the requirement that $\lambda_{0}< 4 \pi$, leading to an upper bound according to Eq.~(\ref{cotabeta}). A lower bound also exits and is set by examining the radiatively induced contributions to $ m_{1} $ and $ m_{2} $ in the model, which suggests that $\tan \beta>1$~\cite{JHEP09-2010}

\begin{eqnarray}\label{cotabeta}
 1 < \text{tan}\ \beta < \sqrt{ \frac{2+2 \sqrt{\big(1-\frac{m^{2}_{h_0} }{m^{2}_{A_0}} \big) \big(1-\frac{m^{2}_{h_0} }{4 \pi v^{2}}\big) } }{ \frac{m^{2}_{h_0}}{m^{2}_{A_0}} \big(1+ \frac{m^{2}_{A_0}- m^{2}_{h_0}}{4 \pi v^{2}}  \big) } -1 }.
\end{eqnarray}

\noindent
From this inequality, we can find the range of allowed values for the parameter $\tan \beta$. In particular, for $m_{A_{0}}=1000$ GeV, it is obtained that $1 < \tan \beta < 10.45$.
\vspace{0.1cm}

\noindent \textbf{ The mixing angle $\theta_{g}$}:
The gauge couplings $g_{A}$ and  $g_{B}$, associated with the $SU(2)_{LA}$ and $SU(2)_{LB}$ gauge bosons,
can be parametrized in a more phenomenological fashion in terms of a mixing angle $\theta_{g}$ and the  $SU(2)_{L}$ gauge coupling:
$\tan \theta_{g}=g_{A}/g_{B}$  and $g=g_{A} g_{B}/\sqrt{g^{2}_{A}+ g^{2}_{B}} $. For simplicity, it is assumed that $\tan \theta_{g}=1$~\cite{PhenomenologyBLH},
which implies that the gauge couplings  $g_{A}$ and  $g_{B}$ are equal. The $g_{A,B}$ values are generated using the restriction $g=0.6525$.
\vspace{0.1cm}

\noindent \textbf{ Symmetry breaking scale $f$}:
The BLHM features a global $  SO(6)_{A}\times SO(6)_{B} $ symmetry that is broken to a diagonal $ SO(6)_{V} $ at a scale $f\sim \mathcal{O} $(TeV) when a nonlinear sigma field, $\sum$, develop a VEV. Bounds on the $f$ scale arise when $\tan \beta$ limits, fine-tuning constraints on the heavy quark masses, and experimental restrictions from producing heavy quarks are considered. Refs.~\cite{Godfrey:2012tf} and ~\cite{Kalyniak}  establish that $f\in (700,3000)$ GeV.
\vspace{0.1cm}

\noindent \textbf{ Symmetry breaking scale $F$}:
A second global symmetry of the $SU(2)_{C}\times SU(2)_{D}$ form is also present in the BLHM and is broken to a diagonal $SU(2)$ at a scale $F>f$ when a second nonlinear sigma field, $\Delta$, develops a VEV.
Due to the characteristics of the BLHM, the energy scale $F$ acquires sufficiently large values compared to the $f$ scale. The purpose is to ensure that the new gauge bosons are much heavier than the exotic quarks. In this way, $F\in [3000,6000]$ GeV~\cite{JHEP09-2010,Kalyniak}.

To predict the estimates of the AMDM and AWMDM of the top-quark, in Table~\ref{parametervalues}, we summarize the values used for the parameters involved in our analysis.

\begin{table}[H]
\caption{Values assigned to the free parameters involved in our numerical analysis in the context of the BLHM.
\label{parametervalues}}
\centering
\begin{tabular}{|c | c|}
\hline
$ \textbf{Parameter} $  & \hspace{1.5cm}    $ \textbf{Value} $  \\
\hline
\hline
$ m_{A_{0}}  $  & \hspace{1.5cm}  $ 1000\  \text{GeV} $  \\
\hline
 $m_{\eta^{0}}  $  & \hspace{1.5cm}   $ 100\  \text{GeV} $  \\
\hline
$ \tan \beta $  & \hspace{1.5cm}   $ 3 $  \\
\hline
$ \tan \theta_{g} $  & \hspace{1.5cm}   $ 1 $  \\
\hline
$ f $  & \hspace{1.5cm}  $ [1000, 3000]\   \text{GeV} $  \\
\hline
$ F $  & \hspace{1.5cm}  $ [3000, 6000] \ \text{GeV} $  \\
\hline
\end{tabular}
\end{table}

\subsection{AMDM of the top-quark at the $ y_2 > y_3 $ scenario}
\label{AMDMk1}

In order to obtain the AMDM of the top-quark, we implement the amplitudes represented by Eqs.~(\ref{amplitudesSV1})-(\ref{amplitudesSV5}) in Mathematica's Feyncalc environment~\cite{Mertig:1990an}, and the integrals involved are solved using the Passarino-Veltman (PV) reduction scheme.  In this environment, Gordon's identity is used to eliminate the terms proportional to $(p_1 + p_2)^{\mu}$. After this, the form factors proportional to the $\sigma^{\mu \nu} q_\nu$ tensor are extracted which in turn generates the AMDM $a_{t}$ of the top-quark through the relation $a_{t}= -2 m_t F^{\gamma}_{M}(q^{2 })$.  At this stage, the analytical expressions obtained for $a_t$ involve the PV scalar functions, and these functions are evaluated using the Package-X package~\cite{Patel:2015tea}. In this paper, we do not report the analytical expressions for $a_{t}$ because they are extensive expressions therefore, we report only our numerical results. It is also important to mention that all the contributions we have found for $a_{t}$ are free of ultraviolet divergences.

\begin{figure}[H]
\subfloat[]{\includegraphics[width=8.0cm]{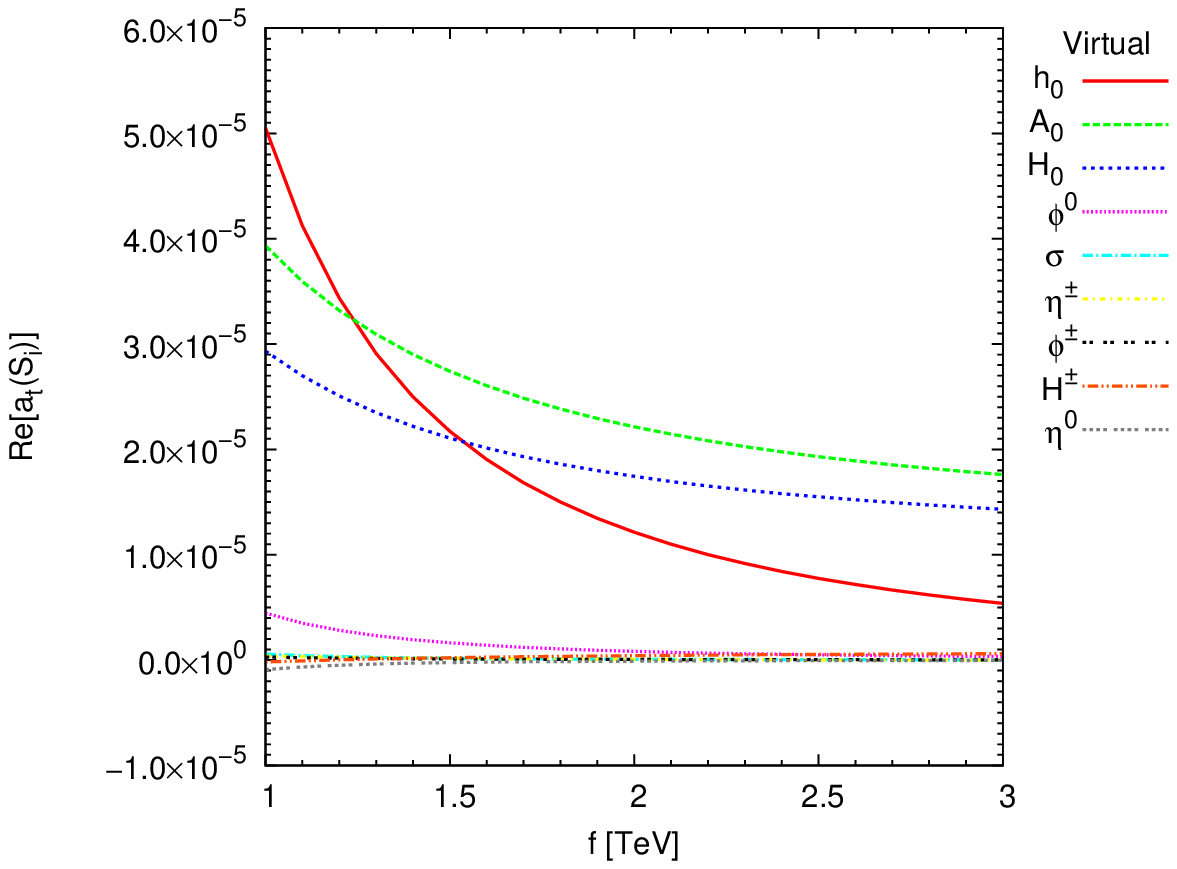}}
\subfloat[]{\includegraphics[width=8.0cm]{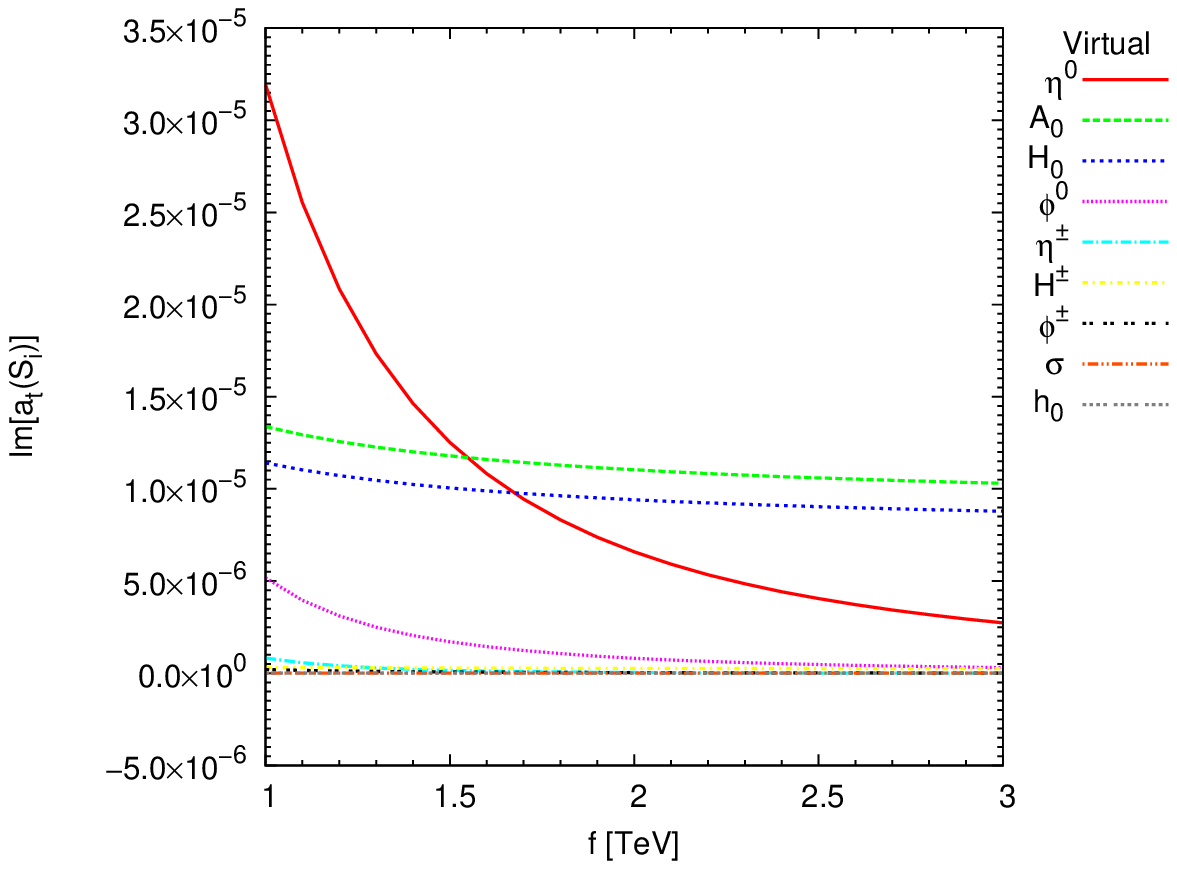}}
\caption{ \label{Si2} Individual scalar contributions to $a_{t}$. The plots are obtained with the fixed value of  $F=4000\, \text{GeV}$. The values provided in Table~\ref{parametervalues} are used for the remaining model parameters.
 a) Re($a_{t}$). b) Im($a_{t}$).}
\end{figure}

As mentioned above, the Feynman diagrams (see Fig.~\ref{dipolo}) contributing to the AMDM of the top are classified for analysis according to the contributions generated by the particles circulating in the $\gamma \bar{t}t$ vertex loop: scalar contributions $[a_{t}]^{S_i}$, vector contributions $[a_{t}]^{V_i} $ and scalar-vector contributions $   [a_{t}]^{S_i - V_i} $.
In this way, we discuss the partial contributions generated by the scalar bosons. Fig.~\ref{Si2} shows the behavior of the real and imaginary parts of $a_t$ as a function of the scale of the new physics, $f$. This energy scale takes values from 1000 to 3000 GeV. We find from Fig.~\ref{Si2}(a) that the largest positive contributions are generated by the scalars $h_0$ and $A_0$: $\text{Re}[a_t (h_0)]= [5.05, 3.43] \times 10^{-5} $ for the interval $f=[1000,1200]$ GeV and $\text{Re}[a_t (A_0)]= ( 3.43, 1.93] \times 10^{-5} $ for $f=(1200,3000]$ GeV. In contrast, the $\eta^{0}$ scalar provides the smallest negative contributions, $\text{Re}[a_t (\eta^{0})]= -[8.97\times 10^{-7},5.74 \times 10^{-8}] $.
Concerning Fig.~\ref{Si2}(b), we can appreciate here that the largest positive curves are generated by the $\eta^{0}$ and $A_0$ scalars:   $\text{Im}[a_t (\eta^{0})]= [3.19, 1.18] \times 10^{-5} $ for $f=[1000, 1540]$ GeV and  $\text{Im}[a_t (A_0)]= (1.18, 1.03] \times 10^{-5} $ for the interval $f=( 1540, 3000]$ GeV. The most suppressed curve is given by the scalar $h_0$, which provides a zero imaginary contribution.

We now turn to the analysis of the vector and scalar-vector contributions. These contributions originate from the Feynman diagrams represented by Figs.~\ref{dipolo}(b)-\ref{dipolo}(c) and~\ref{dipolo}(d)-\ref{dipolo}(e), respectively. Before discussing the contributions mentioned above, it is pertinent to recall that Table~\ref{parametervalues} values are assigned to the free parameters involved in our numerical analysis. The parameter $\tan \theta$ is of particular interest in this paper since, as in Refs.~\cite{Cruz-Albaro:2022kty,Cruz-Albaro:2022lks,PhenomenologyBLH}, it is assumed for simplicity that $\tan \theta =1$ which implies that $\sin \theta_g$ and $\cos \theta_g$ are equal (see Table~\ref{2boson-scalar} in Appendix~\ref{reglasFeynman}). This condition leads to the non-contribution of certain triangular diagrams in Figs.~\ref{dipolo}(d) and~\ref{dipolo}(e); this is due to the cancellation of the Feynman rules for the $A_\gamma W^{+} \phi^{-}$ and $A_\gamma W^{+} \eta^{-}$ vertices, as well as their respective Hermitian conjugates.

\begin{figure}[H]
\subfloat[]{\includegraphics[width=8.0cm]{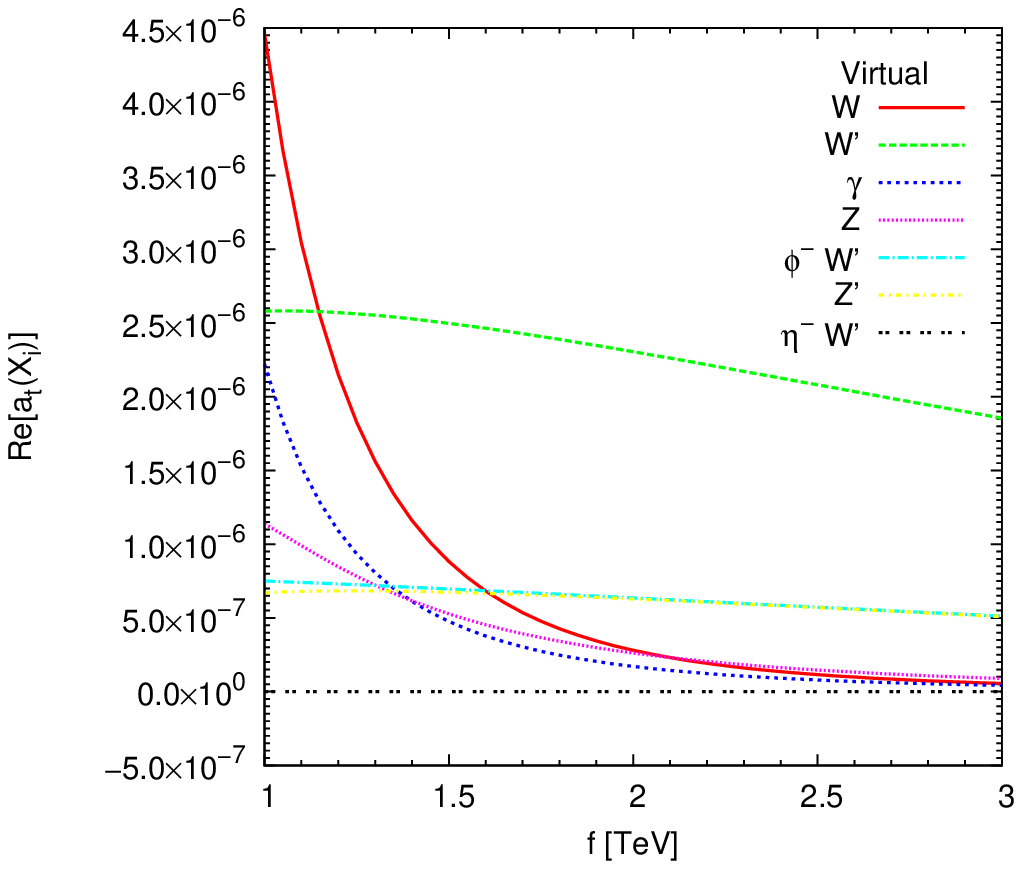}}
\subfloat[]{\includegraphics[width=8.0cm]{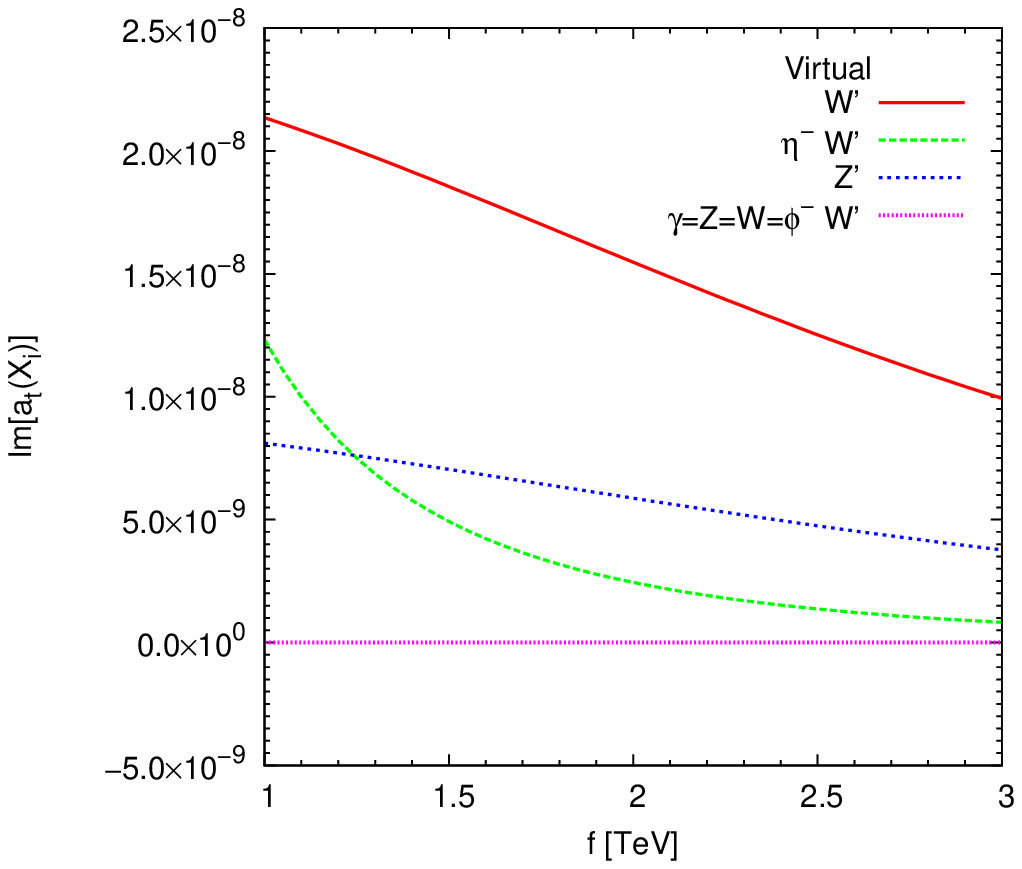}}
\caption{ \label{Vi2} Individual vector and scalar-vector contributions to $a_{t}$. The plots are obtained with the fixed value of  $F=4000\, \text{GeV}$. The values provided in Table~\ref{parametervalues} are used for the remaining model parameters
 a) Re($a_{t}$). b) Im($a_{t}$).}
\end{figure}

In the left plot of Fig.~\ref{Vi2} we can appreciate the individual contributions to the real part of the AMDM of the top-quark generated by the vector and scalar-vector bosons. In this figure, the main partial contributions are given by the vector bosons $W$ and $W'$, $\text{Re}[a_t (W)]= [4.46, 2.64] \times 10^{-6} $ for the interval $f=[1000,1140]$ GeV and $\text{Re}[a_t (W')]= ( 2.64, 2.08] \times 10^{-6} $ for $f=(1140,3000]$ GeV. While the suppressed partial contribution is provided by $\eta^{-} W'$, $\text{Re}[a_t (\eta^{-} W')]=0 $.
The right plot of Fig.~\ref{Vi2} shows the curves contributing to the imaginary part of $a_t$. In this case, the dominant curve is generated by the gauge boson $W'$ with $\text{Im}[a_t ( W')]=[2.13, 1.25]\times 10^{-8} $ for the established analysis range of the energy scale $f$.
Regarding the suppressed curves that generate $ \gamma $, $Z$, $ W $ and $ \phi^{-} W' $, all of them contribute with $\text{Im}[a_t ( \gamma )]=\text{Im}[a_t (Z )]=\text{Im}[a_t ( W )]=\text{Im}[a_t ( \phi^{-}W' )]= 0$.
For a further appreciation of the numerical contributions generated by the particles that circulate in the $\gamma \bar{t}t$ vertex loop, we provide in Tables~\ref{parcial1-2}-\ref{parcial3-2} of Appendix~\ref{numericasparciales} all partial contributions to $a_t$.

\begin{figure}[H]
\subfloat[]{\includegraphics[width=8.0cm]{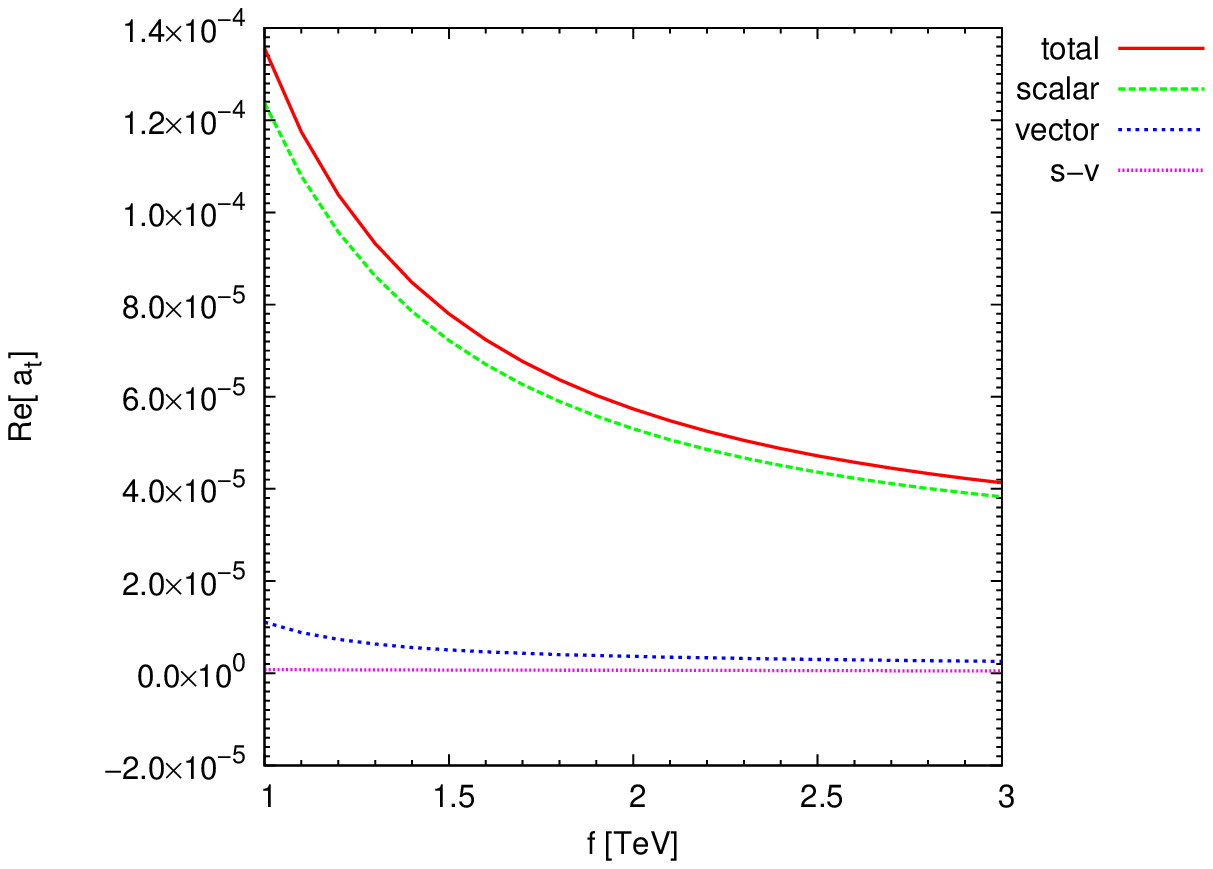}}
\subfloat[]{\includegraphics[width=8.0cm]{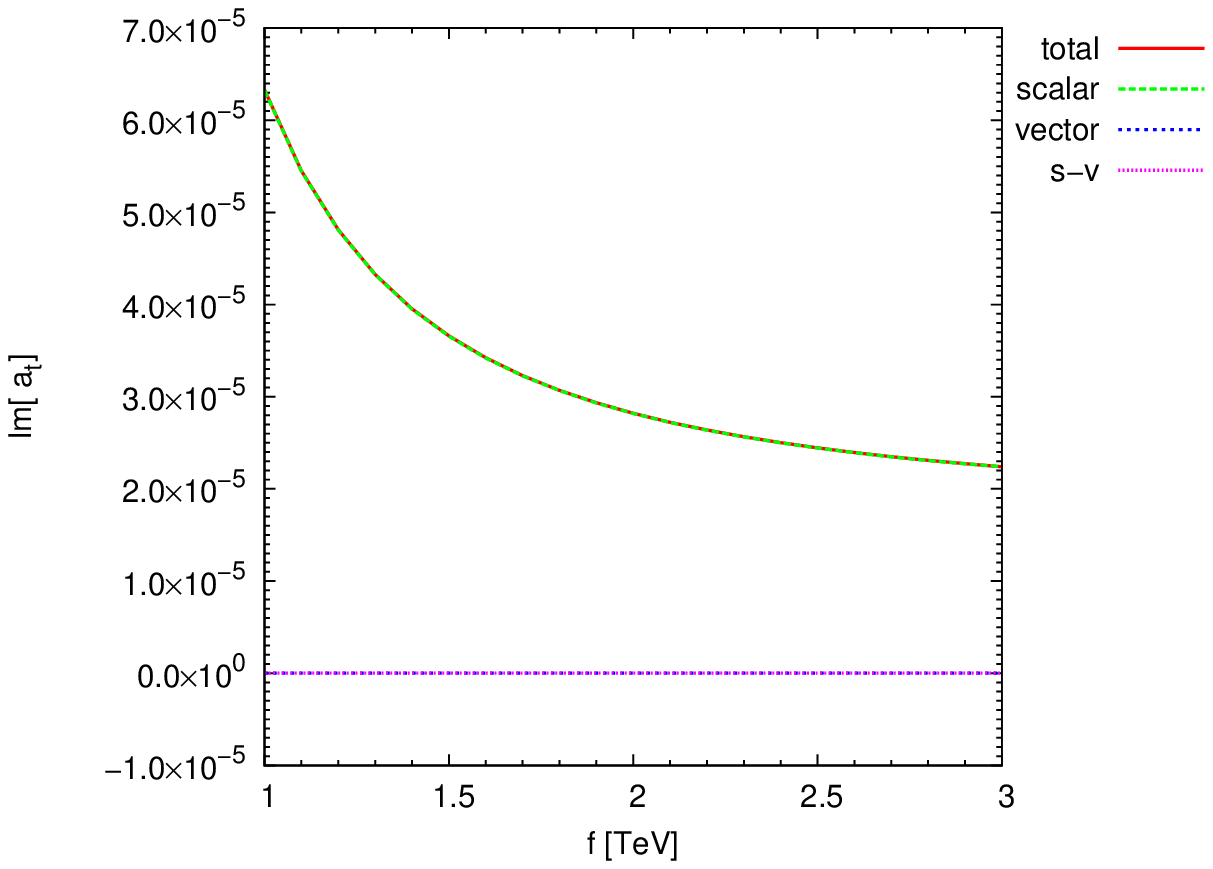}}
\caption{ \label{FS2} Scalar, vector, scalar-vector (s-v) and total contributions  to $a_{t}$. The plots are obtained with the fixed value of  $F=4000\, \text{GeV}$. The values provided in Table~\ref{parametervalues} are used for the remaining model parameters.
 a) Re($a_{t}$). b) Im($a_{t}$).}
\end{figure}

We now present a study of the curves that result from the sum of the partial contributions presented above. In Fig.~\ref{FS2}(a), we can see that the curves $\text{Re}[a_{t}(\text{scalar} )] $,  $\text{Re}[a_{t}(\text{vector})] $ and $\text{Re}[a_{t}(\text{s-v} )] $ contribute positively to the $\text{Re}[a_{t}(\text{total})] $ curve. However, the magnitude of the scalar contribution dominates concerning the vector and scalar-vector contributions. That is, $\text{Re}[a_{t}(\text{total})] $ receives significant contributions from the scalar sector compared to the other two sectors. The numerical estimates obtained for the four sectors are $\text{Re}[a_{t}(\text{scalar} )] = [1.24 \times 10^{-4}, 3.83 \times 10^{-5}]$,  $\text{Re}[a_{t}(\text{vector})] = [1.11 \times 10^{-5}, 2.55 \times 10^{-6} ]$, $\text{Re}[a_{t}(\text{s-v} )]= [7.51, 5.12] \times 10^{-7}$ and $\text{Re}[a_{t}(\text{total})] = [1.36 \times 10^{-4}, 4.13 \times 10^{-5}]$ for $f \in [1000,3000]$ GeV.
Concerning Fig.~\ref{FS2}(b), again, the scalar sector contributes more to the total contribution. In this case, both contributions acquire values of the same magnitude. The remaining sectors interfere very weakly with the total contribution. The numerical data for these contributions are $\text{Im}[a_{t}(\text{scalar} )] = [6.32, 2.24 ]\times 10^{-5}$,  $\text{Im}[a_{t}(\text{vector})] = [2.88, 1.37] \times 10^{-8} $, $\text{Im}[a_{t}(\text{s-v} )]= [1.23 \times 10^{-8},8.21 \times 10^{-10}]$ and $\text{Im}[a_{t}(\text{total})] = [6.33,  2.24] \times 10^{-5}$.
The values of the total contribution to $a_t$ are listed in Table~\ref{GEta1002}.

\begin{figure}[H]
\subfloat[]{\includegraphics[width=8.0cm]{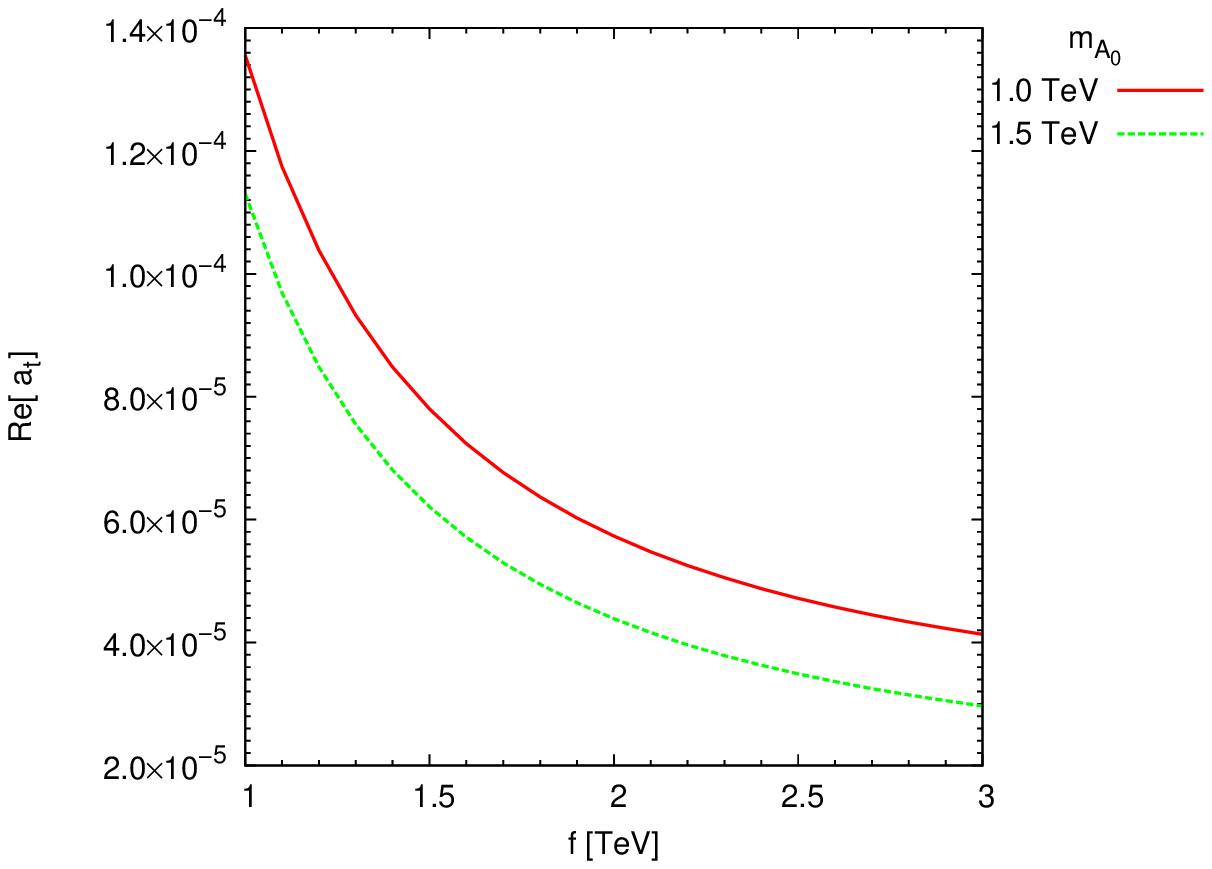}}
\subfloat[]{\includegraphics[width=8.0cm]{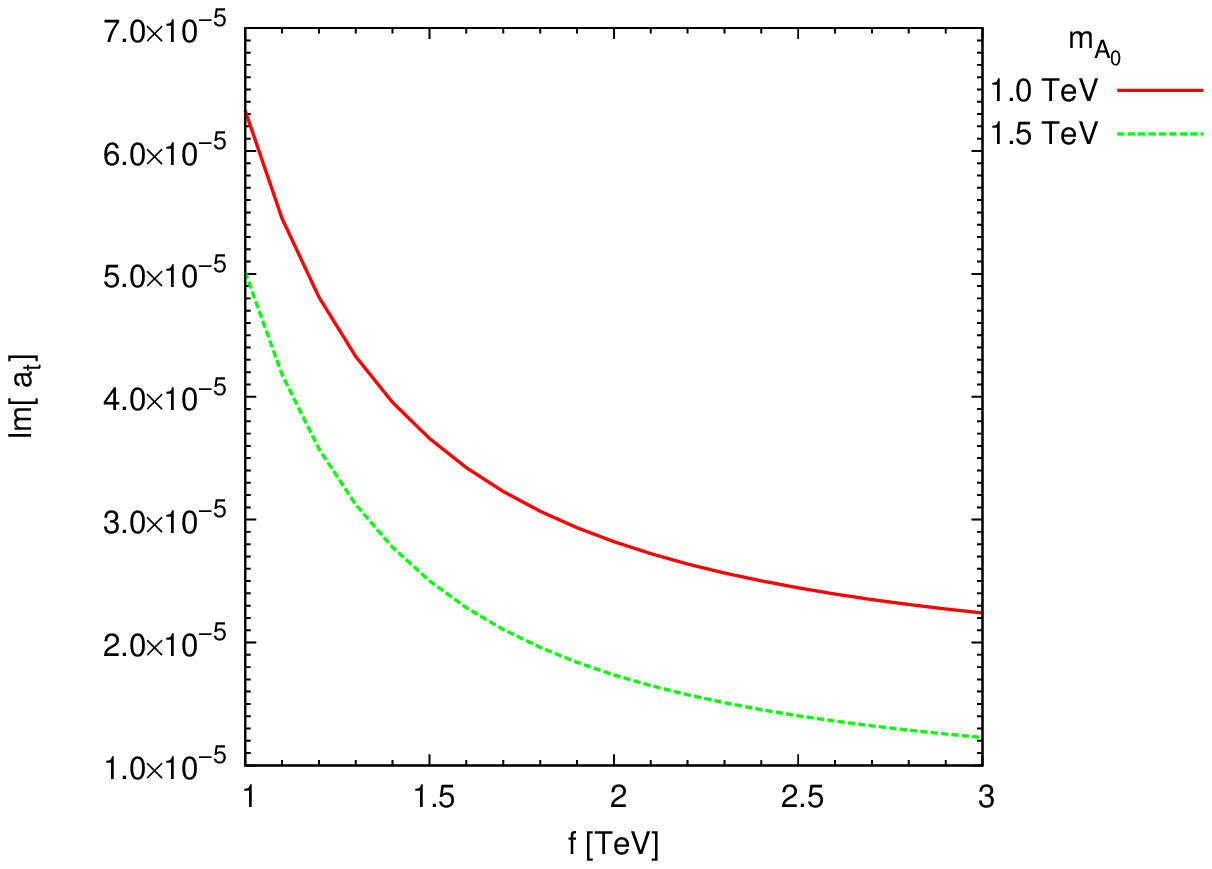}}
\caption{ \label{FmA2} Total contribution to $a_{t}$  for different values of the mass of $A_0$. The plots are obtained with the fixed value of  $F=4000\, \text{GeV}$. The values provided in Table~\ref{parametervalues} are used for the remaining model parameters.
a) Re($a_{t}$). b) Im($a_{t}$).}
\end{figure}

\begin{figure}[H]
\subfloat[]{\includegraphics[width=8.0cm]{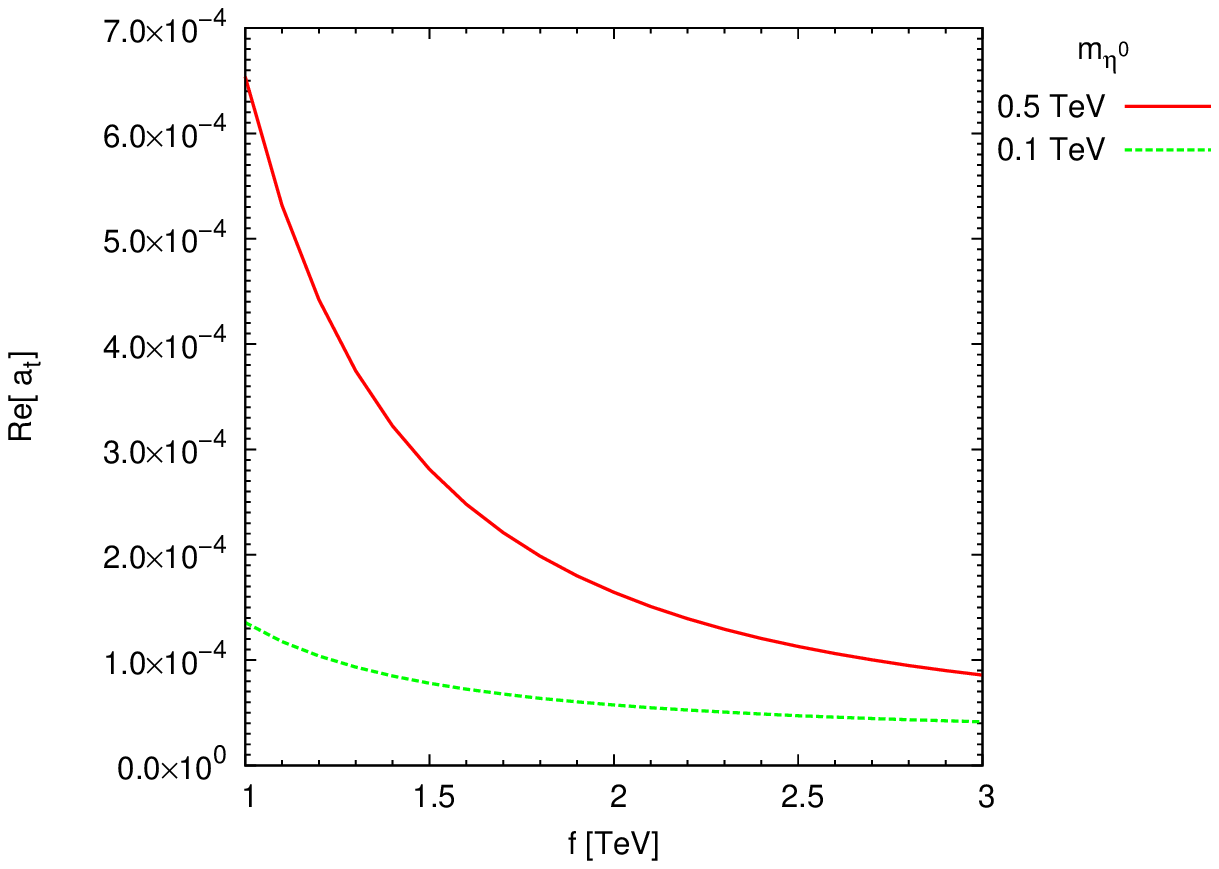}}
\subfloat[]{\includegraphics[width=8.0cm]{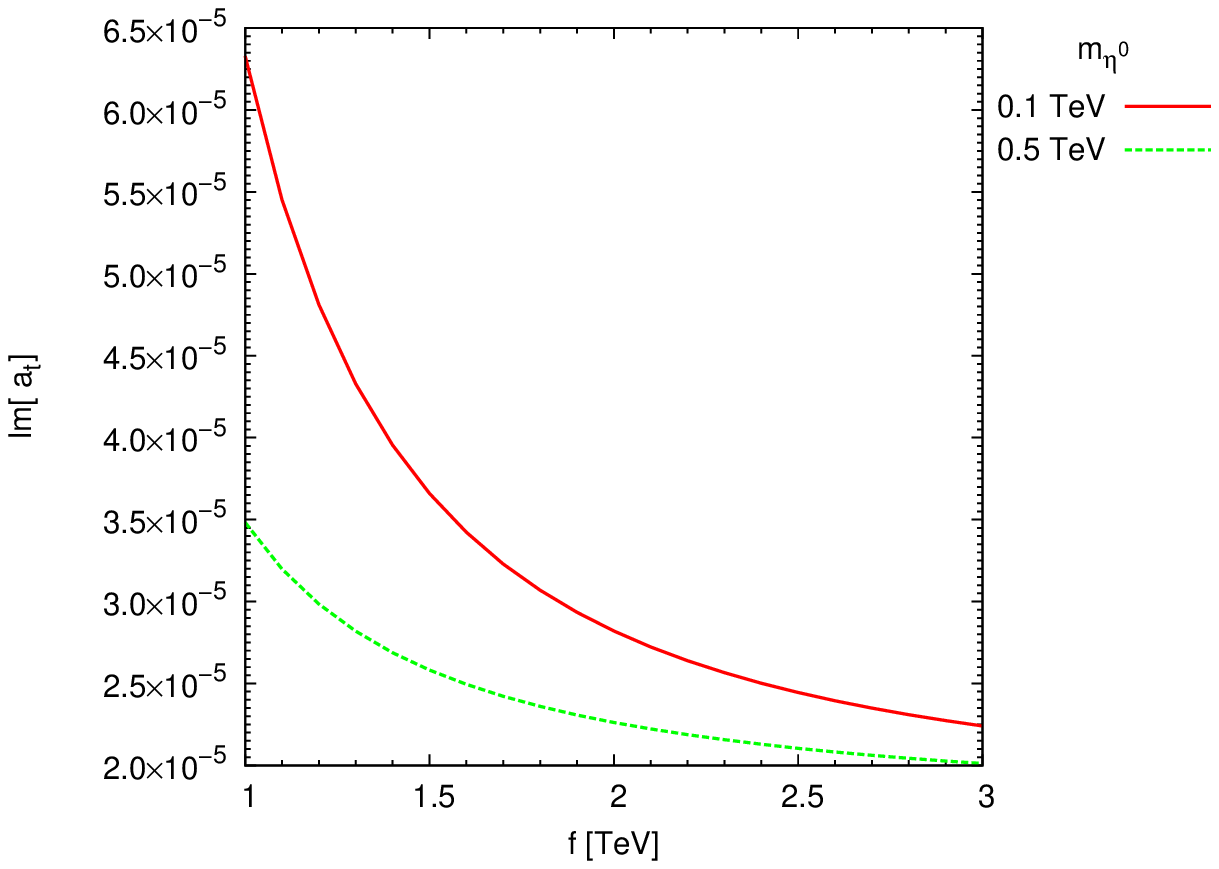}}
\caption{ \label{Fmeta2} Total contribution to $a_{t}$  for different values of the mass of $\eta^{0}$. The plots are obtained with the fixed value of  $F=4000\, \text{GeV}$. The values provided in Table~\ref{parametervalues} are used for the remaining model parameters. a) Re($a_{t}$). b) Im($a_{t}$).}
\end{figure}

The masses of the neutral scalars $A_0$ and $\eta^{0} $ are also input parameters of the BLHM. Thus, we propose to measure the sensitivity of the AMDM $a_t$ of the top-quark by choosing certain values for the masses of the $A_0$ and $\eta^{0} $ states. For this purpose, we pick the values of $m_{A_0}=1000, 1500$ GeV, while for the scalar $\eta^{0}$ we choose $m_{\eta^{0}}=100, 500$ GeV.  With these values for $m_{A_0}$ and $m_{\eta^{0}}$ we generate Figs.~\ref{FmA2} and~\ref{Fmeta2} which show the dependence of $a_t$ vs. $f$.  From Figs.~\ref{FmA2}(a) and~\ref{FmA2}(b) we observe that slightly larger curves are obtained when $m_{A_0}=1000$ GeV compared to $m_{A_0}=1500$ GeV. The numerical data show that $\text{Re}[a_{t}(\text{m}_{A_0}=1000\, \text{GeV})] \sim \text{Re}[a_{t}(\text{m}_{A_0}=1500\, \text{GeV})] \sim 10^{-4}-10^{-5}$ and $\text{Im}[a_{t}(\text{m}_{A_0}=1000\, \text{GeV})] \sim \text{Im}[a_{t}(\text{m}_{A_0}=1500\, \text{GeV})] \sim 10^{-5}$.
We now describe Figs.~\ref{Fmeta2}(a) and~\ref{Fmeta2}(b), in these plots, we can see that for the real part of $a_t$ the largest contributions arise when $m_{\eta^{0}}=500$  GeV: $\text{Re}[a_{t}(\text{m}_{\eta^{0}}=100\, \text{GeV})] \sim \text{Re}[a_{t}(\text{m}_{\eta^{0}}=500\, \text{GeV})] \sim 10^{-4}-10^{-5}$. With respect to the imaginary part of $a_t$ the opposite happens, when $m_{\eta^{0}}=100$ GeV slightly larger curves arise than $m_{\eta^{0}}=500$ GeV: $\text{Im}[a_{t}(\text{m}_{\eta^{0}}=100\, \text{GeV})] \sim \text{Im}[a_{t}(\text{m}_{\eta^{0}}=500\, \text{GeV})] \sim 10^{-5}$.
We can infer from the figures and numerical estimations that $a_t $ depends strongly on the energy scale $ f $. This affirmation is greatly appreciated in the figures generated for the cases analyzed. The curves suffer appreciable changes in the range of analysis of the scale $ f $. For the different choices in the input values of $m_{A_0}$ and $m_{\eta^{0}}$, these parameters are also related to $ a_t $ since the curves drawn show a pretty clear difference between one curve and the other.
For the different cases discussed here, in Tables~\ref{GEta1002}-\ref{mA15002}, we provide the numerical values that $a_t$ acquires in the range of values of the energy scale $f$.

\begin{figure}[H]
\subfloat[]{\includegraphics[width=8.0cm]{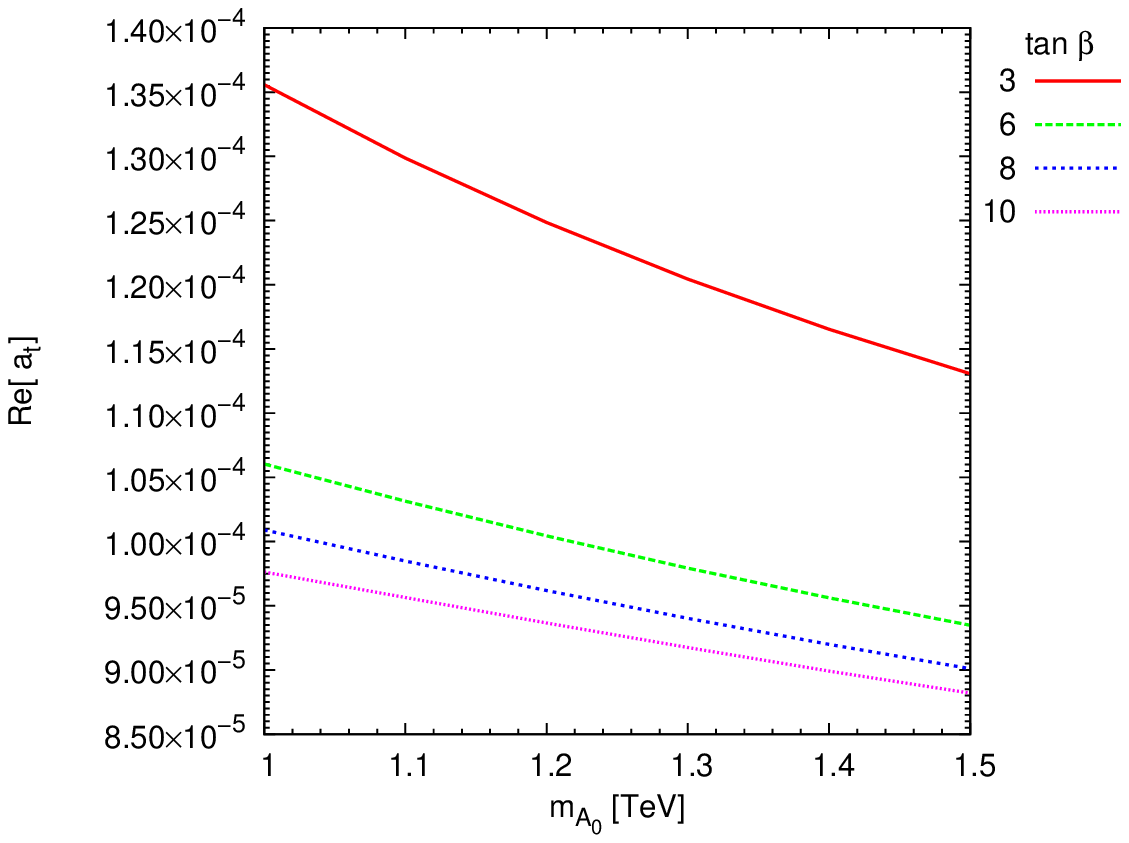}}
\subfloat[]{\includegraphics[width=8.0cm]{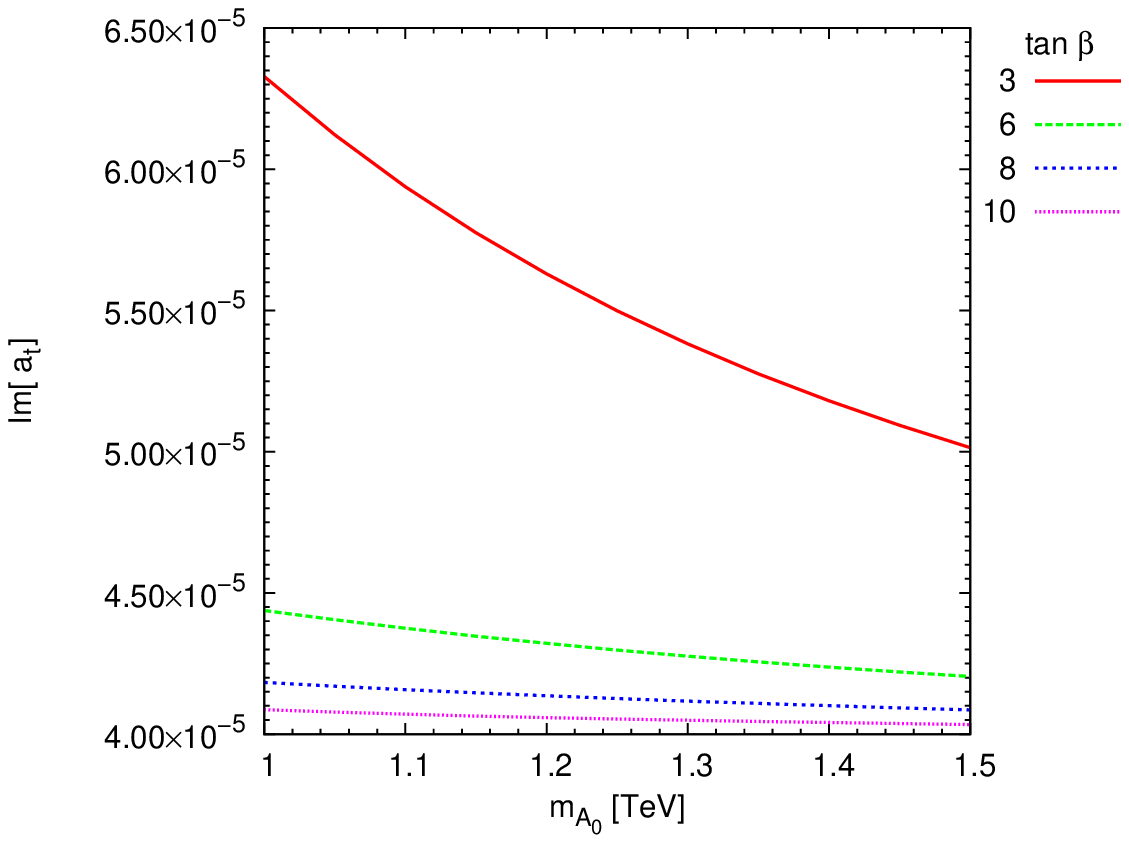}}
\caption{ \label{Ftan2} Total contribution to $a_{t}$ for different values of $\tan \beta$. The plots are obtained with the fixed values of  $f=1000\, \text{GeV}$ and $F=4000\, \text{GeV}$. The values provided in Table~\ref{parametervalues} are used for the remaining model parameters.
a) Re($a_{t}$). b) Im($a_{t}$).}
\end{figure}

We investigate the phenomenological details associated with the increase of $m_{A_0}$ vs. $\text{Re}[a_t]$ and $m_{A_0}$ vs. $\text{Im}[a_t]$. According to Eq.~(\ref{cotabeta}),  the mass of the pseudoscalar $A_0$ is an important input parameter because it sets the range of allowed values for $\tan \beta$, i.e., for $m_{A_0}=1000$ GeV and $m_{A_0}=1500$ GeV the respective intervals of values generated for the $\tan \beta$ parameter are $\tan \beta \in (1, 10.45)$ and $\tan \beta \in (1, 11.99)$.
 In order to evaluate the numerical contributions of $a_t$, we propose to vary $m_{A_0}$ from 1000 GeV to 1500 GeV, and to generate different curves. We take some fixed values for $\tan \beta$ in the allowed value space, $\tan \beta=3,6,8,10$.
 In Fig.~\ref{Ftan2}(a) we appreciate the dependence of $\text{Re}[a_t]$ on $m_{A_0}$, we observe that the main signal is reached for $\tan \beta=3$, while the lowest signal is obtained when $\tan \beta=10$, $ \text{Re}[a_t]=[1.36,1.13]\times 10^{-4} $ and $ \text{Re}[a_t]=[9.76,8.82]\times 10^{-5} $.
 For the remaining curves, $ \text{Re}[a_t]\sim  10^{-4}-10^{-5} $ when $m_{A_0} \in [1000,1500]$ GeV.
  Next we analyze Fig.~\ref{Ftan2}(b) which shows the dependence of $\text{Im}[a_t]$ on $m_{A_0}$, again, the highest signal is obtained for $\tan \beta=3$ and the lowest signal when $\tan \beta=10$.  In this case, all generated curves acquire values of the same order of magnitude, $ 10^{-5} $. According to our numerical predictions,  $\text{Re}[a_t]$  and $\text{Im}[a_t]$  show a clear dependence on the $m_{A_0}$ and $\tan \beta$ parameters. Although $\text{Im}[a_t]$ presents a slightly smaller sensitivity to changes in the $\tan \beta$ parameter since the numerical values obtained by $\text{Im}[a_t]$ are of the same order of magnitude for different choices in the values of $\tan \beta$.

\begin{figure}[H]
\subfloat[]{\includegraphics[width=8.0cm]{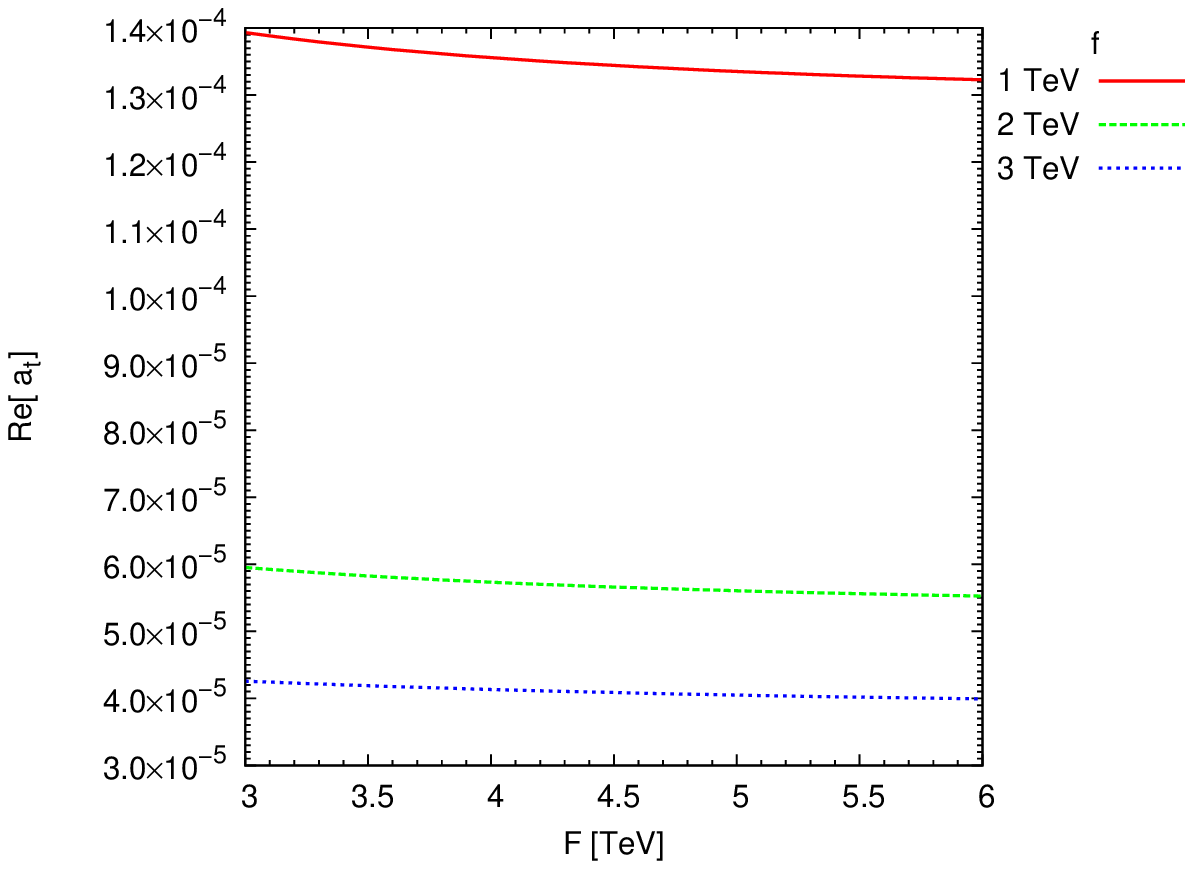}}
\subfloat[]{\includegraphics[width=8.0cm]{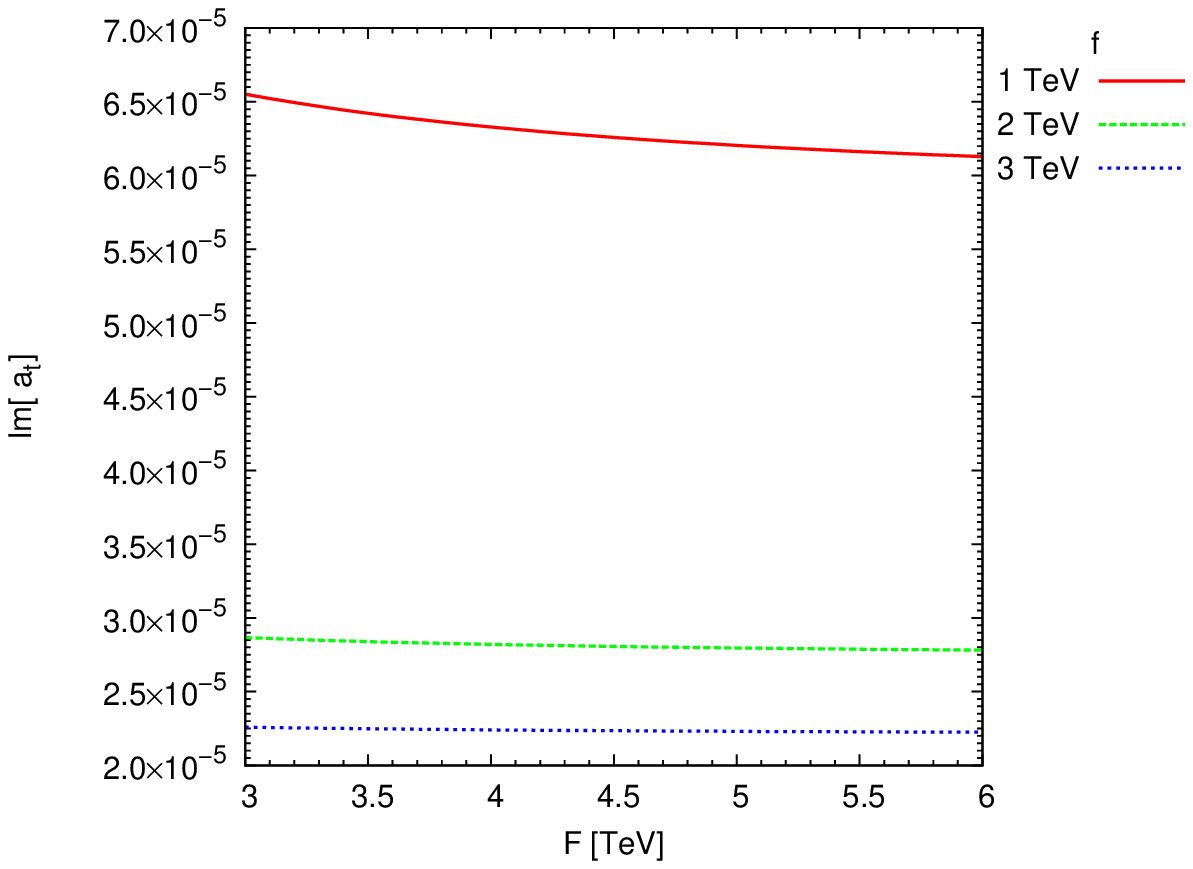}}
\caption{ \label{Ffi2} Total contribution to $a_{t}$ for different values of the energy scale $f$.  The values provided in Table~\ref{parametervalues} are used for the remaining model parameters. a) Re($a_{t}$). b) Im($a_{t}$).}
\end{figure}

The sensitivity of the AMDM $a_t$ of the top-quark has so far been measured by varying the first symmetry-breaking scale $f$. However, due to the characteristics of the BLHM, it also depends on the second symmetry-breaking scale $F$.
Therefore, it is pertinent to examine the dependence of $a_t$ on the $F$ energy scale at the interval set above, $F=[3000, 6000]$ GeV.
In Fig.~\ref{Ffi2}, we show the level of sensitivity exhibited by $a_t$ when varying the $F$ scale while keeping the $f$ scale fixed. The values assigned to $f$ are 1000, 2000, and 3000 GeV.
For the three different energy scales, we find from Fig.~\ref{Ffi2}(a) that the numerical predictions on $\text{Re}[a_t] $ are $\text{Re}[a_t]=[1.39, 1.32]\times 10^{-4} $, $\text{Re}[a_t]=[5.95, 5.52]\times 10^{-5} $ and $\text{Re}[a_t]=[4.26, 3.99] \times 10^{-5} $, respectively.
While the imaginary part of $a_t$, shown in Fig.~\ref{Ffi2}(b), we find $\text{Im}[a_t]=[6.55, 6.13]\times 10^{-5} $, $\text{Im}[a_t]=[2.87, 2.78]\times 10^{-5} $ and $\text{Im}[a_t]=[2.26, 2.22] \times 10^{-5} $.
Based on the numerical contributions found, the dominant contributions arise for small values of the energy scale $f$, particularly when $f=1000$ GeV. It is important to note that $a_t$ depends slightly on the $F$ scale; this is because the curves generated for both $\text{Re}[a_t]$ and $\text{Im}[a_t]$ do not suffer such drastic changes over the range of the $F$ scale.

\begin{figure}[H]
\subfloat[]{\includegraphics[width=8.0cm]{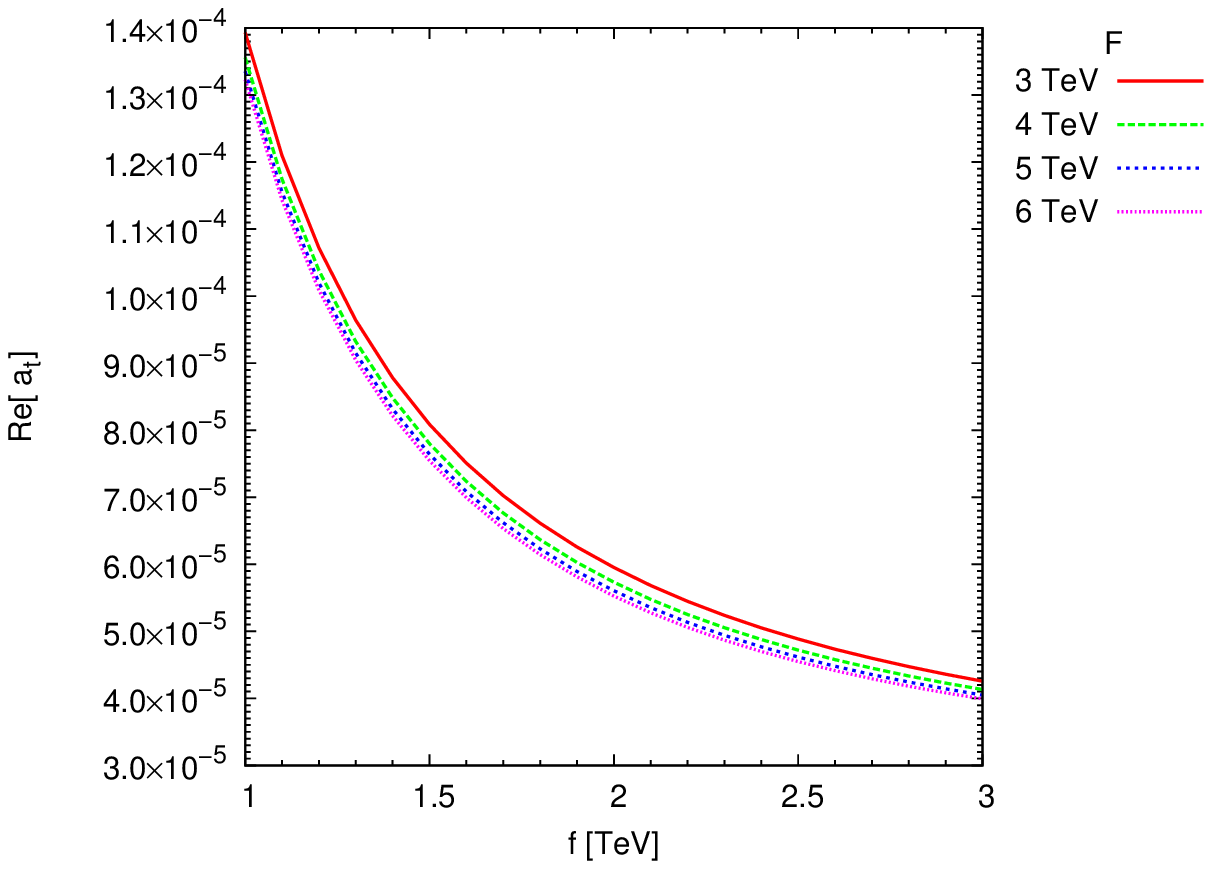}}
\subfloat[]{\includegraphics[width=8.0cm]{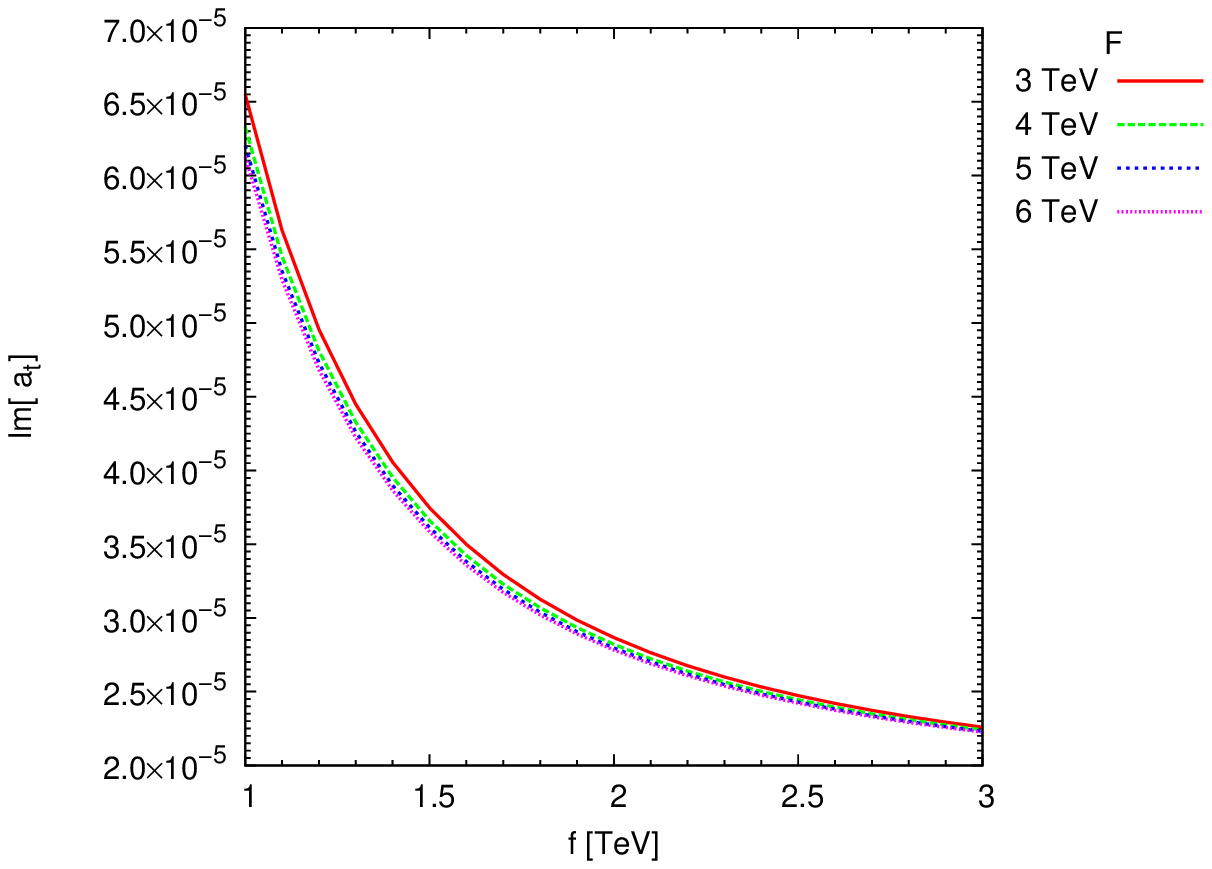}}
\caption{ \label{FFi2} Total contribution to $a_{t}$ for different values of the energy scale $F$.  The values provided in Table~\ref{parametervalues} are used for the remaining model parameters. a) Re($a_{t}$). b) Im($a_{t}$).}
\end{figure}

Finally, we discuss the dependence of $a_t$ on the energy scale $f$ while keeping the second energy scale $F$ fixed. The values assigned to $F$ are 3000, 4000, 5000, and 6000 GeV, and with these fixed values, we generate the different curves shown in Figs.~\ref{FFi2}(a) and~\ref{FFi2}(b).
On the left and right side of Fig.~\ref{FFi2} we find that the slightly larger contributions are obtained for $F=3000$ GeV:  $\text{Re}[a_t]=[1.39\times 10^{-4}, 4.88 \times 10^{-5} ]$ and $\text{Im}[a_t]=[6.55, 2.47]\times 10^{-5} $.
For the rest of the curves, we find that the numerical contributions are $\text{Re}[a_t] \sim 10^{-4}-10^{-5}$ and $\text{Im}[a_t] \sim 10^{-5}$.
For this case, we observe that $a_t$ depends strongly on the energy scale $f$ since the curves generated for the real and imaginary part of the AMDM of the top-quark suffer appreciable changes in the analysis interval set for the scale $f$, $a_t$ has a decrease of at most one order of magnitude as $f$ increases up to 3000 GeV.
In contrast, $a_t$ shows a minor sensitivity to changes in the $F$ energy scale as long as it is between the allowed intervals.

\begin{table}[H]
\caption{Expected sensitivity limits on the $a_t$ in the context of the BLHM with $\sqrt{q^{2}}=500\ \text{GeV}$,
\ $\text{m}_{A_0}=1000\ \text{GeV}$, $\text{m}_{\eta_0}=100\ \text{GeV} $, $F=4000\ \text{GeV}$ and $f=1, 1.5, 2, 2.5, 3\ \text{TeV}$
are represented. All new contributions are considered, scalar bosons, vector bosons, scalar-vector, and heavy quarks.
\label{GEta1002}}
\centering
\begin{tabular}{|c|c|}
\hline
\hline
\multicolumn{2}{|c|}{$\sqrt{q^{2}}=500\ \text{GeV}$,\ $\text{m}_{A_0}=1000\ \text{GeV}$, $\text{m}_{\eta_0}=\bf{ 100}\ \text{GeV} $,\ $F=4000\ \text{GeV}$}\\
\hline
 $f\ [\text{TeV}]$  & $(a_{t})^{\rm total} $  \\
\hline
\hline
$1_\cdot0$  & $ 1.35 \times 10^{-4} + 6.33  \times 10^{-5}\ i $  \\
\hline
$1_\cdot5 $  & $ 7.80 \times 10^{-5} + 3.66 \times 10^{-5}\ i$  \\
\hline
$2_\cdot0 $  & $ 5.73 \times 10^{-5} + 2.82 \times 10^{-5}\ i$  \\
\hline
$2_\cdot5 $  & $  4.72 \times 10^{-5} + 2.44  \times 10^{-5}\ i$  \\
\hline
$3_\cdot0 $  & $  4.13 \times 10^{-5} + 2.24 \times 10^{-5}\ i$  \\
\hline
\end{tabular}
\end{table}

\begin{table}[H]
\caption{Expected sensitivity limits on the $a_t$ in the context of the BLHM with $\sqrt{q^{2}}=500\ \text{GeV}$,
\ $\text{m}_{A_0}=1000\ \text{GeV}$, $\text{m}_{\eta_0}=500\ \text{GeV} $, $F=4000\ \text{GeV}$ and $f=1, 1.5, 2, 2.5, 3\ \text{TeV}$
are represented. All new contributions are considered, scalar bosons, vector bosons, scalar-vector, and heavy quarks.
\label{GEta5002}}
\centering
\begin{tabular}{|c|c|}
\hline
\hline
\multicolumn{2}{|c|}{$\sqrt{q^{2}}=500\ \text{GeV}$,\ $\text{m}_{A_0}=1000\ \text{GeV}$, $\text{m}_{\eta_0}=\bf{ 500}\ \text{GeV} $,\ $F=4000\ \text{GeV}$}\\
\hline
 $f\ [\text{TeV}]$  & $(a_{t})^{\rm total} $  \\
\hline
\hline
$1_\cdot0$  & $ 6.54 \times 10^{-4} + 3.48  \times 10^{-5}\ i $  \\
\hline
$1_\cdot5 $  & $ 2.81 \times 10^{-4} + 2.58 \times 10^{-5}\ i$  \\
\hline
$2_\cdot0 $  & $ 1.64 \times 10^{-4} + 2.26 \times 10^{-5}\ i$  \\
\hline
$2_\cdot5 $  & $ 1.13 \times 10^{-4} + 2.10 \times 10^{-5}\ i$  \\
\hline
$3_\cdot0 $  & $ 8.57 \times 10^{-5} + 2.01 \times 10^{-5}\ i$  \\
\hline
\end{tabular}
\end{table}

\begin{table}[H]
\caption{Expected sensitivity limits on the $a_t$ in the context of the BLHM with $\sqrt{q^{2}}=500\ \text{GeV}$,
\ $\text{m}_{A_0}=1 500\ \text{GeV}$, $\text{m}_{\eta_0}=100\ \text{GeV}$, $F=4000\ \text{GeV}$ and $f=1, 1.5, 2, 2.5, 3\ \text{TeV}$
are represented. All new contributions are considered, scalar bosons, vector bosons, scalar-vector, and heavy quarks.
\label{mA15002}}
\centering
\begin{tabular}{|c|c|}
\hline
\hline
\multicolumn{2}{|c|}{$\sqrt{q^{2}}=500\ \text{GeV}$,\ $\text{m}_{A_0}=\bf{1500}\ \text{GeV}$, $\text{m}_{\eta_0}=100\ \text{GeV}$,  $\ F=4000\ \text{GeV}$}\\
\hline
 $f\ [\text{TeV}]$  & $(a_{t})^{\rm total} $  \\
\hline
\hline
$1_\cdot0$  & $ 1.13 \times 10^{-4} + 5.01 \times 10^{-5}\ i$  \\
\hline
$1_\cdot5 $  & $  6.21 \times 10^{-5} + 2.50 \times 10^{-5}\ i$  \\
\hline
$2_\cdot0 $  & $ 4.39  \times 10^{-5}+ 1.74 \times 10^{-5}\ i$  \\
\hline
$2_\cdot5 $  & $ 3.49 \times 10^{-5} + 1.40  \times 10^{-5}\ i$  \\
\hline
$3_\cdot0 $  & $ 2.97  \times 10^{-5}+ 1.23 \times 10^{-5}\ i$  \\
\hline
\end{tabular}
\end{table}

\subsection{AWMDM of the top-quark at the  $ y_2 > y_3 $ scenario }

We now investigate the contribution of the new physics of the BLHM to the AWMDM $a^{W}_t$ of the top-quark. One difference concerning the AMDM of the top-quark is that it receives additional contributions due to the $Z$ and $Z'$ gauge bosons coupling with the Higgs bosons, $H_i \equiv h_0, H_0$ (see Fig.~\ref{dipoloweak}).
From the amplitudes given by Eqs.~(\ref{amplitudesSV1})-(\ref{amplitudesSV6}), in the same way as was done to find $a_t$, the form factors proportional to the $\sigma^{\mu \nu} q_\nu$ tensor that contains in code form $a^{W}_t$ are obtained.
It should be stressed that the results obtained for $a^{W}_t$ are also finite.

\begin{figure}[H]
\subfloat[]{\includegraphics[width=8.0cm]{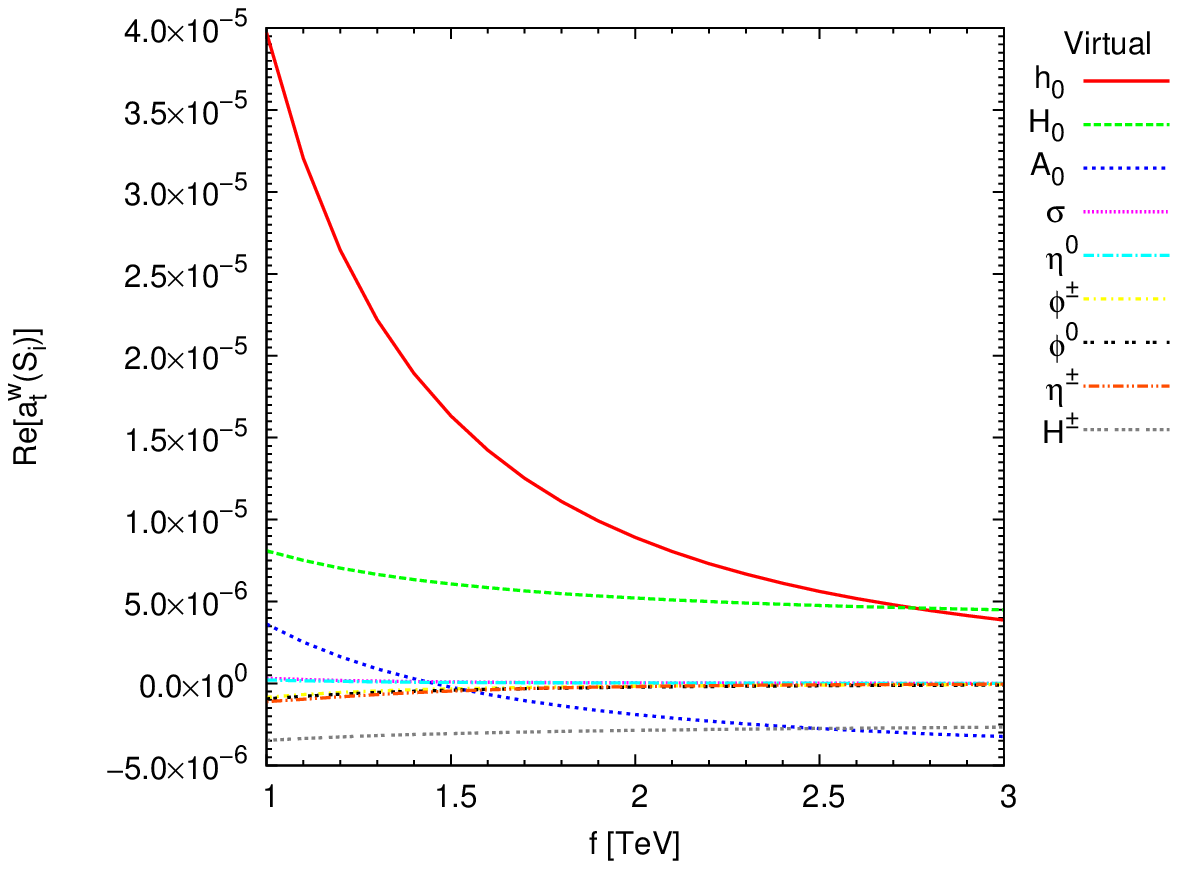}}
\subfloat[]{\includegraphics[width=8.0cm]{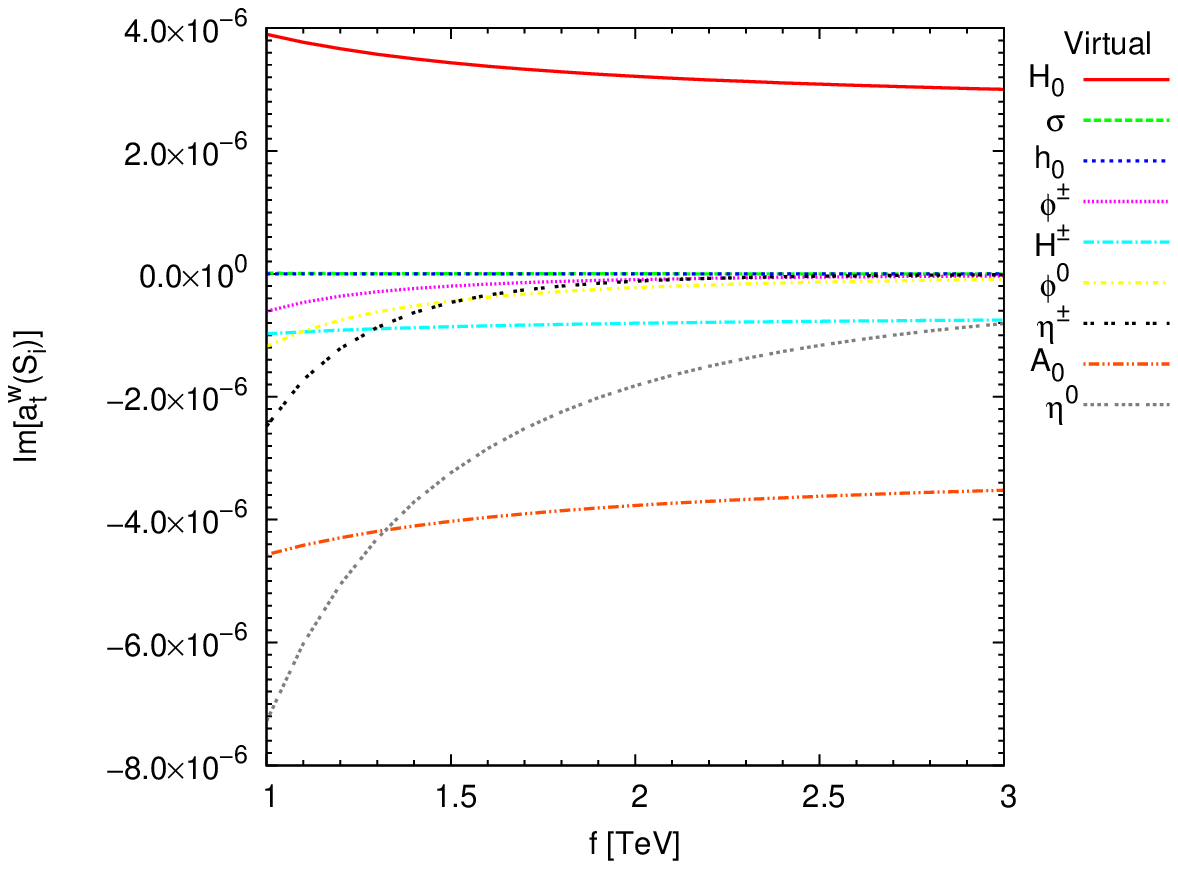}}
\caption{ \label{Si-z2}
 Individual scalar contributions to $a^{W}_{t}$.
The plots are obtained with the fixed value of  $F=4000\, \text{GeV}$. The values provided in Table~\ref{parametervalues} are used for the remaining model parameters.
a) Re($a^{W}_{t}$). b) Im($a^{W}_{t}$).
}
\end{figure}

We again begin by analyzing the magnitude of each partial contribution to the AWMDM of the top-quark. In this way, in Fig.~\ref{Si-z2}, we observe the individual contributions of the scalar bosons to $a^{W}_t$, which develops the real and imaginary parts.
From the left plot of Fig.~\ref{Si-z2}, it is appreciated that the dominant partial contribution arises when the mediator particle in the loop of the vertex $Z\bar{t}t$ is the scalar $h_0$ (SM Higgs boson), while the suppressed partial contribution is generated by the charged scalar $H^{\pm}$:  $\text{Re}[a^{W}_t (h_0)]= [3.97\times 10^{-5}, 3.87 \times 10^{-6}] $ and  $\text{Re}[a^{W}_t (H^{\pm})]= -[3.47, 2.67] \times 10^{-6} $.
Following the right plot of Fig.~\ref{Si-z2}, the main partial contribution is generated by the neutral scalar $H_0$ providing a numerical estimate of $\text{Im}[a^{W}_t (H_0)]= [3.90, 3.00] \times 10^{-6} $  in the analysis range for the new physical scale $f$, $f \in [1000,3000]$ GeV. Complementarily, the curve that represents the lower partial contribution is generated by the scalars $\eta^{0}$ and $A_0$: $\text{Im}[a^{W}_t (\eta^{0})]= - [7.28, 4.31] \times 10^{-6} $ for the interval $f=[1000, 1300]$ GeV  and $\text{Im}[a^{W}_t (A_0)]= - (4.31,3.52] \times 10^{-6} $ when $f=(1300,3000]$ GeV.

\begin{figure}[H]
\subfloat[]{\includegraphics[width=8.0cm]{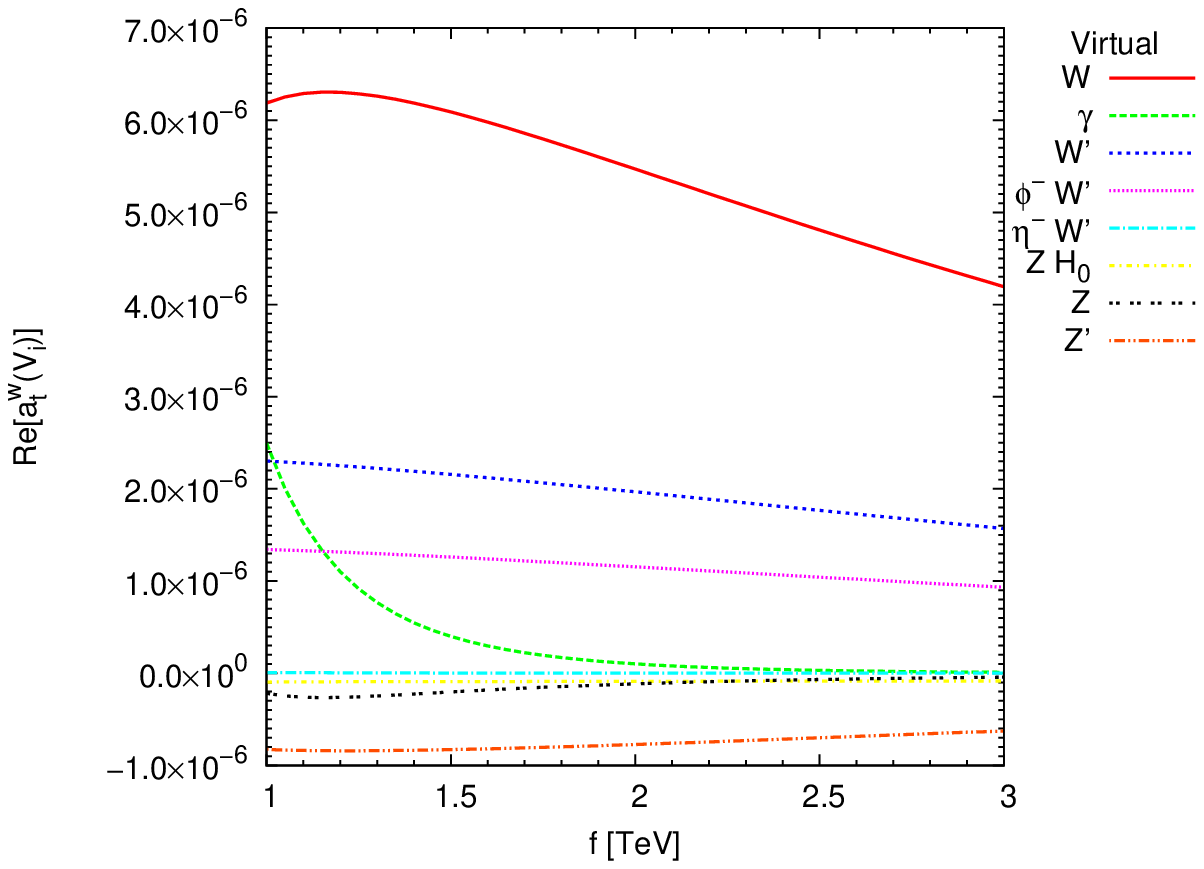}}
\subfloat[]{\includegraphics[width=8.0cm]{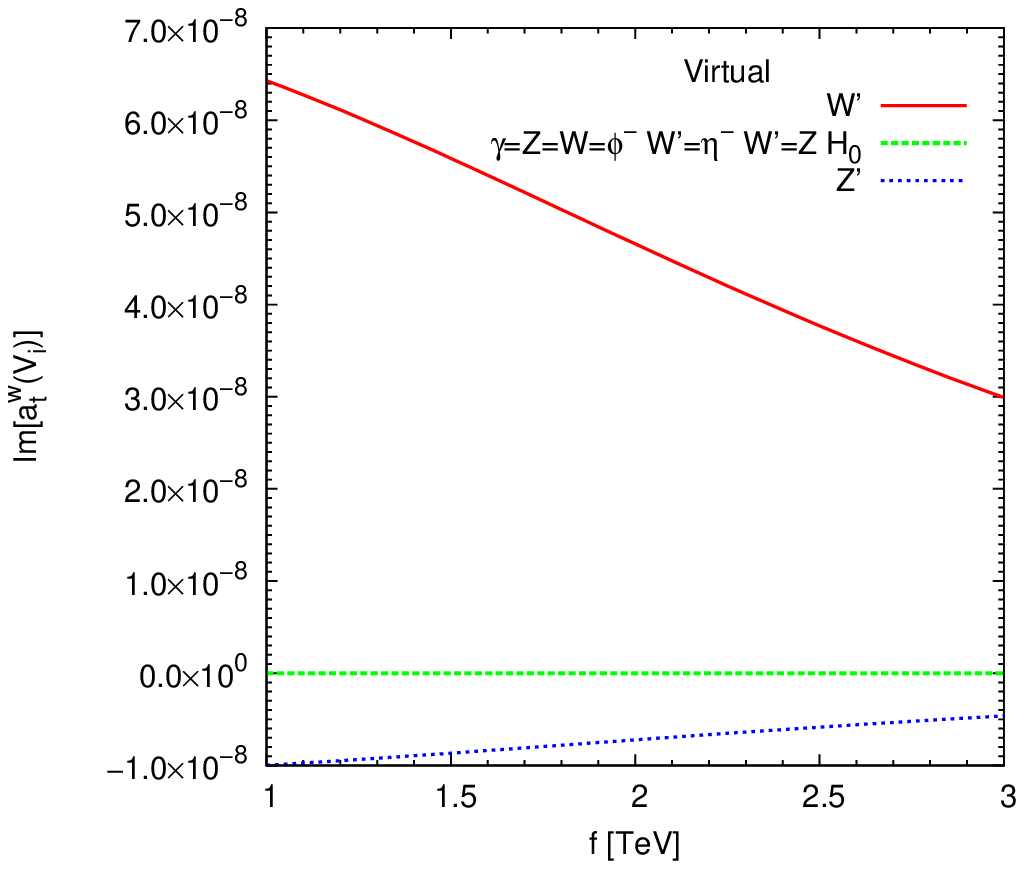}}
\caption{ \label{Vi-z2}
 Individual vector and scalar-vector contributions to $a^{W}_{t}$.
The plots are obtained with the fixed value of  $F=4000\, \text{GeV}$. The values provided in Table~\ref{parametervalues} are used for the remaining model parameters.
a) Re($a^{W}_{t}$). b) Im($a^{W}_{t}$).
}
\end{figure}

We performed the numerical estimates of the vector and scalar-vector contributions to $a^{W}_{t}$. We show in Fig.~\ref{Vi-z2} the behavior of the different curves representing the partial contributions to the AWMDM of the top-quark. However, as in Subsection~\ref{AMDMk1}, in the $\tan \theta_g=1$ scenario which implies that $\sin \theta_g = \cos \theta_g$ leads to the cancellation of some Feynman diagrams in Fig.~\ref{dipoloweak}, this is because the Feynman rules for vertices $ Z W^{+}\phi^{-} $, $ Z \phi^{+} W^{-}$, $ Z W^{+}\eta^{-} $, $ Z \eta^{+} W^{-}$ and $Z_\mu Z'_\nu H_i$ are made zero by the condition mentioned above (see Table~\ref{2boson-scalar} in Appendix~\ref{reglasFeynman}).
The remaining Feynman diagrams provide non-zero partial contributions to $a^{W}_{t}$, these contributions are illustrated in Figs.~\ref{Vi-z2}(a) and~\ref{Vi-z2}(b).
In these figures, we can see that the dominant curves are generated by the gauge bosons $W$ and $W'$, that is,  $\text{Re}[a^{W}_t (W)]= [6.19, 4.19] \times 10^{-6} $ and $\text{Im}[a^{W}_t (W')]= [6.43, 2.99] \times 10^{-8} $ in the allowed $f$ interval.
On the other hand, the suppressed curves are reached when the mediator particle at the corresponding vertex is the gauge boson $Z'$,  $\text{Re}[a^{W}_t (Z')]= - [8.27, 6.26] \times 10^{-7} $ and $\text{Im}[a^{W}_t (Z')]= -[9.98, 4.64] \times 10^{-9} $.
We provide in Tables~\ref{CN1-z2}-\ref{CN3-z2} of Appendix~\ref{numericasparciales} all partial contributions of the different particles that contribute to $a^{W}_t$.

\begin{figure}[H]
\subfloat[]{\includegraphics[width=8.0cm]{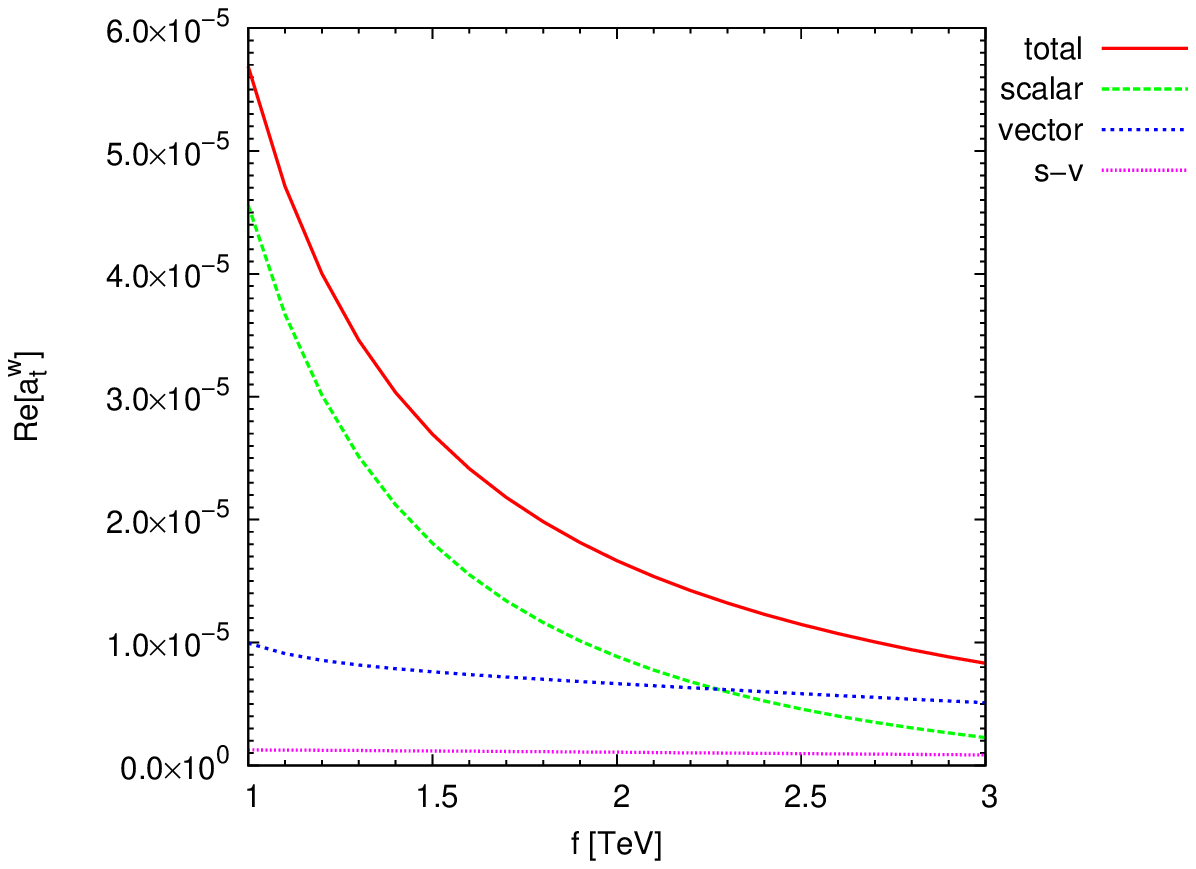}}
\subfloat[]{\includegraphics[width=8.0cm]{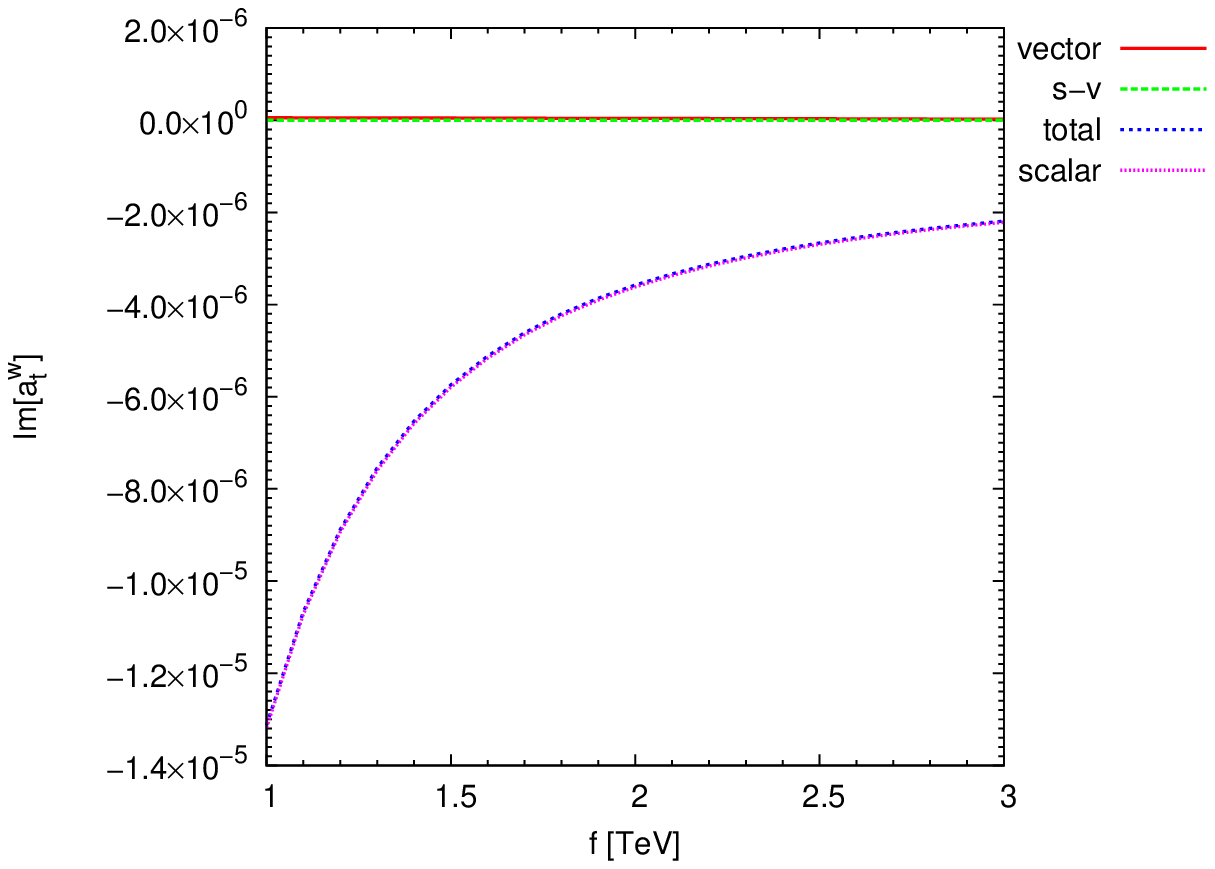}}
\caption{ \label{total-z2} Scalar, vector, scalar-vector and total contributions  to $a^{W}_{t}$. The plots are obtained with the fixed value of  $F=4000\, \text{GeV}$. The values provided in Table~\ref{parametervalues} are used for the remaining model parameters.
 a) Re($a^{W}_{t}$). b) Im($a^{W}_{t}$).}
\end{figure}

We analyze the total contributions of each sector to the AWMDM of the top: the scalar, vector, and scalar-vector sectors. The sum of these three sectors represents the total contribution of the new physics to $a^{W}_t$. In Fig.~\ref{total-z2}, we can observe the behavior of the real and imaginary parts of $a^{W}_t$ for the different sectors when the scale $ f $ takes values from 1000 to 3000 GeV.
The real part of $a^{W}_t$ (see Fig.~\ref{total-z2}(a)) shows that the most significant contribution to the total contribution comes from the scalar contribution followed by the vector contribution. While the scalar-vector sector contributes in a smaller proportion than the other sectors. The numerical estimates found for these sectors are $\text{Re}[a^{W}_t (\text{scalar})]=  [4.56 \times 10^{-5}, 2.26 \times 10^{-6}] $, $\text{Re}[a^{W}_t (\text{vector})]=  [9.94, 5.10] \times 10^{-6} $, $\text{Re}[a^{W}_t (\text{s-v})]= [1.25 \times 10^{-6}, 8.49 \times 10^{-7}] $ and $\text{Re}[a^{W}_t (\text{total})]=  [5.69 \times 10^{-5}, 8.30 \times 10^{-6}] $.
For the imaginary part of $a^{W}_t$ (see Fig.~\ref{total-z2}(b)), we find that the only positive contribution comes from the vector sector, while the contribution that results from the scalars generates a negative contribution. As for the scalar-vector sector, it provides a zero contribution to $\text{Im}(a^{W}_t)$ over the whole allowed interval of the energy scale $f$.
More specifically, we provide the numerical values found for the imaginary part of $a^{W}_t$:
$\text{Im}[a^{W}_t (\text{scalar})]= - [1.32 \times 10^{-5}, 2.21 \times 10^{-6}] $, $\text{Im}[a^{W}_t (\text{vector})]=  [5.43, 2.53] \times 10^{-8} $, $\text{Im}[a^{W}_t (\text{s-v})]= 0 $ and $\text{Im}[a^{W}_t (\text{total})]=  -[1.31 \times 10^{-5}, 2.19 \times 10^{-6}] $.
Considering the numerical magnitudes of each sector, $|\text{Im}[a^{W}_t (\text{scalar})|$ contributes in greater proportion to $|\text{Im}[a^{W}_t (\text{total})]|$.
For more information on the sensitivity of $a^{W}_t$ in the space of allowed values for the $ f $ scale, Table~\ref{ZmEta100-z2} presents some of the values that $a^{W}_t$ acquires.

\begin{figure}[H]
\subfloat[]{\includegraphics[width=8.0cm]{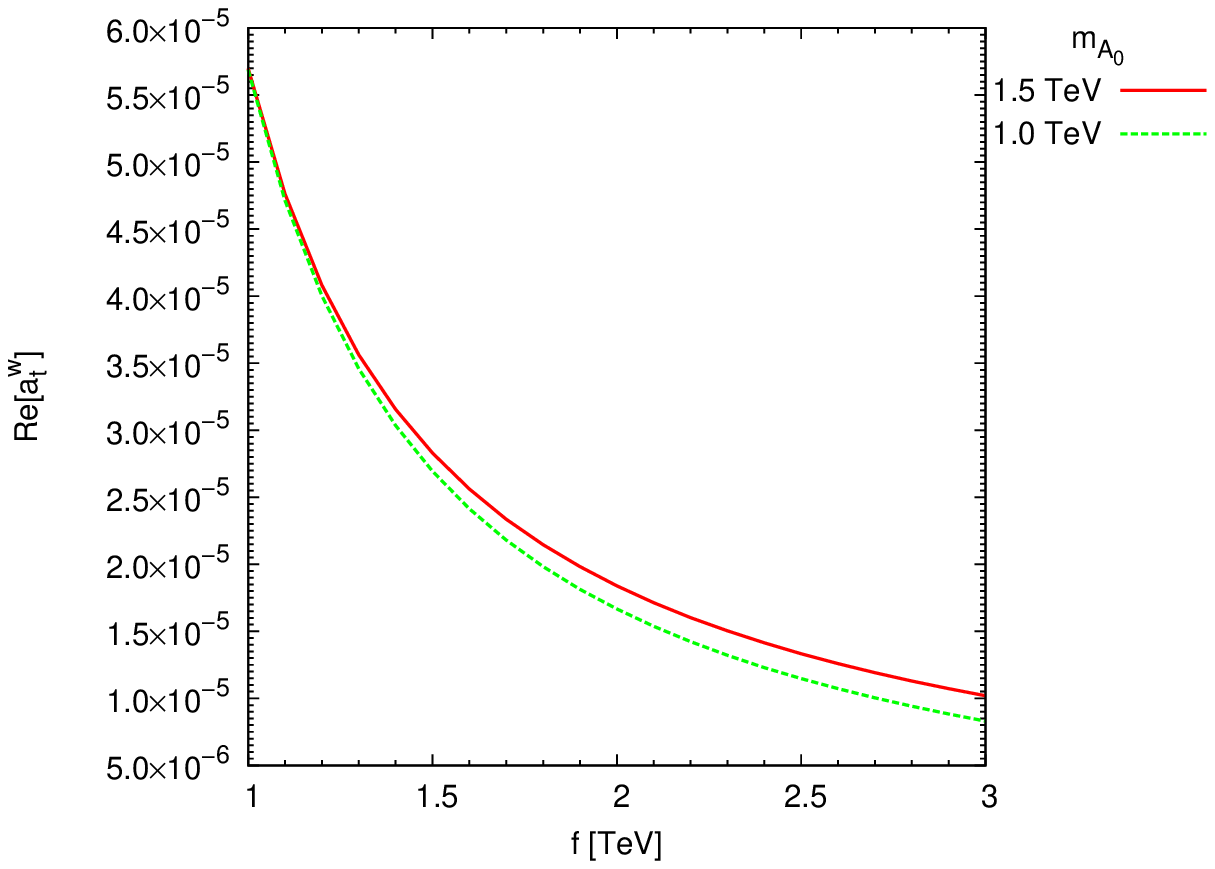}}
\subfloat[]{\includegraphics[width=8.0cm]{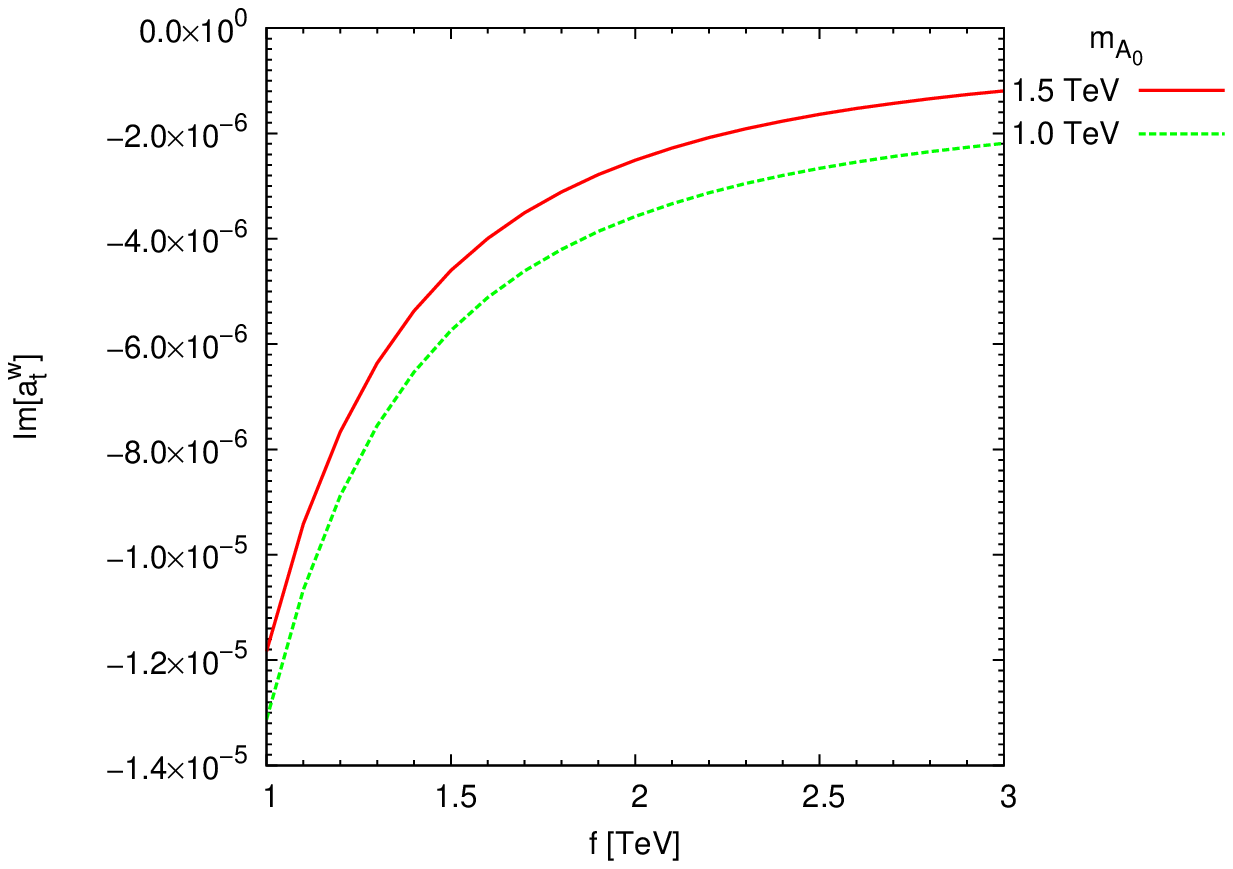}}
\caption{ \label{FmA-z2} Total contribution to $a^{W}_{t}$  for different values of the mass of $A_0$. The plots are obtained with the fixed value of  $F=4000\, \text{GeV}$. The values provided in Table~\ref{parametervalues} are used for the remaining model parameters.
a) Re($a^{W}_{t}$). b) Im($a^{W}_{t}$).}
\end{figure}

Fig.~\ref{FmA-z2} shows the different curves that represent the total contribution to $a^{W}_{t}$ as a function of the energy scale $ f $. The curves are generated for two different values of the mass of the scalar $ A_0 $, $ m_{A_0}=1000 $ GeV and  $m_{A_0}=1500 $ GeV.
We first examine the real part of $a^{W}_{t}$. From Fig.~\ref{FmA-z2}(a), we can see that the two curves seem to follow the same trajectory. However, the difference between one curve and the other is slightly noticeable around the energy scale $ f=2500 $ GeV; this is also seen in the numerical magnitudes:
$\text{Re}[a^{W}_{t}(\text{m}_{A_0}=1000\, \text{GeV})]=[5.69 \times 10^{-5}, 8.30 \times 10^{-6}] $ and  $ \text{Re}[a^{W}_{t}(\text{m}_{A_0}=1500\, \text{GeV})] =[ 5.70, 1.02]\times 10^{-5}$.
Continuing with the imaginary part of $a^{W}_{t}$, Fig.~\ref{FmA-z2}(b) shows again two curves whose differences between one and the other are more appreciable in the whole analysis interval of scale $ f $. In this case the following numerical values of $\text{Im}[a^{W}_{t}]$ are obtained:
$\text{Im}[a^{W}_{t}(\text{m}_{A_0}=1000\ \text{GeV})]=- [1.31 \times 10^{-5}, 2.19 \times 10^{-6}] $ and  $ \text{Im}[a^{W}_{t}(\text{m}_{A_0}=1500\, \text{GeV})] =-[1.18 \times 10^{-5}, 1.19 \times 10^{-6}]$.
The AWMDM $a^{W}_{t}$ of the top-quark is sensitive to changes in $m_{A_0}$ since the generated curves show an appreciable difference between one and the other, an order of magnitude at most.

\begin{figure}[H]
\subfloat[]{\includegraphics[width=8.0cm]{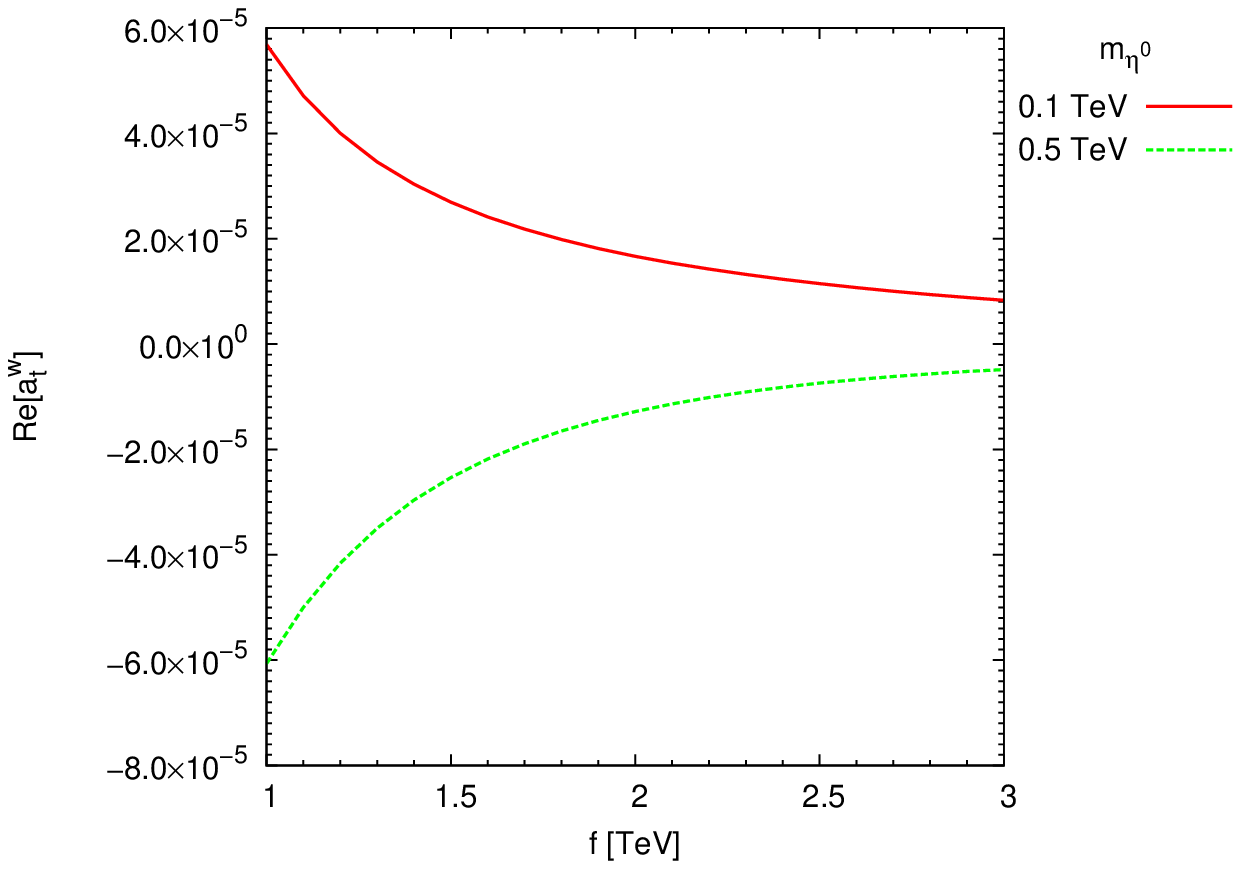}}
\subfloat[]{\includegraphics[width=8.0cm]{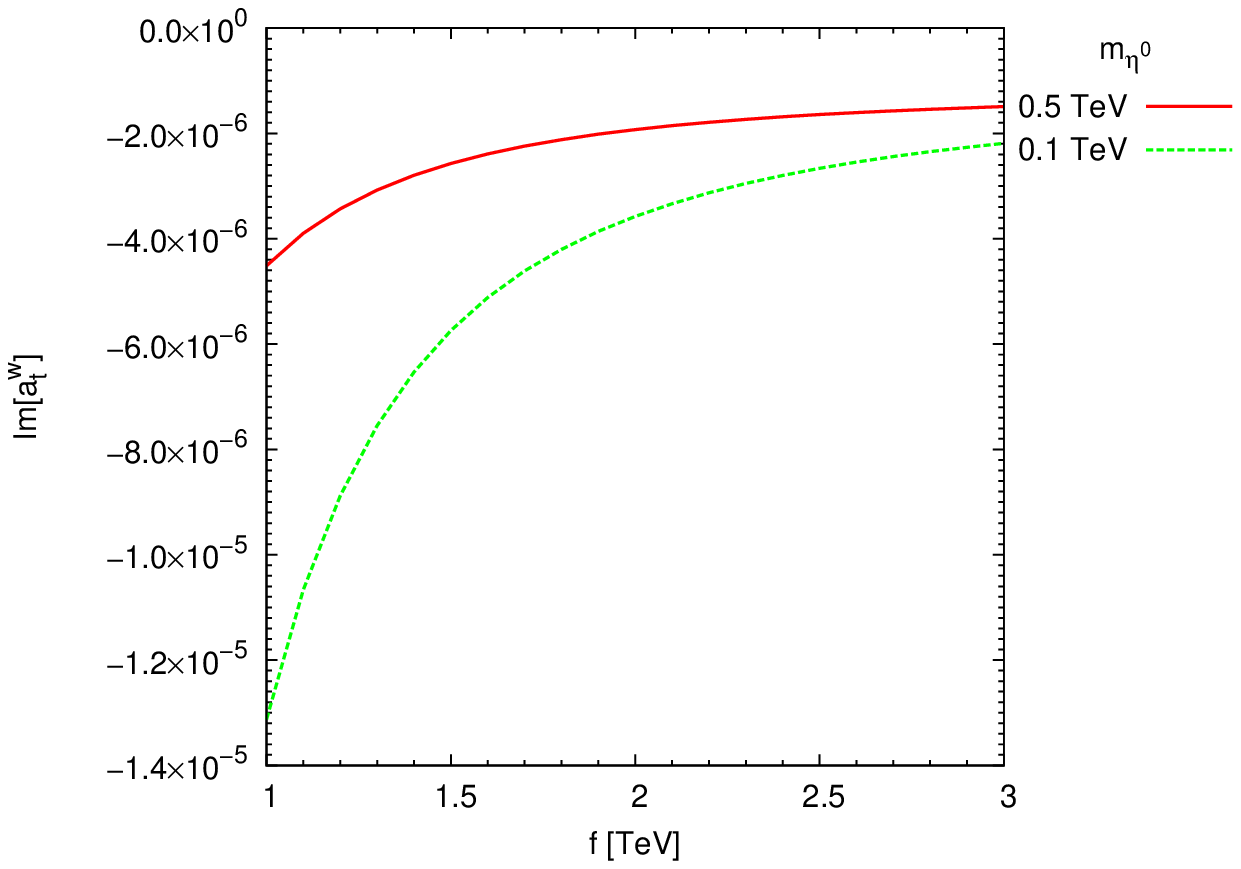}}
\caption{ \label{Fmeta-z2} Total contribution to $a^{W}_{t}$  for different values of the mass of $\eta^{0}$. The plots are obtained with the fixed value of  $F=4000\, \text{GeV}$. The values provided in Table~\ref{parametervalues} are used for the remaining model parameters. a) Re($a^{W}_{t}$). b) Im($a^{W}_{t}$).}
\end{figure}

Using the same approach as that used for the study of the sensitivity of $a^{W}_{t}$ to changes in $m_{A_0}$, we now study the sensitivity of $a^{W}_{t}$ concerning some changes in $m_{\eta^{0}}$ values, $m_{\eta^{0}}=100$ GeV and $m_{\eta^{0}}=500$ GeV.
With these fixed values of the mass of the scalar $ \eta^{0} $, we generate the curves depicted in Figs.~\ref{Fmeta-z2}(a) and~\ref{Fmeta-z2}(b), these are the result of varying the function $a^{W}_{t}$ vs. $ f $.
We provide the numerical estimates we found for the real and imaginary part of $a^{W}_{t}$:
 $ \text{Re}[a^{W}_{t}(\text{m}_{\eta^{0}}=100\, \text{GeV})] = [5.69 \times 10^{-5}, 8.30 \times 10^{-6}]$,  $ \text{Re}[a^{W}_{t}(\text{m}_{\eta^{0}}=500\ \text{GeV})] = -[6.07 \times 10^{-5}, 4.83 \times 10^{-6}]$, $ \text{Im}[a^{W}_{t}(\text{m}_{\eta^{0}}=100\ \text{GeV})] = - [1.31 \times 10^{-5}, 2.19 \times 10^{-6}]$ and $ \text{Im}[a^{W}_{t}(\text{m}_{\eta^{0}}=500\ \text{GeV})] = - [4.51, 1.49] \times 10^{-6}$.
Taking the absolute value of the numerical values, we conclude that the real part of $a^{W}_{t}$ does not show a drastic difference between one curve and the other. In contrast, the imaginary part does show a more significant
numerical difference in the contributions of the first curve concerning the second curve since around $f=1000$ GeV, there is a difference in values of up to an order of magnitude.
The function $a^{W}_{t}$ shows a dependence on the parameter $ m_{\eta^{0}} $.
In Tables~\ref{ZmEta100-z2}-\ref{mA1500debil-z2} we list the values of $a^{W}_{t}$ in relation to the different choices in  $m_{A_0}$ and $m_{\eta^{0}}$.

\begin{figure}[H]
\subfloat[]{\includegraphics[width=8.0cm]{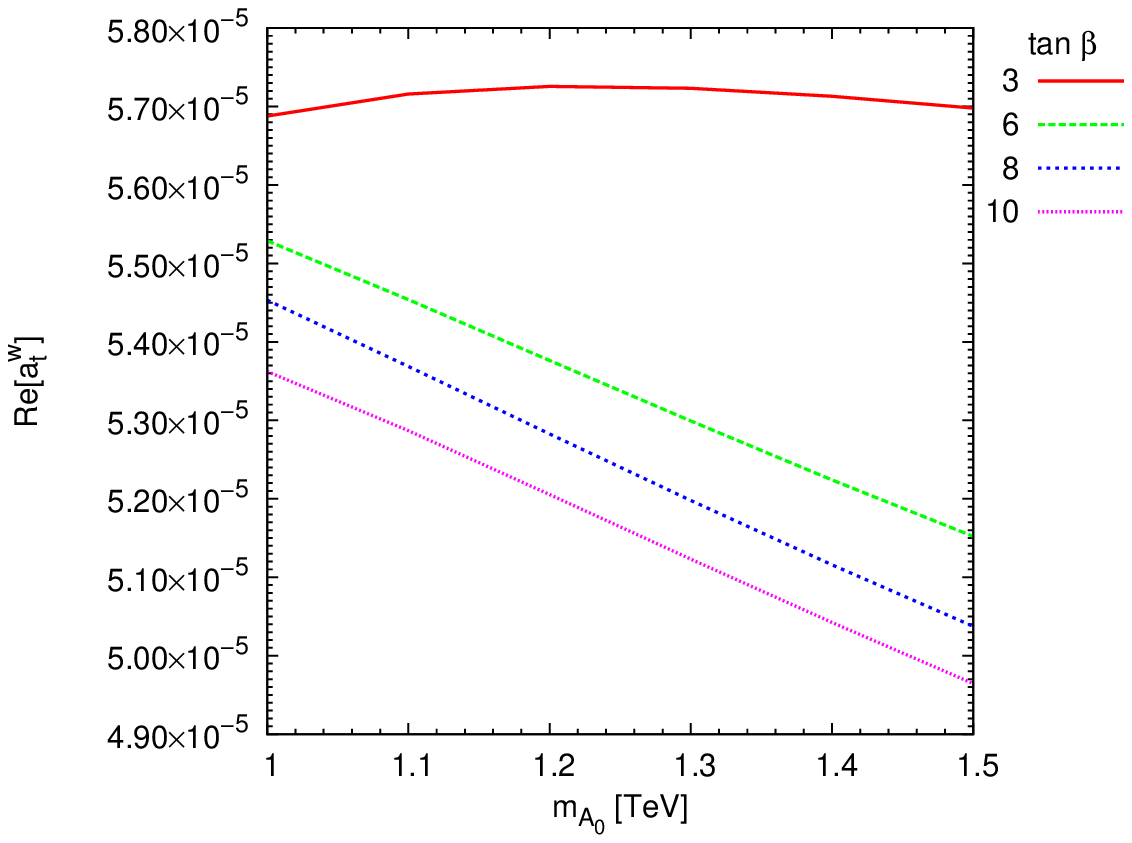}}
\subfloat[]{\includegraphics[width=8.0cm]{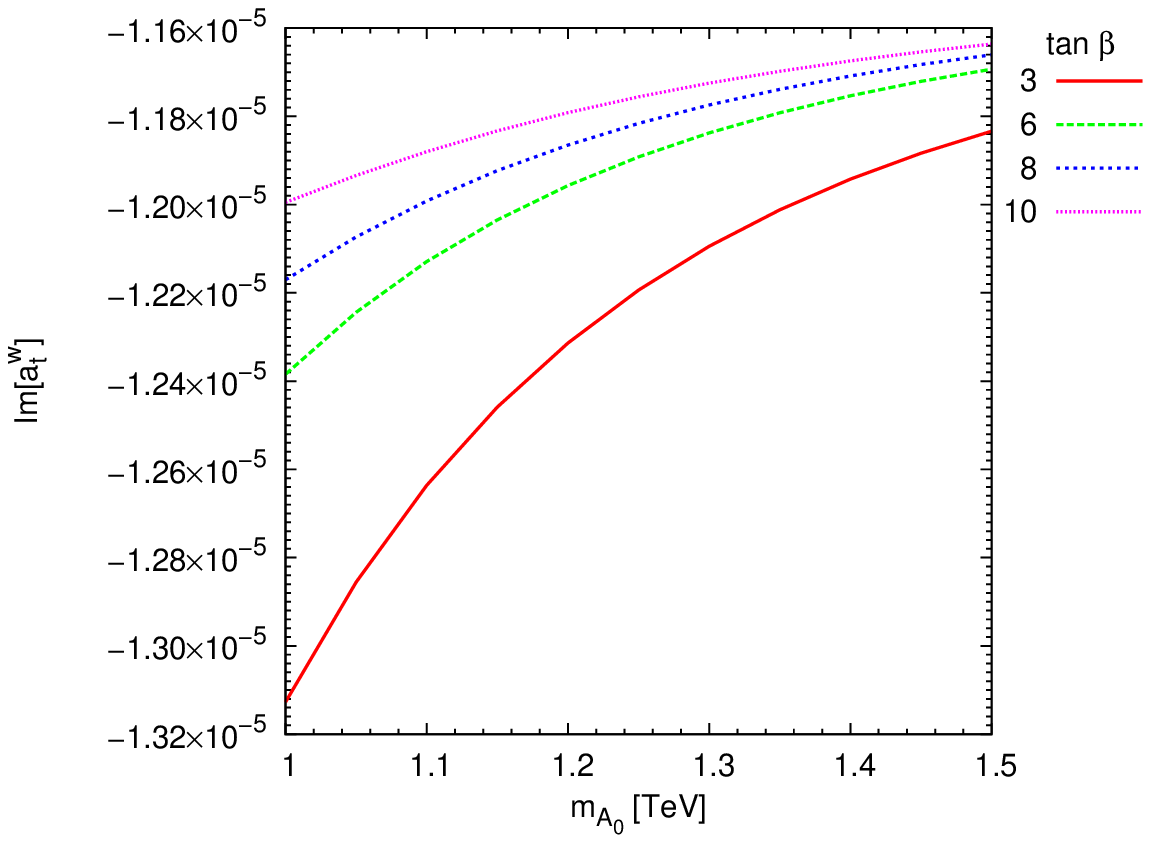}}
\caption{ \label{Ftan-z2} Total contribution to $a^{W}_{t}$ for different values of $\tan \beta$. The plots are obtained with the fixed values of  $f=1000\, \text{GeV}$ and $F=4000\, \text{GeV}$. The values provided in Table~\ref{parametervalues} are used for the remaining model parameters.
a) Re($a^{W}_{t}$). b) Im($a^{W}_{t}$).}
\end{figure}

We are also interested in investigating the $a^{W}_{t}$  vs.  $m_{A_0}$  dependence while the two energy scales $ f $ and $ F $ are fixed. As discussed above, the parameter  $m_{A_0}$  assigns the range of values to $ \tan \beta $; this lets us know that  $m_{A_0}$  and $ \tan \beta $ are strongly linked to each other. It is, therefore, convenient to study these variables. In Figs.~\ref{Ftan-z2}(a) and ~\ref{Ftan-z2}(b),  we show the real and imaginary contributions to $a^{W}_{t}$, the different curves are generated for different choices in the values of $ \tan \beta $.
 From Fig.~\ref{Ftan-z2}(a), we can appreciate that all contributions acquire values of the same order of magnitude, $ 10^{-5} $. However, the curve that provides the most significant contributions is reached when $ \tan \beta =3$,  $ \text{Re}[a^{W}_{t}] = [5.69, 5.70] \times 10^{-5}$ for $m_{A_0}=[1000, 1500]$ GeV.
 Following Fig.~\ref{Ftan-z2}(b), the imaginary part of $a^{W}_{t}$, we observe that all the negative contributions also take values of the same order of magnitude, $ 10^{-5} $. The absolute value of these numerical estimates indicates that the main curve is reached when $ \tan \beta =3$,  $|\text{Im}[a^{W}_{t}]| = [1.31, 1.18] \times 10^{-5}$.
 Based on the figures and numerical estimates, $a^{W}_{t}$ shows an appreciable dependence on the input parameter  $m_{A_0}$. Concerning $ \tan  \beta $,
 the sensitivity of $a^{W}_{t}$ to a change in the values of $ \tan \beta $ is minor.

\begin{figure}[H]
\subfloat[]{\includegraphics[width=8.0cm]{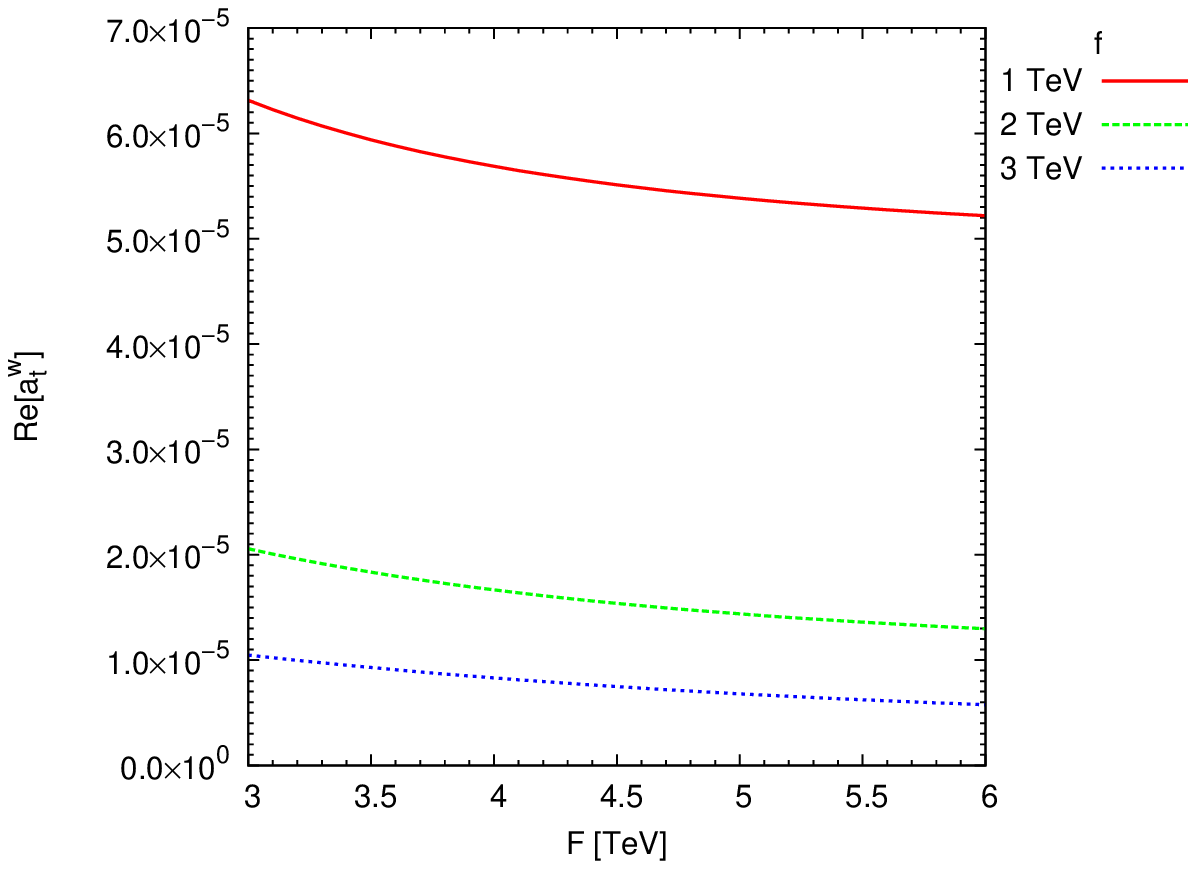}}
\subfloat[]{\includegraphics[width=8.0cm]{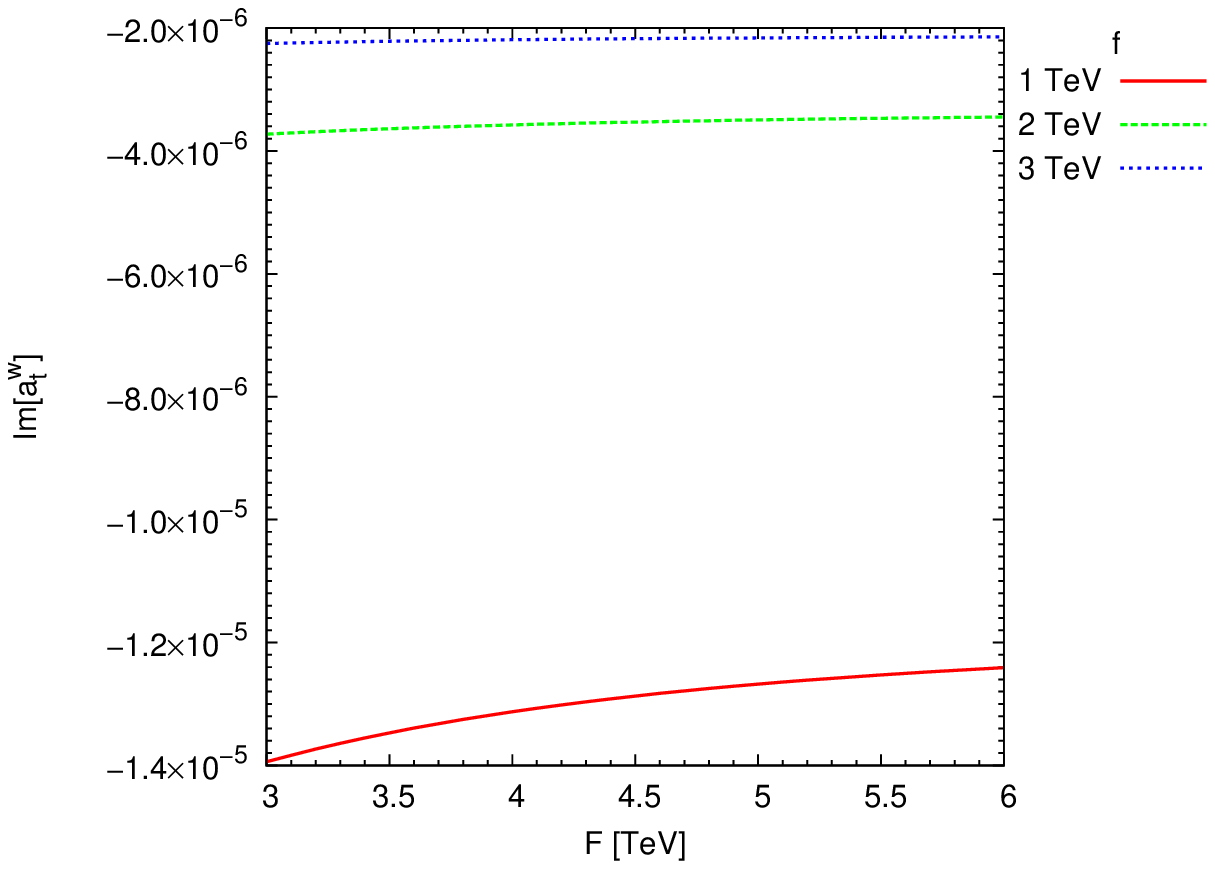}}
\caption{ \label{Ffix} Total contribution to $a^{W}_{t}$ for different values of the energy scale $f$.  The values provided in Table~\ref{parametervalues} are used for the remaining model parameters. a) Re($a^{W}_{t}$). b) Im($a^{W}_{t}$).}
\end{figure}

Let us now examine the behavior of the AWMDM of the top-quark versus variations of the energy scales $ F $ or $ f $.
We first discuss the behavior of $a^{W}_{t}$ as a function of the $ F $ scale, while the other $ f $ scale, we assign the fixed values of $ f=1000, 2000, 3000 $ GeV. With these input values, the curves shown in Figs.~\ref{Ffix}(a) and~\ref{Ffix}(b) are generated, which represent the real and imaginary part of $a^{W}_{t}$, respectively.
From Fig.~\ref{Ffix}(a), we can see that the curve that gives the largest contributions in the $ F $ scale study interval arises for $f=1000$ GeV, the rest of the curves are slightly more suppressed compared to the main curve:
$\text{Re}[a^{W}_t(f=1000\ \text{GeV})]=[6.31, 5.22] \times 10^{-5}$,
 $\text{Re}[a^{W}_t(f=2000\ \text{GeV})]=[2.06, 1.30] \times 10^{-5}$ and $\text{Re}[a^{W}_t(f=3000\ \text{GeV})]=[1.05  \times 10^{-5}, 5.76 \times 10^{-6}]$.
Contrary to what happened with the real part, the imaginary part of $a^{W}_{t}$ (see Fig.~\ref{Ffix}(b)) acquires negative values, and these contributions are more suppressed compared to the numerical predictions of the real part of $a^{W}_{t}$:
$\text{Im}[a^{W}_t(f=1000\ \text{GeV})]=- [1.39, 1.24] \times 10^{-5}$, $\text{Im}[a^{W}_t(f=2000\ \text{GeV})]= - [3.73, 3.45] \times 10^{-6}$ and $\text{Im}[a^{W}_t(f=3000\ \text{GeV})]= -[2.25, 2.14] \times 10^{-6}$.
Considering the absolute value of the numerical estimates, the main curve is also reached for $f=1000$ GeV.
 According to our analysis, $a^{W}_{t}$ depends slightly on the $ F $ scale because the plotted curves do not undergo such appreciable changes over the whole $ F $ scale interval.

\begin{figure}[H]
\subfloat[]{\includegraphics[width=8.0cm]{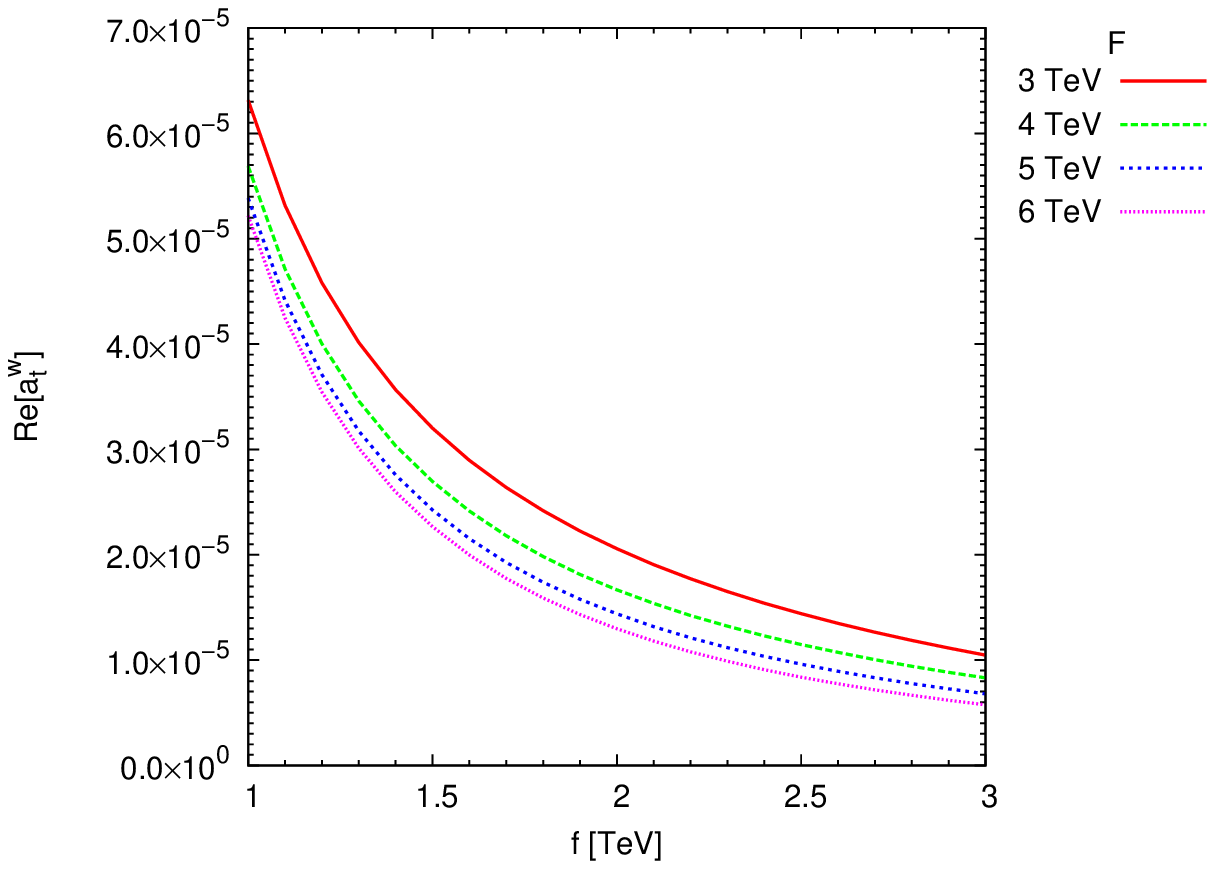}}
\subfloat[]{\includegraphics[width=8.0cm]{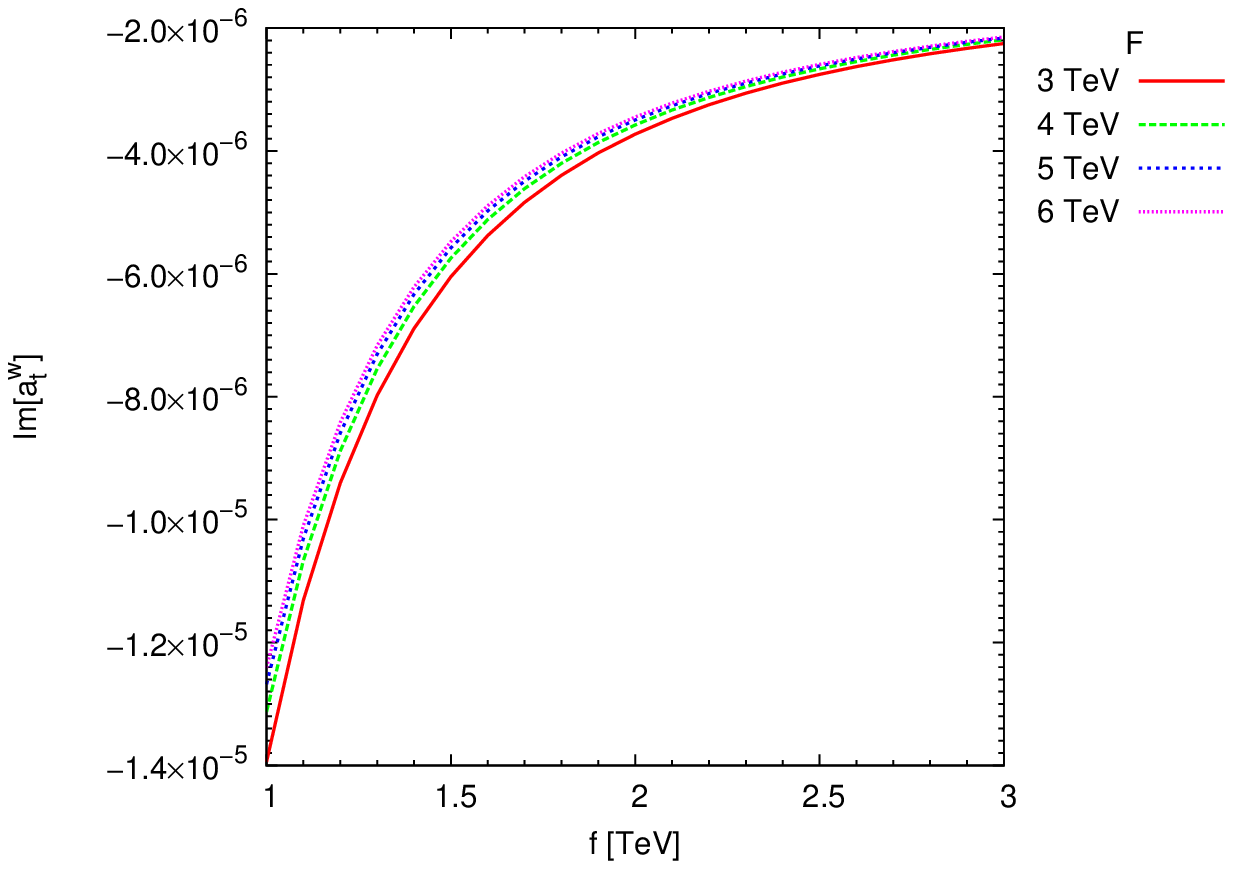}}
\caption{ \label{Fi-z2} Total contribution to $a^{W}_{t}$ for different values of the energy scale $F$.  The values provided in Table~\ref{parametervalues} are used for the remaining model parameters. a) Re($a^{W}_{t}$). b) Im($a^{W}_{t}$).}
\end{figure}

We are adopting the same approach for analyzing of the dependence of $a^{W}_{t}$ vs. $ F $. We are now interested in
studyng the dependence of $a^{W}_{t}$ on the $ f $ scale while using fixed values of the $ F $ scale, $ F= 3000, 4000, 5000, 6000 $ GeV.  In this way, in the interval set for the scale $ f $ and with the input values of $ F $, the contributions to the real and imaginary part of $ a^{W}_{t} $ are generated in Fig.~\ref{Fi-z2}.
In the left plot of Fig.~\ref{Fi-z2}, we observe the real contributions to $a^{W}_{t}$, and these curves are very close to each other over the whole interval $ f $. However, the curve that offers the slightly more significant contributions arises for $F=3000$ GeV,  $\text{Re}[a^{W}_t(F=3000\ \text{GeV})]=[6.31, 1.05] \times 10^{-5}$.
The other curves generate contributions on the order of $10^{-5}-10^{-6}  $.
The right plot in Fig.~\ref{Fi-z2} shows the imaginary contributions to $a^{W}_{t}$; the contributions are negative, and again, the curves are so close to each other. Nevertheless, the curve with the largest contributions to $ a^{W}_{t} $, based on the absolute value of the numerical predictions, is obtained when $F=3000$ GeV:
$|\text{Im}[a^{W}_t(F=3000\ \text{GeV})]|=[1.39\times  10^{-5}, 2.25\times 10^{-6}]$.
The rest of the curves get slightly lower values than the main curve; these numerical values are in the order of $ 10^{-5}-10^{-6} $.
We have found that the $a^{W}_{t}$ function decreases at most one order of magnitude as $f$ increases up to 3000 GeV. Therefore, $a^{W}_{t}$ shows a stronger dependence on the $ f $ energy scale than the $F$.

\begin{table}[H]
\caption{Expected sensitivity limits on the $a^W_t$ in the context of the BLHM with $\sqrt{q^{2}}=500\ \text{GeV}$,
\ $\text{m}_{A_0}=1000\ \text{GeV}$, $\text{m}_{\eta_0}=100\ \text{GeV} $, $F=4000\ \text{GeV}$ and $f=1, 1.5, 2, 2.5, 3\ \text{TeV}$
are represented. All new contributions are considered, scalar bosons, vector bosons, scalar-vector, and heavy quarks.
\label{ZmEta100-z2}}
\centering
\begin{tabular}{|c|c|}
\hline
\hline
\multicolumn{2}{|c|}{$\sqrt{q^{2}}=500\ \text{GeV}$,\ $\text{m}_{A_0}=1000\ \text{GeV}$, $\text{m}_{\eta_0}=\bf{ 100}\ \text{GeV} $,\ $F=4000\ \text{GeV}$}\\
\hline
 $f\ [\text{TeV}]$  & $(a^{W}_{t})^{\rm total} $  \\
\hline
\hline
$1_\cdot0$  & $ 5.69 \times 10^{-5} - 1.31 \times 10^{-5}\ i $  \\
\hline
$1_\cdot5 $  & $ 2.69 \times 10^{-5} - 5.74 \times 10^{-6}\ i$  \\
\hline
$2_\cdot0 $  & $ 1.67 \times 10^{-5} - 3.57 \times 10^{-6}\ i$  \\
\hline
$2_\cdot5 $  & $ 1.15 \times 10^{-5} - 2.66 \times 10^{-6}\ i$  \\
\hline
$3_\cdot0 $  & $ 8.30 \times 10^{-6} - 2.19 \times 10^{-6}\ i$  \\
\hline
\end{tabular}
\end{table}

\begin{table}[H]
\caption{Expected sensitivity limits on the $a^W_t$ in the context of the BLHM with $\sqrt{q^{2}}=500\ \text{GeV}$,
\ $\text{m}_{A_0}=1000\ \text{GeV}$, $\text{m}_{\eta_0}=500\ \text{GeV} $, $F=4000\ \text{GeV}$ and $f=1, 1.5, 2, 2.5, 3\ \text{TeV}$
are represented. All new contributions are considered, scalar bosons, vector bosons, scalar-vector, and heavy quarks.
\label{ZmEta500-z2}}
\centering
\begin{tabular}{|c|c|}
\hline
\hline
\multicolumn{2}{|c|}{$\sqrt{q^{2}}=500\ \text{GeV}$,\ $\text{m}_{A_0}=1000\ \text{GeV}$, $\text{m}_{\eta_0}=\bf{ 500}\ \text{GeV} $,\ $F=4000\ \text{GeV}$}\\
\hline
 $f\ [\text{TeV}]$  & $(a^{W}_{t})^{\rm total} $  \\
\hline
\hline
$1_\cdot0$  & $ - 6.07 \times 10^{-5} - 4.51 \times 10^{-6}\ i $  \\
\hline
$1_\cdot5 $  & $ - 2.53 \times 10^{-5} - 2.57 \times 10^{-6}\ i$  \\
\hline
$2_\cdot0 $  & $ - 1.28 \times 10^{-5} -  1.93 \times 10^{-6}\ i$  \\
\hline
$2_\cdot5 $  & $ - 7.42 \times 10^{-6} - 1.64 \times 10^{-6}\ i$  \\
\hline
$3_\cdot0 $  & $ - 4.83 \times 10^{-6} -  1.49 \times 10^{-6}\ i$  \\
\hline
\end{tabular}
\end{table}

\begin{table}[H]
\caption{Expected sensitivity limits on the $a^{W}_t$ in the context of the BLHM with $\sqrt{q^{2}}=500\ \text{GeV}$,
\ $\text{m}_{A_0}=1 500\ \text{GeV}$, $\text{m}_{\eta_0}=100\ \text{GeV}$, $F=4000\ \text{GeV}$ and $f=1, 1.5, 2, 2.5, 3\ \text{TeV}$
are represented. All new contributions are considered, scalar bosons, vector bosons, scalar-vector, and heavy quarks.
\label{mA1500debil-z2}}
\centering
\begin{tabular}{|c|c|}
\hline
\hline
\multicolumn{2}{|c|}{$\sqrt{q^{2}}=500\ \text{GeV}$,\ $\text{m}_{A_0}=\bf{1500}\ \text{GeV}$, $\text{m}_{\eta_0}=100\ \text{GeV}$,  $\ F=4000\ \text{GeV}$}\\
\hline
 $f\ [\text{TeV}]$  & $(a^{W}_{t})^{\rm total} $  \\
\hline
\hline
$1_\cdot0$  & $ 5.70  \times 10^{-5} - 1.18 \times 10^{-5}\ i$  \\
\hline
$1_\cdot5 $  & $ 2.83 \times 10^{-5} - 4.60 \times 10^{-6}\ i$  \\
\hline
$2_\cdot0 $  & $ 1.84 \times 10^{-5} - 2.51 \times 10^{-6}\ i$  \\
\hline
$2_\cdot5 $  & $ 1.33 \times 10^{-5}  - 1.64 \times 10^{-6}\ i$  \\
\hline
$3_\cdot0 $  & $ 1.02 \times 10^{-5} - 1.19 \times 10^{-6}\ i$  \\
\hline
\end{tabular}
\end{table}

\section{Conclusions}
\label{conclusions}

We explored the AMDM $ a_t $ and AWMDM $ a^{W}_t $ of the top-quark  in the two diagonalization schemes to Yukawa couplings that the BLHM offers.
 The different diagonalization schemes arise when $y_2=y_3$; in this case, the masses of the heavy quarks $T$ and $T_5$  degenerate to lowest order.
In the first scenario, $y_2>y_3$, the mass splitting between the exotic quarks $T_5$ and $T_6$ is relatively tiny and leads to the decays of $T_5$ being predominantly to SM particles. For the second scenario, $y_2<y_3$, the mass splitting between $T_5$ and $T_6$ is large, which increases the decay modes available for the $T_5$ quark through decay cascades to non-SM particles~\cite{Godfrey:2012tf,PhenomenologyBLH,Cruz-Albaro:2022kty}.
Due to the phenomenological implications of each scenario motivates us to calculate  $ a_t $ and $ a^{W}_t $  in the two sample regions of parameter space.

We calculate the effects induced by the new BLHM particles at the one-loop level for $ a_t $ and $ a^{W}_t $ and find that if the particles are produced at $  \sqrt{q^2}=500$ GeV, their numerical effects on $ a_t $ and $ a^{W}_t $ show us that in the two diagonalization schemes, the contributions due to the scalars provide the most significant contributions to both $ a_t $ and $ a^{W}_t $.
The top-quark numerical magnitudes obtained for the AMDM and AWMDM are slightly higher in the $y_2 < y_3$ scenario than
in the $y_2 > y_3$ scenario. In particular, when $f=1000$ GeV and $F=3000$ GeV, the total contributions which give the
best sensitivity on the electroweak dipole moments of the top-quark are given in Table~\ref{values-scenarios}.

\begin{table}[H]
\caption{Total contributions that give the best sensitivity on $a_t$ and $a^{W}_t$ at the scenarios $y_2 > y_3$ and $y_2 < y_3$.}
\label{values-scenarios}
\centering
\begin{tabular}{|c | c| c|}
\hline
$ \textbf{Scenario} $  &    $ a_t $ &   $ a^{W}_t $   \\
\hline
\hline
$ y_2 > y_3 $  &  $ 1.39 \times 10^{-4} + i\,  6.55 \times 10^{-5}  $  & $  6.31 \times 10^{-5} - i\,  1.39 \times 10^{-5}  $ \\
\hline
 $ y_2 < y_3  $  &   $ 2.12 \times 10^{-4} + i\,  4.49 \times 10^{-5}   $  & $  2.13 \times 10^{-4} - i\,  1.34 \times 10^{-5}$ \\
\hline
\end{tabular}
\end{table}

\noindent It should be stressed that the numerical estimate reported in the table above for $ a^{W}_t $ in the $y_2 < y_3$ scenario is obtained from the sum of the result obtained in this paper (Subsection~\ref{B2} of Appendix~\ref{y2menor}) and the contributions previously found in Ref.~\cite{Cruz-Albaro:2022kty}.
 Based on the research performed in $ a_t $ and $ a^{W}_t $, we are varying the scales of the new physics $f$ and $F$ in the intervals $f=[1000, 3000]$ GeV and $F=[3000,6000]$ GeV, as well as the mass of the scalar $A_0$ in the range from 1000 GeV to 1500 GeV.
 We find that the highest sensitivity in the AMDM and AWMDM of the top-quark is reached just when the energy scales $f$ and $F$ acquire small values in the intervals established for their analysis, that is, $f=1000$ GeV and $F=3000$ GeV.
 In general, the total contributions to $ a_t $ and $ a^{W}_t $ generate real and imaginary parts that are of the order of $\text{Re}[a_t] \sim 10^{-4}-10^{-5}$, $\text{Im}[a_t] \sim 10^{-5}$, $\text{Re}[a^{W}_t] \sim 10^{-4}-10^{-6}$ and $\text{Im}[a^{W}_t] \sim 10^{-5}-10^{-6}$ when the scales $f \in [1000,3000]$ GeV and $F \in [3000, 6000]$ GeV.

The AMDM and AWMDM of the top-quark show a very high sensitivity at the $ f $ energy scale compared to the $ F $ scale, and even with $m_{A_0}$, the effects are noticeable in the generated curves of $a_t$ vs. $ f $ and $a^{W}_t$ vs. $ f $ as they undergo quite appreciable changes in the range of analysis set for the $ f $ energy scale.

 In the context of the BLHM, the numerical estimates obtained for $ a_t $ and $ a^{W}_t $ are weak concerning
the SM predictions provided in Table~\ref{limites-SM}.
However, they are comparable, in some cases, to the predictions of several extensions of the SM, such as the models with Two Higgs Doublets (2HDM), $ a_t \sim 10^{-3}-10^{-5}$ and $ a^{W}_t \sim 10^{-3}-10^{-4}$~\cite{Bernabeu:1995gs};
the Minimal Supersymmetric Standard Model (MSSM),  $ a^{W}_t \sim 10^{-4}$~\cite{Bartl:1997iq};
the general effective extended model with flavor changing $\bar{f}_i f_j Z'$ vertices, $ a^{W}_t \in [10^{-6}, 10^{-10}]$~\cite{Vivian:2019zfa};
and finally, in the context of the effective Lagrangian approach, $ a_t  \in [-0.79, 1.3] $~\cite{Martinez:1997ww}.
Another study in the SM scenario derived the one-loop level QCD contributions to the $ a_t $ and $ a^{W}_t $ and found that $ a_t =1.53 \times 10^{-2} $ and $ a^{W}_t=5.2 \times 10^{-3} $ for renormalization scale $\mu=m_t=175$ GeV~\cite{Bernreuther:2005gq}.

The couplings of the top-quark to $V$ gauge bosons have not yet been measured directly.  The current data only provide weak constraints, although there are some studies that have indirectly constrained the couplings of the top to electroweak gauge bosons.
There are pretty promising projects at current colliders such as the LHC~\cite{ATLAS-CONF-2012-126,PRL110-2013,EPJC79-2019,JHEP03-2020} or future colliders such as the FCC-hh~\cite{Barletta:2014vea,Koratzinos:2015fya}, ILC~\cite{Behnke:2013xla,Baer:2013cma,Adolphsen:2013kya} and CLIC~\cite{deBlas:2018mhx,Robson:2018enq,Roloff:2018dqu,CLIC:2016zwp,Abramowicz:2016zbo}  that have as an essential part of their physics programs to investigate $\bar{t}tV$ couplings, in particular, the electroweak dipole moments of the top-quark.
In the LHC or FCC experiment, the most promising channel to study and constrain the electroweak couplings of the top-quark is through the $pp \to \bar{t}tV$ process, which produces direct sensitivity without intrinsic dilution by QCD effects.
Concerning leptonic colliders such as ILC and CLIC, the search for electroweak top-quark interactions will focus on the $e^{+}e^{-} \to V^{*} \to \bar{t}t$ process, which is highly sensitive to both $tt\gamma$ and $ttZ$ couplings simultaneously.

We discuss below some phenomenological studies that have attempted to bound the electroweak dipole moments of the top-quark, in particular, the AMDM $ a_t $ and AWMDM $ a^{W}_{t} $.
For instance, the achievable sensitivity at the LHC at 95$\%$ C.L. for $ a_t $ and $ a^{W}_{t} $ are inferred through the $pp \to tj\gamma $ channel for  $ a_t $~\cite{Etesami:2016rwu}, while $ a^{W}_{t} $ is estimated using the  production cross section of the $t\bar{t}ZZ$ process~\cite{Etesami:2017ufk}. The numerical estimates obtained are $ a_t \in (-0.29, 0.29) $ and $ a^{W}_{t} \in [-0.1, 0.09]$.
Additionally, the authors of Ref.~\cite{Koksal:2019gyo} propose the $ pp \to p t \bar{t}X $ process as an excellent
perspective for probing $a_t$ in the post-LHC era.
On the other hand, at the FCC-he with center-of-mass energy $\sqrt{s}=10$ TeV and integrated luminosity of 1000 $\text{fb}^{-1}$, the $e^{-}p \to \bar{t}\nu_ebp$ process gives a sensitivity to $ a_t $ of $a_t \in (-0.1074,0.1033)$ at 95$\%$ C.L.~\cite{Hernandez-Ruiz:2018oxq,Koksal:2019cjn,Billur:2018mwk}.
FCC-hh will provide collisions at a center-of-mass energy of 100 TeV, a factor 7 higher than the LHC. At this energy and with 10 000 fb$^{-1}$ of data, about $10^{8}$ $\bar{t}tZ$ events will be produced. At a rough estimate, the FCC-hh will provide improved limits to $a^{W}_{t}$ by a factor of 3 to 10 compared to the 3000 fb$^{-1}$ LHC~\cite{Barletta:2014vea,Koratzinos:2015fya,Rontsch:2015una}.
The ILC with $\sqrt{s}=500$ GeV and 500 fb$^{-1}$, the limits of $-0.02 \lesssim a^{W}_{t} \lesssim 0.04$ at $95 \%$ C.L. are expected
to be reached. They are derived by exploiting the total cross-section of the top-quark pair production~\cite{Rontsch:2015una}.
For this same collider, the $e^{+}e^{-} \to \bar{t}t  $ channel provides the limits to $a_t \in (-0.002, 0.002)$ at  $68\%$ C.L.~\cite{ECFADESYLCPhysicsWorkingGroup:2001igx}.
Finally, the expected measurement sensitivity for $a_t$ at the CLIC with the $\gamma \gamma \to \bar{t}t$ process is $a_t \in (-0.2203, 0.0020)$ at $95\%$ C.L.~\cite{Billur:2017yds}.

The ILC is one of the leptonic colliders that offers the possibility of extending the LHC top-quark program~\cite{Baer:2013cma} and is one of the most advanced proposals for a linear $e^{+}e^{-}$  collider.
As mentioned above, the ILC is extremely sensitive to electroweak couplings of the top-quark; in addition, it generates simultaneous sensitivity to $\gamma tt$ and $Ztt$ couplings.
The ILC promises to explore top-quark interactions with a very competitive precision, much better than a hadron collider. This can even be appreciated in the numerical estimates of the phenomenological studies performed on the search for $a_t$ and $a^{W}_t$. These obtain improved limits at the ILC compared to the LHC constraints.
For these reasons, the ILC can verify our results in the future because it has the potential to reach the required level of sensitivity.

We present a comprehensive study of the AMDM and AWMDM of the top-quark at the one-loop level in the context of the BLHM. In our study, we investigated the virtual effects induced by the new BLHM particles and obtained bounds on $a_t$ and $a^{W}_t$; these bounds are small compared to SM theoretical predictions but competitive with those reported in Refs.~\cite{Bernabeu:1995gs,Bartl:1997iq,Vivian:2019zfa,Bernreuther:2005gq}. Our results are within the numerical intervals derived from phenomenological studies.
There are currently no experimental measurements of precision in $a_t$ and $a^{W}_t$. However, future experiments are expected to reach the sensitivity of predicted values for observables in the BLHM.
Our work complements other studies performed at one-loop in models with an extended scalar sector.

\vspace{5.0cm}

\begin{center}
{\bf Acknowledgements}
\end{center}

E. C. A. appreciates the post-doctoral stay. A. G. R. and M. A. H. R. acknowledge support from SNI and PROFEXCE (M\'exico).

\vspace{3cm}


\appendix

\vspace{5cm}

\section{Partial contributions of particles that induce the AMDM and AWMDM of the top-quark in the $y_2 > y_3$ scenario  } \label{numericasparciales}

In this Appendix, we present the partial numerical contributions of the particles that induce the AMDM and AWMDM of the top-quark in the $y_2 > y_3$ scenario.

\begin{table}[H]
\caption{The magnitude of the partial contributions to $a_{t}$ of the BLHM.
The data are obtained by fixing the $f$ and $F$ scales, $f=1000$ GeV and $F=4000$ GeV.
The values provided in Table~\ref{parametervalues} are used for the rest of the model parameters.
{\bf abc} denotes the different particles running in the loop of the vertex $\gamma t t$.
\label{parcial1-2}}
\centering
\begin{tabular}{|c|c|}
\hline
\hline
\multicolumn{2}{|c|}{$\sqrt{q^{2}}=500\ \text{GeV}$,\  $f=1 000\ \text{GeV}$, $F=4000\ \text{GeV}$}\\
\hline
 $\text{Couplings}\  \textbf{abc}$  & $\left(a_{t} \right)^{ \textbf{abc} } $  \\
\hline
\hline
$Z' tt $  & $ 7_\cdot 56 \times 10^{-7} + 8_\cdot 11 \times 10^{-9}\ i$  \\
\hline
$W'bb $   & $ 7.74 \times 10^{-7}+ 2.13 \times 10^{-8}\ i $ \\
\hline
$  Z T T$ & $ 3.42 \times 10^{-9} + 0\ i$ \\
\hline
$  \gamma T T$ & $- 4.31 \times 10^{-8} + 0\ i$ \\
\hline
$  Z' T T$ & $ - 3.93 \times 10^{-12} + 0\ i$ \\
\hline
$  Z T_5 T_5$ & $ 1.81 \times 10^{-6} + 0\ i$ \\
\hline
$  \gamma T_5 T_5$ & $ 7.30 \times 10^{-7} + 0\ i$ \\
\hline
$ Z' T_5 T_5$ & $- 3.47 \times 10^{-8} + 0\ i$ \\
\hline
$  Z T_6 T_6$ & $- 3.86 \times 10^{-7} + 0\ i$ \\
\hline
$  \gamma T_6 T_6$ & $ - 7.58 \times 10^{-9} + 0\ i$ \\
\hline
$  Z' T_6 T_6$ & $- 1.00 \times 10^{-8} + 0\ i$ \\
\hline
$  Z T^{2/3} T^{2/3}$ & $ -2.93 \times 10^{-7} + 0\ i$ \\
\hline
$  \gamma T^{2/3} T^{2/3}$ & $ 1.54 \times 10^{-6} + 0\ i$ \\
\hline
$  Z' T^{2/3} T^{2/3}$ & $- 4.15 \times 10^{-8} + 0\ i$ \\
\hline
$  W T^{5/3} T^{5/3}$ & $ 4.46 \times 10^{-6} + 0\ i$ \\
\hline
$  W' T^{5/3} T^{5/3}$ & $ -1.70 \times 10^{-7} + 0\ i$ \\
\hline
$  W' B B$ & $ -3.91 \times 10^{-11} + 0\ i$ \\
\hline
$  W' W' b$ & $ 1.98  \times 10^{-6} + 0\ i$ \\
\hline
$  W' W' B$ & $ 1.37 \times 10^{-11} + 0\ i$ \\
\hline
$\sigma tt $  & $ 1.92 \times 10^{-8} +8.35 \times 10^{-9}\ i$  \\
\hline
$A_{0} tt   $  & $ 1.85 \times 10^{-5} +  1.34 \times 10^{-5}\ i$  \\
\hline
\end{tabular}
\end{table}

\begin{table}[H]
\caption{Continuation of Table~\ref{parcial1-2}.}
\centering
\label{parcial2-2}
\begin{tabular}{|c|c|}
\hline
\hline
\multicolumn{2}{|c|}{$\sqrt{q^{2}}=500\ \text{GeV}$,\  $f=1 000\ \text{GeV}$, $F=4000\ \text{GeV}$}\\
\hline
$H_{0} tt $  & $ 1.59  \times 10^{-5} + 1.14 \times 10^{-5}\ i$  \\
\hline
$\eta^{0} tt  $  & $ - 3.45  \times 10^{-7} + 3.19 \times 10^{-5}\ i$  \\
\hline
$\phi^{0} tt $  & $ 4.82 \times 10^{-6} + 5.17 \times 10^{-6}\ i$  \\
\hline
$H^{\pm} bb  $  & $ 1.15 \times 10^{-6} + 3.19 \times 10^{-7}\ i$  \\
\hline
$\eta^{\pm} bb   $  & $ 3_\cdot 71 \times 10^{-7} + 8_\cdot 21 \times 10^{-7}\ i$  \\
\hline
$\phi^{\pm} bb  $  & $ 2_\cdot 81 \times 10^{-7} + 2_\cdot 00 \times 10^{-7}\ i$  \\
\hline
$\sigma T T  $  & $ - 4.59 \times 10^{-12} + 0\ i$  \\
\hline
$h_{0} TT  $  & $ - 2.17 \times 10^{-6} + 0\ i$  \\
\hline
$H_0 TT  $  & $- 1_\cdot 40 \times 10^{-7} + 0\ i$  \\
\hline
$A_0 TT  $  & $  1.44 \times 10^{-7} + 0\ i$  \\
\hline
$\phi^{0} TT  $  & $ - 2.49  \times 10^{-7} + 0\ i$  \\
\hline
$\eta^{0} TT  $  & $ - 3.35 \times 10^{-7} + 0\ i$  \\
\hline
$\sigma T_5 T_5  $  & $ - 8.15  \times 10^{-9} + 0\ i$  \\
\hline
$h_0 T_5 T_5  $  & $ 3.48 \times 10^{-5} + 0\ i$  \\
\hline
$H_0 T_5 T_5  $  & $ 1.86 \times 10^{-6} + 0\ i$  \\
\hline
$A_0 T_5 T_5  $  & $ -2.70 \times 10^{-6} + 0\ i$  \\
\hline
$\phi^{0} T_5 T_5  $  & $ -6.07 \times 10^{-8} + 0\ i$  \\
\hline
$\eta^{0} T_5 T_5  $  & $ -1.05 \times 10^{-7} + 0\ i$  \\
\hline
$  \sigma T_6 T_6$ & $ 5.50 \times 10^{-7} + 0\ i$ \\
\hline
$h_0 T_6 T_6  $  & $  1.32 \times 10^{-5} + 0\ i$  \\
\hline
$H_0 T_6 T_6  $  & $ 1.45 \times 10^{-5} + 0\ i$  \\
\hline
$A_0 T_6 T_6  $  & $ 2.41 \times 10^{-5} + 0\ i$  \\
\hline
$\phi^{0} T_6 T_6  $  & $ -1.44 \times 10^{-8} + 0\ i$  \\
\hline
$\eta^{0} T_6 T_6  $  & $ -2.65 \times 10^{-8} + 0\ i$  \\
\hline
$  \sigma T^{2/3} T^{2/3}$ & $- 1.27 \times 10^{-9} + 0\ i$ \\
\hline
\end{tabular}
\end{table}

\begin{table}[H]
\caption{Continuation of Table~\ref{parcial2-2}.}
\centering
\label{parcial3-2}
\begin{tabular}{|c|c|}
\hline
\hline
\multicolumn{2}{|c|}{$\sqrt{q^{2}}=500\ \text{GeV}$,\  $f=1 000\ \text{GeV}$, $F=4000\ \text{GeV}$}\\
\hline
$ h_0 T^{2/3} T^{2/3}$ & $ 4.59 \times 10^{-6} + 0\ i$ \\
\hline
$ H_0 T^{2/3} T^{2/3}$ & $ -2.78  \times 10^{-6} + 0\ i$ \\
\hline
$ A_0 T^{2/3} T^{2/3}$ & $ -6.99  \times 10^{-7} + 0\ i$ \\
\hline
$ \phi^{0} T^{2/3} T^{2/3}$ & $ -4.68 \times 10^{-8} + 0\ i$ \\
\hline
$ \eta^{0} T^{2/3} T^{2/3}$ & $ -8.62 \times 10^{-8} + 0\ i$ \\
\hline
$ H^{\pm} T^{5/3} T^{5/3}$ & $ -1.32 \times 10^{-6} + 0\ i$ \\
\hline
$ H^{\pm} BB $ & $ 1.52 \times 10^{-9} + 0\ i$ \\
\hline
$ \eta^{-} W' b $ & $ 0 + 1.23 \times 10^{-8}\ i$ \\
\hline
$ \phi^{-} W' b $ & $ 7.51 \times 10^{-7}+ 0\ i$ \\
\hline
\end{tabular}
\end{table}

\begin{table}[H]
\caption{The magnitude of the partial contributions to $a^{W}_{t}$ of the BLHM.
The data are obtained by fixing the $f$ and $F$ scales, $f=1000$ GeV and $F=4000$ GeV.
The values provided in Table~\ref{parametervalues} are used for the rest of the model parameters. {\bf abc} denotes the different particles running in the
vertex loop $Ztt$.
\label{CN1-z2}}
\centering
\begin{tabular}{|c|c|}
\hline
\hline
\multicolumn{2}{|c|}{$\sqrt{q^{2}}=500\ \text{GeV}$,\ $f=1 000\ \text{GeV}$, $F=4000\ \text{GeV}$}\\
\hline
 $\text{Couplings} \textbf{ abc}$  & $\left( a^{W}_{t} \right)^{ \textbf{abc} } $  \\
\hline
\hline
$ Z'tt $  & $- 9_\cdot 31\times 10^{-7} - 9_\cdot 98 \times 10^{-9}\ i$  \\
\hline
$ \sigma tt $  & $ 2_\cdot 52\times 10^{-8} + 1_\cdot 09 \times 10^{-8}\ i$  \\
\hline
$H_{0} tt $  & $ 5_\cdot 43 \times 10^{-6} + 3.90 \times 10^{-6}\ i$  \\
\hline
$ A_{0} tt$  & $ -6.31 \times 10^{-6} - 4.57 \times 10^{-6}\ i$  \\
\hline
$ \phi^{0} tt $  & $ -1_\cdot 10 \times 10^{-6} - 1_\cdot 18\times 10^{-6}\ i$  \\
\hline
$ \eta^{0} tt $  & $ 7_\cdot 86 \times 10^{-8} - 7_\cdot 28\times 10^{-6}\ i$  \\
\hline
$ W' bb $  & $ 2_\cdot 33\times 10^{-6} + 6_\cdot 43\times 10^{-8}\ i$  \\
\hline
$H^{\pm} bb $  & $ -3.47 \times 10^{-6} - 9.76 \times 10^{-7}\ i$  \\
\hline
$ \phi^{\pm} bb $  & $ -8_\cdot 46\times 10^{-7} - 6.07 \times 10^{-7}\ i$  \\
\hline
$ \eta^{\pm} bb $  & $ -1_\cdot 12 \times 10^{-6} - 2_\cdot 48\times 10^{-6}\ i$  \\
\hline
$ Z TT $  & $ 4.22 \times 10^{-9} +0\ i $  \\
\hline
$ \gamma TT $  & $ - 5.31 \times 10^{-8} +0\ i $  \\
\hline
$Z' TT $  & $ -4.84 \times 10^{-12}+0\ i $  \\
\hline
$ \sigma TT $  & $ 1.14 \times 10^{-12}+0\ i $  \\
\hline
$ h_{0} TT $  & $ 2.67 \times 10^{-6}+0\ i $  \\
\hline
$ H_{0} TT $  & $ 1_\cdot 72 \times 10^{-7} + 0 i$  \\
\hline
$ A_{0} TT $  & $ -1.78 \times 10^{-7} + 0 i$  \\
\hline
$\phi^{0} TT $  & $ 3.06 \times 10^{-7} + 0\ i$  \\
\hline
$ \eta^{0} TT $  & $ 4.12 \times 10^{-7} +0\ i$  \\
\hline
$ Z T_{5} T_{5} $  & $ -9.94  \times 10^{-7} +0\ i$  \\
\hline
$ \gamma T_{5} T_{5} $  & $ -4.00 \times 10^{-7} +0\ i$  \\
\hline
\end{tabular}
\end{table}

\begin{table}[H]
\caption{Continuation of Table~\ref{CN1-z2}.
\label{CN2-z2}}
\centering
\begin{tabular}{|c|c|}
\hline
\hline
\multicolumn{2}{|c|}{$\sqrt{q^{2}}=500\ \text{GeV}$,\  $f=1 000\ \text{GeV}$, $F=4000\ \text{GeV}$}\\
\hline
\hline
$Z' T_{5} T_{5} $  & $ 1.90 \times 10^{-8} + 0\ i$  \\
\hline
$\sigma T_{5} T_{5} $  & $ -4.47 \times 10^{-9}+0\ i$  \\
\hline
$ h_{0}T_{5} T_{5} $  & $ 1.91 \times 10^{-5} +0\ i$  \\
\hline
$ H_{0} T_{5} T_{5} $  & $ 1.02 \times 10^{-6} +0\ i$  \\
\hline
$ A_{0} T_{5} T_{5} $  & $ -1.48 \times 10^{-6} +0\ i $  \\
\hline
$\phi^{0} T_{5} T_{5} $  & $ -3.33 \times 10^{-8}+0\ i  $  \\
\hline
$\eta^{0} T_{5} T_{5} $  & $ -5.76 \times 10^{-8}+0\ i $  \\
\hline
$ Z T_{6} T_{6} $  & $ 2.12 \times 10^{-7} +0\ i $  \\
\hline
$ \gamma T_{6} T_{6} $  & $4.16 \times 10^{-9}+0\ i $  \\
\hline
$ Z' T_{6}T_{6} $  & $ 5.48 \times 10^{-9} + 0 i$  \\
\hline
$ \sigma T_{6}T_{6} $  & $ 3.01 \times 10^{-7} + 0 i$  \\
\hline
$h_{0} T_{6}T_{6} $  & $ 7.26 \times 10^{-6} + 0\ i$  \\
\hline
$ H_{0} T_{6}T_{6} $  & $ 7.95 \times 10^{-6} +0\ i$  \\
\hline
$ A_{0} T_{6}T_{6} $  & $ 1.32 \times 10^{-5} +0\ i$  \\
\hline
$ \phi^{0} T_{6}T_{6} $  & $ -7.90 \times 10^{-9} +0\ i$  \\
\hline
$\eta^{0} T_{6}T_{6} $  & $ -1.45 \times 10^{-8} + 0\ i$  \\
\hline
$Z T^{2/3} T^{2/3}$  & $ 5.61 \times 10^{-7}+0\ i$  \\
\hline
$ \gamma T^{2/3} T^{2/3} $  & $ 2.94  \times 10^{-6} +0\ i$  \\
\hline
$ Z' T^{2/3} T^{2/3} $  & $ 7.93 \times 10^{-8} +0\ i$  \\
\hline
$ \sigma T^{2/3} T^{2/3} $  & $ -2.95 \times 10^{-9} +0\ i $  \\
\hline
$h_{0} T^{2/3} T^{2/3} $  & $ 1.07 \times 10^{-5}+0\ i  $  \\
\hline
$H_{0} T^{2/3} T^{2/3} $  & $ -6.48 \times 10^{-6}+0\ i $  \\
\hline
$ A_{0}T^{2/3} T^{2/3} $  & $ -1.63 \times 10^{-6} +0\ i $  \\
\hline
$ \phi^{0} T^{2/3} T^{2/3} $  & $ -1.09 \times 10^{-7}+0\ i $  \\
\hline
$ \eta^{0} T^{2/3} T^{2/3} $  & $ -2.01 \times 10^{-7} + 0 i$  \\
\hline
\end{tabular}
\end{table}

\begin{table}[H]
\caption{Continuation of Table~\ref{CN2-z2}.
\label{CN3-z2}}
\centering
\begin{tabular}{|c|c|}
\hline
\hline
\multicolumn{2}{|c|}{$\sqrt{q^{2}}=500\ \text{GeV}$,\ $f=1 000\ \text{GeV}$, $F=4000\ \text{GeV}$}\\
\hline
\hline
$ W T^{5/3} T^{5/3} $  & $ -7.30 \times 10^{-7} + 0 i$  \\
\hline
$W' T^{5/3} T^{5/3}  $  & $ -2.79 \times 10^{-8} + 0\ i$  \\
\hline
$ H^{\pm} T^{5/3} T^{5/3} $  & $ 3.34 \times 10^{-9} +0\ i$  \\
\hline
$ W' BB $  & $ 1.19 \times 10^{-11} +0\ i$  \\
\hline
$H^{\pm} BB $  & $ -6.85 \times 10^{-9} + 0\ i$  \\
\hline
$W' W' b $  & $ 3.61 \times 10^{-6} + 0\ i$  \\
\hline
$W' W' B $  & $ 3.31 \times 10^{-6} + 0\ i$  \\
\hline
$\eta^{-} W' b $  & $ 6.75  \times 10^{-9} + 0\ i$  \\
\hline
$\phi^{-} W' b $  & $ 1.34 \times 10^{-6} + 0\ i$  \\
\hline
$ Z H_{0} t $  & $ -9.53  \times 10^{-8} + 0\ i$  \\
\hline
\end{tabular}
\end{table}

\vspace{5cm}

\section{AMDM and AWMDM of the top-quark at the $y_{2} < y_{3} $ scenario} \label{y2menor}

By the characteristics of the BLHM, its fermionic sector generates two study regions where the given expressions for the masses of the six heavy quarks are valid (see Eqs.~(\ref{mT})-(\ref{mB})), outside these regions the masses of the exotic $T$ and $T_5$ quarks are degenerate at lowest order. Thus, in Section~\ref{resultadosNum}, the AMDM and AWMDM of the top-quark in the region $y_2 > y_3$ were analyzed. To avoid redundancies, we study in this Appendix the AMDM and AWMDM of the top in the region $y_2 < y_3$. It is essential to mention that in our previous work~\cite{Cruz-Albaro:2022kty}, the AWMDM of the top in the $y_2 < y_3$ region has been extensively discussed. Some processes need to be considered for that study; therefore, we complete our analysis by providing the missing processes in this Appendix.

\subsection{AMDM of the top-quark }
\label{B1}

\begin{figure}[H]
\subfloat[]{\includegraphics[width=8.0cm]{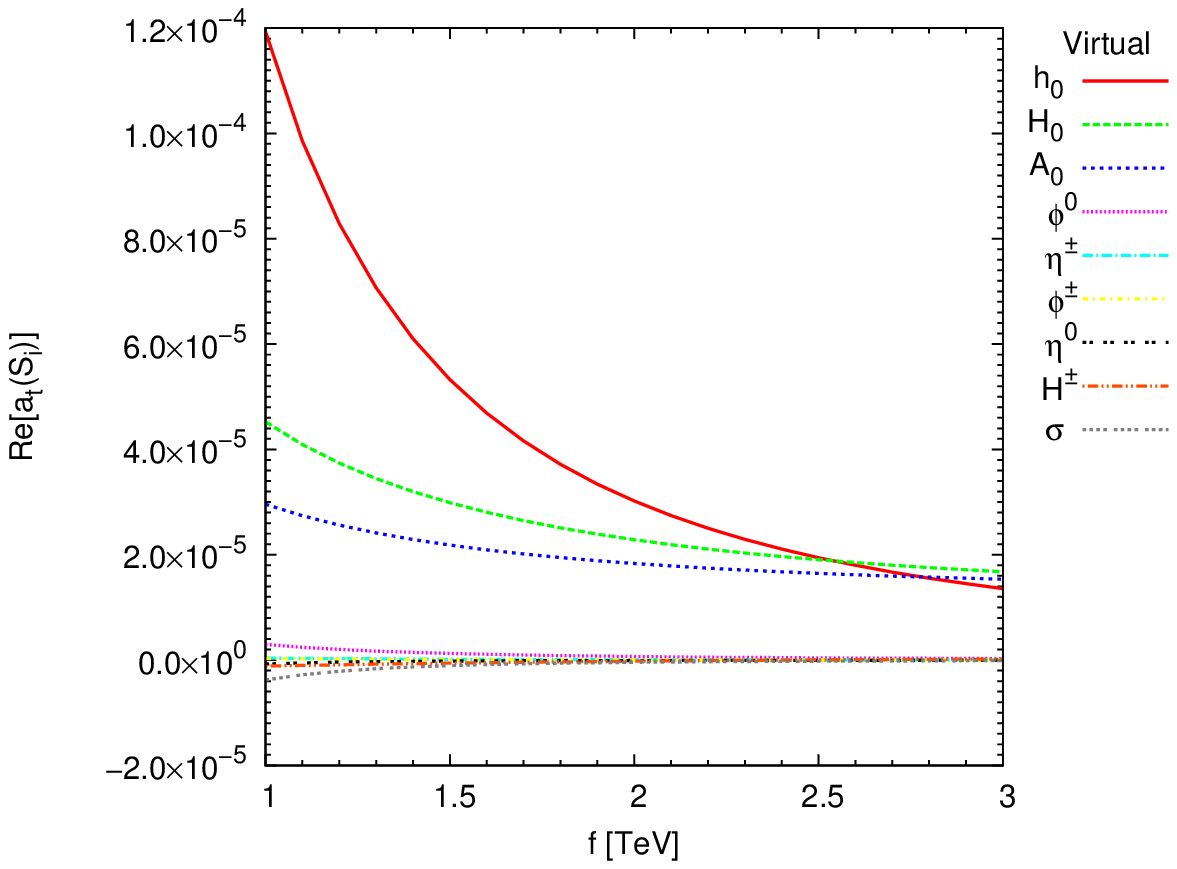}}
\subfloat[]{\includegraphics[width=8.0cm]{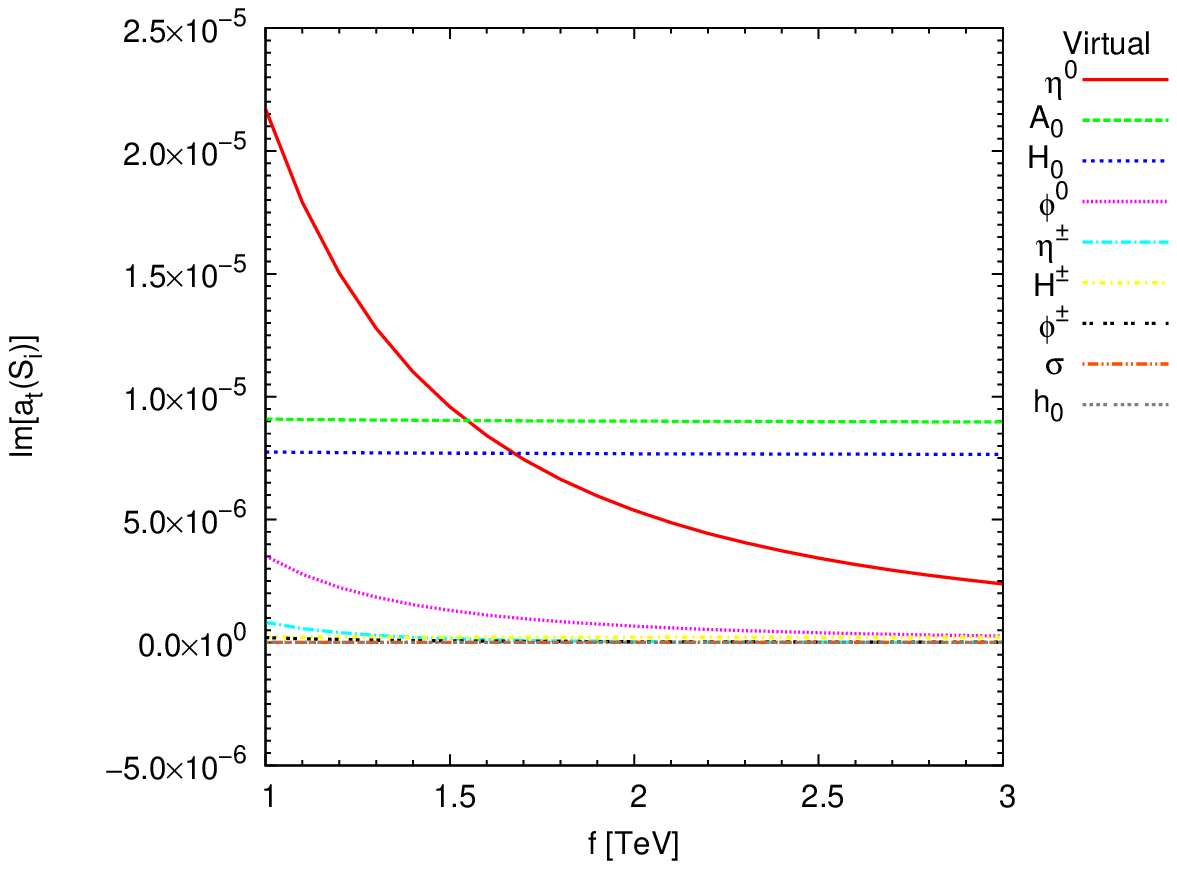}}
\caption{ \label{Si} Individual scalar contributions to $a_{t}$. The plots are obtained with the fixed value of  $F=4000\, \text{GeV}$. The values provided in Table~\ref{parametervalues} are used for the remaining model parameters.
 a) Re($a_{t}$). b) Im($a_{t}$).}
\end{figure}

The Feynman diagrams that contribute to the AMDM of the top-quark in the $y_2 < y_3$  scenario are the same Feynman diagrams as those shown in Fig.~\ref{dipolo}.
In this subsection, we discuss all new physics contributions arising from loops involving new particles, again, for study are cataloged according to the particles circulating in the corresponding vertex loop:
scalar contributions, vector contributions, and scalar-vector contributions.
We will first present the partial contributions due to scalars (see Fig.~\ref{dipolo}(a)).
Thus, in Fig.~\ref{Si}, we show all the individual contributions of the scalars that contribute to $ a_t $, which generate real and imaginary parts.
From Fig.~\ref{Si}(a), we can appreciate that the curve that provides the largest contributions in the whole analysis interval of the energy scale $ f $ are generated by the scalars $h_0$ and $H_0$, their numerical magnitudes are
$ \text{Re}[a_t (h_0)]=  [1.19 \times 10^{-4}, 1.88 \times 10^{-5}]$ for the interval $ f \in [1000, 2540] $ GeV
and
$ \text{Re}[a_t (H_0)]=  (1.88, 1.68] \times 10^{-5}$ when $f \in (2540,3000]$ GeV, respectively.
On the opposite side, the scalar $ \sigma $ generates the curve with the smallest contributions,
$ \text{Re}[a_t (\sigma)]= - [3.78 \times 10^{-6}, 1.25 \times 10^{-7}]$.
Following Fig.~\ref{Si}(b), we observe that the curves generated by the scalars $\eta^{0}$ and $A_0$ limit the curve that provides the large contributions in the scale interval $ f $, that is,
$ \text{Im}[a_t (\eta^{0})]=  [2.17 \times 10^{-5}, 9.09 \times 10^{-6}]$ when $f=[1000,1540]$ GeV and
$ \text{Im}[a_t (A_{0})]= (9.09, 8.98] \times 10^{-6}$ for $f=(1540,3000]$ GeV.
The curve with the menor contributions is given by the scalar $h_0$, $ \text{Im}[a_t (h_{0})]= 0$.

\begin{figure}[H]
\subfloat[]{\includegraphics[width=8.0cm]{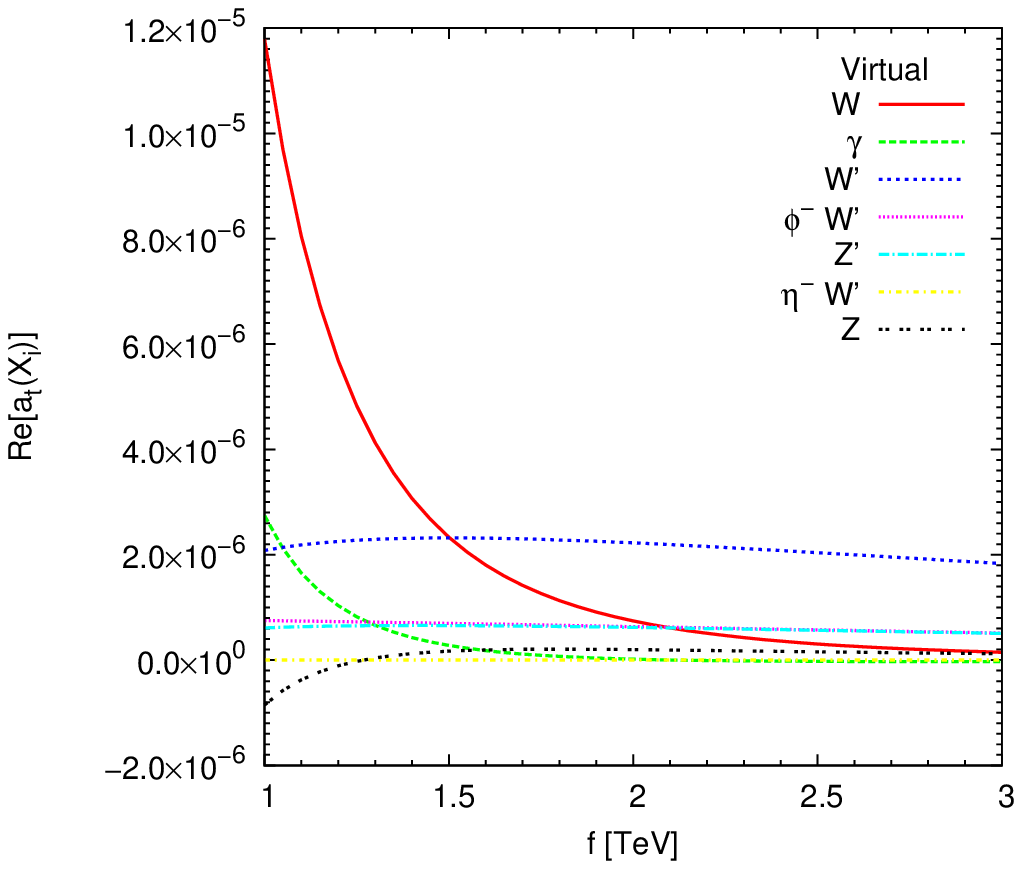}}
\subfloat[]{\includegraphics[width=8.0cm]{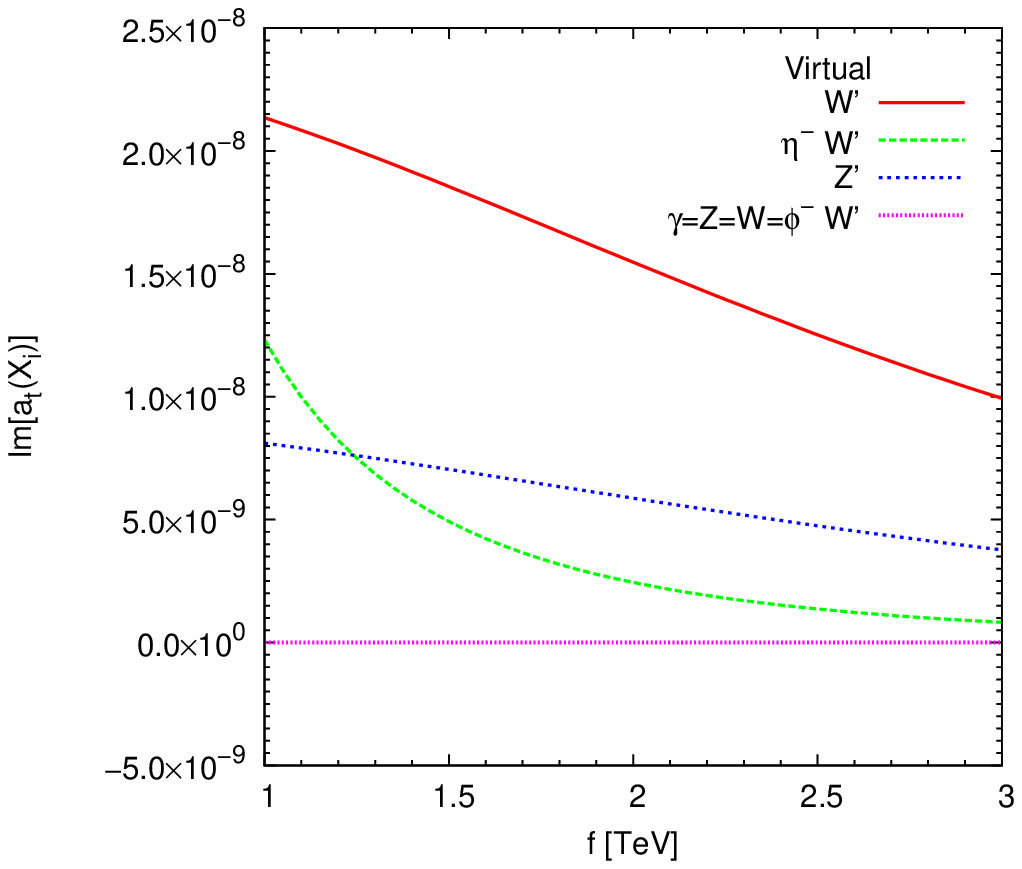}}
\caption{ \label{Vi} Individual vector and scalar-vector contributions to $a_{t}$. The plots are obtained with the fixed value of  $F=4000\, \text{GeV}$. The values provided in Table~\ref{parametervalues} are used for the remaining model parameters
 a) Re($a_{t}$). b) Im($a_{t}$).}
\end{figure}

Regarding vector and scalar-vector contributions, these contributions arise from the Feynman diagrams depicted in Figs.~\ref{dipolo}(b)-\ref{dipolo}(e).
 As in Subsubsection~\ref{AMDMk1}, some Feynman diagrams are cancelled out due to the choice of the $\tan \theta_g=1$ scenario. Therefore, in Figs.~\ref{Vi}(a) and~\ref{Vi}(b), we plot the contributions from the Feynman diagrams that give non-zero contributions.
Fig.~\ref{Vi}(a) shows the real part of the partial contributions to $ a_t $ due to scalar and scalar-vector. From this figure, we find that two curves generate the most significant contributions in the allowed space of scale $ f $. Such contributions arise due to the vector bosons $W$ and $W'$ which play the role of mediator particles in the $\gamma \bar{t}t$ vertex loops:
$ \text{Re}[a_t (W)]=  [1.18 \times 10^{-5}, 2.33 \times 10^{-6}]$ when $ f \in [1000, 1500] $ GeV and
$ \text{Re}[a_t (W')]=  (2.33, 1.83] \times 10^{-6}$ for the interval $ f= (1500, 3000] $ GeV.
The curve that gives the null contributions arises for $\eta^{-}W'^{+}  $.
 With respect to the imaginary part of $ a_t $, Fig.~\ref{Vi}(b), the main curve is again generated by the vector boson $ W' $:
$ \text{Im}[a_t (W')]=  [2.13 \times 10^{-8}, 9.93 \times 10^{-9}]$ for the interval $ f= [1000, 3000] $ GeV.
The null curve is generated by the $ \gamma $, $ Z $, $ W $ and $ \phi^{-}W'^{+} $ particles.
For more information, in Tables~\ref{parcial1}-\ref{parcial3}, we provide the numerical magnitudes of all partial contributions to $ a_t $.

\begin{figure}[H]
\subfloat[]{\includegraphics[width=8.0cm]{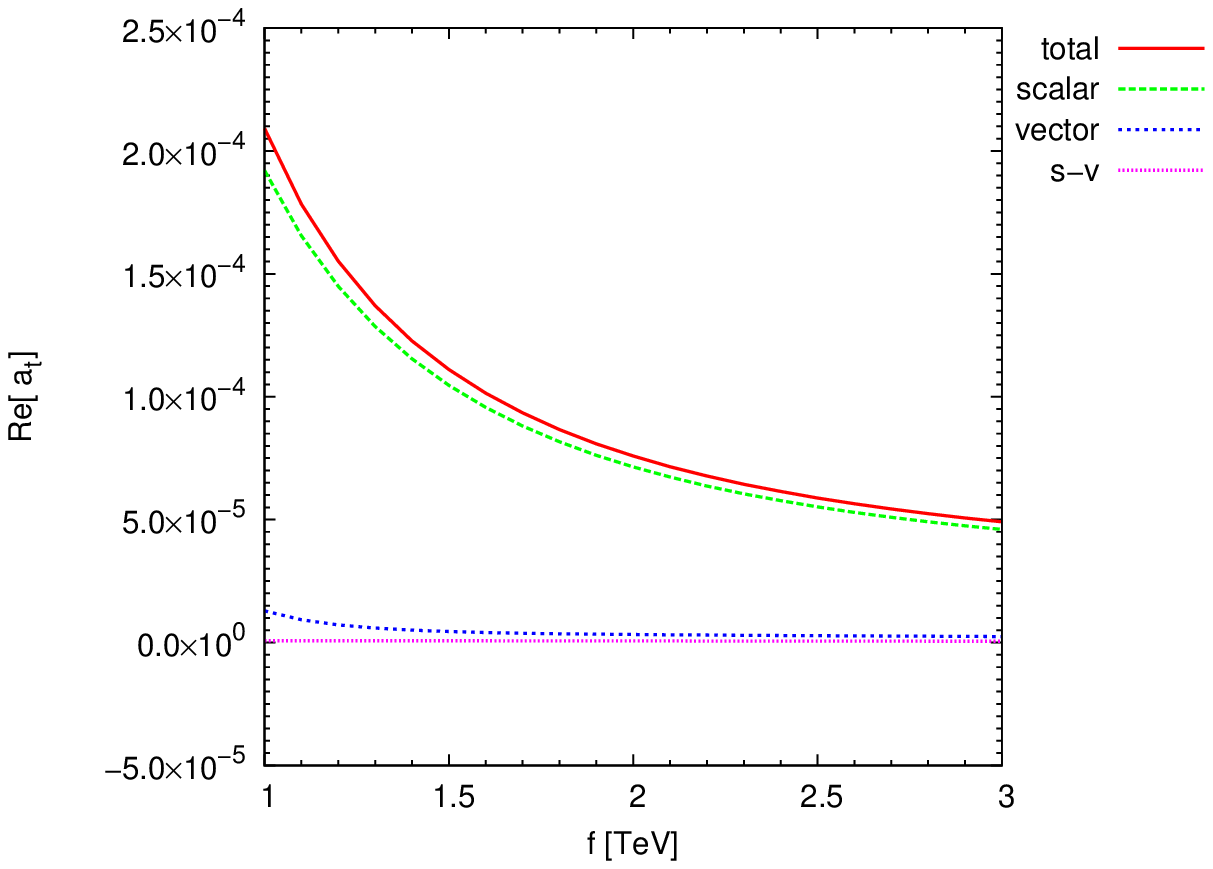}}
\subfloat[]{\includegraphics[width=8.0cm]{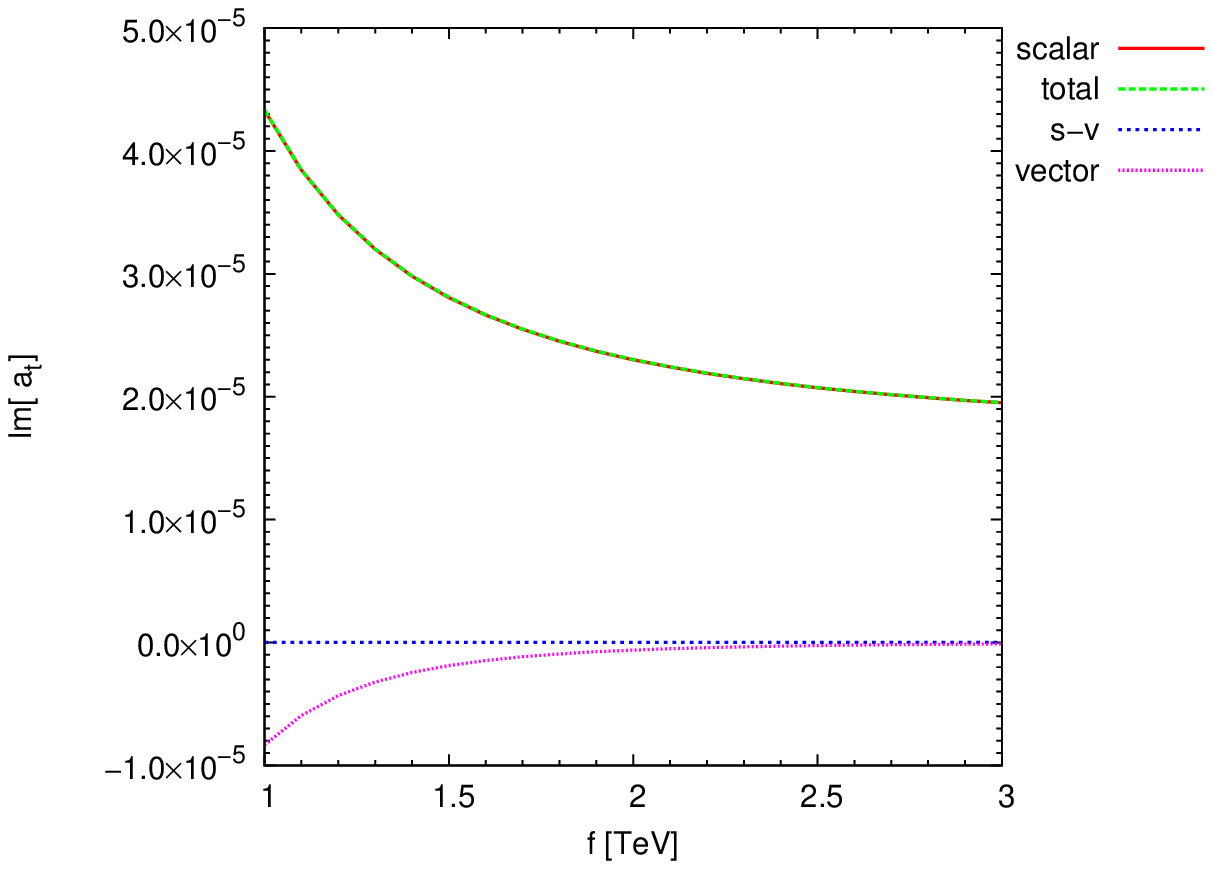}}
\caption{ \label{FS} Scalar, vector, scalar-vector and total contributions  to $a_{t}$. The plots are obtained with the fixed value of  $F=4000\, \text{GeV}$. The values provided in Table~\ref{parametervalues} are used for the remaining model parameters.
 a) Re($a_{t}$). b) Im($a_{t}$).}
\end{figure}

We show in Fig.~\ref{FS} the different curves representing the sum of all the partial contributions of each sector: the scalar, vector, and scalar-vector contributions. The sums of these contributions, in turn, generate the total contribution to $ a_t $.
We first discuss the real contributions to $ a_t $ represented in Fig.~\ref{FS}(a). In this figure, we can appreciate that the curve that provides the significant contributions to the total contribution comes from the contributions due to the scalars. The vector and scalar-vector contributions contribute in smaller proportion to the total contribution:
$\text{Re}[a_t(\text{scalar})]= [1.92 \times 10^{-4}, 4.60 \times 10^{-5}]$,
$\text{Re}[a_t(\text{vector})]= [1.29 \times 10^{-5}, 2.42 \times 10^{-6}]$,
$\text{Re}[a_t(\text{s-v})]= [7.51, 5.12 ] \times 10^{-7}$ and
$\text{Re}[a_t(\text{total })]= [2.09 \times 10^{-4}, 4.91 \times 10^{-5}]$.
We secondly analyze the imaginary contributions to $ a_t $ produced by the different sectors involved. From Fig.~\ref{FS}(b), we observe that the curve with the largest contributions comes from the scalar contributions. Therefore, this sector contributes the largest proportion to the total contribution, whereas the vector and scalar-vector contributions are rather suppressed. The numerical estimates obtained for each sector are
$\text{Im}[a_t(\text{scalar})]= [4.33, 1.95] \times 10^{-5}$,
$\text{Im}[a_t(\text{vector})]= - [8.35 \times 10^{-6}, 1.17 \times 10^{-7}]$,
$\text{Im}[a_t(\text{s-v})]= [1.23 \times 10^{-8}, 8.21  \times 10^{-10}]$ and
$\text{Im}[a_t(\text{total })]=  [4.33, 1.95] \times 10^{-5}$.

\begin{figure}[H]
\subfloat[]{\includegraphics[width=8.0cm]{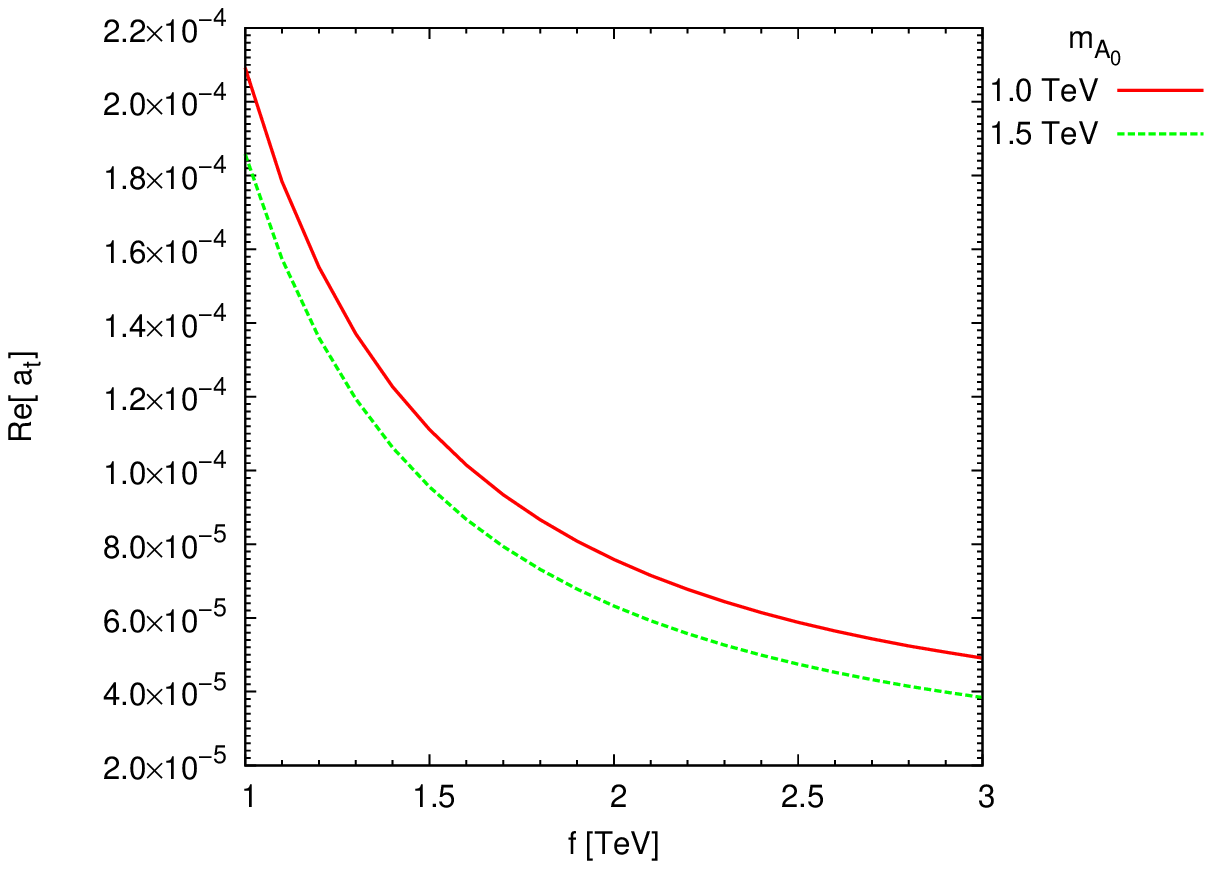}}
\subfloat[]{\includegraphics[width=8.0cm]{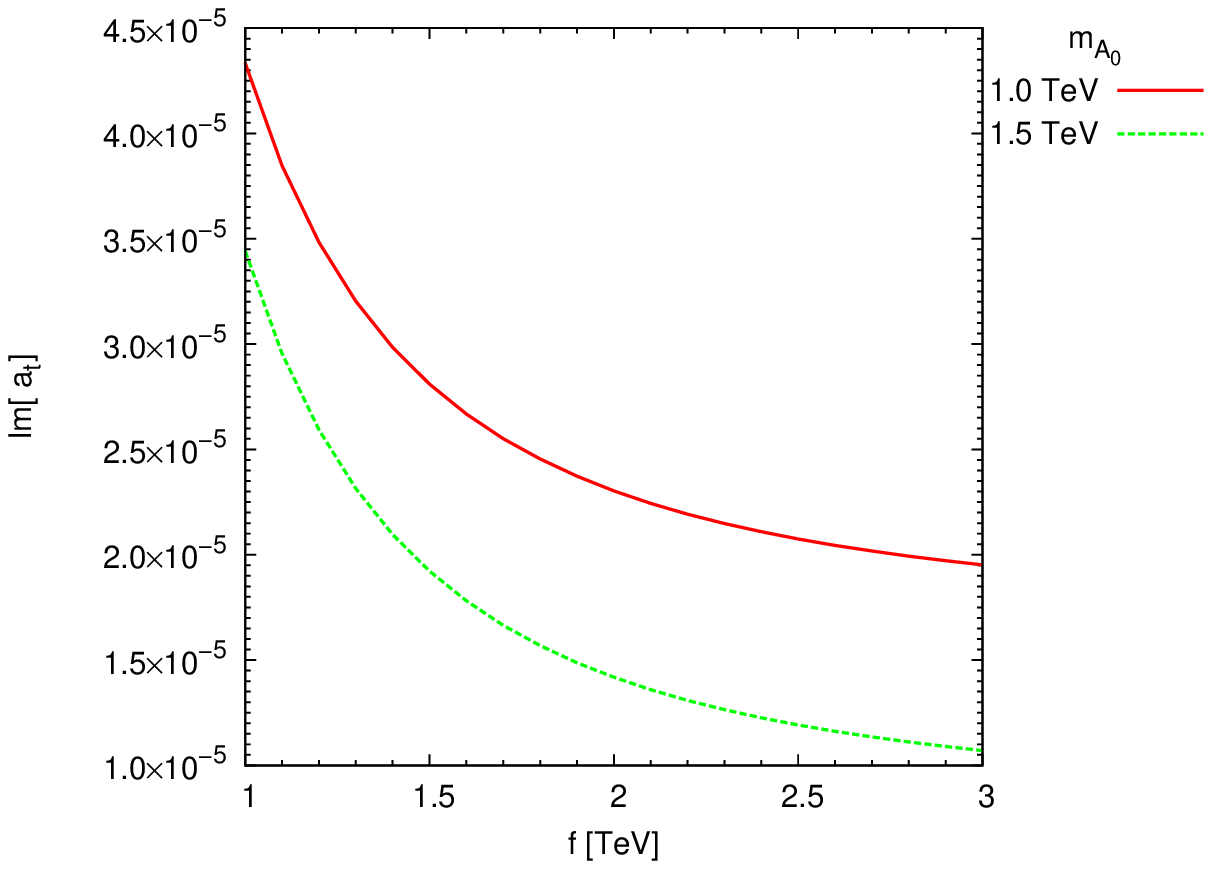}}
\caption{ \label{FmA-gy2menor} Total contribution to $a_{t}$  for different values of the mass of $A_0$. The plots are obtained with the fixed value of  $F=4000\, \text{GeV}$. The values provided in Table~\ref{parametervalues} are used for the remaining model parameters.
a) Re($a_{t}$). b) Im($a_{t}$).}
\end{figure}

\begin{figure}[H]
\subfloat[]{\includegraphics[width=8.0cm]{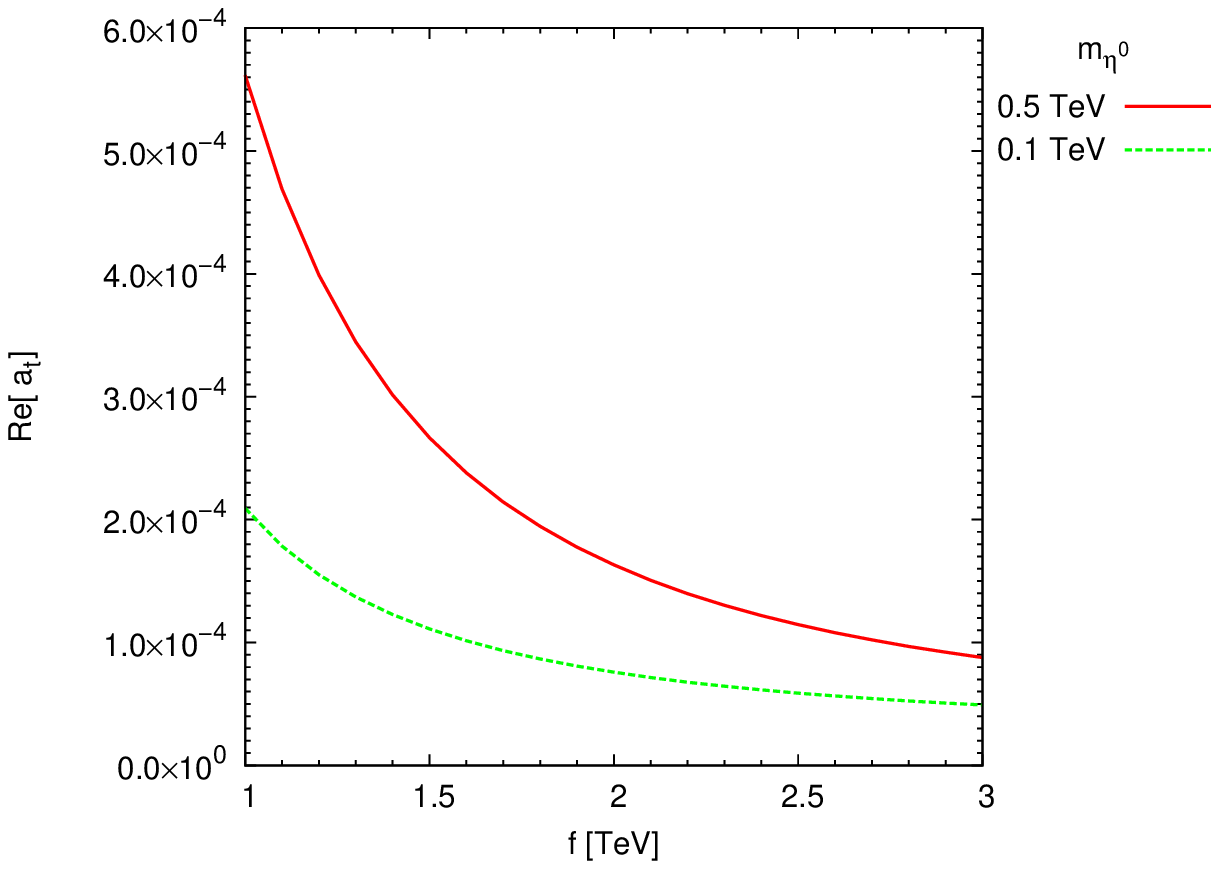}}
\subfloat[]{\includegraphics[width=8.0cm]{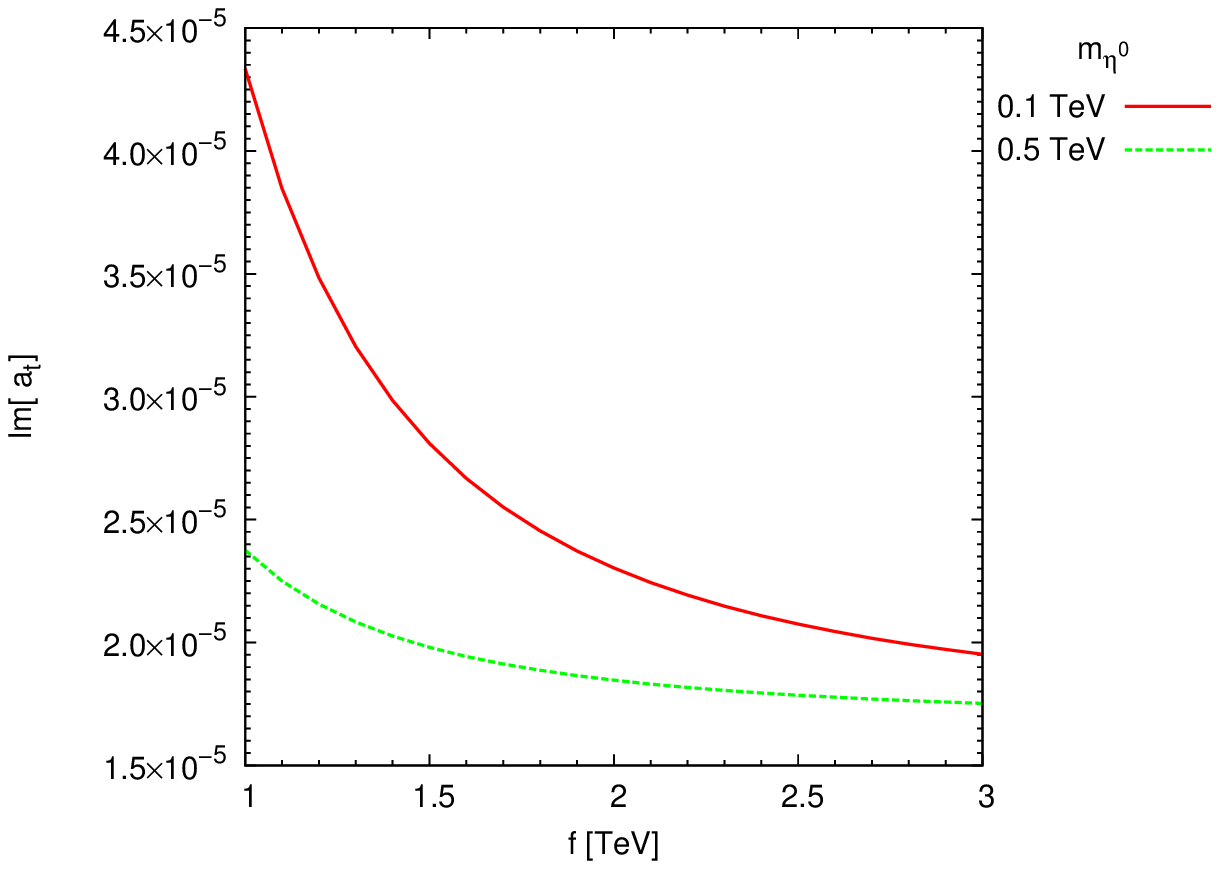}}
\caption{ \label{Fmeta-gy2menor} Total contribution to $a_{t}$  for different values of the mass of $\eta^{0}$. The plots are obtained with the fixed value of  $F=4000\, \text{GeV}$. The values provided in Table~\ref{parametervalues} are used for the remaining model parameters. a) Re($a_{t}$). b) Im($a_{t}$).}
\end{figure}

We now examine the behavior of $ a_t $ vs. $ f $ for specific choices in the input values of the parameters $ m_{A_0} $ and $m_{\eta^{0}}$.
In this way, in Figs.~\ref{FmA-gy2menor} and~\ref{Fmeta-gy2menor} the curves are generated when $ m_{A_0} $ takes values of 1000 and 1500 GeV, while for  $m_{\eta^{0}}$ they acquire values of 100 and 500 GeV.
From Fig.~\ref{FmA-gy2menor}, we can see the contributions to $ a_t $, which present real and imaginary parts. For the real contributions to $ a_t $, both curves acquire values of the order of $ 10^{-4}-10^{-5} $ in the whole study interval of the $ f $ scale. The imaginary part of $ a_t $ also the two curves involved provide contributions of the order of $ 10^{-5} $.
In these cases, the curves that provide slightly more significant contributions are reached when $ m_{A_0} =1000$ GeV.
Concerning Fig.~\ref{Fmeta-gy2menor}, the real contributions have a decrease of about one order of magnitude as $ f $ increases up to 3000 GeV, i.e., $ 10^{-4}-10^{-5} $. On the other hand, the imaginary contributions take values of the order of $ 10^{-5} $ in the allowed $ f $ space.
Significant contributions arise for the real part of $ a_t $ for large values of $m_{\eta^{0}}$, while the imaginary part does so for smaller values of $m_{\eta^{0}}$.
 The different scenarios studied show us that $ a_t $ depends on the choice of the values of the parameters $ m_{A_0} $ and $m_{\eta^{0}}$, as well as on the energy scale $ f $ (see Tables~\ref{GEta100}-\ref{mA1500}).

\begin{figure}[H]
\subfloat[]{\includegraphics[width=8.0cm]{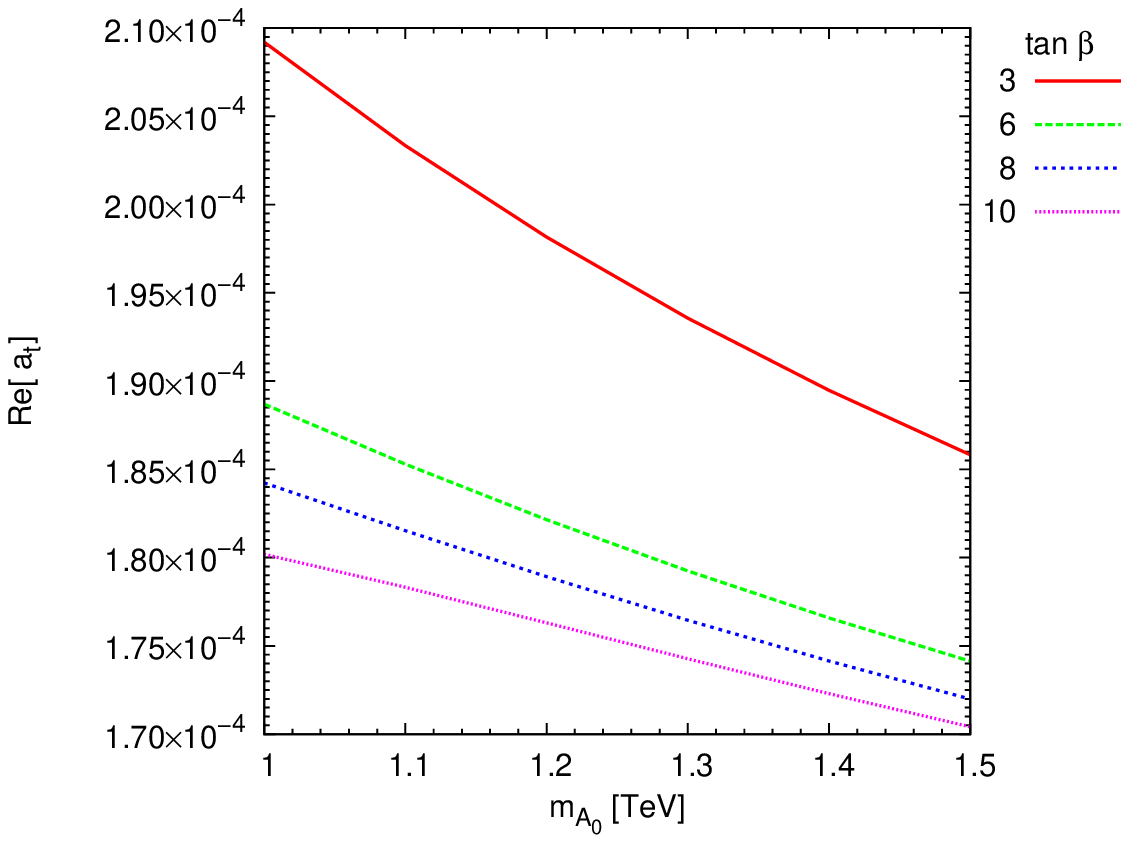}}
\subfloat[]{\includegraphics[width=8.0cm]{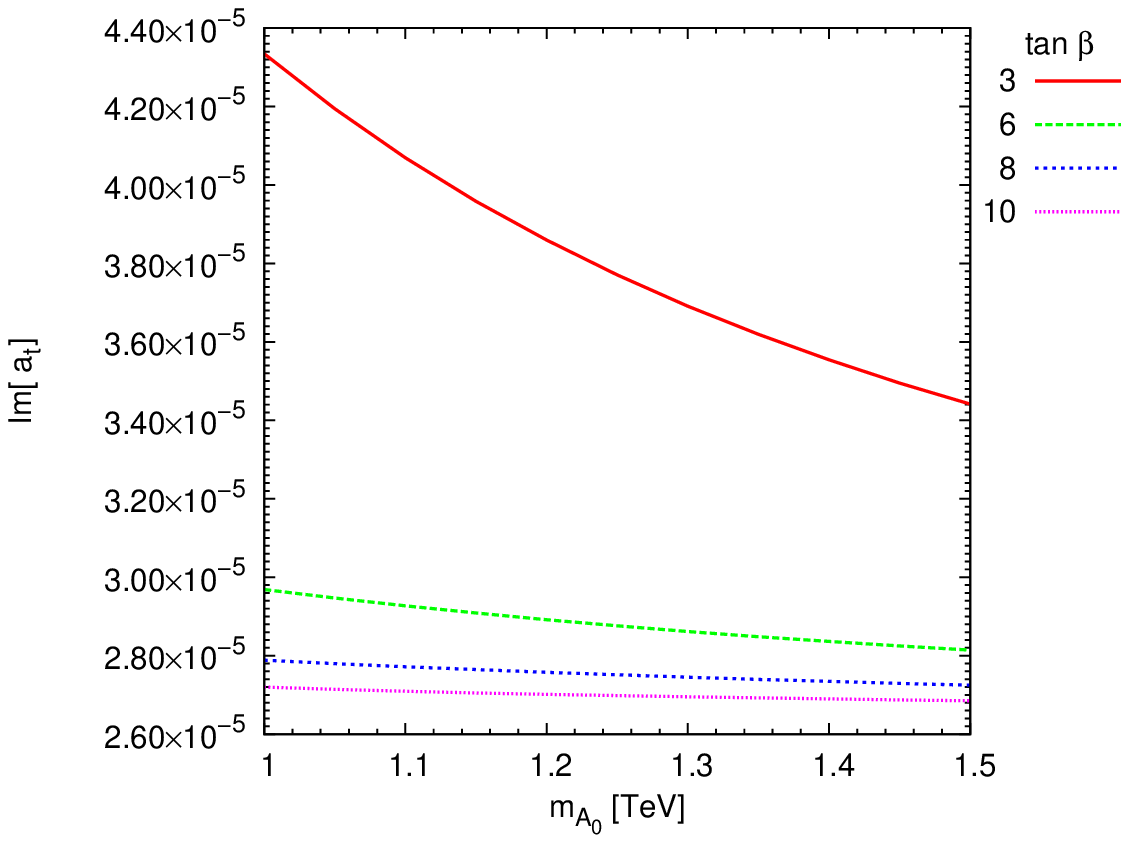}}
\caption{ \label{Ftan-y2menor-gamma} Total contribution to $a_{t}$ for different values of $\tan \beta$. The plots are obtained with the fixed values of  $f=1000\, \text{GeV}$ and $F=4000\, \text{GeV}$. The values provided in Table~\ref{parametervalues} are used for the remaining model parameters.
a) Re($a_{t}$). b) Im($a_{t}$).}
\end{figure}

We now discuss the AMDM $ a_t $ of the top-quark as a function of the mass of the scalar $ A_0 $, different choices in the input values of the parameter $\tan \beta$ generate the different curves shown in Figs.~\ref{Ftan-y2menor-gamma}(a) and \ref{Ftan-y2menor-gamma}(b).
Our selection of values for $\tan \beta=3,6,8$ and $10$ are consistent with the values generated and allowed by Eq.~(\ref{cotabeta}).
In the left plot of Fig.~\ref{Ftan-y2menor-gamma}, the strongest contributions to the real part of $ a_t $ arise when $\tan \beta=3$, $ \text{Re}[a_t ]= [2.09, 1.86] \times 10^{-4} $. The other curves provide weaker contributions, although these values remain in the same order of magnitude as the main curve.
With the right plot of Fig.~\ref{Ftan-y2menor-gamma}, as happened with the real part of $ a_t $, all the generated curves acquire values of the same order of magnitude as the curve that provides the largest contributions,  $ \text{Re}[a_t (\tan \beta=3)]= [4.33, 3.44] \times 10^{-5} $.
There is a strong dependence of $ a_t $ on $ m_{A_0} $ since the values acquired by the AMDM of the top-quark decrease rapidly as $ m_{A_0} $ increases up to 1500 GeV. Concerning the selected input values for $\tan \beta$, larger values of $\tan \beta$ produce weaker contributions to $ a_t $.

\begin{figure}[H]
\subfloat[]{\includegraphics[width=8.0cm]{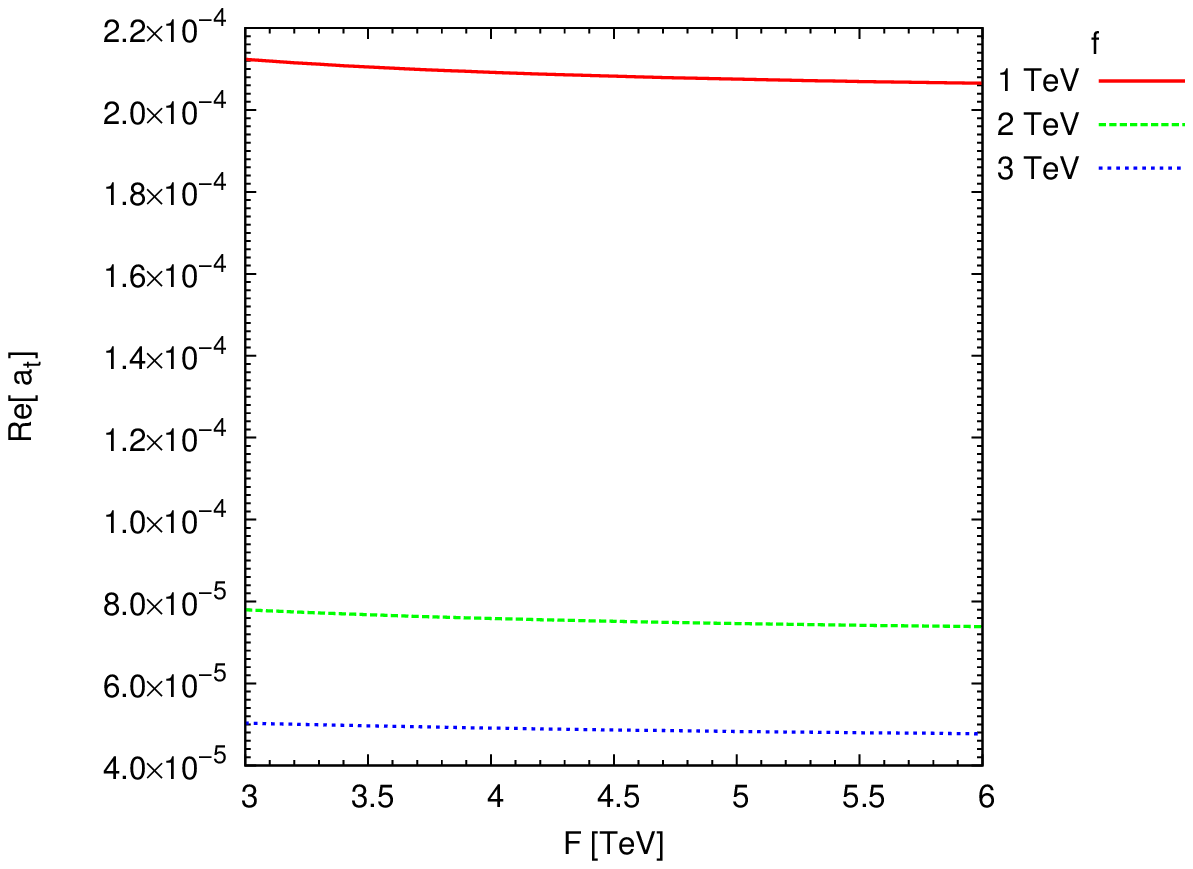}}
\subfloat[]{\includegraphics[width=8.0cm]{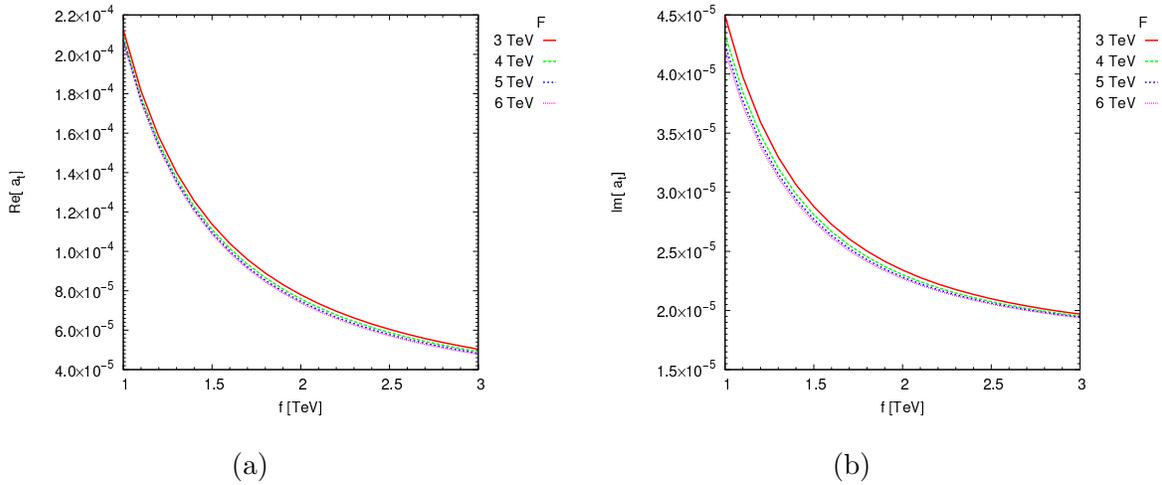}}
\caption{ \label{Ffi-y2menor-gamma} Total contribution to $a_{t}$ for different values of the energy scale $f$.  The values provided in Table~\ref{parametervalues} are used for the remaining model parameters. a) Re($a_{t}$). b) Im($a_{t}$).}
\end{figure}

We continue analyzing the dependence of $ a_t $ on the energy scales $ F $ and $ f $. We first discuss the behavior of $ a_t $ as a function of $ F $ in the interval from 3000 to 6000 GeV, as shown in Fig.~\ref{Ffi-y2menor-gamma}.
The curves are generated for some fixed values of the $ f $ scale, i.e., 1000, 2000, and 3000 GeV. The numerical evaluation shows that for $f=1000$ GeV, we obtain the curves with the largest contributions to the real and imaginary part of $ a_t $ in the analysis range of the $ F $ scale:
$\text{Re}[a_t(f=1000\ \text{GeV})]= [2.12, 2.06] \times 10^{-4}$
and
$\text{Im}[a_t(f=1000\ \text{GeV})]= [4.49, 4.19] \times 10^{-5}$.
More suppressed contributions are obtained for larger values of the $ f $ scale.
As for the remaining curves, these acquire numerical values of the order of magnitude of $ 10^{-5} $; this happens for both the real and the imaginary part of $ a_t $.
According to Figs.~\ref{Ffi-y2menor-gamma}(a) and~\ref{Ffi-y2menor-gamma}(b), we can observe that $ a_t $ shows a weak dependence on the $ F $ scale since the generated curves show small changes in the whole space of analysis allowed for $ F $.

\begin{figure}[H]
\subfloat[]{\includegraphics[width=8.0cm]{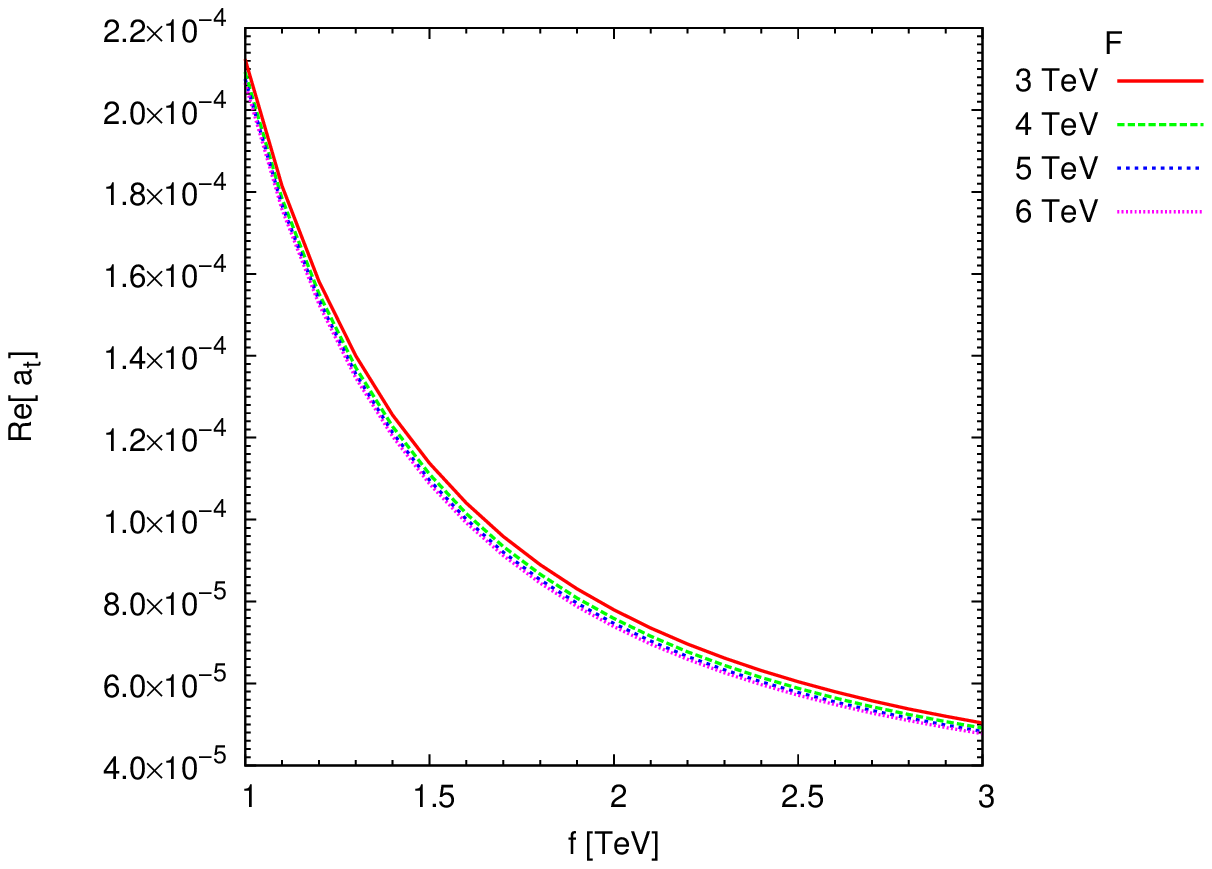}}
\subfloat[]{\includegraphics[width=8.0cm]{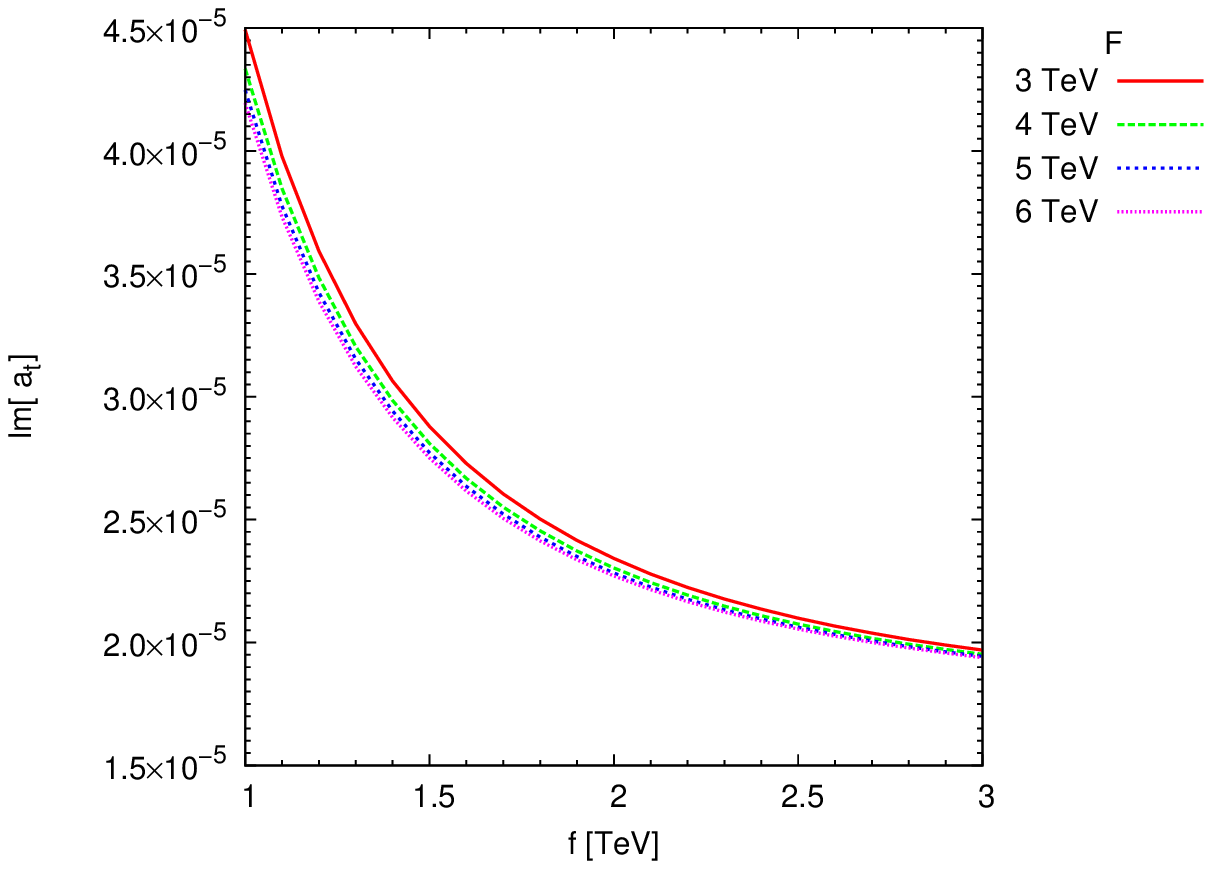}}
\caption{ \label{FFi-y2menor-gamma} Total contribution to $a_{t}$ for different values of the energy scale $F$.  The values provided in Table~\ref{parametervalues} are used for the remaining model parameters. a) Re($a_{t}$). b) Im($a_{t}$).}
\end{figure}

We now discuss secondly the behavior of  $a_t$ vs. $ f $ as shown in Figs.~\ref{FFi-y2menor-gamma}(a) and~\ref{FFi-y2menor-gamma}(b). For the $ F $ scale, we assign fixed values such as 3000, 4000, 5000, and 6000 GeV.
With these input values for $ F $, we generate the curves shown in the abovementioned figures.
Based on the corresponding figures and numerical estimates, we obtain that the largest contributions to the real and imaginary part of $ a_t $ are found when $F=3000$ GeV:
$\text{Re}[a_t(F=3000\ \text{GeV})]= [2.12 \times 10^{-4}, 5.03 \times 10^{-5}]$
and
$\text{Im}[a_t(F=3	000\ \text{GeV})]= [4.49, 1.97] \times 10^{-5}$.
The other curves generate slightly more suppressed contributions than the main contributions for larger values of the $ F $ energy scale.
 In this case, the $ a_t $ function shows a higher dependence on the $ f $ energy scale than the $ F $ scale. This affirmation is best appreciated in the behavior of the rapidly decreasing curves as $ f $ increases.

\begin{table}[H]
\caption{The magnitude of the partial contributions to $a_{t}$ of the BLHM.
The data are obtained by fixing the $f$ and $F$ scales, $f=1000$ GeV and $F=4000$ GeV.
The values provided in Table~\ref{parametervalues} are used for the rest of the model parameters.
{\bf abc} denotes the different particles running in the loop of the vertex $\gamma t t$.
\label{parcial1}}
\centering
\begin{tabular}{|c|c|}
\hline
\hline
\multicolumn{2}{|c|}{$\sqrt{q^{2}}=500\ \text{GeV}$,\  $f=1 000\ \text{GeV}$, $F=4000\ \text{GeV}$}\\
\hline
 $\text{Couplings}\  \textbf{abc}$  & $\left(a_{t} \right)^{ \textbf{abc} } $  \\
\hline
\hline
$Z' tt $  & $ 7_\cdot 56 \times 10^{-7} + 8_\cdot 11 \times 10^{-9}\ i$  \\
\hline
$W'bb $   & $ 7.74 \times 10^{-7}+ 2.13 \times 10^{-8}\ i $ \\
\hline
$  Z T T$ & $2.51 \times 10^{-7} + 0\ i$ \\
\hline
$  \gamma T T$ & $-5.07 \times 10^{-7} + 0\ i$ \\
\hline
$  Z' T T$ & $-3.30 \times 10^{-8} + 0\ i$ \\
\hline
$  Z T_5 T_5$ & $-1.81 \times 10^{-7} + 0\ i$ \\
\hline
$  \gamma T_5 T_5$ & $-8.21 \times 10^{-7} + 0\ i$ \\
\hline
$ Z' T_5 T_5$ & $-9.03\times 10^{-12} + 0\ i$ \\
\hline
$  Z T_6 T_6$ & $-1.46 \times 10^{-7} + 0\ i$ \\
\hline
$  \gamma T_6 T_6$ & $-2.87 \times 10^{-9} + 0\ i$ \\
\hline
$  Z' T_6 T_6$ & $-3.78\times 10^{-9} + 0\ i$ \\
\hline
$  Z T^{2/3} T^{2/3}$ & $-7.75 \times 10^{-7} + 0\ i$ \\
\hline
$  \gamma T^{2/3} T^{2/3}$ & $ 4.07 \times 10^{-6} + 0\ i$ \\
\hline
$  Z' T^{2/3} T^{2/3}$ & $- 1.10 \times 10^{-7} + 0\ i$ \\
\hline
$  W T^{5/3} T^{5/3}$ & $ 1.18 \times 10^{-5} + 0\ i$ \\
\hline
$  W' T^{5/3} T^{5/3}$ & $ -4.50 \times 10^{-7} + 0\ i$ \\
\hline
$  W' B B$ & $ -3.21 \times 10^{-7} + 0\ i$ \\
\hline
$  W' W' b$ & $ 1.98  \times 10^{-6} + 0\ i$ \\
\hline
$  W' W' B$ & $ 1.01 \times 10^{-7} + 0\ i$ \\
\hline
$\sigma tt $  & $ 1_\cdot 49\times 10^{-10} + 6_\cdot 47 \times 10^{-11}\ i$  \\
\hline
$A_{0} tt   $  & $ 1_\cdot 26 \times 10^{-5} +  9_\cdot 09\times 10^{-6}\ i$  \\
\hline
\end{tabular}
\end{table}

\begin{table}[H]
\caption{Continuation of Table~\ref{parcial1}.}
\centering
\label{parcial2}
\begin{tabular}{|c|c|}
\hline
\hline
\multicolumn{2}{|c|}{$\sqrt{q^{2}}=500\ \text{GeV}$,\  $f=1 000\ \text{GeV}$, $F=4000\ \text{GeV}$}\\
\hline
$H_{0} tt $  & $ 1_\cdot 08\times 10^{-5} + 7_\cdot 75\times 10^{-6}\ i$  \\
\hline
$\eta^{0} tt  $  & $ -2_\cdot 34 \times 10^{-7} + 2_\cdot 17 \times 10^{-5}\ i$  \\
\hline
$\phi^{0} tt $  & $ 3_\cdot 28 \times 10^{-6} + 3_\cdot 52 \times 10^{-6}\ i$  \\
\hline
$H^{\pm} bb  $  & $ 7_\cdot 81\times 10^{-7} + 2_\cdot 16\times 10^{-7}\ i$  \\
\hline
$\eta^{\pm} bb   $  & $ 3_\cdot 71 \times 10^{-7} + 8_\cdot 21 \times 10^{-7}\ i$  \\
\hline
$\phi^{\pm} bb  $  & $ 2_\cdot 81 \times 10^{-7} + 2_\cdot 00 \times 10^{-7}\ i$  \\
\hline
$\sigma TT  $  & $ -8_\cdot 48 \times 10^{-9} + 0\ i$  \\
\hline
$h_{0} TT  $  & $ 2_\cdot 78 \times 10^{-5} + 0\ i$  \\
\hline
$H_0 TT  $  & $ 1_\cdot 47 \times 10^{-6} + 0\ i$  \\
\hline
$A_0 TT  $  & $ -2_\cdot 06 \times 10^{-6} + 0\ i$  \\
\hline
$\phi^{0} TT  $  & $ -2_\cdot 89 \times 10^{-7} + 0\ i$  \\
\hline
$\eta^{0} TT  $  & $ -5_\cdot 01 \times 10^{-7} + 0\ i$  \\
\hline
$\sigma T_5 T_5  $  & $ -1_\cdot 13 \times 10^{-8} + 0\ i$  \\
\hline
$h_0 T_5 T_5  $  & $ -3_\cdot 80 \times 10^{-7} + 0\ i$  \\
\hline
$H_0 T_5 T_5  $  & $ -2_\cdot 42 \times 10^{-8} + 0\ i$  \\
\hline
$A_0 T_5 T_5  $  & $ 2_\cdot 24 \times 10^{-8} + 0\ i$  \\
\hline
$\phi^{0} T_5 T_5  $  & $ -2_\cdot 03 \times 10^{-11} + 0\ i$  \\
\hline
$\eta^{0} T_5 T_5  $  & $ -2_\cdot 73 \times 10^{-11} + 0\ i$  \\
\hline
$  \sigma T_6 T_6$ & $-3.75 \times 10^{-6} + 0\ i$ \\
\hline
$h_0 T_6 T_6  $  & $ 1_\cdot 62 \times 10^{-5} + 0\ i$  \\
\hline
$H_0 T_6 T_6  $  & $ 3_\cdot 16 \times 10^{-5} + 0\ i$  \\
\hline
$A_0 T_6 T_6  $  & $ 1_\cdot 28 \times 10^{-5} + 0\ i$  \\
\hline
$\phi^{0} T_6 T_6  $  & $ -5_\cdot 46 \times 10^{-9} + 0\ i$  \\
\hline
$\eta^{0} T_6 T_6  $  & $ -1_\cdot 00 \times 10^{-8} + 0\ i$  \\
\hline
$  \sigma T^{2/3} T^{2/3}$ & $- 3.36 \times 10^{-9} + 0\ i$ \\
\hline
$ h_0 T^{2/3} T^{2/3}$ & $ 7.57 \times 10^{-5} + 0\ i$ \\
\hline
$ H_0 T^{2/3} T^{2/3}$ & $ 1.40 \times 10^{-6} + 0\ i$ \\
\hline
\end{tabular}
\end{table}

\begin{table}[H]
\caption{Continuation of Table~\ref{parcial2}.}
\centering
\label{parcial3}
\begin{tabular}{|c|c|}
\hline
\hline
\multicolumn{2}{|c|}{$\sqrt{q^{2}}=500\ \text{GeV}$,\  $f=1 000\ \text{GeV}$, $F=4000\ \text{GeV}$}\\
\hline
$ A_0 T^{2/3} T^{2/3}$ & $ 6.32 \times 10^{-6} + 0\ i$ \\
\hline
$ \phi^{0} T^{2/3} T^{2/3}$ & $ -6.52 \times 10^{-9} + 0\ i$ \\
\hline
$ \eta^{0} T^{2/3} T^{2/3}$ & $ -1.20 \times 10^{-8} + 0\ i$ \\
\hline
$ H^{\pm} T^{5/3} T^{5/3}$ & $ -2.09 \times 10^{-6} + 0\ i$ \\
\hline
$ H^{\pm} BB $ & $ 1.58 \times 10^{-7} + 0\ i$ \\
\hline
$ \eta^{-} W' b $ & $ 0+ 1.23 \times 10^{-8}\ i$ \\
\hline
$ \phi^{-} W' b $ & $ 7.51 \times 10^{-7}+ 0\ i$ \\
\hline
\end{tabular}
\end{table}

\begin{table}[H]
\caption{Expected sensitivity limits on the $a_t$ in the context of the BLHM with $\sqrt{q^{2}}=500\ \text{GeV}$,
\ $\text{m}_{A_0}=1000\ \text{GeV}$, $\text{m}_{\eta_0}=100\ \text{GeV} $, $F=4000\ \text{GeV}$ and $f=1, 1.5, 2, 2.5, 3\ \text{TeV}$
are represented. All new contributions are considered, scalar bosons, vector bosons, scalar-vector, and heavy quarks.
\label{GEta100}}
\centering
\begin{tabular}{|c|c|}
\hline
\hline
\multicolumn{2}{|c|}{$\sqrt{q^{2}}=500\ \text{GeV}$,\ $\text{m}_{A_0}=1000\ \text{GeV}$, $\text{m}_{\eta_0}=\bf{ 100}\ \text{GeV} $,\ $F=4000\ \text{GeV}$}\\
\hline
 $f\ [\text{TeV}]$  & $(a_{t})^{\rm total} $  \\
\hline
\hline
$1_\cdot0$  & $ 2.09 \times 10^{-4} + 4.33  \times 10^{-5}\ i $  \\
\hline
$1_\cdot5 $  & $ 1.11 \times 10^{-4} + 2.81 \times 10^{-5}\ i$  \\
\hline
$2_\cdot0 $  & $ 7.59 \times 10^{-5} + 2.30 \times 10^{-5}\ i$  \\
\hline
$2_\cdot5 $  & $  5.88 \times 10^{-5} + 2.07  \times 10^{-5}\ i$  \\
\hline
$3_\cdot0 $  & $  4.91 \times 10^{-5} + 1.95 \times 10^{-5}\ i$  \\
\hline
\end{tabular}
\end{table}

\begin{table}[H]
\caption{Expected sensitivity limits on the $a_t$ in the context of the BLHM with $\sqrt{q^{2}}=500\ \text{GeV}$,
\ $\text{m}_{A_0}=1000\ \text{GeV}$, $\text{m}_{\eta_0}=500\ \text{GeV} $, $F=4000\ \text{GeV}$ and $f=1, 1.5, 2, 2.5, 3\ \text{TeV}$
are represented. All new contributions are considered, scalar bosons, vector bosons, scalar-vector, and heavy quarks.
\label{GEta500}}
\centering
\begin{tabular}{|c|c|}
\hline
\hline
\multicolumn{2}{|c|}{$\sqrt{q^{2}}=500\ \text{GeV}$,\ $\text{m}_{A_0}=1000\ \text{GeV}$, $\text{m}_{\eta_0}=\bf{ 500}\ \text{GeV} $,\ $F=4000\ \text{GeV}$}\\
\hline
 $f\ [\text{TeV}]$  & $(a_{t})^{\rm total} $  \\
\hline
\hline
$1_\cdot0$  & $ 5.61 \times 10^{-4} + 2.37  \times 10^{-5}\ i $  \\
\hline
$1_\cdot5 $  & $ 2.67 \times 10^{-4} + 1.98 \times 10^{-5}\ i$  \\
\hline
$2_\cdot0 $  & $ 1.63 \times 10^{-4} + 1.85 \times 10^{-5}\ i$  \\
\hline
$2_\cdot5 $  & $ 1.14 \times 10^{-4} + 1.78 \times 10^{-5}\ i$  \\
\hline
$3_\cdot0 $  & $ 8.78 \times 10^{-5} + 1.75 \times 10^{-5}\ i$  \\
\hline
\end{tabular}
\end{table}

\begin{table}[H]
\caption{Expected sensitivity limits on the $a_t$ in the context of the BLHM with $\sqrt{q^{2}}=500\ \text{GeV}$,
\ $\text{m}_{A_0}=1 500\ \text{GeV}$, $\text{m}_{\eta_0}=100\ \text{GeV}$, $F=4000\ \text{GeV}$ and $f=1, 1.5, 2, 2.5, 3\ \text{TeV}$
are represented. All new contributions are considered, scalar bosons, vector bosons, scalar-vector, and heavy quarks.
\label{mA1500}}
\centering
\begin{tabular}{|c|c|}
\hline
\hline
\multicolumn{2}{|c|}{$\sqrt{q^{2}}=500\ \text{GeV}$,\ $\text{m}_{A_0}=\bf{1500}\ \text{GeV}$, $\text{m}_{\eta_0}=100\ \text{GeV}$,  $\ F=4000\ \text{GeV}$}\\
\hline
 $f\ [\text{TeV}]$  & $(a_{t})^{\rm total} $  \\
\hline
\hline
$1_\cdot0$  & $ 1.86 \times 10^{-4} + 3.44 \times 10^{-5}\ i$  \\
\hline
$1_\cdot5 $  & $  9.55 \times 10^{-5} + 1.92 \times 10^{-5}\ i$  \\
\hline
$2_\cdot0 $  & $ 6.32  \times 10^{-5}+ 1.42 \times 10^{-5}\ i$  \\
\hline
$2_\cdot5 $  & $ 4.74 \times 10^{-5} + 1.19 \times 10^{-5}\ i$  \\
\hline
$3_\cdot0 $  & $ 3.84  \times 10^{-5}+ 1.07 \times 10^{-5}\ i$  \\
\hline
\end{tabular}
\end{table}

\subsection{AWMDM of the top-quark  }
\label{B2}

In this Subsection, we briefly present the analysis of the processes that are not considered in our previous work~\cite{Cruz-Albaro:2022kty} related to the weak dipole moments of the top-quark in the $y_2 < y_3$ diagonalization scheme at the BLHM. In our first mentioned work, only processes mediated by scalar bosons in the $Z\bar{t}t$ vertex loop are investigated. However, processes mediated by vector particles and scalar-vector particles that could potentially generate contributions more significant than the scalar contributions are also generated.

\begin{figure}[H]
\subfloat[]{\includegraphics[width=8.0cm]{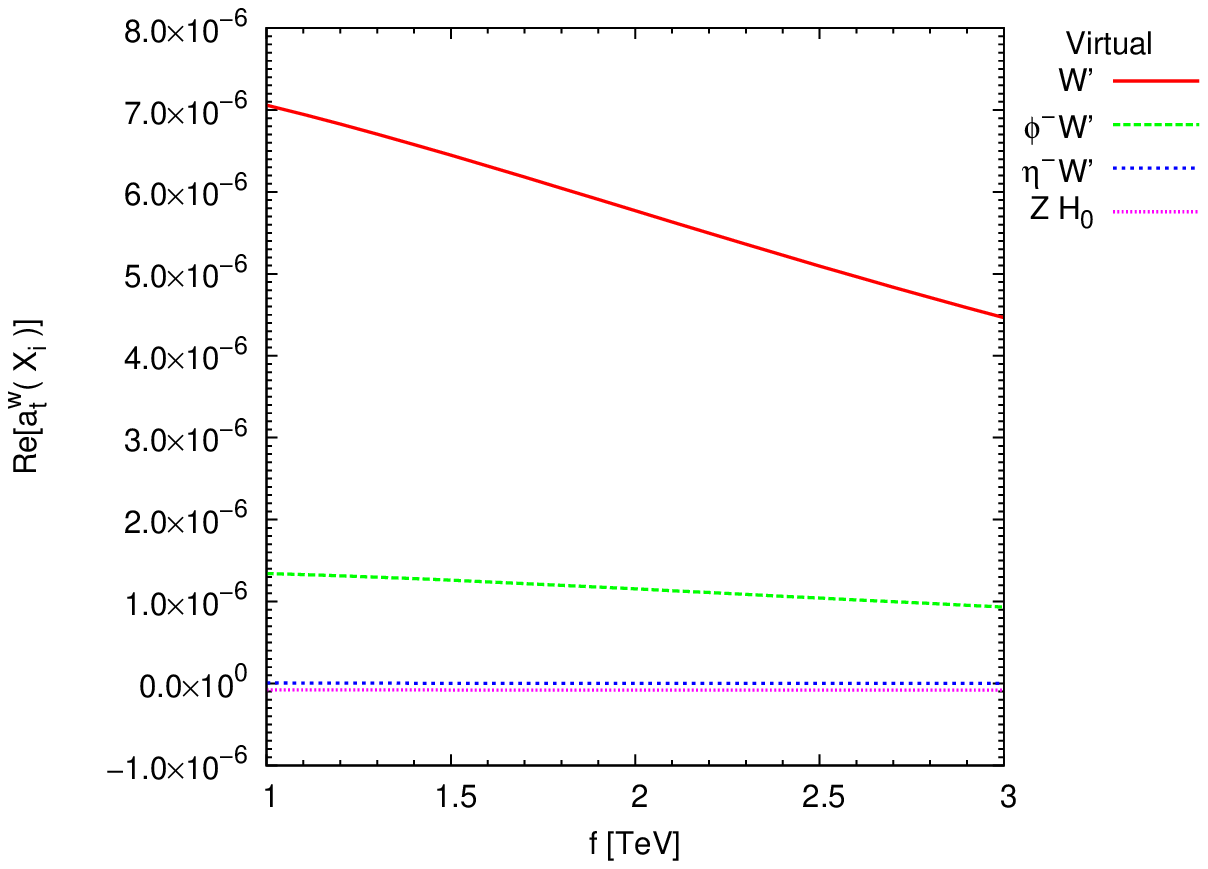}}
\subfloat[]{\includegraphics[width=8.0cm]{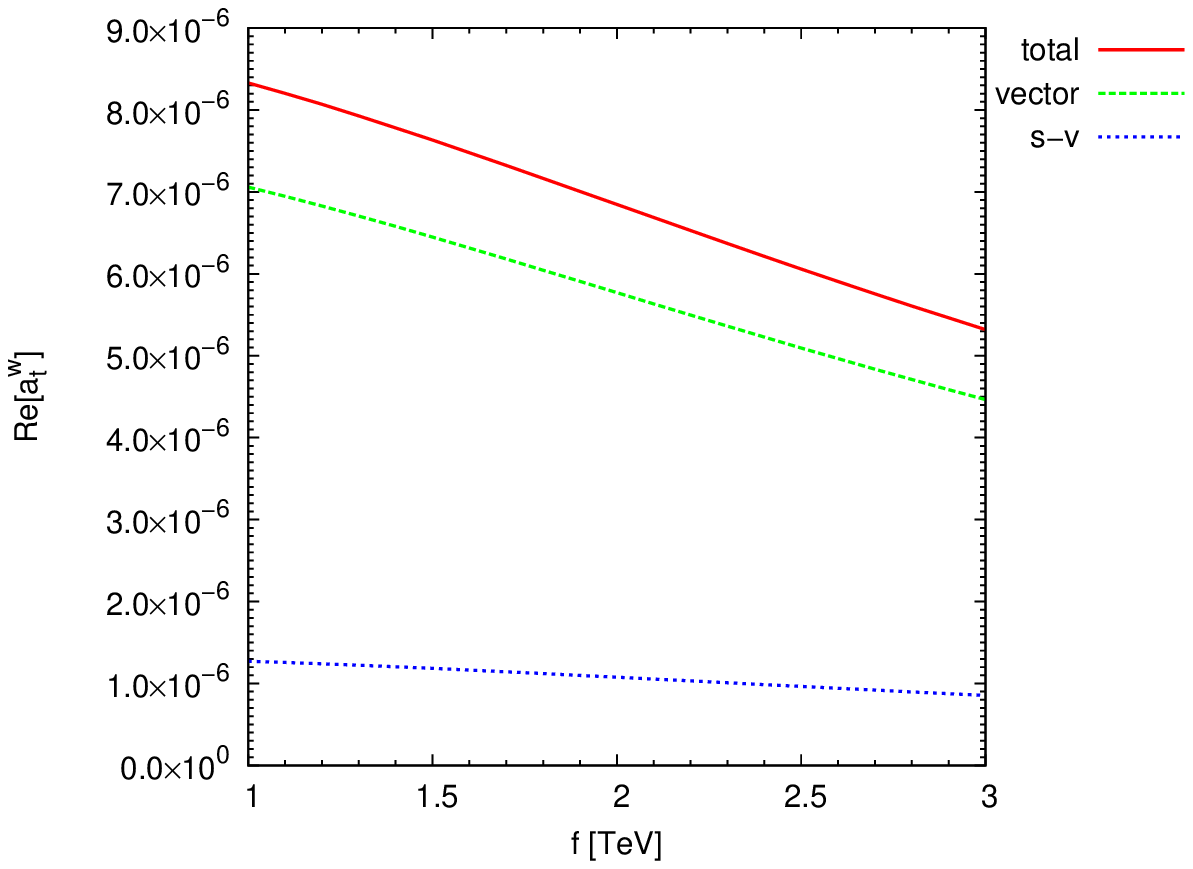}}
\caption{ \label{parcialZ}
a) Individual vector and scalar-vector contributions to Re($a^{W}_{t}$).
b) Scalar-vector (s-v), vector, and total contributions to Re($a^{W}_{t}$).
The plots are obtained with the fixed value of  $F=4000\, \text{GeV}$. The values provided in Table~\ref{parametervalues} are used for the remaining model parameters.
}
\end{figure}

Following the same approach as in the previous Subsection, we present in Fig.~\ref{parcialZ}(a) the individual contributions to the AWMDM $ a^{W}_{t} $ of the top-quark of the different particles circulating in the loop of the corresponding vertex.
In this case, the generated partial contributions provide only real contributions to $ a^{W}_{t} $. The curve that provides the largest contributions arises when the mediator particle is the exotic vector boson $W'$, $ \text{Re}[a^{W}_{t}]=[7.06, 4.46]\times 10^{-6} $ for the interval $f=[1000, 3000]$ GeV. The scalar-vector contributions $ \phi^{-} W'$, $ \eta^{-} W' $ and $Z H_0$   provide weaker contributions than the main curve. In Table~\ref{parcialweakZtt}, we provide the numerical magnitudes of all partial contributions.
Fig.~\ref{parcialZ}(b) shows the contributions due to the scalar and scalar-vector contributions, as well as the total contribution that results from the sum of the scalar and scalar-vector contributions.
This figure shows that the total contribution receives significant contributions from the vector contribution, followed by the scalar-vector contribution.
The numerical estimates obtained for these sectors in the energy scale range $ f $ are $ \text{Re}[a^{W}_t (\text{vector})]=[7.06, 4.46] \times 10^{-6} $, $ \text{Re}[a^{W}_t (\text{s-v})]=[1.27 \times 10^{-6}, 8.54\times 10^{-7}] $ and $ \text{Re}[a^{W}_t (\text{total})]=[8.33, 5.32] \times 10^{-6} $.

\begin{figure}[H]
\subfloat[]{\includegraphics[width=8.0cm]{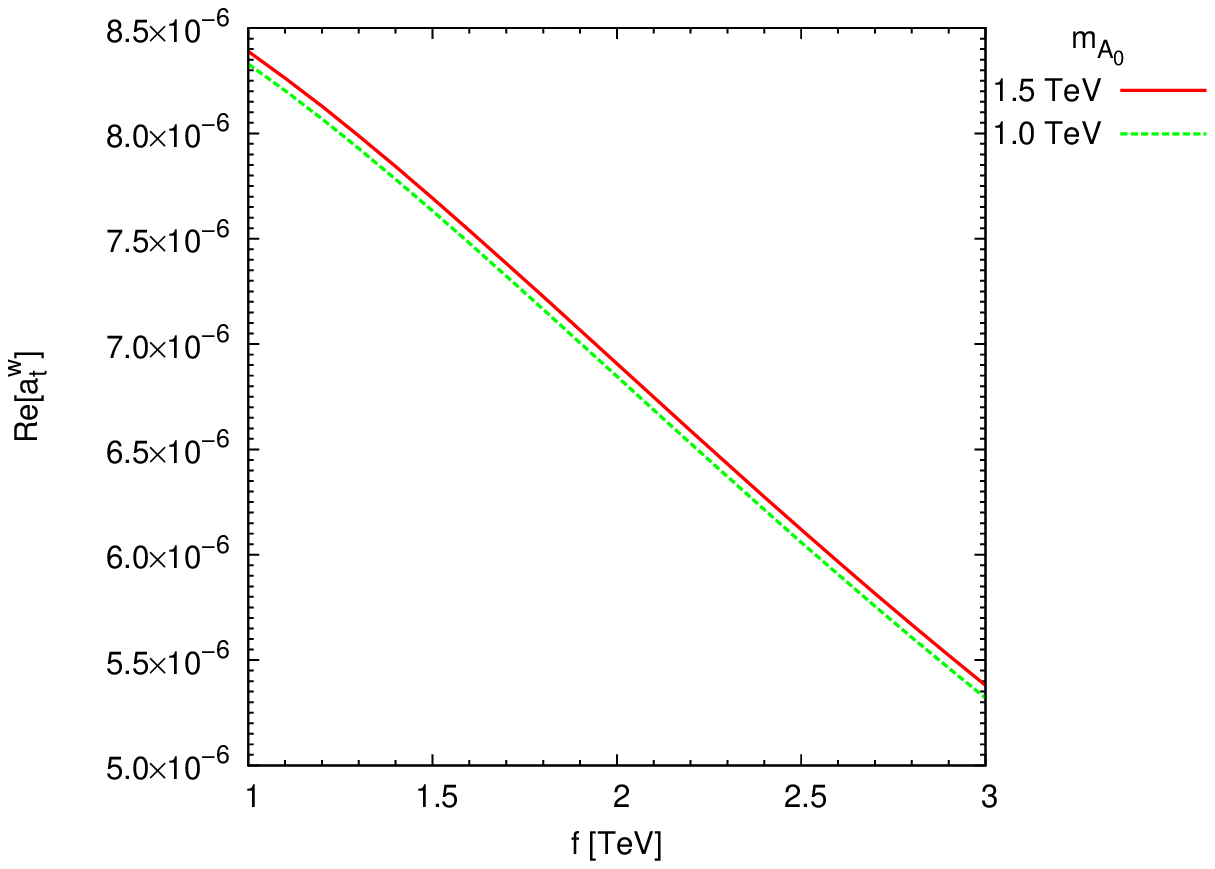}}
\subfloat[]{\includegraphics[width=8.0cm]{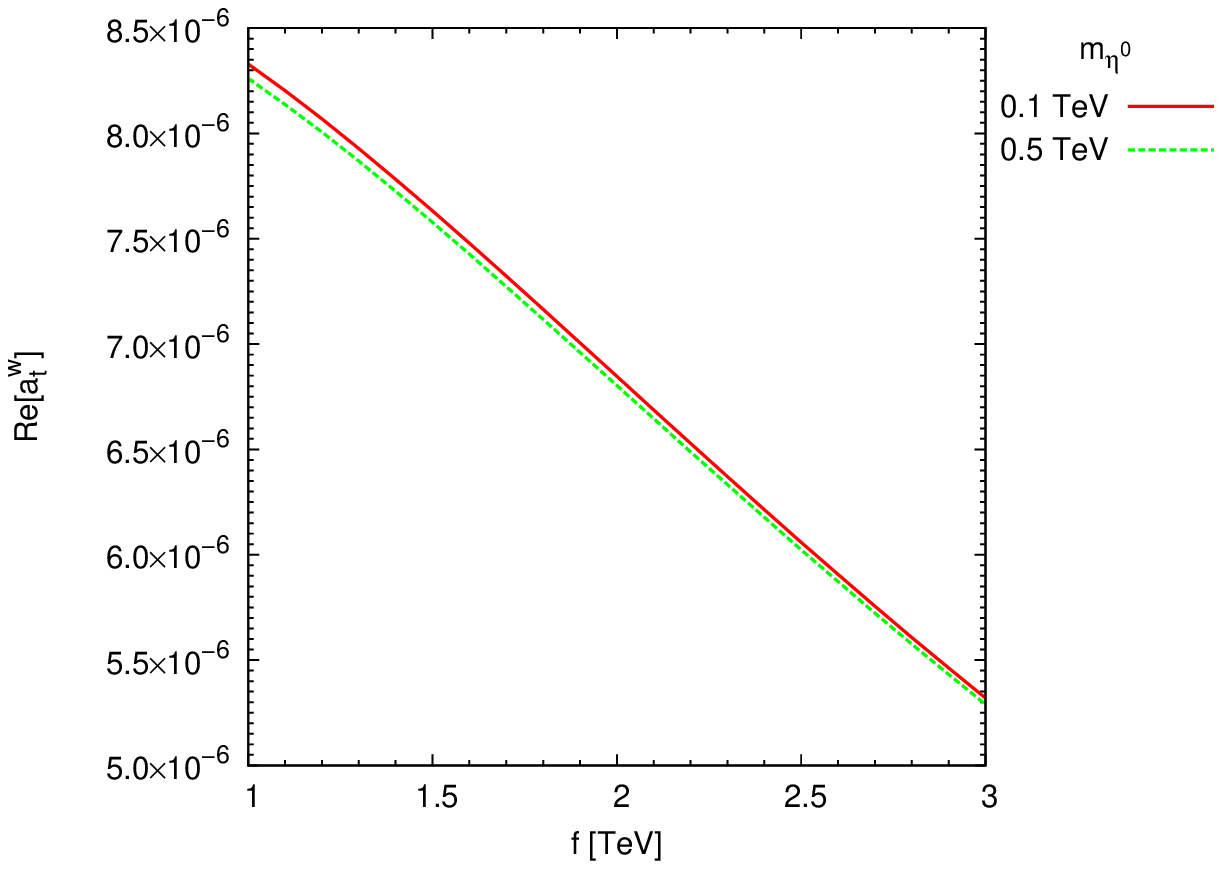}}
\caption{ \label{FmA} a) Total contribution to  Re($a^{W}_{t}$) for different values of the mass of $A_0$. b) Total contribution to  Re($a^{W}_{t}$) for different values of the mass of $\eta^{0}$. The plots are obtained with the fixed value of  $F=4000\, \text{GeV}$. The values provided in Table~\ref{parametervalues} are used for the remaining model parameters.
}
\end{figure}

We investigated the sensitivity of the total contribution to $ a^{W}_t $ while selecting certain fixed values to the input parameters $ m_{A_0} $ and $ m_{\eta^{0}} $.  In Figs.~\ref{FmA}(a) and~\ref{FmA}(b) we find the behavior of $ a^{W}_t $ for different choices in the masses of the scalars $ A_0 $ and $ \eta^{0} $, $ m_{A_0}=1000, 1500 $ GeV and $ m_{\eta^{0}} =100, 500$ GeV.
From Fig.~\ref{FmA}(a), we appreciate that the curves generated for $ a^{W}_t $ depend slightly on $ m_{A_0}$ since the curves, in the scale analysis interval $ f $, are quite close to each other. Both curves obtain contributions of the same magnitude, $ 10^{-6} $.
Concerning Fig.~\ref{FmA}(b), $ a^{W}_t $ also shows a slight sensitivity to any election in the input values set for $ m_{\eta^{0}}$, the two curves provide contributions of the same order of magnitude, that is, $ 10^{-6} $.
For more information on the numerical contributions generated for the different cases analyzed, we provide the numerical values in Tables~\ref{ZmEta100}-\ref{mA1500debil}.

We examine the dependence of $a^{W}_{t}$ as a function of the mass of the scalar $ A_0 $, which will take values from 1000 to 1500 GeV. The different curves are obtained by assigning fixed values to $ \tan \beta $, which are 3, 6, 8, and 10. All four curves acquire values of the same order of magnitude over the whole study interval for $ m_{A_0} $. The most significant contributions to $a^{W}_{t}$, in the study interval for  $ m_{A_0} $, are generated when $ \tan \beta =10$, while more suppressed values are obtained for small values of $ \tan \beta $.

\begin{figure}[H]
\centering
\subfloat[]{\includegraphics[width=8.0cm]{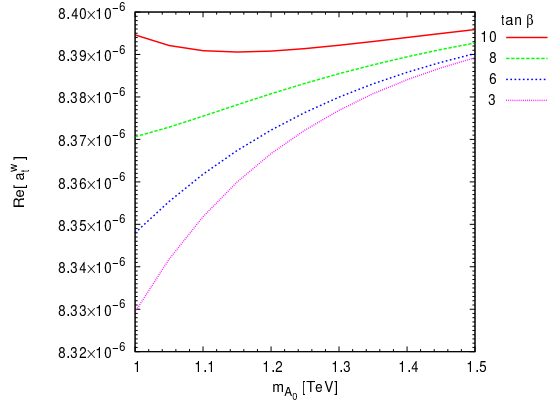}}
\caption{ \label{Ftan} Total contribution to Re($a^{W}_{t}$) for different values of $\tan \beta$.
The plot is obtained with the fixed values of  $f=1000\, \text{GeV}$ and $F=4000\, \text{GeV}$. The values provided in Table~\ref{parametervalues} are used for the remaining model parameters.}
\end{figure}

\begin{figure}[H]
\subfloat[]{\includegraphics[width=8.0cm]{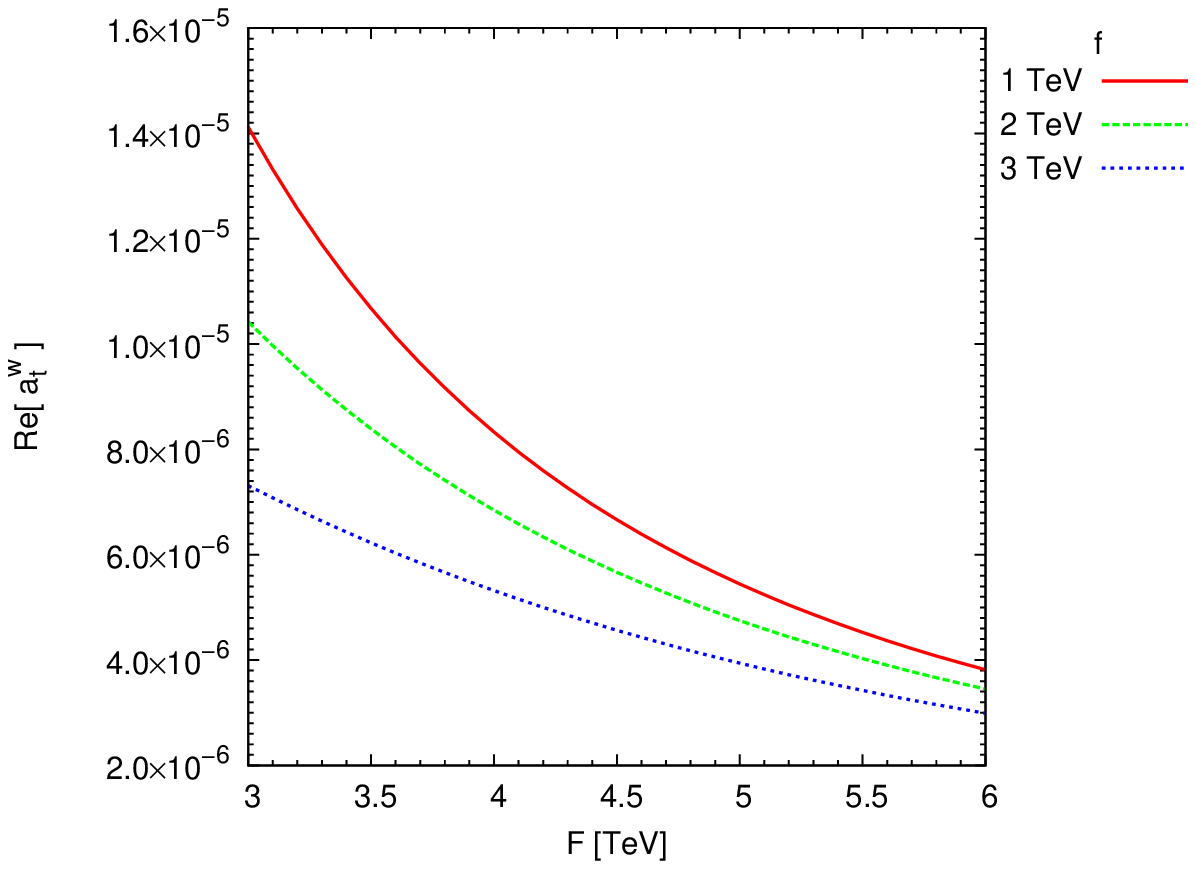}}
\subfloat[]{\includegraphics[width=8.0cm]{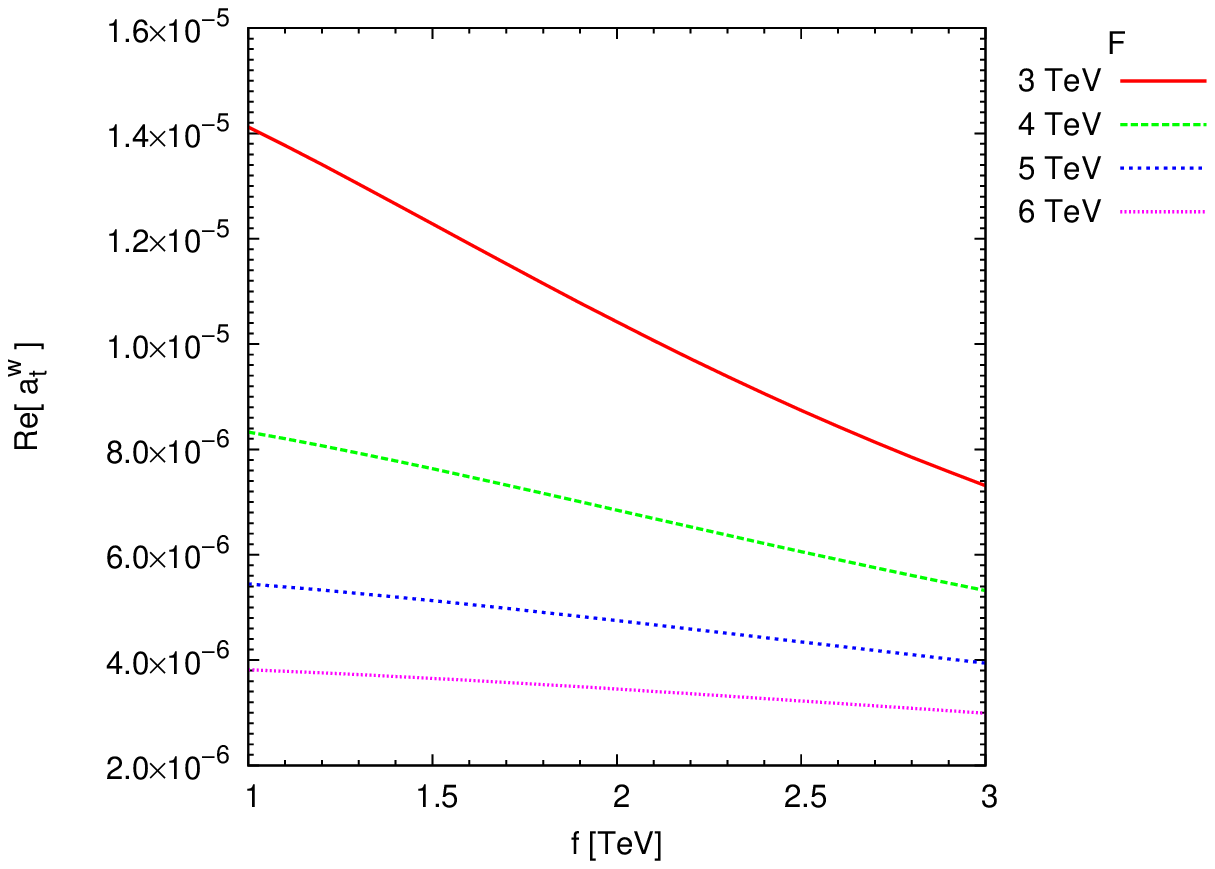}}
\caption{ \label{FFi} a) Total contribution to Re($a^{W}_{t}$) for different values of the energy scale $f$.  b) Total contribution to Re($a^{W}_{t}$) for different values of the energy scale $F$. The values provided in Table~\ref{parametervalues} are used for the remaining model parameters.
}
\end{figure}

Finally, we discuss the behavior of $\text{Re}[a^{W}_t]$ as a function of scale $ F $ or $ f $ as shown in Figs.~\ref{FFi}(a) and~\ref{FFi}(b).
From Fig.~\ref{FFi}(a), we find that $\text{Re}[a^{W}_t]$ decreases by at most one order of magnitude as $ F $ increases up to 6000 GeV. In particular, the curve that gives slightly larger values is obtained when $f=1000$ GeV, $\text{Re}[a^{W}_t]=[1.41 \times 10^{-5}, 3.81 \times 10^{-6}]$.
Regarding Fig.~\ref{FFi}(b), in this case, we find that the significant contributions occur when $F=3000$ GeV, the corresponding curve acquires values of the order of $10^{-5}$ to $10^{-6}$ when the scale $f \in [1000, 3000]$ GeV. The rest of the curves acquire suppressed values, $ 10^{-6} $.
The numerical evaluations indicate that $\text{Re}[a^{W}_t]$ depend on the energy scales $ F $ and $ f $; this is because the different curves generated all show appreciable changes in the intervals established for their respective analyses.

\begin{table}[H]
\caption{The magnitude of the partial contributions to $a^{W}_{t}$ of the BLHM.
The data are obtained by fixing the $f$ and $F$ scales, $f=1000$ GeV and $F=4000$ GeV.
The values provided in Table~\ref{parametervalues} are used for the rest of the model parameters.
{\bf abc} denotes the different particles running in the loop of the vertex $Z t t$.
\label{parcialweakZtt}}
\centering
\begin{tabular}{|c|c|}
\hline
\hline
\multicolumn{2}{|c|}{$\sqrt{q^{2}}=500\ \text{GeV}$,\  $f=1 000\ \text{GeV}$, $F=4000\ \text{GeV}$}\\
\hline
 $\text{Couplings}\  \textbf{abc}$  & $\left(a^{W}_{t} \right)^{ \textbf{abc} } $  \\
\hline
\hline
$W'W' b$  & $ 3_\cdot 61 \times 10^{-6} + 0\ i$  \\
\hline
$W'W' B$   & $ 3.45 \times 10^{-6}+ 0\ i $ \\
\hline
$ \eta^{-} W'^{+} b$ & $ 6.75 \times 10^{-9} + 0\ i$ \\
\hline
$ \phi^{-} W'^{+} b$ & $ 1.34 \times 10^{-6} + 0\ i$ \\
\hline
$  Z H_0 t$ & $- 7.86 \times 10^{-8} + 0\ i$ \\
\hline
\end{tabular}
\end{table}

\begin{table}[H]
\caption{Expected sensitivity limits on the $a^W_t$ in the context of the BLHM with $\sqrt{q^{2}}=500\ \text{GeV}$,
\ $\text{m}_{A_0}=1000\ \text{GeV}$, $\text{m}_{\eta_0}=100\ \text{GeV} $, $F=4000\ \text{GeV}$ and $f=1, 1.5, 2, 2.5, 3\ \text{TeV}$
are represented. All contributions are considered, vector bosons and scalar-vector.
\label{ZmEta100}}
\centering
\begin{tabular}{|c|c|}
\hline
\hline
\multicolumn{2}{|c|}{$\sqrt{q^{2}}=500\ \text{GeV}$,\ $\text{m}_{A_0}=1000\ \text{GeV}$, $\text{m}_{\eta_0}=\bf{ 100}\ \text{GeV} $,\ $F=4000\ \text{GeV}$}\\
\hline
 $f\ [\text{TeV}]$  & $(a^{W}_{t})^{\rm total} $  \\
\hline
\hline
$1_\cdot0$  & $ 8.33 \times 10^{-6} + 0\ i $  \\
\hline
$1_\cdot5 $  & $ 7.63 \times 10^{-6} + 0\ i$  \\
\hline
$2_\cdot0 $  & $ 6.85 \times 10^{-6} + 0\ i$  \\
\hline
$2_\cdot5 $  & $ 6.06 \times 10^{-6} + 0\ i$  \\
\hline
$3_\cdot0 $  & $ 5.32 \times 10^{-6} + 0\ i$  \\
\hline
\end{tabular}
\end{table}

\begin{table}[H]
\caption{Expected sensitivity limits on the $a^W_t$ in the context of the BLHM with $\sqrt{q^{2}}=500\ \text{GeV}$,
\ $\text{m}_{A_0}=1000\ \text{GeV}$, $\text{m}_{\eta_0}=500\ \text{GeV} $, $F=4000\ \text{GeV}$ and $f=1, 1.5, 2, 2.5, 3\ \text{TeV}$
are represented. All contributions are considered, vector bosons and scalar-vector.
\label{ZmEta500}}
\centering
\begin{tabular}{|c|c|}
\hline
\hline
\multicolumn{2}{|c|}{$\sqrt{q^{2}}=500\ \text{GeV}$,\ $\text{m}_{A_0}=1000\ \text{GeV}$, $\text{m}_{\eta_0}=\bf{ 500}\ \text{GeV} $,\ $F=4000\ \text{GeV}$}\\
\hline
 $f\ [\text{TeV}]$  & $(a^{W}_{t})^{\rm total} $  \\
\hline
\hline
$1_\cdot0$  & $ 8.26 \times 10^{-6} + 0\ i $  \\
\hline
$1_\cdot5 $  & $ 7.58 \times 10^{-6} + 0\ i$  \\
\hline
$2_\cdot0 $  & $ 6.80 \times 10^{-6} + 0\ i$  \\
\hline
$2_\cdot5 $  & $ 6.02 \times 10^{-6} + 0\ i$  \\
\hline
$3_\cdot0 $  & $ 5.29 \times 10^{-6} + 0\ i$  \\
\hline
\end{tabular}
\end{table}

\begin{table}[H]
\caption{Expected sensitivity limits on the $a^{W}_t$ in the context of the BLHM with $\sqrt{q^{2}}=500\ \text{GeV}$,
\ $\text{m}_{A_0}=1 500\ \text{GeV}$, $\text{m}_{\eta_0}=100\ \text{GeV}$, $F=4000\ \text{GeV}$ and $f=1, 1.5, 2, 2.5, 3\ \text{TeV}$
are represented. All contributions are considered, vector bosons and scalar-vector.
\label{mA1500debil}}
\centering
\begin{tabular}{|c|c|}
\hline
\hline
\multicolumn{2}{|c|}{$\sqrt{q^{2}}=500\ \text{GeV}$,\ $\text{m}_{A_0}=\bf{1500}\ \text{GeV}$, $\text{m}_{\eta_0}=100\ \text{GeV}$,  $\ F=4000\ \text{GeV}$}\\
\hline
 $f\ [\text{TeV}]$  & $(a^{W}_{t})^{\rm total} $  \\
\hline
\hline
$1_\cdot0$  & $ 8.39 \times 10^{-6} + 0\ i$  \\
\hline
$1_\cdot5 $  & $ 7.69 \times 10^{-6} + 0\ i$  \\
\hline
$2_\cdot0 $  & $ 6.91 \times 10^{-6}+ 0\ i$  \\
\hline
$2_\cdot5 $  & $ 6.12 \times 10^{-6} + 0\ i$  \\
\hline
$3_\cdot0 $  & $ 5.38 \times 10^{-6}+ 0\ i$  \\
\hline
\end{tabular}
\end{table}

\vspace{5cm}

\section{The Feynman rules for the BLHM} \label{reglasFeynman}

In this Appendix, we present the Feynman rules for the BLHM involved in our calculation for the AMDM and AWMDM of the top-quark.
It is convenient to define the following useful notation:

\begin{eqnarray}
c_{\beta} &=& \cos \beta, \\
 s_{\beta} &=& \sin \beta, \\
s_{\alpha} &=& \sin \alpha, \\
c_{\alpha} &=& \cos \alpha, \\
c_{g} &=& \cos \theta_g, \\
s_{g} &=& \sin \theta_g, \\
x_{s} &=& \frac{1}{2 \cos \theta_{W} } \sin \theta_{g} \cos \theta_{g} (\sin^{2} \theta_{g}-\cos^{2} \theta_{g} ), \\
y_{t,b }&=& \frac{m_{t,b}}{v \sin \beta} \left(1-\frac{v^{2}}{3 f^{2}} \right)^{-1/2}, \\
g_{Y} &=& g'.
\end{eqnarray}

\begin{table}[H]
\caption{Self-couplings of gauge bosons in the BLHM.
\label{3boson}}
\begin{tabular}{|p{5.2cm} p{10.8cm}|}
\hline
\hline
\textbf{Vertex} &   \hspace{3.5cm} \textbf{Couplings} \\
\hline
\hline
$Z_{\mu}(q) W'^{-}_{\alpha}(k) W'^{+}_{\beta}(p) $      &  $\ i g\, c_{W} \left[ \delta_{\beta \mu} \left( p_{\alpha}- q_{\alpha} \right) + \delta_{\alpha \mu} \left( q_{\beta}- k_{\beta}  \right)  + \delta_{\alpha \beta} \left( k_{\mu}- p_{\mu} \right) \right] \hfill \break - \frac{g\, v^{2} x_s (2\, c^{2}_{g} -2\, s^{2}_{g}- c_{g} s_{g} c_{W}(2 c_{W}+1) )}{2 c_{g} s_{g} (f^{2}+F^{2}) }   [ \delta_{\beta \mu} \left( p_{\alpha}- q_{\alpha} \right) + \delta_{\alpha \mu} \left( q_{\beta}- k_{\beta}  \right)    \hfill \break + \delta_{\alpha \beta} \left( k_{\mu}- p_{\mu} \right) ] $ \\
\hline
\hline
$A_{\mu}(q) W'^{-}_{\alpha}(k) W'^{+}_{\beta}(p) $     & $\  i g\, s_{W} \left[ \delta_{\alpha \beta} \left( k_{\mu} - p_{\mu} \right)  + \delta_{\beta \mu } \left(  p_{\alpha} - q_{\alpha} \right) + \delta_{\alpha \mu} \left( q_{\beta} - k_{\beta} \right) \right] \hfill \break - \frac{g\,  v^{2} c_{W} s_{W} x_{s}  }{ (f^{2}+ F^{2})}  \left[ \delta_{\alpha \beta} \left( k_{\mu} - p_{\mu} \right)  + \delta_{\beta \mu } \left(  p_{\alpha} - q_{\alpha} \right) + \delta_{\alpha \mu} \left( q_{\beta} - k_{\beta} \right) \right] $ \\
\hline
\hline
\end{tabular}
\end{table}

\begin{table}[H]
\caption{Three-point couplings of two gauge bosons to a scalar in the BLHM.
\label{2boson-scalar}}
\begin{tabular}{|p{3.4cm}  p{12.8cm}|}
\hline
\hline
\textbf{Vertex} &    \hspace{3.5cm} \textbf{Couplings} \\
\hline
\hline
$Z Z' h_{0} $   &   $ -\frac{g
   s_{W} v \left(c_{g}^2-s_{g}^2\right)
   \left(g^2+g'^2\right) \sin (\alpha +\beta )}{2 c_{g}
    s_{g} g' }+  \frac{g s_{W} v^3
   \left(c_{g}^2-s_{g}^2\right) \left(g^2+g'^2\right)
   \sin (\alpha +\beta )}{6 c_{g} s_{g} g' f^2 }  \hfill \break  + \frac{v^3
   x_{s} \sin (\alpha +\beta ) \left(c_{g}^2 g g'
   s_{W} \left(g^2+g'^2\right)+2 c_{g} s_{g}
   \left(g^4 s_{W}^2+g^2 g'^2 \left(2
   s_{W}^2+1\right)+g'^4 s_{W}^2\right)-g g'
   s_{g}^2 s_{W} \left(g^2+g'^2\right)\right)}{2
   c_{g}  s_{g}g'^2 \left(f^2+F^2\right)} $ \\
\hline
\hline
$Z Z H_{0} $  &  $    \frac{s_{W}^2 v
   \left(g^2+g'^2\right)^2 \cos (\alpha +\beta )}{2 g'^2}  - \frac{s_{W}^2 v^3
   \left(g^2+g'^2\right)^2 \cos (\alpha +\beta )}{6
   g'^2 f^2 } \hfill \break -\frac{s_{W} v^3 x_{s} \left(g^2+g'^2\right) \cos
   (\alpha +\beta ) \left(c_{g}^2 (-g) g'+c_{g}
   s_{g} s_{W} \left(g^2+g'^2\right)+g g'
   s_{g}^2\right)}{2 c_{g}  s_{g}g'^2
   \left(f^2+F^2\right)}  $ \\
\hline
\hline
$ Z Z' H_{0} $    & $ -\frac{g
   s_{W} v \left(c_{g}^2-s_{g}^2\right)
   \left(g^2+g'^2\right) \cos (\alpha +\beta )}{2 c_{g}
    s_{g} g' }+\frac{g s_{W} v^3
   \left(c_{g}^2-s_{g}^2\right) \left(g^2+g'^2\right)
   \cos (\alpha +\beta )}{6 c_{g} s_{g} g' f^2  } \hfill \break + \frac{v^3
   x_{s} \cos (\alpha +\beta ) \left(c_{g}^2 g g'
   s_{W} \left(g^2+g'^2\right)+2 c_{g} s_{g}
   \left(g^4 s_{W}^2+g^2 g'^2 \left(2
   s_{W}^2+1\right)+g'^4 s_{W}^2\right)-g g'
   s_{g}^2 s_{W} \left(g^2+g'^2\right)\right)}{2
   c_{g}  s_{g} g'^2 \left(f^2+F^2\right)} $ \\
\hline
\hline
$  A_\gamma W^{+} \eta^{-} $   & $ \frac{ g^3 s_{W}^2 v^4 x_{s}
   \left(c_{g}^2+s_{g}^2\right)}{8 c_{g}s_{g}  g' f
    \left(f^2+F^2\right)}  $ \\
\hline
\hline
$  A_\gamma W'^{+} \eta^{-}  $  & $-\frac{ g^2 s_{W} v^2
   \left(c_{g}^2+s_{g}^2\right)}{8 c_{g} s_{g} f} + \frac{ g^3 s_{W}^2 v^4 x_{s}
   \left(c_{g}^2+s_{g}^2\right)}{16 c_{g}s_{g} g'
   f  \left(f^2+F^2\right)}  $ \\
\hline
\hline
$ A_\gamma W^{+} \phi^{-}   $  & $ \frac{ f g^3 s_{W}^2 v^2 x_{s}
   \left(c_{g}^2+s_{g}^2\right)}{2 c_{g}
   s_{g} g' \left(f^2+F^2\right)}-\frac{ g^3 s_{W}^2 v^4
   x_{s} \left(c_{g}^2+s_{g}^2\right)}{8 c_{g} s_{g} g'
    f \left(f^2+F^2\right)}  $ \\
\hline
\hline
$  A_\gamma W'^{+} \phi^{-}  $  & $-\frac{ f g^2 s_{W}
   \left(c_{g}^2+s_{g}^2\right)}{2 c_{g} s_{g}} +  \frac{ g^2 s_{W} v^2
   \left(c_{g}^2+s_{g}^2\right)}{8 c_{g}
   s_{g} f }+\frac{ f g^3 s_{W}^2 v^2 x_{s}
   \left(c_{g}^2+s_{g}^2\right)}{4 c_{g}
   s_{g} g' \left(f^2+F^2\right)} -\frac{ g^3 s_{W}^2 v^4 x_{s}
   \left(c_{g}^2+s_{g}^2\right)}{16 c_{g}s_{g} g'
    f \left(f^2+F^2\right)} $ \\
\hline
\hline
$  Z W^{+} \eta^{-} $  & $ -\frac{ g^2 s_{W}^2 v^4 x_{s}
   \left(c_{g}^2+s_{g}^2\right)}{8 c_{g} s_{g} f
   \left(f^2+F^2\right)}  $ \\
\hline
\hline
$  Z W'^{+} \eta^{-} $  & $ \frac{ g g' s_{W} v^2
   \left(c_{g}^2+s_{g}^2\right)}{8 c_{g}
   s_{g} f }-\frac{ g s_{W} v^4 x_{s}  \left(c_{g}^2+s_{g}^2\right)
   (g s_{W}+g')}{16 c_{g}  s_{g} f
   \left(f^2+F^2\right)}  $ \\
\hline
\hline
$  Z W^{+} \phi^{-} $  & $ \frac{ f g^2 v^2 x_{s}
   \left(c_{g}^2+s_{g}^2\right) \left(g^2
   s_{W}^2-g'^2\right)}{2 c_{g} g'^2 s_{g}
   \left(f^2+F^2\right)}  -\frac{ g^2 v^4 x_{s}  \left(c_{g}^2+s_{g}^2\right) \left(2 g^2
   s_{W}^2+g'^2 \left(s_{W}^2-2\right)\right)}{8
   c_{g}s_{g}  g'^2  f \left(f^2+F^2\right)}   $ \\
\hline
\hline
$  Z W'^{+} \phi^{-} $  & $-\frac{ f g^3 s_{W}
   \left(c_{g}^2+s_{g}^2\right)}{2 c_{g}s_{g}  g' }   +\frac{ g s_{W}
   v^2  \left(c_{g}^2+s_{g}^2\right) \left(2
   g^2+g'^2\right)}{8 c_{g} s_{g} g'  f }+\frac{ f
   g^3 s_{W} v^2 x_{s} \left(c_{g}^2+s_{g}^2\right) (g
   s_{W}+g')}{4 c_{g} s_{g} g'^2
   \left(f^2+F^2\right)} \hfill \break -\frac{ g s_{W} v^4 x_{s}  \left(c_{g}^2+s_{g}^2\right)
   \left(2 g^2+g'^2\right) (g s_{W}+g')}{16 c_{g}
   s_{g} g'^2 f \left(f^2+F^2\right)} $ \\
\hline
\hline
\end{tabular}
\end{table}

\begin{table}[H]
\caption{Three-point couplings of a gauge boson to two scalars in the BLHM.
\label{1boson-2scalar}}
\begin{tabular}{|p{3.4cm}  p{12.8cm}|}
\hline
\hline
\textbf{Vertex} &    \hspace{3.5cm} \textbf{Couplings} \\
\hline
\hline
$ A_\gamma \eta^{+} \eta^{-}   $  & \hspace{4.5cm} $ 0 $ \\
\hline
\hline
$ A_\gamma \phi^{+} \phi^{-}   $  & \hspace{4.5cm} $ 0 $ \\
\hline
\hline
$ Z \eta^{+} \eta^{-}   $  & \hspace{4.5cm}  $ 0 $ \\
\hline
\hline
$ Z \phi^{+} \phi^{-}   $  &   \hspace{4.5cm} $ 0 $ \\
\hline
\hline
\end{tabular}
\end{table}

\begin{table}[H]
\caption{Scalar-fermion couplings in the BLHM.
\label{1scalar-2fermions-1}}
\begin{tabular}{|p{2.8cm} p{13.4cm}|}
\hline
\hline
\textbf{Vertex} &   \hspace{3.5cm} \textbf{Couplings} \\
\hline
\hline
$A_{0}t\bar{t}$  & $-\frac{3 i c_{\beta} (P_{L}-P_{R}) y_{1} y_{2}
   y_{3}}{\sqrt{y_{1}^2+y_{2}^2}
   \sqrt{y_{1}^2+y_{3}^2}}+\frac{2 i c_{\beta}
   (P_{L}-P_{R}) s_{\beta} v y_{2}
   \left(y_{3}^2-2 y_{1}^2\right) y_{3}}{f
   \sqrt{y_{1}^2+y_{2}^2} \left(y_{1}^2+y_{3}^2\right)}   \hfill \break +   \frac{3 i c_{\beta} (P_{L}-P_{R}) s_{\beta}^2
   v^2 y_{1} y_{2} \left(4 y_{1}^6+4 \left(2
  y_{2}^2+y_{3}^2\right) y_{1}^4+\left(4 y_{2}^4+8
  y_{3}^2 y_{2}^2-2 \sqrt{y_{1}^2+y_{3}^2}\right)
  y_{1}^2+y_{2}^2 \left(4 y_{2}^2
   y_{3}^2+\sqrt{y_{1}^2+y_{3}^2}\right)\right)
   y_{3}^3}{2 f^2 \left(y_{1}^2+y_{2}^2\right)^{5/2}
   \left(y_{1}^2+y_{3}^2\right)^{5/2}} $ \\
\hline
\hline
$A_{0} t\bar{T}$  & $ \frac{i c_{\beta}
   P_{R} \left(2 y_{1}^2-y_{2}^2\right)
   y_{3}}{\sqrt{y_{1}^2+y_{2}^2}
   \sqrt{y_{1}^2+y_{3}^2}}+\frac{i c_{\beta}
   s_{\beta} v y_{1} \left(P_{L} y_{2} \left(2
   y_{1}^4+\left(2 y_{2}^2+5 y_{3}^2\right) y_{1}^2-4
   y_{2}^2 y_{3}^2\right)-2 P_{R} y_{3}
   \left(-y_{1}^4+y_{3}^2 y_{1}^2+y_{2}^4+y_{2}^2
   y_{3}^2\right)\right)}{f
   \left(y_{1}^2+y_{2}^2\right)^{3/2}
   \left(y_{1}^2+y_{3}^2\right)} \hfill \break +  \frac{3 i c_{\beta} P_{R} s_{\beta}^2 v^2
   y_{1}^2 \left(4 y_{1}^6+4 \left(2
   y_{2}^2+y_{3}^2\right) y_{1}^4+\left(4 y_{2}^4+8
   y_{3}^2 y_{2}^2-2 \sqrt{y_{1}^2+y_{3}^2}\right)
   y_{1}^2+y_{2}^2 \left(4 y_{2}^2
   y_{3}^2+\sqrt{y_{1}^2+y_{3}^2}\right)\right)
   y_{3}^3}{2 f^2 \left(y_{1}^2+y_{2}^2\right)^{5/2}
   \left(y_{2}^2-y_{3}^2\right)
   \left(y_{1}^2+y_{3}^2\right)^{3/2}} $ \\
\hline
\hline
$A_{0} t\bar{T}_{5}$  & $  -\frac{i c_{\beta}
   P_{L} \left(2 y_{1}^2-y_{3}^2\right)
   y_{2}}{\sqrt{y_{1}^2+y_{2}^2}
   \sqrt{y_{1}^2+y_{3}^2}} +
   \frac{i c_{\beta} P_{R}
   s_{\beta} v y_{1} y_{3}}{2 f
   \left(y_{1}^2+y_{2}^2\right)^2
   \left(y_{1}^2+y_{3}^2\right)^{3/2}
   \left(y_{3}^2-y_{2}^2\right)} \bigg(4 \left(2
   y_{2}^2+y_{3}^2\right) y_{1}^6 \hfill \break +2 \left(8
   y_{2}^4-y_{3}^2 y_{2}^2+2 y_{3}^4\right)
   y_{1}^4   +2 \left(y_{2}^2-y_{3}^2\right)  \left(4
   y_{2}^4-4 y_{3}^2
   y_{2}^2+\sqrt{y_{1}^2+y_{3}^2}\right) y_{1}^2 \hfill \break -10
   y_{2}^6 y_{3}^2-y_{2}^4
   \left(\sqrt{y_{1}^2+y_{3}^2}-4
   y_{3}^4\right)+y_{2}^2 y_{3}^2
   \sqrt{y_{1}^2+y_{3}^2} \bigg)  \hfill \break +  \frac{3 i c_{\beta} s_{\beta}^2 v^2 y_{1}^2
   y_{3} \left(3 P_{L} y_{2} y_{3}
   \left(y_{1}^2+y_{3}^2\right)-2 P_{R}
   \sqrt{y_{1}^2+y_{2}^2} \left(y_{1}^2-2
   y_{3}^2\right) \sqrt{y_{1}^2+y_{3}^2}\right)
   y_{2}^2}{f^2 \sqrt{y_{1}^2+y_{2}^2}
   \left(y_{2}^2-y_{3}^2\right)
   \left(y_{1}^2+y_{3}^2\right)^{5/2}}  $ \\
\hline
\hline
$A_{0} t\bar{T}_{6}$  & $ \frac{2 i P_{L}
   s_{\beta} y_{1}y_{2}}{\sqrt{y_{1}^2+y_{2}^2}} -\frac{i v P_{R} y_{1}
   y_{3}}{2f} \hfill \break \times \frac{\left(4 y_{1}^6+4 \left(2
   y_{2}^2+y_{3}^2\right) y_{1}^4+\left(4 y_{2}^4+8
   y_{3}^2 y_{2}^2-2 \sqrt{y_{1}^2+y_{3}^2}\right)
   y_{1}^2+y_{2}^2 \left(4 y_{2}^2
   y_{3}^2+\sqrt{y_{1}^2+y_{3}^2}\right)\right)
   s_{\beta}^2+8 c_{\beta}^2
   \left(y_{1}^2+y_{2}^2\right)^2
   \left(y_{1}^2+y_{3}^2\right) }{
   \left(y_{1}^2+y_{2}^2\right)^2
   \left(y_{1}^2+y_{3}^2\right)^{3/2}} \hfill \break +  \frac{i c_{\beta}^2 s_{\beta} y_{3} \left(3
   P_{L} y_{1} y_{2} y_{3}+4 P_{R}
   \sqrt{y_{1}^2+y_{2}^2}
   \left(y_{3}^2-y_{1}^2\right)\right) }{f^2
   \sqrt{y_{1}^2+y_{2}^2}
   \left(y_{1}^2+y_{3}^2\right)}   $ \\
\hline
\hline
$A_{0} t\bar{T}^{2/3}$  & $   - \frac{2 i c_{\beta}
   P_{R} y_{1} y_{3}}{\sqrt{y_{1}^2+y_{3}^2}} -\frac{i c_{\beta}
   s_{\beta} \left(P_{L} y_{1} y_{2} \left(4
   y_{1}^2-11 y_{3}^2\right)+2 P_{R}
   \sqrt{y_{1}^2+y_{2}^2} y_{3}
   \left(y_{1}^2-y_{3}^2\right)\right) v}{f
   \sqrt{y_{1}^2+y_{2}^2}
   \left(y_{1}^2+y_{3}^2\right)} \hfill \break -\frac{i c_{\beta} P_{R} s_{\beta}^2 y_{1}
   y_{3} \left(4 y_{1}^6+4 \left(2
   y_{2}^2+y_{3}^2\right) y_{1}^4+\left(4 y_{2}^4+8
   y_{3}^2 y_{2}^2-2 \sqrt{y_{1}^2+y_{3}^2}\right)
   y_{1}^2+y_{2}^2 \left(4 y_{2}^2
   y_{3}^2+\sqrt{y_{1}^2+y_{3}^2}\right)\right) v^2}{2 f^2
   \left(y_{1}^2+y_{2}^2\right)^2
   \left(y_{1}^2+y_{3}^2\right)^{3/2}} $ \\
\hline
\hline
\end{tabular}
\end{table}

\begin{table}[H]
\caption{Continuation of Table~\ref{1scalar-2fermions-1}.
\label{1scalar-2fermions-2}}
\begin{tabular}{|p{1.9cm} p{14.3cm}|}
\hline
\hline
\textbf{Vertex} &   \hspace{5.5cm} \textbf{Couplings} \\
\hline
\hline
$h_{0} t \bar{T}$  & $ \frac{c_{\alpha}
   P_{R} \left(y_{2}^2-2 y_{1}^2\right)
   y_{3}}{\sqrt{y_{1}^2+y_{2}^2}
   \sqrt{y_{1}^2+y_{3}^2}} + \frac{c_{\alpha}
   s_{\beta} y_{1} \left(2 P_{R} y_{3}
   \left(-y_{1}^4+y_{3}^2 y_{1}^2+y_{2}^4+y_{2}^2
   y_{3}^2\right)+P_{L} y_{2} \left(2 y_{1}^4+\left(2
   y_{2}^2+5 y_{3}^2\right) y_{1}^2-4 y_{2}^2
   y_{3}^2\right)\right) v}{f
   \left(y_{1}^2+y_{2}^2\right)^{3/2}
   \left(y_{1}^2+y_{3}^2\right)} $ \\
\hline
\hline
$h_{0} t \bar{T}_{5}$  & $ \frac{c_{\alpha}
   P_{L} y_{2} \left(y_{3}^2-2
   y_{1}^2\right)}{\sqrt{y_{1}^2+y_{2}^2}
   \sqrt{y_{1}^2+y_{3}^2}} -\frac{c_{\alpha}
   P_{R}v s_{\beta} y_{1} y_{3}}{2f
   \left(y_{1}^2+y_{2}^2\right)^2
   \left(y_{1}^2+y_{3}^2\right)^{3/2}
   \left(y_{3}^2-y_{2}^2\right)}  \bigg(4 \left(2
   y_{2}^2+y_{3}^2\right) y_{1}^6 \hfill \break +2 \left(8
   y_{2}^4-y_{3}^2 y_{2}^2+2 y_{3}^4\right)
   y_{1}^4   +2 \left(y_{2}^2-y_{3}^2\right) (4
   y_{2}^4-4 y_{3}^2
   y_{2}^2+\sqrt{y_{1}^2+y_{3}^2}) y_{1}^2-10
   y_{2}^6 y_{3}^2-y_{2}^4
   (\sqrt{y_{1}^2+y_{3}^2}-4
   y_{3}^4 )   +y_{2}^2 y_{3}^2
   \sqrt{y_{1}^2+y_{3}^2}\bigg)  $ \\
\hline
\hline
$h_{0} t \bar{T}_{6}$  & $ -\frac{2 P_{L}
   s_{\alpha} y_{1} y_{2}}{\sqrt{y_{1}^2+y_{2}^2}} +\frac{ v P_{R} y_{1}
   y_{3}}{2 f \left(y_{1}^2+y_{3}^2\right)^{3/2}} \bigg(8 c_{\alpha} c_{\beta}
   \left(y_{1}^2+y_{3}^2\right)  \hfill \break  +  \frac{s_{\alpha}
   s_{\beta} \left(4 y_{1}^6+4 \left(2
   y_{2}^2+y_{3}^2\right) y_{1}^4+2 \left(2 y_{2}^4+4
   y_{3}^2 y_{2}^2+\sqrt{y_{1}^2+y_{3}^2}\right)
   y_{1}^2+4 y_{2}^4 y_{3}^2-y_{2}^2
   \sqrt{y_{1}^2+y_{3}^2}\right)}{\left(y_{1}^2+y_{2}^
   2\right)^2}\bigg) \hfill \break
    -\frac{ v^2}{3 f^2
   \sqrt{y_{1}^2+y_{2}^2}
   \left(y_{1}^2+y_{3}^2\right)}  \bigg(12 P_{L} s_{\alpha} y_{1} y_{2}
   \left(y_{1}^2+y_{3}^2\right) c_{\beta}^2+2
   c_{\alpha} s_{\beta} \big(P_{L} y_{1}
   y_{2} \left(4 y_{1}^2-5 y_{3}^2\right)  \hfill \break  -3 P_{R}
   \sqrt{y_{1}^2+y_{2}^2} y_{3} \left(y_{1}^2-3
   y_{3}^2\right) \big) c_{\beta} +s_{\alpha}
   s_{\beta}^2 \big(P_{L} y_{1} y_{2} \left(4
   y_{1}^2-5 y_{3}^2\right) \hfill \break   +6 P_{R}
   \sqrt{y_{1}^2+y_{2}^2} y_{3}
   \left(y_{1}^2+y_{3}^2\right)\big)\bigg) $ \\
\hline
\hline
$h_{0} t \bar{T}^{2/3}$  & $ -\frac{2 c_{\alpha}
   P_{R} y_{1} y_{3}}{\sqrt{y_{1}^2+y_{3}^2}} + \frac{\left(2
   c_{\beta} P_{L} s_{\alpha} y_{1} y_{2}
   \left(y_{1}^2+y_{3}^2\right)+c_{\alpha}
   s_{\beta} \left(2 P_{R} \sqrt{y_{1}^2+y_{2}^2}
   y_{3} \left(y_{3}^2-y_{1}^2\right)+P_{L} \left(7
   y_{1} y_{2} y_{3}^2-2 y_{1}^3
   y_{2}\right)\right)\right) v}{f \sqrt{y_{1}^2+y_{2}^2}
   \left(y_{1}^2+y_{3}^2\right)} \hfill \break
    -\frac{P_{R} y_{1} y_{3}v^2}{6 f^2
   \left(y_{1}^2+y_{2}^2\right)^2
   \left(y_{1}^2+y_{3}^2\right)^{3/2}} \bigg(6 c_{\alpha} \hfill \break \times
   \left(2 y_{1}^6+2 \left(2 y_{2}^2+y_{3}^2\right)
   y_{1}^4+2 \left(y_{2}^4+2 y_{3}^2
   y_{2}^2+\sqrt{y_{1}^2+y_{3}^2}\right) y_{1}^2 +2
   y_{2}^4 y_{3}^2-y_{2}^2
   \sqrt{y_{1}^2+y_{3}^2}\right) s_{\beta}^2 \break +  c_{\beta} s_{\alpha} \Big(4 y_{1}^6+4 \left(2 y_{2}^2+y_{3}^2\right) y_{1}^4+\left(4
   y_{2}^4+8 y_{3}^2 y_{2}^2+6
   \sqrt{y_{1}^2+y_{3}^2}\right) y_{1}^2+4 y_{2}^4
   y_{3}^2  \hfill \break  -3 y_{2}^2 \sqrt{y_{1}^2+y_{3}^2}\Big)
   s_{\beta} - 4 c_{\alpha} c_{\beta}^2
   \left(y_{1}^2+y_{2}^2\right)^2
   \left(y_{1}^2+y_{3}^2\right)\bigg)  $ \\
\hline
\hline
$H_{0} t\bar{t}$  & $  \frac{3
   s_{\alpha} y_{1} y_{2}
   y_{3}}{\sqrt{y_{1}^2+y_{2}^2}
   \sqrt{y_{1}^2+y_{3}^2}}  -\frac{2
    s_{\alpha} s_{\beta} y_{2}
   y_{3} \left(y_{3}^2-2 y_{1}^2\right) v}{f
   \sqrt{y_{1}^2+y_{2}^2}
   \left(y_{1}^2+y_{3}^2\right)}   - \frac{ y_{1} y_{2} y_{3} v^2}{2 f^2
   \left(y_{1}^2+y_{2}^2\right)^{5/2}
   \left(y_{1}^2+y_{3}^2\right)^{5/2}}
   \bigg(s_{\alpha}  \Big(14 y_{1}^8 +4 \left(7
   y_{2}^2+4 y_{3}^2\right) y_{1}^6 \hfill \break +2 \left(7
   y_{2}^4+16 y_{3}^2 y_{2}^2+y_{3}^4+2
   \sqrt{y_{1}^2+y_{3}^2}\right) y_{1}^4  -2 \big(-8
   y_{3}^2 y_{2}^4  +\left(\sqrt{y_{1}^2+y_{3}^2}-2
   y_{3}^4\right) y_{2}^2  \hfill \break  + y_{3}^2
   \sqrt{y_{1}^2+y_{3}^2}\big) y_{1}^2 +y_{2}^2
   y_{3}^2 \left(2 y_{2}^2 y_{3}^2+\sqrt{y_{1}^2+y_{3}^2}\right)\Big)
   s_{\beta}^2   -2 c_{\alpha} c_{\beta}
   \left(y_{1}^2+y_{3}^2\right) \hfill \break \times \left(6 y_{1}^6+6 \left(2
   y_{2}^2+y_{3}^2\right) y_{1}^4+2 \left(3 y_{2}^4+6
   y_{3}^2 y_{2}^2+\sqrt{y_{1}^2+y_{3}^2}\right)
   y_{1}^2+6 y_{2}^4 y_{3}^2-y_{2}^2
   \sqrt{y_{1}^2+y_{3}^2}\right) s_{\beta} \hfill \break   +2
   c_{\beta}^2 s_{\alpha}
   \left(y_{1}^2+y_{2}^2\right)^2
   \left(y_{1}^2+y_{3}^2\right)^2\bigg)   $ \\
\hline
\hline
\end{tabular}
\end{table}

\begin{table}[H]
\caption{Continuation of Table~\ref{1scalar-2fermions-2}.
\label{1scalar-2fermions-3}}
\begin{tabular}{|p{1.9cm} p{14.3cm}|}
\hline
\hline
\textbf{Vertex} &   \hspace{5.5cm} \textbf{Couplings} \\
\hline
\hline
$H_{0} t\bar{T}$  & $ \frac{P_{R}
   s_{\alpha} \left(2 y_{1}^2-y_{2}^2\right)
   y_{3}}{\sqrt{y_{1}^2+y_{2}^2}
   \sqrt{y_{1}^2+y_{3}^2}} -\frac{s_{\alpha}
   s_{\beta} y_{1} \left(2 P_{R} y_{3}
   \left(-y_{1}^4+y_{3}^2 y_{1}^2+y_{2}^4+y_{2}^2
   y_{3}^2\right)+P_{L} y_{2} \left(2 y_{1}^4+\left(2
   y_{2}^2+5 y_{3}^2\right) y_{1}^2-4 y_{2}^2
   y_{3}^2\right)\right) v}{f
   \left(y_{1}^2+y_{2}^2\right)^{3/2}
   \left(y_{1}^2+y_{3}^2\right)} \hfill \break - \frac{P_{R} y_{3} v^2}{6 f^2
   \left(y_{1}^2+y_{2}^2\right)^{5/2}
   \left(y_{2}^2-y_{3}^2\right)
   \left(y_{1}^2+y_{3}^2\right)^{3/2}} \bigg(3 s_{\alpha} \Big(4 \left(2
   y_{2}^2+y_{3}^2\right) y_{1}^8+2 \left(5 y_{2}^4+11
   y_{3}^2 y_{2}^2  +2 y_{3}^4\right) y_{1}^6 \hfill \break +2 \left(-2
   y_{2}^6+13 y_{3}^2 y_{2}^4+\left(7
   y_{3}^4+\sqrt{y_{1}^2+y_{3}^2}\right) y_{2}^2  +2
   y_{3}^2 \sqrt{y_{1}^2+y_{3}^2}\right)
   y_{1}^4  \hfill \break +y_{2}^4 \left(-6 y_{2}^4+2 y_{3}^2
   y_{2}^2+16 y_{3}^4-3 \sqrt{y_{1}^2+y_{3}^2}\right)
   y_{1}^2-y_{2}^4 \left(y_{3}^2-y_{2}^2\Big)
   \left(\sqrt{y_{1}^2+y_{3}^2}-6 y_{2}^2
   y_{3}^2\right)\right) s_{\beta}^2+c_{\alpha}
   c_{\beta} \left(y_{3}^2-y_{2}^2\right) \bigg(20
   y_{1}^8   +4 \left(6 y_{2}^2+5 y_{3}^2\right)
   y_{1}^6+6 \left(-2 y_{2}^4+4 y_{3}^2
   y_{2}^2+\sqrt{y_{1}^2+y_{3}^2}\right)
   y_{1}^4 \hfill \break -\left(16 y_{2}^6+12 y_{3}^2 y_{2}^4+9
   \sqrt{y_{1}^2+y_{3}^2} y_{2}^2\right) y_{1}^2-16
   y_{2}^6 y_{3}^2+3 y_{2}^4
   \sqrt{y_{1}^2+y_{3}^2}\bigg) s_{\beta} \hfill \break +2
   c_{\beta}^2 s_{\alpha} \left(2
   y_{1}^2-y_{2}^2\right)
   \left(y_{1}^2+y_{2}^2\right)^2
   \left(y_{2}^2-y_{3}^2\right)
   \left(y_{1}^2+y_{3}^2\right)\bigg)  $ \\
\hline
\hline
$H_{0} t\bar{T}_{5}$  & $ \frac{P_{L}
   s_{\alpha} y_{2} \left(2
   y_{1}^2-y_{3}^2\right)}{\sqrt{y_{1}^2+y_{2}^2}
   \sqrt{y_{1}^2+y_{3}^2}} + \frac{P_{R} v
   s_{\alpha} s_{\beta} y_{1} y_{3}}{2f
   \left(y_{1}^2+y_{2}^2\right)^2
   \left(y_{1}^2+y_{3}^2\right)^{3/2}
   \left(y_{3}^2-y_{2}^2\right)} \hfill \break \times  \bigg(4
   \left(2 y_{2}^2+y_{3}^2\right) y_{1}^6+2 \left(8
   y_{2}^4-y_{3}^2 y_{2}^2+2 y_{3}^4\right)
   y_{1}^4+2 \left(y_{2}^2-y_{3}^2\right) \left(4
   y_{2}^4-4 y_{3}^2
   y_{2}^2+\sqrt{y_{1}^2+y_{3}^2}\right) y_{1}^2-10
   y_{2}^6 y_{3}^2-y_{2}^4
   \left(\sqrt{y_{1}^2+y_{3}^2}-4
   y_{3}^4\right)+y_{2}^2 y_{3}^2
   \sqrt{y_{1}^2+y_{3}^2}\bigg)  + \frac{v^2}{3 f^2 \sqrt{y_{1}^2+y_{2}^2}
   \left(y_{2}^2-y_{3}^2\right) \left(y_{1}^2+y_{3}^2\right)^{5/2}} \hfill \break \times \bigg(3 s_{\alpha} y_{3} \Big(2 P_{R} \sqrt{y_{1}^2+y_{2}^2} \sqrt{y_{1}^2+y_{3}^2}
   \left(-4 y_{3}^4+10 y_{2}^2 y_{3}^2 +y_{1}^2
   \left(y_{2}^2-4 y_{3}^2\right)\right) y_{1}^2 \hfill \break +P_{L}
   y_{2} y_{3} \left(3 y_{1}^2+y_{2}^2-y_{3}^2\right) \left(2
   y_{1}^4+y_{3}^2 y_{1}^2-y_{3}^4\right)\Big)
   s_{\beta}^2    +2 c_{\alpha} c_{\beta} P_{L}
   y_{2}   \left(y_{1}^2+y_{3}^2\right)^2 \hfill \break \times \left(4
   y_{3}^4-4 y_{2}^2 y_{3}^2+y_{1}^2 \left(2
   y_{2}^2-11 y_{3}^2\right)\right) s_{\beta}   +c_{\beta}^2 P_{L} s_{\alpha} y_{2}
   \left(y_{2}^2-y_{3}^2\right)
   \left(y_{1}^2+y_{3}^2\right)^2 \left(4 y_{1}^2+7
   y_{3}^2\right)\bigg) $ \\
\hline
\hline
$H_{0} t\bar{T}_{6}$  & $ -\frac{2 c_{\alpha} P_{L} y_{1} y_{2}}{\sqrt{y_{1}^2+y_{2}^2}}  +\frac{P_{R}v y_{1}
   y_{3}}{2 f \left(y_{1}^2+y_{3}^2\right)^{3/2}} \hfill \break \times \left(\frac{c_{\alpha} s_{\beta} \left(4
   y_{1}^6+4 \left(2 y_{2}^2+y_{3}^2\right) y_{1}^4+2
   \left(2 y_{2}^4+4 y_{3}^2
   y_{2}^2+\sqrt{y_{1}^2+y_{3}^2}\right) y_{1}^2+4
   y_{2}^4 y_{3}^2-y_{2}^2
   \sqrt{y_{1}^2+y_{3}^2}\right)}{\left(y_{1}^2+y_{2}^
   2\right)^2}-8 c_{\beta} s_{\alpha}
   \left(y_{1}^2+y_{3}^2\right)\right)  \hfill \break - \frac{v^2}{3
   f^2 \sqrt{y_{1}^2+y_{2}^2}
   \left(y_{1}^2+y_{3}^2\right)} \bigg(2 c_{\beta} s_{\alpha} s_{\beta}
   \left(3 P_{R} \sqrt{y_{1}^2+y_{2}^2} y_{3}
   \left(y_{1}^2-3 y_{3}^2\right)+P_{L} \left(5 y_{1}
   y_{2} y_{3}^2-4 y_{1}^3
   y_{2}\right)\right)+c_{\alpha} \left(12 P_{L}
   y_{1} y_{2} \left(y_{1}^2+y_{3}^2\right)
   c_{\beta}^2+s_{\beta}^2 \left(P_{L} y_{1}
   y_{2} \left(4 y_{1}^2-5 y_{3}^2\right)+6 P_{R}
   \sqrt{y_{1}^2+y_{2}^2} y_{3}
   \left(y_{1}^2+y_{3}^2\right)\right)\right)\bigg)  $ \\
\hline
\hline
$H_{0} t\bar{T}^{2/3}$  & $ \frac{2 P_{R}
   s_{\alpha} y_{1}
   y_{3}}{\sqrt{y_{1}^2+y_{3}^2}} +\frac{\left(2
   c_{\alpha} c_{\beta} P_{L} y_{1} y_{2}
   \left(y_{1}^2+y_{3}^2\right)+s_{\alpha}
   s_{\beta} \left(P_{L} y_{1} y_{2} \left(2
   y_{1}^2-7 y_{3}^2\right)+2 P_{R}
   \sqrt{y_{1}^2+y_{2}^2} y_{3}
   \left(y_{1}^2-y_{3}^2\right)\right)\right) v}{f
   \sqrt{y_{1}^2+y_{2}^2}
   \left(y_{1}^2+y_{3}^2\right)} \hfill \break  -\frac{P_{R} y_{1} y_{3} v^2}{6 f^2
   \left(y_{1}^2+y_{2}^2\right)^2
   \left(y_{1}^2+y_{3}^2\right)^{3/2}} \bigg(-6 s_{\alpha}
   \Big(2 y_{1}^6+2 \left(2 y_{2}^2+y_{3}^2\right)
   y_{1}^4  +2 \left(y_{2}^4+2 y_{3}^2
   y_{2}^2+\sqrt{y_{1}^2+y_{3}^2}\right) y_{1}^2 \hfill \break +2
   y_{2}^4 y_{3}^2-y_{2}^2
   \sqrt{y_{1}^2+y_{3}^2}\Big) s_{\beta}^2+c_{\alpha} c_{\beta} \Big(4 y_{1}^6+4
   \left(2 y_{2}^2+y_{3}^2\right) y_{1}^4+\left(4
   y_{2}^4+8 y_{3}^2 y_{2}^2+6
   \sqrt{y_{1}^2+y_{3}^2}\right) y_{1}^2 \hfill \break  +4 y_{2}^4
   y_{3}^2-3 y_{2}^2 \sqrt{y_{1}^2+y_{3}^2}\Big)
   s_{\beta}+4 c_{\beta}^2 s_{\alpha}
   \left(y_{1}^2+y_{2}^2\right)^2
   \left(y_{1}^2+y_{3}^2\right)\bigg)  $ \\
 \hline
 \hline
\end{tabular}
\end{table}

\begin{table}[H]
\caption{Continuation of Table~\ref{1scalar-2fermions-3}.
\label{1scalar-2fermions-4}}
\begin{tabular}{|p{2.0cm} p{14.2cm}|}
\hline
\hline
\textbf{Vertex} &   \hspace{5.5cm} \textbf{Couplings} \\
\hline
\hline
$\sigma t\bar{t}$  & $ -\frac{3 c_{\beta}
    v y_{1} y_{2} y_{3}}{\sqrt{2} f
   \sqrt{y_{1}^2+y_{2}^2} \sqrt{y_{1}^2+y_{3}^2}} + \frac{\sqrt{2} c_{\beta} s_{\beta} v^2 y_{2} y_{3} \left(2 y_{1}^2+5
   y_{3}^2\right)}{f^2 \sqrt{y_{1}^2+y_{2}^2}
   \left(y_{1}^2+y_{3}^2\right)} $ \\
\hline
\hline
$\sigma t\bar{T}$  & $  \frac{c_{\beta}
   P_{R} \left(y_{2}^2-2 y_{1}^2\right) y_{3}
   v}{\sqrt{2} f \sqrt{y_{1}^2+y_{2}^2}
   \sqrt{y_{1}^2+y_{3}^2}} + \frac{c_{\beta} s_{\beta} y_{1} \left(2 P_{R}
   \left(y_{1}^2+y_{2}^2\right) y_{3} \left(3
   y_{1}^2+y_{2}^2+5 y_{3}^2\right)+P_{L} \left(-2
   y_{2} y_{1}^4+\left(13 y_{2} y_{3}^2-2
   y_{2}^3\right) y_{1}^2+4 y_{2}^3
   y_{3}^2\right)\right) v^2}{\sqrt{2} f^2
   \left(y_{1}^2+y_{2}^2\right)^{3/2}
   \left(y_{1}^2+y_{3}^2\right)}  $ \\
\hline
\hline
$\sigma t\bar{T}_{5}$  & $  \frac{c_{\beta} P_{L} v y_{2} \left(2 y_{1}^2+5
   y_{3}^2\right)}{\sqrt{2} f \sqrt{y_{1}^2+y_{2}^2}
   \sqrt{y_{1}^2+y_{3}^2}}-\frac{c_{\beta} P_{R}
   s_{\beta} v^2 y_{1} y_{3} }{2 \sqrt{2} f^2} \hfill \break \times \frac{ \left(12 y_{3}^2
   y_{1}^6+6 y_{3}^2 \left(y_{2}^2+2 y_{3}^2\right)
   y_{1}^4+2 \left(y_{3}^2-y_{2}^2\right) \left(12
   y_{2}^2 y_{3}^2+\sqrt{y_{1}^2+y_{3}^2}\right)
   y_{1}^2-18 y_{2}^6 y_{3}^2+y_{2}^4 \left(12
   y_{3}^4+\sqrt{y_{1}^2+y_{3}^2}\right)-y_{2}^2
   y_{3}^2 \sqrt{y_{1}^2+y_{3}^2}\right)}{
   \left(y_{1}^2+y_{2}^2\right)^2
   \left(y_{1}^2+y_{3}^2\right)^{3/2}
   \left(y_{3}^2-y_{2}^2\right)} $ \\
\hline
\hline
$\sigma t\bar{T}_{6}$  & $   \frac{\sqrt{2}
   s_{\beta}v}{f} \left(2 P_{R} y_{3}-\frac{P_{L}
   y_{1} y_{2} \left(y_{1}^2-2
   y_{3}^2\right)}{\sqrt{y_{1}^2+y_{2}^2}
   \left(y_{1}^2+y_{3}^2\right)}\right)  + \frac{P_{R} y_{1} y_{3}v^2}{2 \sqrt{2} f^2} \hfill \break \times \frac{ \left(\left(4 y_{1}^6+4
   \left(2 y_{2}^2+y_{3}^2\right) y_{1}^4+2 \left(2
   y_{2}^4+4 y_{3}^2
   y_{2}^2+\sqrt{y_{1}^2+y_{3}^2}\right) y_{1}^2+4
   y_{2}^4 y_{3}^2-y_{2}^2
   \sqrt{y_{1}^2+y_{3}^2}\right) s_{\beta}^2+8
   c_{\beta}^2 \left(y_{1}^2+y_{2}^2\right)^2
   \left(y_{1}^2+y_{3}^2\right)\right)}{
   \left(y_{1}^2+y_{2}^2\right)^2
   \left(y_{1}^2+y_{3}^2\right)^{3/2}}  $ \\
\hline
\hline
$\sigma t\bar{T}^{2/3}$  & $  -\frac{\sqrt{2} c_{\beta} P_{R} v y_{1} y_{3}}{f
   \sqrt{y_{1}^2+y_{3}^2}} + \frac{c_{\beta} s_{\beta} v^2 \left(2 P_{R}
   \sqrt{y_{1}^2+y_{2}^2} y_{3}
   \left(y_{3}^2-y_{1}^2\right)+P_{L} y_{1} y_{2}
   \left(4 y_{1}^2+19 y_{3}^2\right)\right)}{\sqrt{2} f^2
   \sqrt{y_{1}^2+y_{2}^2}
   \left(y_{1}^2+y_{3}^2\right)} $ \\
\hline
\hline
$\phi^{0} t\bar{t}$  & $ \frac{3 i (P_{L}-P_{R}) s_{\beta} v y_{1}
   y_{2} y_{3}}{2 f \sqrt{y_{1}^2+y_{2}^2}
   \sqrt{y_{1}^2+y_{3}^2}}-\frac{i (P_{L}-P_{R})
   s_{\beta}^2 v^2 y_{2} y_{3} \left(y_{3}^2-2
   y_{1}^2\right)}{f^2 \sqrt{y_{1}^2+y_{2}^2}
   \left(y_{1}^2+y_{3}^2\right)} $ \\
\hline
\hline
$\phi^{0} t\bar{T}$  & $  \frac{i P_{R}
   s_{\beta} \left(2 y_{1}^2+5 y_{2}^2\right) y_{3}
   v}{2 f \sqrt{y_{1}^2+y_{2}^2}
   \sqrt{y_{1}^2+y_{3}^2}} \hfill \break + \frac{i y_{1} \left(4 P_{L} y_{2}
   \left(y_{1}^2+y_{2}^2\right)
   \left(y_{1}^2+y_{3}^2\right) c_\beta^2+s_{\beta}^2 \left(2 P_{R} y_{3}
   \left(-y_{1}^4+y_{3}^2 y_{1}^2+y_{2}^4+y_{2}^2
   y_{3}^2\right)+P_{L} y_{2} \left(2 y_{1}^4+\left(2
   y_{2}^2-7 y_{3}^2\right) y_{1}^2+2 y_{2}^2
   y_{3}^2\right)\right)\right) v^2}{2 f^2
   \left(y_{1}^2+y_{2}^2\right)^{3/2}
   \left(y_{1}^2+y_{3}^2\right)} $ \\
\hline
\hline
$\phi^{0} t\bar{T}_{5}$  & $ \frac{i P_{L}
   s_{\beta} y_{2} \left(2 y_{1}^2-y_{3}^2\right)
   v}{2 f \sqrt{y_{1}^2+y_{2}^2}
   \sqrt{y_{1}^2+y_{3}^2}}  +  \frac{3 i P_{R} s_{\beta}^2 y_{1} y_{3}}{4 f^2v^2
   \left(y_{1}^2+y_{2}^2\right)^2
   \left(y_{1}^2+y_{3}^2\right)^{3/2}
   \left(y_{3}^2-y_{2}^2\right)}  \Big(4
   y_{3}^2 y_{1}^6 +2 \left(2 y_{3}^4+7 y_{2}^2
   y_{3}^2\right) y_{1}^4 \hfill \break  +\left(16 y_{3}^2 y_{2}^4+2
   \left(4 y_{3}^4+\sqrt{y_{1}^2+y_{3}^2}\right)
   y_{2}^2-2 y_{3}^2 \sqrt{y_{1}^2+y_{3}^2}\right)
   y_{1}^2  +6 y_{2}^6 y_{3}^2  \hfill \break  -y_{2}^4
   \left(\sqrt{y_{1}^2+y_{3}^2}-4
   y_{3}^4\right)  +y_{2}^2 y_{3}^2
   \sqrt{y_{1}^2+y_{3}^2}\Big)  $ \\
\hline
\hline
$\phi^{0} t\bar{T}_{6}$  & $ \frac{i c_{\beta} P_{L} v y_{1} y_{2}}{f
   \sqrt{y_{1}^2+y_{2}^2}}-\frac{i c_{\beta} P_{R}
   s_{\beta} v^2 y_{1} y_{3} \left(4 y_{1}^6+4
   \left(2 y_{2}^2+y_{3}^2\right) y_{1}^4+2 \left(2
   y_{2}^4+4 y_{3}^2
   y_{2}^2+\sqrt{y_{1}^2+y_{3}^2}\right) y_{1}^2+4
   y_{2}^4 y_{3}^2-y_{2}^2
   \sqrt{y_{1}^2+y_{3}^2}\right)}{4 f^2
   \left(y_{1}^2+y_{2}^2\right)^2
   \left(y_{1}^2+y_{3}^2\right)^{3/2}} $ \\
\hline
\hline
$\phi^{0} t\bar{T}^{2/3}$  & $ \frac{i P_{R} s_{\beta} v y_{1} y_{3} \left(2
   y_{1}^6+2 \left(2 y_{2}^2+y_{3}^2\right)
   y_{1}^4+\left(2 y_{2}^4+4 y_{3}^2 y_{2}^2-2
   \sqrt{y_{1}^2+y_{3}^2}\right) y_{1}^2+y_{2}^2
   \left(2 y_{2}^2
   y_{3}^2+\sqrt{y_{1}^2+y_{3}^2}\right)\right)}{2 f
   \left(y_{1}^2+y_{2}^2\right)^2
   \left(y_{1}^2+y_{3}^2\right)^{3/2}} \hfill \break -\frac{i v^2 \left(2
   P_{L} y_{1} y_{2} \left(y_{1}^2+y_{3}^2\right)
   c_{\beta}^2+s_{\beta}^2 \left(2 P_{R}
   \sqrt{y_{1}^2+y_{2}^2} y_{3}
   \left(y_{3}^2-y_{1}^2\right)+P_{L} \left(7 y_{1}
   y_{2} y_{3}^2-2 y_{1}^3
   y_{2}\right)\right)\right)}{2 f^2
   \sqrt{y_{1}^2+y_{2}^2}
   \left(y_{1}^2+y_{3}^2\right)} $ \\
   \hline
   \hline
\end{tabular}
\end{table}

\begin{table}[H]
\caption{Continuation of Table~\ref{1scalar-2fermions-4}.
\label{1scalar-2fermions-5}}
\begin{tabular}{|p{2.0cm} p{14.2cm}|}
\hline
\hline
\textbf{Vertex} &   \hspace{5.5cm} \textbf{Couplings} \\
\hline
\hline
$\eta^{0} t\bar{t}$  & $ -\frac{3 i
   (P_{L}-P_{R}) s_{\beta} v y_{1} y_{2}
   y_{3}}{2 f \sqrt{y_{1}^2+y_{2}^2}
   \sqrt{y_{1}^2+y_{3}^2}} + \frac{i (P_{L}-P_{R}) s_{\beta}^2 v^2 y_{2}
   y_{3} \left(y_{3}^2-2 y_{1}^2\right)}{f^2
   \sqrt{y_{1}^2+y_{2}^2}
   \left(y_{1}^2+y_{3}^2\right)} $ \\
\hline
\hline
$\eta^{0} t\bar{T}$  & $ -\frac{i P_{R}
   s_{\beta} \left(2 y_{1}^2+5 y_{2}^2\right) y_{3}
   v}{2 f \sqrt{y_{1}^2+y_{2}^2}
   \sqrt{y_{1}^2+y_{3}^2}} \hfill \break -\frac{i y_{1} \left(4 P_{L} y_{2}
   \left(y_{1}^2+y_{2}^2\right)
   \left(y_{1}^2+y_{3}^2\right) c^{2}_{\beta} +s_{\beta}^2 \left(2 P_{R} y_{3}
   \left(-y_{1}^4+y_{3}^2 y_{1}^2+y_{2}^4+y_{2}^2
   y_{3}^2\right)+P_{L} y_{2} \left(2 y_{1}^4+\left(2
   y_{2}^2-7 y_{3}^2\right) y_{1}^2+2 y_{2}^2
   y_{3}^2\right)\right)\right) v^2}{2 f^2
   \left(y_{1}^2+y_{2}^2\right)^{3/2}
   \left(y_{1}^2+y_{3}^2\right)} $ \\
\hline
\hline
$\eta^{0} t\bar{T}_{5}$  & $ -\frac{i P_{L}
   s_{\beta} y_{2} \left(2 y_{1}^2-y_{3}^2\right)
   v}{2 f \sqrt{y_{1}^2+y_{2}^2}
   \sqrt{y_{1}^2+y_{3}^2}}  -\frac{3 i P_{R} s_{\beta}^2 y_{1} y_{3} v^2 }{4 f^2    \left(y_{1}^2+y_{2}^2\right)^2
   \left(y_{1}^2+y_{3}^2\right)^{3/2}
   \left(y_{3}^2-y_{2}^2\right)}  \Big(4
   y_{3}^2 y_{1}^6+2 \left(2 y_{3}^4+7 y_{2}^2
   y_{3}^2\right) y_{1}^4 \hfill \break +\left(16 y_{3}^2 y_{2}^4+2
   \left(4 y_{3}^4+\sqrt{y_{1}^2+y_{3}^2}\right)
   y_{2}^2-2 y_{3}^2 \sqrt{y_{1}^2+y_{3}^2}\right)
   y_{1}^2+6 y_{2}^6 y_{3}^2 \hfill \break -y_{2}^4
   \left(\sqrt{y_{1}^2+y_{3}^2}-4
   y_{3}^4\right)+y_{2}^2 y_{3}^2
   \sqrt{y_{1}^2+y_{3}^2}\Big)  $ \\
\hline
\hline
$\eta^{0} t\bar{T}_{6}$  & $ -\frac{i c_{\beta
   } P_{L} v y_{1} y_{2}}{f
   \sqrt{y_{1}^2+y_{2}^2}} + \frac{i c_{\beta} P_{R} s_{\beta} v^2 y_{1}
   y_{3} \left(4 y_{1}^6+4 \left(2
   y_{2}^2+y_{3}^2\right) y_{1}^4+2 \left(2 y_{2}^4+4
   y_{3}^2 y_{2}^2+\sqrt{y_{1}^2+y_{3}^2}\right)
   y_{1}^2+4 y_{2}^4 y_{3}^2-y_{2}^2
   \sqrt{y_{1}^2+y_{3}^2}\right)}{4 f^2
   \left(y_{1}^2+y_{2}^2\right)^2
   \left(y_{1}^2+y_{3}^2\right)^{3/2}} $ \\
\hline
\hline
$\eta^{0} t\bar{T}^{2/3}$  & $ -\frac{i P_{R}
   s_{\beta} v y_{1} y_{3} \left(2 y_{1}^6+2
   \left(2 y_{2}^2+y_{3}^2\right) y_{1}^4+\left(2
   y_{2}^4+4 y_{3}^2 y_{2}^2-2
   \sqrt{y_{1}^2+y_{3}^2}\right) y_{1}^2+y_{2}^2
   \left(2 y_{2}^2
   y_{3}^2+\sqrt{y_{1}^2+y_{3}^2}\right)\right)}{2 f
   \left(y_{1}^2+y_{2}^2\right)^2
   \left(y_{1}^2+y_{3}^2\right)^{3/2}} \hfill \break + \frac{i v^2 \left(2 P_{L} y_{1} y_{2}
   \left(y_{1}^2+y_{3}^2\right) c^{2}_\beta
   +s_{\beta}^2 \left(2 P_{R}
   \sqrt{y_{1}^2+y_{2}^2} y_{3}
   \left(y_{3}^2-y_{1}^2\right)+P_{L} \left(7 y_{1}
   y_{2} y_{3}^2-2 y_{1}^3
   y_{2}\right)\right)\right)}{2 f^2
   \sqrt{y_{1}^2+y_{2}^2}
   \left(y_{1}^2+y_{3}^2\right)} $ \\
\hline
\hline
$H^{+} \bar{t} b$  & $\sqrt{2} c_{\beta}
   P_{R} \text{yb}  -\frac{3 \sqrt{2} c_{\beta} P_{L} y_{1} y_{2}
   y_{3}}{\sqrt{y_{1}^2+y_{2}^2}
   \sqrt{y_{1}^2+y_{3}^2}}+\frac{2 \sqrt{2} c_{\beta}
   P_{L} s_{\beta} v y_{2} \left(y_{3}^2-2
   y_{1}^2\right) y_{3}}{f \sqrt{y_{1}^2+y_{2}^2}
   \left(y_{1}^2+y_{3}^2\right)}$ \\
\hline
\hline
$H^{+} \bar{T}^{5/3} t$  & $  \frac{2 \sqrt{2} c_{\beta} P_{R} y_{1}
   y_{3}}{\sqrt{y_{1}^2+y_{3}^2}}+\frac{2 \sqrt{2}
   c_{\beta} P_{R} s_{\beta} v
   \left(y_{1}^2-y_{3}^2\right) y_{3}}{f
   \left(y_{1}^2+y_{3}^2\right)}  $ \\
\hline
\hline
$H^{-} \bar{B} t$  & $ \frac{\sqrt{2} c_\beta P_{R} \left(y_{2}^2-2 y_{1}^2\right)
   y_{3}}{\sqrt{y_{1}^2+y_{2}^2}
   \sqrt{y_{1}^2+y_{3}^2}} + \frac{2 \sqrt{2} c_{\beta} P_{R} s_{\beta} v
   y_{1} \left(-y_{1}^2+y_{2}^2+y_{3}^2\right)
   y_{3}}{f \sqrt{y_{1}^2+y_{2}^2}
   \left(y_{1}^2+y_{3}^2\right)}  $ \\
\hline
\hline
$\phi^{+} \bar{t} b$  & $  \frac{i s_{\beta} v}{\sqrt{2} f} \left(\frac{3 P_{L} y_{1}
   y_{2} y_{3}}{\sqrt{y_{1}^2+y_{2}^2}
   \sqrt{y_{1}^2+y_{3}^2}}-P_{R}
   y_b \right)-\frac{i \sqrt{2} P_{L}
   s_{\beta}^2 v^2 y_{2} y_{3} \left(y_{3}^2-2
   y_{1}^2\right)}{f^2 \sqrt{y_{1}^2+y_{2}^2}
   \left(y_{1}^2+y_{3}^2\right)} $ \\
\hline
\hline
$\phi^{+} \bar{T}^{5/3} t$  & $ -\frac{i P_{R}
   s_{\beta} v y_{1} y_{3} \left(2 y_{1}^6+2
   \left(2 y_{2}^2+y_{3}^2\right) y_{1}^4+\left(2
   y_{2}^4+4 y_{3}^2 y_{2}^2-2
   \sqrt{y_{1}^2+y_{3}^2}\right) y_{1}^2+y_{2}^2
   \left(2 y_{2}^2
   y_{3}^2+\sqrt{y_{1}^2+y_{3}^2}\right)\right)}{\sqrt{2}
   f \left(y_{1}^2+y_{2}^2\right)^2
   \left(y_{1}^2+y_{3}^2\right)^{3/2}} \hfill \break  + \frac{i \sqrt{2} P_{R} s_{\beta}^2 v^2 y_{3}
   \left(y_{3}^2-y_{1}^2\right)}{f^2
   \left(y_{1}^2+y_{3}^2\right)} $ \\
\hline
\hline
$\phi^{-} \bar{B} t$  & $  \frac{i P_{R} s_{\beta} v \left(2 y_{1}^2+5
   y_{2}^2\right) y_{3}}{\sqrt{2} f
   \sqrt{y_{1}^2+y_{2}^2}
   \sqrt{y_{1}^2+y_{3}^2}}-\frac{i \sqrt{2} P_{R}
   s_{\beta}^2 v^2 y_{1} y_{3}
   \left(y_{1}^2-y_{2}^2-y_{3}^2\right)}{f^2
   \sqrt{y_{1}^2+y_{2}^2}
   \left(y_{1}^2+y_{3}^2\right)} $ \\
\hline
\hline
$ \eta^{+} b\bar{t} $  & $ i \left(-\frac{3 P_{L} s_{\beta} v y_{2}
   y_{3} y_{1}^3}{\sqrt{2} Y_{1} Y_{2}^{3} }-\frac{3 P_{L}
   s_{\beta} v y_{2} y_{3}^3 y_{1}}{\sqrt{2} Y_{1} Y_{2}^{3}}+\frac{P_{R}
   s_{\beta} v y_b}{\sqrt{2}}\right) \left( \frac{1}{f} \right) $  \\
\hline
\hline
\end{tabular}
\end{table}

\begin{table}[H]
\caption{Gauge boson-fermion couplings in the BLHM.
\label{1boson-2fermions-1}}
\begin{tabular}{|p{2.0cm} p{14.2cm}|}
\hline
\hline
\textbf{Vertex} &   \hspace{5.5cm} \textbf{Couplings} \\
\hline
\hline
$\gamma \bar{T} T$  & $\frac{2}{3} i g s_{W} \gamma ^{\mu
   }  +  \frac{2 i g s_{W}  v^2}{3
   f^2} \bigg( c_{\beta}^2  \gamma ^{\mu }.P_{R}   +\frac{y_{1}^2 \left(4 c_{\beta}^2
   \left(y_{2}^2-y_{3}^2\right)^2+s_{\beta}^2 \left(2
   y_{2}^2+y_{3}^2\right)^2\right) \gamma ^{\mu
   }.P_{L}}{\left(y_{1}^2+y_{2}^2\right)
   \left(y_{2}^2-y_{3}^2\right)^2}+\frac{s_{\beta}^2
   \left(y_{2}^2-2 y_{1}^2\right)^2 y_{3}^2 \gamma ^{\mu
   }.P_{R}}{\left(y_{1}^2+y_{2}^2\right)^2
   \left(y_{1}^2+y_{3}^2\right)}\bigg) $ \\
\hline
\hline
$\gamma \bar{T_{5}} T_{5}$  & $ \frac{2}{3} i g s_{W}
  \gamma ^{\mu }+  \frac{3 i g s_{W} s_{\beta}^2 v^2 y_{1}^2
   \left(\left(6 y_{3}^4-2 y_{2}^2 y_{3}^2+y_{1}^2
   \left(y_{2}^2+2 y_{3}^2\right)\right) \gamma ^{\mu
   }.P_{L}-\left(y_{2}^2-4 y_{3}^2\right)
   \left(y_{1}^2+y_{3}^2\right) \gamma ^{\mu
   }.P_{R}\right) y_{2}^2}{2 f^2
   \left(y_{2}^2-y_{3}^2\right)^2
   \left(y_{1}^2+y_{3}^2\right)^2} $ \\
\hline
\hline
$\gamma \bar{T_{6}} T_{6}$  & $ \frac{2}{3}
   i g s_{W} \gamma ^{\mu } +  \frac{i c_{\beta}^2 g s_{W} \gamma ^{\mu } \left(4 P_{L}+P_{R}\right) v^2}{2 f^2} $ \\
\hline
\hline
$\gamma \bar{T }^{2/3} T^{2/3}$  & $\frac{2}{3} i g s_{W} \gamma ^{\mu
   } +  \frac{2 i g s_{W} s_{\beta}^2   \gamma ^{\mu }\left(
P_{L}+4 P_{R}\right) v^2}{3
   f^2} $ \\
\hline
\hline
$\gamma \bar{T }^{5/3} T^{5/3}$  & $ \frac{5}{3} i g s_{W} \gamma ^{\mu
   } $ \\
\hline
\hline
$\gamma \bar{t} T$  & $ -\frac{i g s_{W}
   y_{1} y_{2} \gamma ^{\mu }.P_{L}}{6
   \left(y_{1}^2+y_{2}^2\right)}  -\frac{i g s_{W}
   s_{\beta} \left(2 y_{1}^2-y_{2}^2\right) y_{3}
   \left(8 y_{1}^4+8 \left(y_{2}^2+y_{3}^2\right)
   y_{1}^2+8 y_{2}^2 y_{3}^2-3
   \sqrt{y_{1}^2+y_{3}^2}\right) \gamma ^{\mu }.P_{R}
   v}{12 f \left(y_{1}^2+y_{2}^2\right)^2
   \left(y_{1}^2+y_{3}^2\right)^{3/2}} \hfill \break +  \frac{2 i g s_{W} y_{1} y_{2} \left(4
   \left(y_{2}^2-y_{3}^2\right)
   \left(y_{1}^2+y_{3}^2\right) \text{c$\beta
   $}^2+s_{\beta}^2 \left(2 y_{1}^2-y_{3}^2\right)
   \left(2 y_{2}^2+y_{3}^2\right)\right) \gamma ^{\mu
   }.P_{L} v^2}{3 f^2 \left(y_{1}^2+y_{2}^2\right)
   \left(y_{2}^2-y_{3}^2\right)
   \left(y_{1}^2+y_{3}^2\right)}   $ \\
\hline
\hline
$\gamma \bar{t} T_{5}$  & $  \frac{i g s_{W} s_{\beta} v}{6
   f}  \left(\frac{y_{2}
   \left(y_{3}^4-y_{2}^2 y_{3}^2+y_{1}^2 \left(2
   y_{2}^2+y_{3}^2\right)\right) \gamma ^{\mu
   }.P_{L}}{\sqrt{y_{1}^2+y_{2}^2}
   \left(y_{2}^2-y_{3}^2\right)
   \left(y_{1}^2+y_{3}^2\right)}+\frac{8 y_{3} \gamma
   ^{\mu }.P_{R}}{\sqrt{y_{1}^2+y_{3}^2}}\right) \hfill \break -\frac{i g s_{W} s_{\beta}^2 v^2 y_{1} y_{3} }{4 f^2}
   \left(\frac{8}{y_{1}^2+y_{3}^2}-\frac{\left(2
   y_{1}^2-y_{2}^2\right) \left(y_{2}^2+2
   y_{3}^2\right)}{\left(y_{1}^2+y_{2}^2\right)^2
   \left(y_{1}^2+y_{3}^2\right)^{3/2}
   \left(y_{3}^2-y_{2}^2\right)}\right) \gamma ^{\mu
   }.P_{R}  $ \\
\hline
\hline
$\gamma \bar{t} T_{6}$  & $   \frac{i c_{\beta} g s_{W} v y_{2} \gamma ^{\mu
   }.P_{L}}{3 f \sqrt{y_{1}^2+y_{2}^2}}-\frac{i
   c_{\beta} g s_{W} s_{\beta} v^2 \left(2
   y_{1}^2-y_{2}^2\right) y_{3} \gamma ^{\mu
   }.P_{R}}{4 f^2 \left(y_{1}^2+y_{2}^2\right)^2
   \left(y_{1}^2+y_{3}^2\right)} $ \\
\hline
\hline
$\gamma \bar{t} T^{2/3}$  & $ -\frac{7 i g s_{W}
   s_{\beta} v y_{3} \gamma ^{\mu }.P_{R}}{3 f
   \sqrt{y_{1}^2+y_{3}^2}} + \frac{2 i g s_{W} s_{\beta}^2 v^2 \left(y_{2}
   \left(y_{3}^2-2 y_{1}^2\right) \gamma ^{\mu }.P_{L}+4
   y_{1} \sqrt{y_{1}^2+y_{2}^2} y_{3} \gamma ^{\mu
   }.P_{R}\right)}{3 f^2 \sqrt{y_{1}^2+y_{2}^2}
   \left(y_{1}^2+y_{3}^2\right)} $ \\
\hline
\hline
$Z \bar{T} T$  & $ \frac{i g \left(3 c_{W}^2-s_{W}^2\right) \gamma ^{\mu } }{6 c_{W}}-\frac{2 i
   g s_{W}^2 v^2 y_{1}^2 \left(4 c_{\beta}^2
   \left(y_{2}^2-y_{3}^2\right)^2+s_{\beta}^2 \left(2
   y_{2}^2+y_{3}^2\right)^2\right)\gamma ^{\mu }  }{3 c_{W} f^2
   \left(y_{1}^2+y_{2}^2\right)
   \left(y_{2}^2-y_{3}^2\right)^2} $ \\
\hline
\hline
$Z \bar{T_{5}} T_{5}$  & $ -\frac{2 i g s_{W}^2
   \gamma ^{\mu } }{3 c_{W}} \hfill \break  +  \frac{3 i g s_{\beta}^2 v^2 y_{1}^2 y_{2}^2
   \left(\left(c_{W}^2 \left(5 y_{3}^4-2 y_{2}^2
   y_{3}^2+y_{1}^2 \left(y_{2}^2+2
   y_{3}^2\right)\right)-s_{W}^2 y_{3}^4\right) \gamma
   ^{\mu }.P_{L}-c_{W}^2 \left(y_{2}^2-4
   y_{3}^2\right) \left(y_{1}^2+y_{3}^2\right) \gamma
   ^{\mu }.P_{R}\right)}{2 c_{W} f^2
   \left(y_{2}^2-y_{3}^2\right)^2
   \left(y_{1}^2+y_{3}^2\right)^2} $ \\
\hline
\hline
$Z \bar{T_{6}} T_{6}$  & $ -\frac{2 i g s_{W}^2 \gamma ^{\mu }}{3 c_{W}} + \frac{i c_{W} c_{\beta}^2 g v^2 \gamma ^{\mu }  \left(4 P_{L}+ P_{R}\right)}{2 f^2} $ \\
\hline
\hline
$Z \bar{T }^{2/3} T^{2/3}$  & $ -\frac{i g \left(3 c_{W}^2+7 s_{W}^2\right)
   \gamma ^{\mu }}{6 c_{W}} -\frac{2 i g s_{W}^2 s_{\beta}^2 \gamma ^{\mu } \left(P_{L}+4 P_{R}\right) v^2}{3 c_{W}
   f^2} $ \\
\hline
\hline
$Z \bar{T }^{5/3} T^{5/3}$  & $   \frac{i g \left(3 c_{W}^2-7 s_{W}^2\right)\gamma ^{\mu } }{6 c_{W}}  $ \\
\hline
\hline
$Z \bar{t} T$  & $   \frac{i g y_{1} y_{2} s_{W}^2
   \gamma ^{\mu }.P_{L} }{6 c_{W}
   \left(y_{1}^2+y_{2}^2\right)}+\frac{i g s_{\beta} v
   \left(2 y_{1}^2-y_{2}^2\right) y_{3} \left(3
   \sqrt{y_{1}^2+y_{3}^2} c_{W}^2+8 s_{W}^2
   \left(y_{1}^2+y_{2}^2\right)
   \left(y_{1}^2+y_{3}^2\right)\right) \gamma ^{\mu
   }.P_{R}}{12 c_{W} f \left(y_{1}^2+y_{2}^2\right)^2
   \left(y_{1}^2+y_{3}^2\right)^{3/2}} \hfill \break -\frac{2 i g v^2 y_{1} y_{2} \left(4
   \left(y_{2}^2-y_{3}^2\right)
   \left(y_{1}^2+y_{3}^2\right) \text{c$\beta
   $}^2+s_{\beta}^2 \left(2 y_{1}^2 -y_{3}^2\right)
   \left(2 y_{2}^2+y_{3}^2\right)\right)  s_{W}^2 \gamma ^{\mu
   }.P_{L}}{3 c_{W} f^2
   \left(y_{1}^2+y_{2}^2\right)
   \left(y_{2}^2-y_{3}^2\right)
   \left(y_{1}^2+y_{3}^2\right)} $ \\
\hline
\hline
$Z \bar{t} T_{5}$  & $  -\frac{i g s_{\beta} v }{6 c_{W} f} \Big(\frac{8 s_{W}^2 \gamma ^{\mu
   }.P_{R}
   y_{3}^3}{\left(y_{1}^2+y_{3}^2\right)^{3/2}}+\frac{8
   s_{W}^2 y_{1}^2 \gamma ^{\mu }.P_{R}
   y_{3}}{\left(y_{1}^2+y_{3}^2\right)^{3/2}} \hfill \break +\frac{\text{
   y2} \left(3 \left(2 y_{1}^2-y_{3}^2\right)
   \left(y_{2}^2-y_{3}^2\right) c_{W}^2+s_{W}^2
   \left(4 y_{3}^4-4 y_{2}^2 y_{3}^2+y_{1}^2 \left(8
   y_{2}^2-5 y_{3}^2\right)\right)\right) \gamma ^{\mu
   }.P_{L}}{\sqrt{y_{1}^2+y_{2}^2}
   \left(y_{2}^2-y_{3}^2\right)
   \left(y_{1}^2+y_{3}^2\right)}\Big) $ \\
\hline
\hline
$Z \bar{t} T_{6}$  & $ -\frac{i c_{\beta} g
   \left(3 c_{W}^2+4 s_{W}^2\right) y_{2} v \gamma ^{\mu
   }.P_{L} }{3 c_{W} f \sqrt{y_{1}^2+y_{2}^2}}  -\frac{i c_{W} c_{\beta} g s_{\beta} \left(2
   y_{1}^2-y_{2}^2\right) y_{3}v^2 \gamma ^{\mu }.P_{R}
   }{4 f^2 \left(y_{1}^2+y_{2}^2\right)^2
   \left(y_{1}^2+y_{3}^2\right)} $ \\
\hline
\hline
$Z \bar{t } T^{2/3}$  & $ -\frac{i g \left(3
   c_{W}^2-4 s_{W}^2\right) s_{\beta} y_{3} v \gamma
   ^{\mu }.P_{R} }{3 c_{W} f \sqrt{y_{1}^2+y_{3}^2}} -\frac{2 i g s_{W}^2 s_{\beta}^2 v^2 \left(y_{2}
   \left(y_{3}^2-2 y_{1}^2\right) \gamma ^{\mu }.P_{L}+4
   y_{1} \sqrt{y_{1}^2+y_{2}^2} y_{3} \gamma ^{\mu
   }.P_{R}\right) }{3 c_{W} f^2
   \sqrt{y_{1}^2+y_{2}^2}
   \left(y_{1}^2+y_{3}^2\right)} $ \\
\hline
\hline
\end{tabular}
\end{table}

\begin{table}[H]
\caption{Continuation of Table~\ref{1boson-2fermions-1}.
\label{1boson-2fermions-2}}
\begin{tabular}{|p{2.0cm} p{14.2cm}|}
\hline
\hline
\textbf{Vertex} &   \hspace{5.5cm} \textbf{Couplings} \\
\hline
\hline
$Z' \bar{t} t$  & $  \frac{1}{2} i g \gamma ^{\mu }.P_{L}-\frac{i g s_{\beta}^2
   v^2 y_{3}^2 }{8 f^2 \left(y_{1}^2+y_{3}^2\right)^2} \left(\frac{\left(y_{2}^2-2
   y_{1}^2\right)^2}{\left(y_{1}^2+y_{2}^2\right)^4}-16
   \left(y_{1}^2+y_{3}^2\right)\right) \gamma ^{\mu
   }.P_{R} $ \\
\hline
\hline
$Z' \bar{t} T$  & $ -\frac{i g s_{\beta} v \left(2 y_{1}^2-y_{2}^2\right)
   y_{3} \gamma ^{\mu }.P_{R}}{4 f
   \left(y_{1}^2+y_{2}^2\right)^2
   \left(y_{1}^2+y_{3}^2\right)} $ \\
\hline
\hline
$Z' \bar{t} T_{5}$  & $ -\frac{i g s_{\beta} v y_{2} \left(2
   y_{1}^2-y_{3}^2\right) \gamma ^{\mu }.P_{L}}{2 f
   \sqrt{y_{1}^2+y_{2}^2} \left(y_{1}^2+y_{3}^2\right)} + \frac{i g s_{\beta}^2 v^2 y_{1} y_{3} }{4 f^2}
   \left(\frac{8}{y_{1}^2+y_{3}^2}-\frac{\left(2
   y_{1}^2-y_{2}^2\right) \left(y_{2}^2+2
   y_{3}^2\right)}{\left(y_{1}^2+y_{2}^2\right)^2
   \left(y_{1}^2+y_{3}^2\right)^{3/2}
   \left(y_{3}^2-y_{2}^2\right)}\right) \gamma ^{\mu
   }.P_{R}  $ \\
\hline
\hline
$Z' \bar{t } T_{6}$  & $ -\frac{i c_{\beta} g v
   y_{2} \gamma ^{\mu }.P_{L}}{f
   \sqrt{y_{1}^2+y_{2}^2}} + \frac{i c_{\beta} g s_{\beta} v^2 \left(2
   y_{1}^2-y_{2}^2\right) y_{3} \gamma ^{\mu }.P_{R}}{4
   f^2 \left(y_{1}^2+y_{2}^2\right)^2
   \left(y_{1}^2+y_{3}^2\right)} $ \\
\hline
\hline
$Z' \bar{t } T^{2/3}$  & $  \frac{i g s_{\beta} v y_{3} \gamma ^{\mu }.P_{R}}{f
   \sqrt{y_{1}^2+y_{3}^2}} $ \\
\hline
\hline
$W'^{+} \bar{t} b $  & $  \frac{ig \gamma^{\mu}.P_{L} }{\sqrt{2}} $ \\
\hline
\hline
$W'^{-} \bar{ b} t $  & $ \frac{i g \gamma ^{\mu }.P_{L}}{\sqrt{2}}  $ \\
\hline
\hline
\end{tabular}
\end{table}

\newpage

\end{document}